\pdfoutput=1
\documentclass[11pt]{article}
\usepackage{jheppub}
\notoc
\usepackage{graphicx} 
\usepackage{graphbox}
\usepackage[dvipsnames]{xcolor} 
\usepackage{tikz}
\usepackage{tikz-cd}
\usepackage{dirtytalk}
\usepackage{diagbox}
\usepackage{dsfont}
\usepackage{amsmath}
\usepackage{nccmath}
\usepackage{amsthm}
\usepackage{float}
\usepackage{bbm, dsfont}
\usepackage[normalem]{ulem}
\usepackage{booktabs}
\usepackage{array}
\usepackage{slashed}
\usepackage{mathtools}
\usepackage{cancel}
\usepackage[normalem]{ulem}
\usepackage{cleveref}
\usepackage{orcidlink}
\usepackage{enumerate}
\usepackage{fontawesome}
\usepackage{adjustbox}
\usepackage{tcolorbox}
\usepackage{ulem}
\usepackage{dsfont}
\usepackage{shuffle}

\usetikzlibrary{arrows.meta,positioning,calc}

\def\beq{\begin{equation}}
\def\eeq{\end{equation}}
\def\ie{\begin{equation}\begin{aligned}}
\def\fe{\end{aligned}\end{equation}}

\newcommand{\tvec}{\vec{p}}

\newcommand{\rmax}{r^{\rm max}}

\newcommand{\sv}{\text{sv}}
\newcommand{\eqv}{\text{eqv}}
\newcommand{\II}{\mathbb{I}}
\newcommand{\mm}{\mathfrak{m}}
\newcommand{\dr}{\mathfrak{dr}}

\newcommand{\ep}{\epsilon}
\newcommand{\bep}{\boldsymbol{\varepsilon}}
\newcommand{\est}{\text{G}}
\newcommand{\SLZ}{\mathrm{SL}_2(\mathbb{Z})}
\newcommand{\ANN}[1]{\mathbb{A}(\ep;#1)}
\newcommand{\AN}[1]{\mathbb{A}_{\SLZ}(\ep;#1)}
\newcommand{\IN}[1]{\mathbb I_{\SLZ}(\ep;#1)}
\newcommand{\INN}[1]{\mathbb I(\ep;#1)}
\newcommand{\INeqv}[1]{\mathbb I_{\SLZ}^\eqv(\ep;#1)}
\newcommand{\INNeqv}[1]{\mathbb I^\eqv(\ep;#1)}
\newcommand{\TcocN}{\mathbb{T}_{\SLZ}(\ep)}
\newcommand{\ScocN}{\mathbb{S}_{\SLZ}(\ep)}

\newcommand{\INNsv}[1]{\mathbb{I}^\sv(\ep;#1)}
\newcommand{\epdem}[1]{\ep^{\rm dem}_{#1}}

\definecolor{basisblue}{HTML}{1D4ED8}
\definecolor{basisred}{HTML}{C62828}
\definecolor{loopgreen}{HTML}{16803C}

\def\nn{\nonumber}

\def\ad{\mathrm{ad}}
\def\dd{\text{d}}
\def\mot{\mathfrak{m}}
\def\BF{{\rm BF}}

\newcommand{\MMV}[2]{\mathrm{m}\! \left[\begin{smallmatrix}#1\\#2\end{smallmatrix}\right]}
\newcommand{\MMVo}[3]{\mathrm{m}\! \left[\begin{smallmatrix}#1\\#2\\#3\end{smallmatrix}\right]}

\newcommand{\ee}[3]{{\cal E}\! \left[\begin{smallmatrix}#1\\#2\end{smallmatrix};#3\right]}

\newcommand{\eez}[4]{{\cal E}\! \left[\begin{smallmatrix}#1\\#2\\#3\end{smallmatrix};#4\right]}

\newcommand{\eezeqv}[4]{{\cal E}^{\eqv}\! \left[\begin{smallmatrix}#1\\#2\\#3\end{smallmatrix};#4\right]}

\newcommand{\Gcyc}{{\mathbb G}_1}
\newcommand{\egen}{\text{e}}

\newcommand{\agen}{a}
\newcommand{\bgen}{b}

\def\Im{{\rm Im \,}}
\def\GG{{\rm G}}
\def\dumz{{\cal Z}}
\newcommand{\per}{\mathsf{p}}

\usepackage[bbgreekl]{mathbbol}
\DeclareSymbolFontAlphabet{\mathbbm}{bbold}
\DeclareSymbolFontAlphabet{\mathbb}{AMSb}

\title{Equivariant iterated Eisenstein integrals for congruence subgroups of level 2}

\author{Emiel Claasen$^a$, Mehregan Doroudiani$^b$, Claude Duhr$^c$, Axel Kleinschmidt$^a$, Franca Lippert$^d$, Oliver Schlotterer$^{e,f}$, Yoann Sohnle$^{e,g}$}

\vspace{8mm}
\affiliation[a]{Max-Planck-Institut f\"ur Gravitationsphysik (Albert-Einstein-Institut), Am M\"uhlenberg 1, 14476 Potsdam, Germany}
\affiliation[b]{School of Physics \& Astronomy, University of Southampton, University Road, SO17 1BJ, Southampton, U.K.}
\affiliation[c]{Bethe Center for Theoretical Physics and Cluster of Excellence ``Color meets Flavor'', Universit\"at Bonn, 53115 Bonn, Germany}
\affiliation[d]{Bethe Center for Theoretical Physics, Universit\"at Bonn, 53115 Bonn, Germany}
\affiliation[e]{Department of Physics and Astronomy, Uppsala University, Box 516, 75120 Uppsala, Sweden}
\affiliation[f]{Department of Mathematics, Centre for Geometry and Physics, Uppsala University, Box 480, 75106 Uppsala, Sweden}
\affiliation[g]{SLAC National Accelerator Laboratory, Stanford University, Stanford, CA 94309, USA}

\emailAdd{emiel.claasen@aei.mpg.de}
\emailAdd{m.doroudiani@soton.ac.uk}
\emailAdd{cduhr@uni-bonn.de}
\emailAdd{axel.kleinschmidt@aei.mpg.de}
\emailAdd{flippert@uni-bonn.de}
\emailAdd{oliver.schlotterer@physics.uu.se}
\emailAdd{ysohnle@stanford.edu}

\vspace{5mm}

\abstract{The modular properties of holomorphic Eisenstein series for $\SLZ$ and its congruence subgroups do not carry over to their primitives. Instead, the transformation of iterated Eisenstein integrals under congruence subgroups of level $N$ introduces inhomogeneous terms comprising (among others) cyclotomic multiple zeta values of level $N$. We combine iterated Eisenstein integrals for the congruence subgroups of level $N=2$ of arbitrary iteration depth with their complex conjugates and (single-valued) level-2 cyclotomic multiple zeta values such that they define non-holomorphic modular forms. Our construction of these so-called equivariant iterated Eisenstein integrals for $\Gamma(2)$ and $\Gamma_0(2)$ proceeds at the level of generating series that take values in an infinite-dimensional (pro-unipotent) Lie group and closely follow Brown's analogous construction in the $\SLZ$ case. The required insertions of cyclotomic multiple zeta values are controlled by zeta generators for level 2 that normalise the corresponding (pro-nilpotent) Lie algebra of the generating series and whose Lie-bracket relations we determine systematically. Upon restriction to Eisenstein series of modular weight~2, we recover single-valued multiple polylogarithms on the genus 0 modular curve $\Gamma(2)\backslash \mathbb{H}$.
}

\preprint{UUITP--29/26, BONN-TH-2026-23}

\begin{document}

\maketitle

\newpage
\setcounter{tocdepth}{2}
\tableofcontents

\vspace{10mm}
\hrule
\vspace{5mm}

\section{Introduction}

A fruitful exchange of ideas between quantum field theory, string theory, algebraic geometry and number theory revolves around iterated integrals. In particle physics and gravity, state-of-the-art computations of scattering amplitudes at the precision level of current and near-future experiments require computational control over increasingly challenging Feynman integrals. Their complexity level can be conveniently organized according to their hidden geometry in suitable parameter spaces. In this way, their mathematical backbone can often be mapped to iterated integrals on moduli- or configuration spaces of (punctured) Riemann surfaces or higher-dimensional varieties \cite{Bourjaily:2022bwx}.

The modern algebraic-geometry and number-theory literature offers a wealth of techniques for integration on different geometries and powerful concepts to organize the resulting spaces of functions and numbers. Flat connections associated with Riemann surfaces with values in suitable Lie algebras fuel the construction of function spaces of (higher-genus) polylogarithms that close under integration \cite{DHoker:2025szl}. Moreover, the theory of motives allows us to further dissect the numbers appearing through the monodromies or special values of polylogarithms, and their links can sometimes be expressed in terms of a coaction \cite{brown2017notes}.  

Scattering amplitudes in string theory offer particularly clean showcases of and testing grounds for many of these mathematical concepts \cite{Berkovits:2022ivl}. For instance, different classes of polylogarithms adapted to the sphere, the torus or more general Riemann surfaces are naturally disentangled by the topological expansion of string amplitudes.
The formulation of perturbative string amplitudes in terms of a worldsheet conformal field theory provides preferred integration kernels on Riemann surfaces that yield concrete realisations of the abstract Lie-algebraic structures appearing in the algebraic-geometry literature.

The modular graph forms \cite{DHoker:2015gmr, DHoker:2015wxz, DHoker:2016mwo, Gerken:review, DHoker:2024book} in the low-energy expansion of genus-1 string amplitudes stimulated Brown's construction of iterated integrals over holomorphic Eisenstein series \cite{brown2017multiple, Brown:2017qwo, Brown:2017qwo2}, which close under the action of the full modular group $\SLZ$. Their homogeneous $\SLZ$ transformation matches that of the flat connection which generates the iterated Eisenstein integrals and is therefore referred to as \emph{equivariant}. 
Brown's construction was made explicit \cite{Dorigoni:2024oft} in terms of generating series in certain infinite-dimensional (pro-unipotent) Lie groups. A first class of their Lie-algebra generators, known as Tsunogai derivations \cite{Tsunogai}, is associated with holomorphic Eisenstein series for $\SLZ$. Additionally, so-called zeta generators \cite{EnriquezEllAss,tsunogai11,Brown:depth3,Schneps:2015mzv,hain_matsumoto_2020} control the appearance of multiple zeta values (MZVs) in the equivariant completion of iterated Eisenstein integrals through their Lie-bracket relations with Tsunogai derivations~\cite{Dorigoni:2024oft}. 

The equivariant iterated Eisenstein integrals are closely related to single-valued integrals \cite{Brown:2017qwo2}. They 
offer concrete representatives of de Rham uplifts of these integrals (assuming that such an uplift exists, see~\cite{Matthes:talk} for a discussion of motivic elliptic MZVs), which in turn play an important role in constructing a coaction on iterated Eisenstein integrals~\cite{Kleinschmidt:2025dtk}. Understanding the Tsunogai derivations, zeta generators and equivariant iterated Eisenstein integrals is thus important if one wants to understand the mathematical structures underlying this class of special functions.

Iterated Eisenstein integrals are not only important in the context of string theory, but they also arise in the computation of multiloop Feynman integrals. It was first observed in~\cite{adams2018feynmanintegralsiteratedintegrals} in the context of the two-loop equal-mass sunrise integral that certain classes of Feynman integrals can be expressed in terms of iterated Eisenstein integrals. For Feynman integrals, however, Eisenstein series for the full modular group $\SLZ$ are not enough, but instead one needs to consider congruence subgroups of $\SLZ$. Hence, if we want to harvest the mathematical advances in the understanding of iterated Eisenstein integrals for $\SLZ$, it is essential to extend the known results from $\SLZ$ to the congruence subgroups, including the appropriate generalisations of the Tsunogai derivations and zeta generators. 

The extension to genuine congruence subgroups $\Gamma$ introduces novel challenges absent for the full modular group: For example, there may be more than one Eisenstein series of a given weight, including weight 2. Moreover, at least for normal subgroups, the equivariant integrals now come equipped with an action of the quotient group $\SLZ/\Gamma$, leading to transformation properties under $\SLZ$. 
Finally, the inhomogeneous terms in the modular transformation of individual iterated Eisenstein integrals now include cyclotomic generalisations of MZVs, so their equivariant completions require more efforts.

In a companion paper~\cite{wip:PRL}, we argue that these structures naturally arise from studying so-called canonical differential equations for Feynman integrals, first introduced in~\cite{Henn:2013pwa} in the context of Feynman integrals that evaluate to multiple polylogarithms~\cite{Goncharov:1998kja}, and later extended to cases that involve iterated Eisenstein integrals~\cite{Adams:2018yfj,Broedel:2018rwm,Gorges:2023zgv,Duhr:2025lbz,e-collaboration:2025frv,Bree:2025tug}. In particular, in~\cite{wip:PRL} we argue that the matrices defining the canonical differential equations satisfied by Feynman integrals that evaluate to iterated Eisenstein integrals provide explicit, finite-dimensional representations of the relevant extensions of Tsunogai derivations and zeta generators to congruence subgroups.

In this paper we take a complementary approach, and we initiate the study of iterated Eisenstein integrals, their underlying Lie-algebraic structures and equivariant analogues. We focus here on the simplest class of congruence subgroups, namely those of level 2, with a special focus on the principal congruence subgroup $\Gamma(2)\subset \SLZ$. An example of a canonical differential equation satisfied by a hypergeometric function leading to iterated Eisenstein integrals for level 2 was presented in~\cite{Broedel:2018rwm}. 
We emphasise that our results synergize with those of~\cite{wip:PRL}. Indeed, on the one hand we draw inspiration from the explicit finite-dimensional representations encountered from canonical differential equations to construct explicit formulas for congruence subgroups of level 2. On the other hand, the results of this paper are independent of any finite-dimensional representation, and thus provide the algebraic backbone underlying canonical differential equations associated with iterated Eisenstein integrals of level 2. Our results thus also provide the first steps to constructing a coaction for iterated Eisenstein integrals associated with congruence subgroups~\cite{wip:coactionlong}.

Our construction of equivariant iterated Eisenstein integrals for $\Gamma(2)$ draws heavily 
from flat connections on (the universal cover of) the moduli space of twice-punctured tori. For Brown's
construction in the ${\rm SL}_2(\mathbb Z)$ case, the Tsunogai derivations are contained in
the flat Calaque-Enriquez-Etingof (CEE) connection \cite{KZB}, which also governs the construction of
elliptic polylogarithms \cite{Levrac, BrownLev,  Broedel:2014vla, Broedel:2017kkb, Enriquez:2023emp}. The $\Gamma(2)$ analogues of the Tsunogai derivations in turn arise
from the flat ellipsitomic connections of Calaque-Gonzalez (CG) \cite{Calaque:2019ell, Calaque:2020ell}, which incorporate torsion points
on the torus and are governed by congruence subgroups of  ${\rm SL}_2(\mathbb Z)$.
The origin of the $\Gamma(2)$ derivations from the CG connections leads us to derive
infinite families of relations analogous to those of the Tsunogai derivations \cite{LNT, Pollack, Broedel:2015hia}. 

The flat CEE connection also links the Tsunogai derivations with zeta generators \cite{EnriquezEllAss,tsunogai11,Brown:depth3,Schneps:2015mzv,hain_matsumoto_2020}, the 
second Lie-algebra ingredient of equivariant iterated integrals of ${\rm SL}_2(\mathbb Z)$ \cite{Dorigoni:2024oft}.
In the $\Gamma(2)$ case, we leave a direct study of zeta generators from the CG connection
and their links with \cite{Hopper:2022ell, Hirose:2023} to future work. Instead, we reverse the logic and construct zeta generators for level 2 by imposing equivariance conditions in the following way:
By assuming the existence of equivariant iterated Eisenstein integrals for $\Gamma(2)$ and an
explicit construction analogous to the ${\rm SL}_2(\mathbb Z)$ case \cite{Dorigoni:2024oft}, we derive concrete Lie-series
expansions of zeta generators and their bracket relations as necessary conditions. Conversely, we verify
the necessary conditions for the equivariance properties of this construction to be sufficient for a variety of cases,\footnote{The reach of our explicit tests can be quantified through the total modular weight of the Eisenstein series
in the integration kernels which matches the \textit{degree} of the associated derivations. We have verified
 our construction of zeta generators to produce the expected equivariant iterated Eisenstein integrals 
 for $\Gamma(2)$ for each term of their generating series up to and including degree 10.} namely that our
expansions of zeta 
generators indeed lead to the expected families of equivariant iterated Eisenstein integrals for~$\Gamma(2)$.

Later parts of this work are dedicated to two noteworthy corollaries of our main results: 
one on retrieving single-valued multiple polylogarithms on the modular curve $Y_{\Gamma(2)} = \Gamma(2)\backslash \mathbb{H} $ from Eisenstein series of modular weight 2,
and another one on extracting equivariant iterated Eisenstein integrals for $\Gamma_0(2)$ (or other congruence subgroups of level 2) from those of~$\Gamma(2)$.

The level-2 results of this paper are intended to furnish the groundwork for future generalisations to equivariant iterated Eisenstein integrals for more general congruence subgroups
(explaining the detailed presentation with numerous examples). This will not only feed into applications to elliptic Feynman integrals, but it will also set the stage for equivariant iterated integrals for Riemann surfaces of genus $g\geq 2$ -- both for the modular group ${\rm Sp}_{2g}(\mathbb Z)$ and its congruence subgroups. In the long run, equivariant iterated integrals on Riemann surfaces are hoped to serve as prototypes for generalisations to higher-dimensional varieties including the K3 surface and Calabi-Yau manifolds in recent studies of Feynman integrals.

\subsection{Overview of main results}
\label{int:main}

In order to facilitate the navigation through this long paper, we shall now gather
more specific pointers to its main results.
\begin{itemize}
\item In~\eqref{eq: N2conn} we introduce a flat connection $\ANN{\tau}$ living on the upper half-plane $\mathbb{H}$ and taking values in a \emph{derivation algebra} $\mathfrak{u}$ generated by (Tsunogai) derivations that are tied to Eisenstein series for $\Gamma(2)$ and, for the purpose of this introduction, are collectively denoted by $\ep$. The connection $\ANN{\tau}$ transforms covariantly under $\Gamma(2)$ and $\SLZ$. 
Its path-ordered exponential is denoted by $\INN{\tau}$ and generates iterated integrals of Eisenstein series for $\Gamma(2)$. Similar to the $\SLZ$ case of~\cite{brown2017multiple, Brown:2017qwo2}, the generating series $\INN{\tau}$ has non-trivial cocycles under $\Gamma(2)$ and $\SLZ$ transformations, shown in (\ref{trfpexp}). Since $\Gamma(2)$ is normal in $\SLZ$, many $\Gamma(2)$ properties can be extended to $\SLZ$ by attaching a further representation of the quotient $\SLZ/\Gamma(2)\cong \mathcal{S}_3$, the symmetric group on three letters. We use this structure to further organise many~equations. 
\item A central proposal of our work is the construction of a generating series $\INNeqv{\tau}$ in~\eqref{eq: eqv2proposal} whose expansion coefficients are \emph{equivariant} versions of iterated Eisenstein integrals for $\Gamma(2)$ that transform 
in the same way as $\ANN{\tau}$ under $\SLZ$. The construction of the series $\INNeqv{\tau}$ relies on the use of additional derivations $\sigma_w$ 
($w\geq 1$ odd) that act on the derivations $\ep \in \mathfrak{u}$ appearing in $\INN{\tau}$ by the adjoint action, and the resulting $[\sigma_w,\ep]$ are Lie series in $\ep$. 
The derivations $\sigma_w$~--~dubbed \textit{zeta generators}~--~control the single-valued cyclotomic multiple zeta values needed for building the equivariant objects in $\INNeqv{\tau}$ out of the non-equivariant ones in $\INN{\tau}$, generalising~\cite{Dorigoni:2024oft}.
The properties of the derivations $\ep\in\mathfrak{u}$ and zeta generators $\sigma_w$ are further main results of this work.
\item A powerful tool for studying the derivations $\ep\in \mathfrak{u}$ comes from an extension of $\ANN{\tau}$ to a flat \emph{ellipsitomic connection} $\mathbb K_\text{CG}(z,\tau)$ in section \ref{sec:fltcg}, studied already by Calaque-Gonzalez (CG)~\cite{Calaque:2019ell,Calaque:2020ell}. On top of $\tau\in\mathbb{H}$, the CG connection also depends on a puncture $z$ on the universal cover of the torus and takes values in a larger algebra that also includes generators associated with homology cycles and 2-torsion points. The properties of this connection 
lead to explicit formulas (\ref{epac00}) to (\ref{eq: epsona}) for the action of $\mathfrak{u}$ on the larger algebra.\footnote{It is this Lie algebra that the $\ep$ are actual (outer) derivations of.}
\item From the explicit description of
$\mathfrak{u}$ we can study the relations obeyed by the generating derivations $\ep$ that enter the original connection $\ANN{\tau}$. We find by low-order investigations and, in complete agreement with expectations from $\SLZ$~\cite{Pollack}, that the algebra $\mathfrak{u}$ is not free, but it satisfies relations related to cusp forms for $\Gamma(2)$ that we call \emph{Pollack relations}. This is the content of section~\ref{sec: pollack}.
\item We also obtain explicit results for the structure of the zeta generators $\sigma_w$. For studying them we investigate the equivariance conditions of the series $\INNeqv{\tau}$ in section~\ref{sec: level2}. We find that many structures and techniques generalise from $\SLZ$ \cite{Dorigoni:2024oft, Dorigoni:2024iyt}. In particular, the generators decompose as $\sigma_w = z_w + \sigma_w^{\rm g}$, where the two parts are called the \emph{arithmetic} and the \emph{geometric} part of the zeta generator, respectively. We find explicit results for the geometric part up to so-called modular depth 2, see (\ref{sysxi2.09}) and (\ref{hwlevel2}), and for the commutators $[\sigma_w, \ep]$ along with a recursive procedure for obtaining higher modular-depth terms. Here \emph{modular depth} refers to the nesting depth of commutators of the $\ep$ that generate $\mathfrak{u}$.  Of particular interest in the context of $\Gamma(2)$ is the existence of the zeta generator $\sigma_1$, which does not exist for $\SLZ$ where only $\sigma_w$ for $w\geq 3$ odd tied to odd zeta values appear. The generator $\sigma_1$ is tied to $\log 2$, and we have a closed and simple formula for it in~\eqref{exactz1} and its brackets in section \ref{sec:z1ep}.
\item We further study the restriction of the function space to weight-2 Eisenstein series in section~\ref{sec:weight2}. Eisenstein series of modular weight 2 integrate to logarithms of a Hauptmodul $\lambda$ of the modular curve $Y_{\Gamma(2)} = \Gamma(2)\backslash \mathbb{H} $. Similarly,  single-valued primitives of weight-2 Eisenstein series for $\Gamma(2)$ are known to yield single-valued logarithms of $\lambda$ \cite{Duhr:2025lvr}. By projecting the  Eisenstein series for $\Gamma(2)$ in our construction of single-valued iterated integrals to modular weight 2, we recover single-valued multiple polylogarithms \cite{svpolylog} of~$\lambda$, see \eqref{eq:pi2.IIsv}. This requires matching the connection and its iterated integrals including the tangential base-point regularisations as well as the emergence of the genus-0 zeta generators. We also present equivariant polylogarithms resulting from the projection of equivariant integrals onto their weight-2 contributions \eqref{defgeqv}. These polylogarithms transform equivariantly under changes of base-point.
\item $\Gamma(2)$ is not the only level-2 congruence subgroup of $\SLZ$ and, as remarked above, other congruence subgroups show up in Feynman integrals. For this reason, we extend our analysis to all congruence subgroups $\Gamma$ of level 2 in section~\ref{sec: extensions}. Our approach rests on work by Hopper \cite{Hopper:2021ell} and leads to a simple compatibility of connections and iterated integrals exhibited in \eqref{eq:piA}. It also clarifies further the structure of the Pollack relations in section \ref{sec: pollack} in relation to whether the $\Gamma(2)$ cusp forms lift to one of the larger congruence subgroups. We outline the general procedure and apply it to explicitly construct the generating series of equivariant iterated Eisenstein integrals for $\Gamma_0(2)$ in section \ref{specg02}.

\end{itemize}
We would like to emphasise that the equivariance properties of the non-holomorphic series $\INNeqv{\tau}$ in~\eqref{eq: eqv2proposal} are conjectural at this point. Still, they are a consequence of a simpler conjecture, namely that the $S$-cocycle of holomorphic iterated Eisenstein integrals takes the factorised form (\ref{eq:Scoc2}) in terms of series in zeta generators. The level-1 analogue \cite{Kleinschmidt:2025dtk} of the factorised $S$-cocycle in turn is established from the results of \cite{brown2017multiple, Schlotterer:2025qjv}.

\medskip
Several appendices contain complementary material and additional details on some of the calculations of this article. In appendix~\ref{sec: falph}, we outline the $f$-alphabet for cyclotomic MZVs at level 2 as they enter in our construction of the generating series $\INNeqv{\tau}$. Appendix~\ref{app:mf2} recalls the ring  of modular forms for $\Gamma(2)$ and lists in particular cusp forms as they arise from related congruence subgroups. Details on Fay identities that we used when proving flatness of the CEE and CG connections can be found in appendix~\ref{apc.sec}. Appendix~\ref{app:mpl} explains how the single-valued polylogarithms at genus 0 that yield cyclotomic level-2 MZVs are connected to our construction and in particular probes the coaction at genus 0. 
In appendix~\ref{sec:fp}, we sketch a partial first-principles derivation of zeta generators based on degeneration limits of the CG connection to genus 0 but leave the subsequent extensions back to elliptic curves with level structure to future work.
Twisted elliptic MZVs are related to our construction in appendix~\ref{app:eMZV}. 
Additional and elaborate examples for the simplified $S$-cocycle, higher-degree terms of zeta generators and equivariant iterated integrals of modular depth four and five are given in appendix~\ref{app: examples}.
The final appendix~\ref{sec:num.mmv} contains details on how we numerically evaluated some of the multiple modular values (a.k.a.\ period integrals) associated with our iterated integrals and how this was used to determine certain exact expressions.

The arXiv submission of this article includes an ancillary file containing the Pollack relations of section~\ref{sec: pollack} (organized by their $\mathcal{S}_3$ representations), the geometric parts $\sigma_w^{\rm g}$ of the zeta generators and the brackets $[z_w,\ep_{k|\mu}]$ of section~\ref{sec: level2}, the logarithm of the series $\mathbb X(\ep)$ in the factorised $S$-cocycle (\ref{ag:Scoc2}), the $f$-alphabet isomorphism $\rho$ of appendix~\ref{app:a.3}, the generating series (\ref{eq:Eeqvdef}) of equivariant iterated Eisenstein integrals and the numerical values of the multiple modular values obtained in appendix~\ref{sec:num.mmv}.

\subsection*{Acknowledgements}

We would like to thank David Broadhurst, Daniele Dorigoni, Abhiram Kidambi, Martin Raum, Leila Schneps, Oliver Schnetz, Aleksander Shmakov for combinations of valuable discussions and collaboration on related topics.
We would like to thank the Erwin Schr\"odinger International Institute for Mathematics and Physics (ESI), University of Vienna (Austria), for the opportunity to participate in the Thematic Programme ``Amplitudes and Algebraic Geometry'' in 2026 where a significant part of this work has been accomplished and for the support given. The research of EC, AK, OS, YS is funded by the European Union under ERC Synergy Grant MaScAmp 101167287, and the research of CD and FL is supported in part by the European Union under ERC Consolidator Grant LoCoMotive 1101043686 and by the Deutsche Forschungsgemeinschaft (DFG, German Research
Foundation) under Germany’s Excellence Strategy – EXC 3107 – Project-ID 53376636. Views and opinions expressed are however those of the author(s) only and do not necessarily reflect those of the European Union or the European Research Council. Neither the European Union nor the granting authority can be held responsible for them. MD was supported by the ERE grant RF\textbackslash ERE\textbackslash 221103 associated with the Royal Society University Research Fellowship Grant URF\textbackslash R1\textbackslash 221236. MD was supported in part by a grant of access to OpenAI models through the ChatGPT for Academic Researchers program.
The research of YS is supported by the Knut and Alice Wallenberg Foundation, Grant KAW 2025.0414.

\subsection*{Tool and computational resource disclosure}

Suggestions by AI tools such as ChatGPT and Claude were used in the research process and for the creation of the figures. However, all the statements of this work, as well as their proofs, are due to the authors. The authors take full responsibility for the content of the publication.

\section{Review of iterated Eisenstein integrals and main construction}
\label{sec:motiv}

In \cite{Brown:2017qwo,Brown:2017qwo2}, a class of non-holomorphic modular forms was constructed from iterated integrals over holomorphic Eisenstein series for the full modular group $\SLZ$. The quest to construct similar objects for congruence subgroups was first explored in \cite{Drewitt:2023con, Bonisch:2024nru, Duhr:2025lvr} for $\Gamma(N)$, but it was limited to iterated integrals of low iteration depth. Our goal is to write generating series of such equivariant iterated Eisenstein integrals for $\Gamma(2)$ at any depth and to provide explicit examples up to depth 5. 
A major innovation of this work is to infer the cyclotomic multiple zeta values needed for the equivariant completions from Lie-algebra elements known as zeta generators. In this section, we review the construction for the full modular group and summarise our main result, namely a generalisation to equivariant iterated Eisenstein integrals for modular forms for $\Gamma(2)$.

\subsection{Review of modular forms for congruence subgroups}
\label{sec:N2conn}

Before we propose a form of the connection that will enter the construction of equivariant iterated Eisenstein integrals for $\Gamma(2)$, we first introduce some general definitions related to modular forms and congruence subgroups (cf.,~e.g., \cite{diamond} for a review). Let $N>0$ be an integer. We denote by $\Gamma(N)$ the principal congruence subgroup of level $N$ of the modular group $\SLZ$. It is given by
\beq
\Gamma(N)=\{\gamma\in \SLZ :\gamma\equiv \mathbbm{1}\bmod N\}\,,
\label{eq:gammaN}
\eeq
and defines a normal subgroup of $\SLZ$. The associated quotient group is isomorphic to $\SLZ/\Gamma(N)\cong{\rm SL}_2(\mathbb{Z}/N\mathbb{Z})$.
The modular group $\SLZ=\Gamma(1)$ is generated by 
the two matrices
\beq
T = \begin{pmatrix}1&1\\0&1\end{pmatrix}\,,\qquad
\qquad S =\begin{pmatrix}0&-1\\1&0\end{pmatrix}\,.
\label{eq: ST}
\eeq
We define an action of $\SLZ$ on $\tau \in \mathbb{H}\cup \mathbb{P}^1(\mathbb{Q})$, where $\mathbb{P}^1(\mathbb{Q})$ can be represented by $\mathbb{Q} \cup \{\infty\}$ and $\mathbb{H}$ is the upper half-plane $\mathbb{H}=\{\tau\in\mathbb{C}:\Im(\tau)>0\}$,
via M\"obius transformations 
\beq
\gamma \tau = \frac{a\tau+b}{c\tau +d}\,,\qquad \text{for} \quad \gamma=\begin{pmatrix}
    a&b\\c&d
\end{pmatrix}\in \SLZ\,.
\label{eq: SLZ}
\eeq
This action on $\tau$ induces an action on functions $f:\mathbb{H}\rightarrow \mathbb{C}$ via $f(\gamma\tau)=f\big(\tfrac{a\tau+b}{c\tau+d}\big)$. 

Let $\Gamma$ denote a general congruence subgroup of $\SLZ$, i.e., any finite-index subgroup of $\SLZ$ containing $\Gamma(N)$ for some $N$. A meromorphic function $f:\mathbb{H}\rightarrow \mathbb{C}$ that transforms as 
\beq 
f(\gamma\tau)=(c\tau+d)^kf(\tau)\,,\qquad \text{for all }\gamma \in \Gamma\,,
\label{eq: modtrafo}
\eeq
defines a weakly modular form of (integer) weight $k$ for the group $\Gamma$.\footnote{Throughout this paper we restrict to modular forms without a multiplier system, i.e.,~the transformation is as stated in~\eqref{eq: modtrafo}. In principle, $\Gamma$ admits non-trivial characters $\chi: \Gamma \to \mathbb{C}^\times$ that could arise in the transformation law, but for our purposes the trivial $\chi=1$ is all that is needed. For the prevalent example $\Gamma=\Gamma(2)$ below the connection does not see non-trivial characters in applications~\cite{wip:PRL}, justifying this restriction.} 
Since $T^N\in \Gamma$ for the $N^{\rm th}$ power of the generator $T$ in (\ref{eq: ST}), $f$ admits a Fourier expansion
\beq
f(\tau)=\sum_{n\in \mathbb{Z}} a_n q_N^n\,,\ \ \ \ \text{with}\, \ \ q_N =e^{\frac{2\pi i}{N}\tau}\,.
\label{eq: Fourexp}
\eeq
Additionally, if $f$ is holomorphic and for all $\gamma\in \Gamma$ the function $f(\gamma\tau)$ has no nonzero negative Fourier coefficients, i.e., $a_{-n}=0$ for all $n\geq 1$, then $f$ defines a holomorphic modular form for $\Gamma$ of weight~$k$ (with trivial character).

We define the cusps of the congruence subgroup $\Gamma$ as the $\Gamma$-orbits in $\mathbb{P}^1(\mathbb{Q})$. A holomorphic modular form $f$ that vanishes at all cusps of $\Gamma$ (i.e., that has vanishing $a_0$ in the Fourier expansion near each cusp; equation~\eqref{eq: Fourexp} is the Fourier expansion near the cusp at infinity) is known as a cusp form for $\Gamma$. The $\mathbb{C}$-linear space of holomorphic modular forms for $\Gamma$ of weight $k$ is denoted $\mathcal{M}_k(\Gamma)$. It can be decomposed into the space of cusp forms $\mathcal{S}_k(\Gamma)$ and the complementary space spanned by Eisenstein series $\mathcal{E}_k(\Gamma)$ (that do not vanish at all cusps) as $\mathcal{M}_k(\Gamma)=\mathcal{E}_k(\Gamma)\oplus\mathcal{S}_k(\Gamma)$. As we will consider iterated integrals built from Eisenstein series, we merely focus on the Eisenstein subspace from now on. For discussions on iterated integrals of cusp forms we refer to \cite{brown2017multiple,
brown2017a, Dorigoni:2021ngn, Dorigoni:2024oft, wip:cusp}.

As a notational remark, in the rest of this paper we discuss and construct objects for $\SLZ$ as well as the congruence subgroups $\Gamma(2)$ in (\ref{eq:gammaN}) and $\Gamma_0(2)=\Gamma_1(2)$ in (\ref{eq: gamma01}) below. 
When using symbolic notation for the group $\Gamma$, the dependence on $\Gamma$ enters via subscripts or superscripts (e.g., for the connection we use $\mathbb{A}_\Gamma$). As the paper mainly concerns $\Gamma(2)$, we often omit the explicit labels referring to $\Gamma(2)$ in order to not clutter the notation (for example $\mathbb{A} = \mathbb{A}_{\Gamma(2)}$).

\subsubsection{\texorpdfstring{Eisenstein series for $\Gamma(N)$ from Kronecker-Eisenstein coefficients}{Eisenstein series for Gamma(N) from Kronecker-Eisenstein coefficients}}

The Eisenstein subspace for $\Gamma(N)$ is generated by the doubly-periodic Kronecker-Eisenstein coefficients 
\beq
f^{(k)}(z,\tau) = - \sum_{\substack{m,n \in \mathbb Z \\ (m,n) \neq (0,0)}} \frac{e^{2\pi i (nu - mv)}}{(m\tau + n)^k} \, , \ \ \ \ z = u\tau + v \, , \ \ \ \ u,v\in \mathbb R \, , \ \ \ \ k\in \mathbb N = \{1,2,\ldots\}\,,
\label{eq: kronEis}
\eeq
evaluated at torsion points $z \in \mathbb Q + \tau \mathbb Q$ on the elliptic curve $\Sigma_\tau = \mathbb{C}/(\mathbb{Z}+\tau\mathbb{Z})$, leading to a level structure $\Gamma(N)$ \cite{BrownLev,paşol2026eisensteinkroneckermodularsymbolbeilinsonkato}  when the least common multiple of the denominators of the torsion point is $N$. The conventional choice of fundamental cell for $\Sigma_\tau$ is attained by restricting the co-moving coordinates to $u,v \in [0,1)$. The Kronecker-Eisenstein coefficients~(\ref{eq: kronEis}) evaluated at torsion points with $u,v \in \tfrac{1}{N}\mathbb{Z}$ produce a spanning set of $\mathcal{E}_k(\Gamma(N))$ and are absolutely convergent to holomorphic functions in $\tau$ for integer $k>2$.
Their Fourier expansions are given by 
\begin{align}
&f^{(k)}(u \tau {+} v,\tau)= \frac{(2\pi i)^k}{k!} B_k(u) \label{eq: FKroneis} \\
&\quad + \frac{(2\pi i)^k}{(k{-}1)!} \bigg[ 
\frac{u^{k-1}}{1-e^{-2\pi i (u\tau+v)}}
+\sum_{n=1}^\infty q_1^n \bigg(
\frac{(u+n)^{k-1}}{q_1^n-e^{-2\pi i (u\tau+v)}}
- \frac{(u-n)^{k-1}}{q_1^n-e^{2\pi i (u\tau+v)}}
\bigg)
\bigg] 
\notag\,,
\end{align}
where $B_k(u)$ denotes the $k^{\rm th}$ Bernoulli polynomial.\footnote{\label{foot:ber}Bernoulli polynomials are generated by $\frac{ t e^{ut}}{e^t-1} = \sum_{k=0}^\infty  B_k(u) \frac{t^k}{k!}$ and reduce to Bernoulli numbers $B_k =~B_k(0)$ at the special value $u=0$, for instance $B_2(u)=u^2-u+\frac{1}{6}$ leading to $B_2=\frac{1}{6}$.} 
The Fourier expansion~\eqref{eq: FKroneis} is convergent also for $k=2$ as long as $z=u\tau{+}v \notin \mathbb{Z}{+}\tau \mathbb{Z}$, and we take this as the definition of $f^{(2)}(z,\tau)$. For general $z$, they carry modular weight $k\geq 2$ under $\gamma=\big( \smallmatrix a & b \\ c & d \endsmallmatrix \big) \in \SLZ$,
\beq
f^{(k)}\bigg( \frac{z}{c \tau + d},\frac{a\tau + b}{c \tau + d} \bigg) = (c \tau + d)^k f^{(k)}(z,\tau)\,,
\label{eq: KEmod}
\eeq
where the $\SLZ$ action $z\rightarrow \frac{z}{c \tau + d}$ transforms the co-moving coordinates $u,v$ according to 
\beq
\begin{pmatrix} v \\ -u \end{pmatrix}
\rightarrow \begin{pmatrix} a & b \\ c& d \end{pmatrix}
\begin{pmatrix} v \\ -u  \end{pmatrix}
\quad\Longleftrightarrow\quad
\begin{pmatrix} u \\  v \end{pmatrix}
\rightarrow \begin{pmatrix} d & -c \\ -b& a \end{pmatrix}
\begin{pmatrix} u \\ v  \end{pmatrix}
\,,
\label{eq: uv}
\eeq
where the  matrix in the second representation is $\big(\smallmatrix   d &-c \\ -b &a  \endsmallmatrix \big) = (\gamma^{-1})^T$. We will mostly encounter $f^{(k)}$
evaluated at fixed non-zero torsion points that are permuted by the action of $\SLZ$, and for $k=2$, one has to exclude the point $z=0$ in~\eqref{eq: KEmod}.

As an example, the Eisenstein subspace $\mathcal{E}_k(\SLZ)$ for the full modular group is one-dimensional for every $k\geq 4$ even and vanishes for $k=2$ and $k$ odd. For $k\geq 4$ even, it is generated by $f^{(k)}(0,\tau)=-\GG_k(\tau)$, where $\GG_k(\tau)$ is the holomorphic Eisenstein series of even weight $k\geq 4$,
\beq
\GG_k(\tau)=\sum_{\substack{m,n \in \mathbb Z \\ (m,n) \neq (0,0)}}\frac{1}{(m\tau+n)^k}=2\zeta_k+\frac{2(2\pi i)^k}{(k{-}1)!}\sum_{n=1}^\infty \sigma_{k-1}(n)q_1^n\,.
\label{eq: Eis}
\eeq
Here $\zeta_k$ is the Riemann zeta function $\zeta(z)=\sum_{n= 1}^\infty n^{-z}$ evaluated at $z=k$, and the coefficients $\sigma_{k-1}(n)$ in the Fourier expansion are given by the (positive) divisor sums $\sigma_{k-1}(n)=\sum_{m|n}m^{k-1}$. The $k=2$ case of \eqref{eq: Eis} needs to be excluded, since
the asymptotics of $f^{(2)}(z,\tau)$ at $|z| \ll 1$ features a contribution $-\frac{\pi}{\Im \tau }\frac{ \bar z}{z}$, which obstructs a well-defined $z \rightarrow 0$ limit. Correspondingly, there exist no holomorphic Eisenstein series for $\SLZ$ of weight 2. Furthermore, the alternating parity $f^{(k)}(-z,\tau)=(-1)^k f^{(k)}(z,\tau)$ readily leads to the vanishing ${\rm G}_{k}(\tau)=0$ for odd $k$ (and also to vanishing $f^{(k)}(z,\tau)$ at half-periods $z \in \{ \frac{1}{2},\frac{\tau}{2},\frac{1{+}\tau}{2} \}$ for odd $k$, see below). Note that both the lattice-sum representation and Fourier expansion of holomorphic Eisenstein series can be obtained by specialising \eqref{eq: kronEis} and \eqref{eq: FKroneis} to $u=v=0$ and using the relation $(2\pi i)^k\tfrac{B_{k}(0)}{k!}=(2\pi i)^k\tfrac{B_{k}}{k!}=-2\zeta_k$ for even $k$, where $B_k$ is the $k^{\rm th}$ Bernoulli number (see footnote~\ref{foot:ber}), and we note $B_1=-\frac12$.

\subsubsection{\texorpdfstring{Properties of $\Gamma(2)$}{Properties of Gamma(2)}}
\label{sec:G2}

Before considering the Eisenstein subspace for $\Gamma(2)$, it is worth noting some properties of $\Gamma(2)$ itself. The generators of $\Gamma(2)$ can be expressed in terms of the generators of level 1 from~\eqref{eq: ST}, yielding the presentation $\Gamma(2)=\langle T^2,ST^2S^{-1},-\mathbbm{1}\rangle$. The set $\mathbb{P}^{1}(\mathbb{Q})$ decomposes into three $\Gamma(2)$-orbits, represented by $0$, $1$, and $\infty$. The quotient group ${\rm SL}_2(\mathbb{Z}/2\mathbb{Z}) \cong \SLZ/\Gamma(2)$ is isomorphic to the symmetric group $\mathcal{S}_3$ of permutations on three letters. There are three irreducible representations of $\mathcal{S}_3$ over $\mathbb{C}$: 
\begin{enumerate}
    \item[(i)]
    the one-dimensional trivial representation that we call $\mathbf{1}$;
    \item[(ii)]
    the one-dimensional alternating representation that we call $\bep$;
    \item[(iii)]
    the two-dimensional standard representation (that can be realised for instance on three points $\{(z_1,z_2,z_3) \in \mathbb{C}^3 \,|\, z_1+z_2+z_3=0\}$) that we call $\mathbf{2}$.
\end{enumerate}
Since $\Gamma(2)$ is normal in $\SLZ$, the space $\mathcal{M}_k(\Gamma(2)) =  \mathcal{E}_k(\Gamma(2))\oplus \mathcal{S}_k(\Gamma(2))$ of modular forms carries an induced action of $\SLZ$.\footnote{\label{foot:normal}This can be argued as follows. Let $\gamma=\begin{psmallmatrix}
    p&q\\r&s
\end{psmallmatrix}\in \Gamma$ and let $g=\begin{psmallmatrix}
    a&b\\c&d
\end{psmallmatrix}\in\SLZ$. Let $f(\tau)\in\mathcal{M}_k(\Gamma)$. Then the condition for having an induced action of $\SLZ$ comes down to $\tilde{f}_g(\tau)\in \mathcal{M}_k(\Gamma)$ or equivalently $\tilde{f}_g(\gamma\tau)=(r\tau+s)^k\tilde{f}_g(\tau)$, where $\tilde{f}_g(\tau) = (c\tau+d)^{-k}f(g\tau)$. Now write $\tilde{f}_g(\gamma\tau)=(c\gamma\tau+d)^{-k}f(g\gamma\tau)=(c\gamma\tau+d)^{-k}f(g\gamma g^{-1}g\tau)$. If $\Gamma$ is normal, we have $\gamma'=\begin{psmallmatrix}
    p' & q'\\r'&s'
\end{psmallmatrix}=g\gamma g^{-1}\in \Gamma$.
This yields then $(c\gamma\tau+d)^{-k}f(g\gamma g^{-1}g\tau)=(c\gamma\tau+d)^{-k}(r'g\tau+s')^kf(g\tau)=(r\tau+s)^k\tilde{f}_g(\tau)$, where in the last equality we used $\gamma'g=g\gamma$ to simplify the factors of automorphy.} 
Moreover, since $\Gamma(2)$ acts via (\ref{eq: modtrafo}) on $\mathcal{M}_k(\Gamma(2))$, this action factors through the quotient $\SLZ/\Gamma(2)\cong\mathcal{S}_3$, so $\mathcal{M}_k(\Gamma(2))$ is a representation of $\mathcal{S}_3$. In general, $\mathcal{S}_3$ mixes the modular forms, but preserves the cuspidal and the Eisenstein subspaces of $\mathcal{M}_k(\Gamma(2))$. Note that because $-\mathbbm{1}\in \Gamma(2)$, we have that $\mathcal{M}_{k}(\Gamma(2))=0$ for odd $k$.
In Appendix~\ref{app:mf2}, we review some general aspects of the theory of holomorphic modular forms for~$\Gamma(2)$.

The dimension of the Eisenstein subspace for $\Gamma(2)$ (with trivial character) is given by  
\begin{align}
    \dim\mathcal{E}_k\big(\Gamma(2) \big)=\begin{cases}
        2\,, \ \ \ \ k=2\,,\\
        3\,, \ \ \ \ k\geq4\, \ \text{even}\,,\\
        0\,, \ \ \ \ k\, \ \text{odd}\,.
    \end{cases}
    \label{eq: dims}
\end{align}
For even $k\geq 4$ this equals the number of cusps of $\Gamma(2)$, while for $k=2$ it drops by one due to a linear relation among the constant terms at the cusps (see also \eqref{funrel.02}). The three cusps can be identified with the torsion points of exact order 2 on the elliptic curve, so that we can construct a basis of the Eisenstein subspace $\mathcal{E}_k(\Gamma(2))$ by evaluating the Kronecker-Eisenstein coefficients from \eqref{eq: kronEis} at $(u,v)\in \{(0,\frac{1}{2}),(\frac{1}{2},\frac{1}{2}),(\frac{1}{2},0)\}$. We shall employ the notation for the torsion points of exact order 2,
\beq
\omega_1 = 0 \, , \ \ \ \
\omega_2 = \frac{1}{2} \, , \ \ \ \
\omega_3 = \frac{1+\tau}{2} \, , \ \ \ \
\omega_4 = \frac{\tau}{2} \,,
\label{eq: torsion}
\eeq
giving us $\{f^{(k)}(\omega_\mu,\tau)\}_{\mu=2,3,4}$ for $k\geq 2$ even as a spanning set of $\mathcal{E}_2(\Gamma(2))$ and a basis for $\mathcal{E}_k(\Gamma(2))$ for $k\geq 4$ even. To single out a basis at $k=2$, we use the relation
\beq
f^{(2)}(\omega_2 ,\tau)
+ f^{(2)}(\omega_3 ,\tau)
+ f^{(2)}(\omega_4 ,\tau)
= 0\,.
\label{funrel.02}
\eeq
We choose to eliminate $f^{(2)}(\omega_2,\tau)$ in favor of the others, so that our choice of basis for $\mathcal{E}_2(\Gamma(2))$ consists of $\{f^{(2)}(\omega_3,\tau),f^{(2)}(\omega_4,\tau)\}$. For $k\geq4$, the $\{f^{(k)}(\omega_\mu,\tau)\}_{\mu=2,3,4}$ already form a basis, but obey an identity that relates them to $\GG_k(\tau)\in \mathcal{E}_k(\SLZ)$ 
\beq
f^{(k)}(\omega_2 ,\tau)
+ f^{(k)}(\omega_3 ,\tau)
+ f^{(k)}(\omega_4 ,\tau)
= \frac{2^{k-2}-1}{2^{k-2}} {\rm G}_k(\tau)\,,
\label{funrel.01}
\eeq
which can be readily verified from the lattice sums in (\ref{eq: kronEis}) and (\ref{eq: Eis}).
Note that $\GG_k(\tau)$ for even $k\geq 4$ are automatically modular forms for $\Gamma(2)$, as $\mathcal{E}_k(\SLZ)\subset \mathcal{E}_k(\Gamma(2))$ for $k\geq 4$.

The induced action of $\SLZ$, or equivalently of $\mathcal{S}_3$, on the Eisenstein subspace $\mathcal{E}_k(\Gamma(2))$ can be inferred from \eqref{eq: uv}, yielding:
\beq
\left( \smallmatrix
f^{(k)}(\omega_2,\tau+1) \\ 
f^{(k)}(\omega_3,\tau+1)\\
f^{(k)}(\omega_4,\tau+1)
\endsmallmatrix \right)
=
\left( \smallmatrix
\big.1 \ &0 &\ 0 \\ 
0 \ &0 &\ 1\\
0 \ &1 &\ 0\big.
\endsmallmatrix \right)
\left( \smallmatrix
f^{(k)}(\omega_2,\tau) \\ 
f^{(k)}(\omega_3,\tau)\\
f^{(k)}(\omega_4,\tau)
\endsmallmatrix \right)
\, , \ \ \ \ 
\left( \smallmatrix
f^{(k)}(\omega_2,-1/\tau) \\ 
f^{(k)}(\omega_3,-1/\tau)\\
f^{(k)}(\omega_4,-1/\tau)
\endsmallmatrix \right)
=
\tau^k \left( \smallmatrix
\big.0 \ &0 &\ 1 \\ 
0 \ &1 &\ 0\\
1 \ &0 &\ 0\big.
\endsmallmatrix \right)
\left( \smallmatrix
f^{(k)}(\omega_2,\tau) \\ 
f^{(k)}(\omega_3,\tau)\\
f^{(k)}(\omega_4,\tau)
\endsmallmatrix \right)\,,
\label{eq: S3}
\eeq
from which one can work out the decomposition into irreducible representations of $\mathcal{S}_3$, giving the irreducible standard representation $\mathbf{2}$ for $k=2$ and a reducible representation for even $k\geq 4$ that decomposes as $\mathbf{1}\oplus \mathbf{2}$. The absence of $\mathbf{1}$ for $k=2$ is directly reflected in the relation \eqref{funrel.02} among $f^{(2)}(\omega_\mu,\tau)$. 
The trivial representation $\mathbf{1}$ for $k\geq 4$ corresponds to the $\SLZ$-modular form $\mathrm{G}_k(\tau)$.

\subsection{\texorpdfstring{Equivariant flat connection for $\Gamma(2)$}{Equivariant flat connection for Gamma(2)}}
\label{sec:2.1.3}

To the end of constructing non-holomorphic modular forms built from homotopy-invariant iterated Eisenstein integrals for $\Gamma(2)$, we need to define a flat connection form that involves the bases of $\mathcal{E}_k(\Gamma(2))$ for all even $k\geq 2$. Inspired by the framework for $\SLZ$ to be reviewed in section \ref{sec:lvl1}, we shall employ the following connection form\footnote{A different version of the same connection was introduced in \cite{Calaque:2019ell} and will be reviewed and related to \eqref{eq: N2conn} in section \ref{sec: extensions}. 
\label{foot:altcon}
}
\beq
\ANN{\tau} =-\dd \tau \sum_{\mu=3}^4 \frac{f^{(2)}(\omega_\mu,\tau)}{2\pi i} \tilde{\ep}_{2|\mu} -  \dd \tau \sum_{k=4}^{\infty} (k{-}1) \sum_{j=0}^{k-2} \frac{(-1)^j}{j!} (2\pi i \tau)^{j} \sum_{\mu = 2}^4 \frac{ f^{(k)}(\omega_\mu,\tau) }{(2\pi i)^{k-1}} \ep_{k | \mu}^{(j)}\,,
\label{eq: N2conn}
\eeq
where the non-commuting letters $\tilde{\ep}_{2|\mu}$ and $\ep_{k|\mu}^{(j)}$ are said to be of degree $2$ and $k$, respectively, and are treated as bookkeeping variables for now; their origin from the ellipsitomic connection of \cite{Calaque:2019ell,Calaque:2020ell} will be expanded upon in section~\ref{sec: Connection}. They are to be thought of as the level 2 analogue of Tsunogai's derivations (see section \ref{sec:lvl1} and \cite{Tsunogai,Pollack}). The $\tilde{\ep}_{2|\mu},\epsilon_{k|\mu}^{(j)}$ of (\ref{eq: N2conn}) are taken to transform in the dual representations of $\mathcal{S}_3$ by converting the $S$- and $T$-transformations (\ref{eq: S3}) of integration kernels into operators $\theta_S$ and $\theta_T$ acting on the $\tilde{\ep}_{2|\mu},\ep_{k|\mu}^{(j)}$. This can be realised by defining
\begin{align}
\theta_T^{-1} \left( \smallmatrix
\ep_{k| 2}^{(j)} \\
\ep_{k| 3}^{(j)}\\
\ep_{k| 4}^{(j)}
\endsmallmatrix \right)\theta_T
&=
\left( \smallmatrix
\big.1 \ &0 &\ 0 \\ 
0 \ &0 &\ 1\\
0 \ &1 &\ 0\big.
\endsmallmatrix \right)
\left( \smallmatrix
\ep_{k| 2}^{(j)} \\
\ep_{k| 3}^{(j)}\\
\ep_{k| 4}^{(j)}
\endsmallmatrix \right)
\, , \ \ \ \
\theta_S^{-1} \left( \smallmatrix
\ep_{k| 2}^{(j)} \\
\ep_{k| 3}^{(j)}\\
\ep_{k| 4}^{(j)}
\endsmallmatrix \right)\theta_S
=
\left( \smallmatrix
\big.0 \ &0 &\ 1 \\ 
0 \ &1 &\ 0\\
1 \ &0 &\ 0\big.
\endsmallmatrix \right)
\left( \smallmatrix
\ep_{k| 2}^{(j)} \\
\ep_{k| 3}^{(j)}\\
\ep_{k| 4}^{(j)}
\endsmallmatrix \right)\,,
\label{eq: theta}
\end{align}
and a separate transformation property for degree $k=2$, which reflects our choice of eliminating $f^{(2)}(\omega_2,\tau)$ from the basis of $\mathcal{E}_2(\Gamma(2))$ in (\ref{eq: N2conn})
\begin{align}
    \theta_T^{-1}\begin{pmatrix}
        \tilde{\epsilon}_{2|3}\\\tilde{\epsilon}_{2|4}
    \end{pmatrix}\theta_T = \begin{pmatrix} 0&1 \\ 1& 0 \end{pmatrix}\begin{pmatrix}
        \tilde{\epsilon}_{2|3}\\\tilde{\epsilon}_{2|4}
    \end{pmatrix}\,,\quad\quad\theta_S^{-1}
\begin{pmatrix}
        \tilde{\epsilon}_{2|3}\\\tilde{\epsilon}_{2|4}
    \end{pmatrix}\theta_S = \begin{pmatrix} 1&-1 \\ 0& -1 \end{pmatrix}\begin{pmatrix}
        \tilde{\epsilon}_{2|3}\\\tilde{\epsilon}_{2|4}
    \end{pmatrix}\,.
    \label{eq: thetatild}
\end{align}
The connection transforms covariantly under the generators of $\SLZ$,
\begin{align}
\ANN{\tau{+}1} &=
U_T^{-1}
\theta_T^{-1} \ANN{\tau} \theta_T U_T \,,\label{eq: N2trafo} \\
\ANN{-\tfrac{1}{\tau}} &=
U_S^{-1}
\theta_{S}^{-1} \ANN{\tau} \theta_S
U_S\,,\notag
\end{align}
and later quantities with the same transformation law are said to be \emph{equivariant}. 
The conjugations by $U_T$ and $U_S$ act on $\tilde{\ep}_{2|\mu}$, $\ep_{k|\mu}^{(j)}$ while keeping the degree $k$ and the index $\mu$ for the torsion point inert.
These transformations are known from level 1~\cite{Dorigoni:2024oft} and change the exponents $j=0,1,\ldots,k{-}2$ of the integration kernels $\tau^j f^{(k)}(\omega_\mu,\tau)$ according to 
\begin{align}
U_S^{-1}
\tilde  \ep_{2| \mu}
U_S = \tilde  \ep_{2| \mu}\, , \ \ \ \ \ \ 
&U_S^{-1}
\ep_{k| \mu}^{(j)}
U_S
= (-1)^j (2\pi i)^{k-2-2j} \frac{j!}{(k{-}j{-}2)!} \ep_{k| \mu}^{(k-2-j)}\,,\label{eq: US}\\
U_T^{-1}
\tilde  \ep_{2| \mu}
U_T = \tilde  \ep_{2| \mu}\, , \ \ \ \ \ \ 
&U_T^{-1}\ep_{k|\mu}^{(j)}U_T=\sum_{p=0}^{k-2-j}\frac{(-2\pi i)^p}{p!}\ep^{(j+p)}_{k|\mu}\,.
\notag
\end{align}
We extend the action in~\eqref{eq: N2trafo} from the generators $S$ and $T$ to all of $\SLZ$  via
\begin{equation}
\label{eq:genact}
    \ANN{\gamma\tau} =
U_
\gamma^{-1}
\theta_\gamma^{-1} \ANN{\tau} \theta_\gamma U_\gamma\,,\quad \gamma \in \SLZ\,,
\end{equation}
where in our convention $U_{\gamma_1\gamma_2} = U_{\gamma_2} U_{\gamma_1}$ and $\theta_{\gamma_1\gamma_2} = \theta_{\gamma_2} \theta_{\gamma_1}$.
Even though the matrices $\theta_\gamma$ are defined for all $\gamma\in\SLZ$, they reduce to a representation of $\mathcal{S}_3$ (up to swapping the order in the matrix product).
Note that  conjugations by $\theta_\gamma$ and $U_{\gamma'}$ commute for any $\gamma,\gamma'\in\SLZ$ in view of their action on different indices of $\tilde{\ep}_{2|\mu}$ and $\ep_{k|\mu}^{(j)}$.

\subsection{Iterated Eisenstein integrals and their cocycles for level~2}
\label{sec:IEI2}

Given the connection in \eqref{eq: N2conn}, one can build a generating series of iterated Eisenstein integrals for $\Gamma(2)$ by taking the path-ordered exponential of the connection,
\begin{align}
\INN{\tau} &= {\rm Pexp} \bigg( \int^{i\infty}_\tau \ANN{\tau_1} \bigg) \label{prevpexp}\\
&= 1+\int^{i\infty}_\tau \ANN{\tau_1}
+\int^{i\infty}_\tau \ANN{\tau_1} \int^{\tau_1}_\tau \ANN{\tau_2} + \ldots \,.
\notag
\end{align}
It satisfies the differential equation \begin{equation}
    \dd \tau \, \partial_\tau \INN{\tau} = - \INN{\tau}\ANN{\tau}\,,
\end{equation}
with the regularised boundary value $\lim_{\tau \rightarrow i \infty} \INN{\tau} = 1$ with the tangential-base-point prescription of \cite{brown2017multiple}.
As will be detailed in section \ref{tau level 2}, this generating series exhibits \emph{cocycles} $\mathbb T = \mathbb C_T$ and $\mathbb S = \mathbb C_S$ under the modular $T$ and $S$ transformations, and thus departs from the simple transformation law (\ref{eq: N2trafo}) of the connection:
\begin{align}
\INN{\gamma \tau} &= {\rm Pexp} \bigg( \int^{i\infty}_{\gamma \tau} \ANN{\tau_1} \bigg)
= {\rm Pexp} \bigg( \int^{\gamma^{-1}(i\infty)}_\tau \ANN{\gamma \tau_2} \bigg)
\label{trfpexp} \\
&= {\rm Pexp} \bigg( \int^{\gamma^{-1}(i\infty)}_{i\infty} \ANN{\gamma \tau_2} \bigg)
\,
{\rm Pexp} \bigg( \int^{i\infty}_\tau \ANN{\gamma \tau_2} \bigg) \notag \\
&=  \mathbb C_\gamma(\ep)  U^{-1}_\gamma
\theta^{-1}_\gamma 
 \INN{\tau} 
\theta_\gamma  U_\gamma\,.
\notag
\end{align}
We have changed integration variable to $\tau_1 = \gamma \tau_2$ in the first line, applied the composition of paths formula for path-ordered exponentials in passing to the second line and introduced the following notation for the cocycle associated with $\gamma \in \SLZ$ in the third line:
\begin{align}
\mathbb C_\gamma(\ep) &= 
 U^{-1}_\gamma
\theta^{-1}_\gamma 
{\rm Pexp} \bigg( \int^{\gamma^{-1}(i\infty)}_{i\infty} \ANN{\tau_2} \bigg)
\theta_\gamma  U_\gamma
= {\rm Pexp} \bigg( \int^{i\infty}_{\gamma(i\infty)} \ANN{\tau_1} \bigg)\,.
\label{defcocyc}
\end{align}
Being a 1-cocycle means that $\mathbb{C}_{\gamma_1\gamma_2} = \mathbb{C}_{\gamma_1}(\gamma_1\cdot \mathbb{C}_{\gamma_2})$ for $\gamma_1,\gamma_2\in \SLZ$ and $\gamma_1 \cdot \mathbb{C}_{\gamma_2}=U_{\gamma_1}^{-1} \theta_{\gamma_1}^{-1} \mathbb{C}_{\gamma_2} \theta_{\gamma_1} U_{\gamma_1}$ is the action~\eqref{eq:genact} of $\SLZ$ on $\ep$-valued series.

The problem to be tackled in this work can be summarised as follows: constructing a generating series $\INNeqv{\tau}$ of iterated Eisenstein integrals for $\Gamma(2)$ from (\ref{prevpexp}) that transforms equivariantly under all of $\gamma \in \SLZ$, namely, which shares the transformation law $\ANN{\gamma\tau}=U_\gamma^{-1}\theta^{-1}_\gamma \ANN{\tau} \theta_\gamma U_\gamma$ of the connection in (\ref{eq: N2trafo}),
\begin{align}  \INNeqv{\gamma\tau}=U_\gamma^{-1}\theta^{-1}_\gamma \INNeqv{\tau} \theta_\gamma U_\gamma\,.
    \label{eq: equiv2}
\end{align}
The cocycle factor $\mathbb C_\gamma$ in the last line of (\ref{trfpexp}) prevents the series $\INN{\tau}$ of (\ref{prevpexp}) from transforming equivariantly by itself.
It is worth emphasizing that, in view of normality of $\Gamma(2)$ in $\SLZ$, we require full $\SLZ$ equivariance instead of just equivariance under transformations in $\Gamma(2)$. Still, $\SLZ$ equivariance of \eqref{eq: equiv2} implies $\Gamma(2)$ equivariance, since the conjugation by $\theta_\gamma$ acts trivially for $\gamma \in \Gamma(2)$, as one can check from compositions of (\ref{eq: theta}) and (\ref{eq: thetatild}).

Before we describe the construction of $\INNeqv{\tau}$ for $\Gamma(2)$ subject to (\ref{eq: equiv2}), we review the analogous construction for the full modular group in the next section. This serves as a stepping stone towards our generalisation to the congruence subgroup $\Gamma(2)$.

\subsection{Review of the level-1 framework}
\label{sec:lvl1}

To motivate the expression \eqref{eq: N2conn} of the connection for $\Gamma(2)$ and the defining property (\ref{eq: equiv2}) of the equivariant version of its path-ordered exponential, we review the respective counterparts at level 1.

\subsubsection{Tsunogai's derivation algebra}
\label{sec:221ts}

The starting point is the following connection form \cite{brown2017multiple, Brown:2017qwo2, Dorigoni:2022npe, Dorigoni:2024oft}
\beq
\AN{\tau} =  \dd \tau \sum_{k=4}^{\infty} (k{-}1) \sum_{j=0}^{k-2} \frac{(-1)^j}{j!} (2\pi i \tau)^{j} \frac{ {\rm G}_{k}(\tau) }{(2\pi i)^{k-1}} \ep_k^{(j)}\,,
\label{eq: N1conn}
\eeq
taking values in Tsunogai's derivation algebra~\cite{Tsunogai}, also called the (pro-nilpotent) algebra of geometric derivations $\mathfrak{u}_{\SLZ}$, which is non-freely generated by elements $\ep_k^{(j)}$ with $k\geq 4$ even and $j=0,1,\ldots,k{-}2$.\footnote{The literature is not uniform whether or not to include $\ep_0$ and/or the central $\ep_2$ defined below in the generators of $\mathfrak{u}_{\SLZ}$. We make the choice not to include them.} Moreover, $\mathfrak{u}_{\SLZ}$ is equipped with an $\mathfrak{sl}_2$ structure, where the $\mathfrak{sl}_2$ algebra is generated by the following derivations (which do not belong to $\mathfrak{u}_{\SLZ}$)
\begin{align}
   &\epsilon_0 \quad (\text{raising operator} )\,,\notag\\
   &\epsilon_0^\vee \quad (\text{lowering operator} )\,,\label{eq: sl2}\\
   &h=[\ep_0,\ep_0^\vee] \quad (\text{Cartan operator} )\,.\notag
\end{align}
These $\mathfrak{sl}_2$ generators 
act via the adjoint action on the $\ep_k^{(j)}$
to yield irreducible  $\mathfrak{sl}_2$ representations  
\beq
\ep_k^{(j)} = {\rm ad}_{\ep_0}^j(\ep_k) 
\, , \ \ \ \ \text{with $j=0,1,\ldots,k{-}2$ and nilpotency}\, \ \
\ep_k^{(k-1)}  = {\rm ad}_{\ep_0}^{k-1}(\ep_k) = 0\,.
\label{eq: adj}
\eeq
The generators $\ep_k^{(j)}$ of degree $k\geq 4$ form $(k{-}1)$-dimensional irreducible representations under this $\mathfrak{sl}_2$ action (see, e.g., section 5.1 of \cite{Dorigoni:2024iyt}),
\beq
 {\rm ad}_{\ep_0} \ep_k^{(j)}=\ep_k^{(j+1)}\, , \ \ \ \  {\rm ad}_{\ep_0^\vee} \ep_k^{(j)}= j (k{-}1{-}j) \ep_k^{(j-1)} \, , \ \ \ \ k\geq 4\,,
 \label{eq: sl2rep}
\eeq
leading to Cartan eigenvalues $[h,\ep^{(j)}_{k}]=(2j{+}2{-}k) \ep^{(j)}_{k}$. 

The structure of the derivation algebra $\mathfrak{u}_{\SLZ}$ follows from the geometric origin of the $\ep_k$ as geometric derivations on the free Lie algebra ${\rm Lie}[a,b]$ associated to the (the pro-unipotent completion of the) fundamental group of the once-punctured torus~\cite{Tsunogai,KZB}, see section \ref{sec:flcee}. 
Let ${\rm Der}^0{\rm Lie}[a,b]$ denote the Lie subalgebra of derivations that annihilate the bracket $[a,b]\in {\rm Lie}[a,b]$. Then the Tsunogai derivations $\ep_k$ (with $k \geq 0$ even)  are defined  
as the elements  of ${\rm Der}^0{\rm Lie}[a,b]$ given by (they only belong to $\mathfrak{u}_{\SLZ}$ for $k\geq 4$), 
\begin{align}
    &[\ep_k,b]={\rm ad}_b^k(a)\,,\ \ \ \ \text{and}\ \ \ \  [\ep_k,a]=\sum_{j=0}^{k/2-1}(-1)^j \big[{\rm ad}_b^j(a),{\rm ad}_b^{k-1-j}(a) \big]\, \ \ \ \text{for}\, \ k\geq 0\,,
    \label{eq: derivations}
\end{align}
which are easily checked to obey $[\ep_k,[a,b]]= 0$,
as a consequence of the Leibniz property.
The above 
actions of $\ep_k$ include that of $\ep_0$, but we write out its properties as well as the definition of $\ep_0^\vee$ separately:
\begin{equation}
    [\ep_0,b]=a\,,\ \ \ \ [\ep_0,a]=0 \,,\ \ \ \ [\ep_0^\vee,b]=0\,,\ \ \ \   [\ep_0^\vee,a]=b\,.
\end{equation}
We note that $[\ep_0,[a,b]]=[\ep^\vee_0,[a,b]]=0$, so that $\ep_0,\ep_0^\vee\in {\rm Der}^0{\rm Lie}[a,b]$.
Note that derivations in ${\rm Der}^0{\rm Lie}[a,b]$ 
are entirely determined by their action on $b$ (or on $a$), see for example Theorem 2.1 of \cite{Schneps:2012}. The derivation $\ep_2$ commutes with all other $\ep_k$ for level 1, and can be consistently quotiented out. This is why it does not appear in the level-1 connection~\eqref{eq: N1conn}. As will be reviewed in section \ref{sec:flcee} below, the bracket relations of $\ep_k$ can be derived as flatness conditions on the Calaque-Enriquez-Etingof connection \cite{KZB}, which produces $\AN{\tau}$ in (\ref{eq: N1conn}) under a suitable restriction and gauge transformation.

We define the algebra $\mathfrak{u}_{\SLZ}$ as the algebra generated by the $\ep_k$ for even $k\geq 4$, together with their $\mathfrak{sl}_2$ images~\eqref{eq: sl2rep}. However, it is not freely generated by the $\ep^{(j)}_k$ ($k\geq 4$), but 
features relations beyond~\eqref{eq: adj}. These relations are known as Pollack relations \cite{Pollack}, such as
\begin{align}
    &0=[\ep_4,\ep_{10}] - 3 [\ep_6,\ep_8]\,,\notag \\
    &0=80[\ep_4^{(1)},\ep_{12}]+16[\ep_{12}^{(1)},\ep_4]-250[\ep_6^{(1)},\ep_{10}]-125[\ep_{10}^{(1)},\ep_6]+280[\ep_8^{(1)},\ep_8]\notag\\&\qquad-462[\ep_4,[\ep_4,\ep_8]]-1725[\ep_6,[\ep_6,\ep_4]]\, ,
    \label{eq: pollack}
\end{align}
and their respective images under $\mathfrak{sl}_2$. 
The existence of relations in $\mathfrak{u}_{\SLZ}$ is known to be associated to the existence of holomorphic cusp forms for $\SLZ$ \cite{Pollack}. More precisely, to each holomorphic cusp form for $\SLZ$, one can associate an infinite sequence of Pollack relations,
and the relative coefficients of the terms of modular depth 2 can be traced back to the period polynomial for the same cusp form.

To a derivation $\delta\in {\rm Der}^0{\rm Lie}[a,b]$, we assign a notion of 
$a$-degree $r$ and $b$-degree $s$ if $[\delta,a]$ is a Lie polynomial of homogeneous degree $r {+} 1$ in $a$ and $s$ in $b$, in which case $[\delta,b]$ is necessarily of $a$-degree $r$ and $b$-degree $s {+} 1$ (unless it vanishes) \cite{KZB}. Accordingly, the $(a,b)$ degrees of $\ep_0$, $\ep_0^\vee$ and $\ep_k^{(j)}$ are $(1,-1)$, $(-1,1)$ and $(j{+}1,k{-}j{-}1)$, respectively. We will also refer to the sum of the $a$- and $b$-degree of a derivation as its \emph{total degree}, leading for instance to the total degree $k$ of $\ep_k^{(j)}$, irrespective of $j$.

We separately assign \emph{modular depth} 1 to $\ep_k$ with $k\geq 4$, but modular depth 0 to the $\mathfrak{sl}_2$ generators $\ep_0,\ep_0^\vee$. In this way,
any $\ep_{k}^{(j)}$ with $k\geq 4$, and in fact the entire connection (\ref{eq: N1conn}), has modular depth one. 
Accordingly, words $\ep_{k_1}^{(j_1)} \ldots \ep_{k_r}^{(j_r)}$ with $k_i\geq 4$ in the universal enveloping algebra of $\mathfrak{u}_{\SLZ}$ have modular depth $r$. This sets the stage for extending the notion of modular depth to iterated Eisenstein integrals, where it counts the number of Eisenstein kernels $\tau^j{\rm G}_k(\tau)$, for instance in the expansion of the path-ordered exponential (\ref{prevpexp}). Pollack relations typically have non-uniform modular depths. For example, the first relation in \eqref{eq: pollack} is said to be of modular depth 2, while the second relation mixes terms of modular depths 2 and 3.

A first important link to the level-2 connection $\ANN{\tau}$ in \eqref{eq: N2conn} 
is that the $\SLZ$ connection in \eqref{eq: N1conn} can be obtained from $\ANN{\tau}$ by formally setting%
\begin{equation}
\label{eq:SL2limit}
\tilde \ep_{2\mid\nu}\mapsto 0 \,, \quad {\rm for} \ \nu=3,4
\, , \ \ \ \ \ \
\ep_{k\mid\mu}^{(j)}\mapsto\tfrac{2^{k-2}}{1-2^{k-2}}\ep_k^{(j)} \,,
\quad\text{for $\mu=2,3,4$}\,,
\end{equation}
and using \eqref{funrel.01}. In fact, this observation will play an important role in section \ref{sec: extensions}, where we argue that, given equivariant iterated Eisenstein integrals for $\Gamma(2)$, one can obtain all level-2 congruence subgroups with the property $\SLZ\supset \Gamma\supset \Gamma(2)$ as a simple corollary. In section~\ref{sec: extensions} we will derive the substitution rule~\eqref{eq:SL2limit} by this method and study in particular the case $\Gamma = \Gamma_0(2) = \Gamma_1(2)$.

\subsubsection{\texorpdfstring{Iterated Eisenstein integrals and cocycles of $\SLZ$}{Iterated Eisenstein integrals and cocycles of SL(2,Z)}}
\label{sec:lvl1coc}

To construct iterated Eisenstein integrals from the connection in \eqref{eq: N1conn}, we take its path-ordered exponential (solving $\dd \tau \,\partial_\tau \IN{\tau} = - \IN{\tau} \AN{\tau}$, in analogy with (\ref{prevpexp})) and expand it in Tsunogai derivations
\begin{align}
\IN{\tau} &= {\rm Pexp} \bigg( \int^{i\infty}_\tau \AN{\tau_1} \bigg) \label{eq: I1} \\
&= 1+ \sum_{k=4}^\infty (k{-}1) \sum_{j=0}^{k-2} \frac{(-1)^j}{j!} 
\ee{j }{k}{\tau} \ep_k^{(j)} \notag \\
&\quad
+ \sum_{k_1,k_2=4}^\infty (k_1{-}1)(k_2{-}1) \sum_{j_1=0}^{k_1-2} \sum_{j_2=0}^{k_2-2} \frac{(-1)^{j_1+j_2}}{j_1! j_2!} 
\ee{j_1 &j_2}{k_1 &k_2}{\tau} \ep_{k_1}^{(j_1)} \ep_{k_2}^{(j_2)} + \ldots
\notag\,,
\end{align}
where the coefficient functions are homotopy-invariant iterated Eisenstein integrals due to flatness of the connection \eqref{eq: N1conn}. They are defined recursively via 
\beq
\ee{j_1 &\ldots &j_r}{k_1 &\ldots &k_r}{\tau} = (2\pi i)^{1+j_r - k_r} \int_\tau^{i\infty} \dd \tau_r \, \tau_r^{j_r} \GG_{k_r}(\tau_r) \ee{j_1 &\ldots &j_{r-1}}{k_1 &\ldots &k_{r-1}}{\tau_r} \,,
\label{eq: calE}
\eeq
together with $\ee{\emptyset}{\emptyset}{\tau}=1$.
We employ tangential base point regularisation for the endpoint divergence at $\tau_r=i\infty$ \cite{brown2017multiple}, leading for instance to $ \int^\tau_{i\infty} \dd \tau_1 (\tau_1)^j = \frac{\tau^{j+1}}{j+1}$ while preserving shuffle relations. 
The number $r$ of integrations (\ref{eq: calE}) is also referred to as the \emph{modular depth} of $\ee{j_1 &\ldots &j_r}{k_1 &\ldots &k_r}{\tau}$, by analogy with assigning modular depth $r$ to $\ep_{k_1}^{(j_1)}\ldots \ep_{k_r}^{(j_r)}$.
Note that the Pollack relations among the derivations $\ep_k^{(j)}$ project the iterated Eisenstein integrals within the expansion of (\ref{eq: I1}) to those linear combinations which can be realised as elliptic MZVs \cite{Broedel:2015hia, Lochak:2020} which are special values of elliptic multiple polylogarithms \cite{Enriquez:Emzv}.

Next, we investigate the modular equivariance (or the lack thereof) of the generating series \eqref{eq: I1}. Rather than deriving all the results for level 1, we simply list them and explicitly derive them for level 2 in section \ref{sec: level2} (which generalises the level-1 derivation). 
By following the steps of (\ref{trfpexp}), we find the $T$- and $S$-transformations \cite{Brown:2017qwo2, Dorigoni:2024oft} 
\begin{align}
    \IN{\tau{+}1}&= \TcocN U_T^{-1}\IN{\tau} U_T
    \label{eq: I1T}\,,\\
    \IN{-\tfrac{1}{\tau}}&=\ScocN U_S^{-1}\IN{\tau} U_S\notag\,,
\end{align}
where the SL$_2$ transformations $U_T$ and $U_S$ act on the superscripts of the $\ep_k^{(j)}$ in the expansion of (\ref{eq: I1}) as seen in (\ref{eq: US}). The cocycles $\TcocN$, $\ScocN$ are defined as in 
(\ref{defcocyc}) (departing from the unifying notation $\mathbb C_\gamma$ and setting $\theta_\gamma \rightarrow 1$), namely
\begin{align}
\TcocN & = {\rm Pexp}\bigg(\int_{i\infty +1}^{i\infty} \AN{\tau}\bigg) \, ,
\label{lv1coc} \\
\ScocN & =
{\rm Pexp}\bigg(\int_0^{i\infty} \AN{\tau}\bigg) \, . \notag
\end{align}
They are evaluated via tangential base point regularisation
and exhibit different number-theoretic properties: In the
notation
\beq
\BF_k = \frac{B_k}{k!}\,,
\label{defbf}
\eeq
for the ubiquitous ratios involving Bernoulli numbers, the $T$-cocycle only receives contributions from the
zero modes 
$2\zeta_k=-(2\pi i)^k \BF_k$ of the holomorphic Eisenstein series (\ref{eq: Eis}) and can be compactly written as~\cite{Brown:2017qwo2, Dorigoni:2024oft}
\begin{align}
\TcocN = e^{2\pi i \mathfrak{N}_{\SLZ}} 
U_T
\,, \ \ \ \ \text{with}\, \ \ \mathfrak{N}_{\SLZ}=-\ep_0+\sum_{k=4}^\infty(k{-}1) \BF_k \,\ep_k \, \ \ \ \text{and}\ \ U_T=e^{2\pi i \ep_0}\,,
    \label{eq: Tcoc1simple}
\end{align}
leading to rational multiples of powers of $2\pi i$ in each term of its expansion in Tsunogai derivations. Accordingly, the modular $T$-transformation of (\ref{eq: I1T}) reduces to $\IN{\tau{+}1}$ $=e^{2\pi i \mathfrak{N}_{\SLZ} }\IN{\tau} U_T$.

The tangentially regulated $S$-cocycle $\mathbb{S}_{\SLZ}(\ep)$ in the second lines of \eqref{eq: I1T} and (\ref{lv1coc}) in turn features a wealth of MZVs in its expansion,
\begin{align}
    &\mathbb{S}_{\SLZ}(\ep) = 1+\sum_{k_1=4}^\infty (k_1{-}1)\sum_{j_1=0}^{k_1-2}\frac{(-1)^{j_1}}{j_1!}(2\pi i)^{j_1+1-k_1}\MMV{j_1}{k_1}\ep_{k_1}^{(j_1)}
    \label{eq: MMVgen}\\
&\quad \quad +\sum_{k_1,k_2=4}^\infty (k_1{-}1)(k_2{-}1) \sum_{j_1=0}^{k_1-2} \sum_{j_2=0}^{k_2-2}\frac{ (-1)^{j_1+j_2} }{j_1!j_2!}(2\pi i)^{j_1+j_2+2-k_1-k_2}\MMV{j_1&j_2}{k_1&k_2}\ep_{k_1}^{(j_1)}\ep_{k_2}^{(j_2)}+\dots\,,\notag
\end{align}
where the coefficients $\MMV{\dots}{\dots}$ are referred to as \emph{multiple modular values} (MMVs) \cite{brown2017multiple} 
\begin{align}
    \MMV{j_1&j_2&\dots&j_r}{k_1&k_2&\dots &k_r}=\int_0^{i\infty}\tau_r^{j_r} \GG_{k_r}(\tau_r)\dd\tau_r\int_{\tau_r}^{i\infty}\dots\int_{\tau_3}^{i\infty}\tau_2^{j_2}\GG_{k_2}(\tau_2)\dd\tau_2\int_{\tau_2}^{i\infty}\tau_1^{j_1}\GG_{k_1}(\tau_1)\dd\tau_1\,.
    \label{eq: mmv}
\end{align}
Individual MMVs at modular depth $r \geq 2$ may evaluate to combinations of MZVs \cite{saad2020multiple, Kleinschmidt:2025dtk} and more general numbers such as non-critical $L$-values of holomorphic cusp forms \cite{brown2017multiple, Brown2019, Dorigoni:2021ngn}. In the expansion (\ref{eq: MMVgen}) of the $S$-cocycle, however, the Pollack relations \cite{Pollack} among $\ep_k^{(j)}$ only retain $\mathbb Q[2\pi i]$-linear combinations of MZVs, as was proven to arbitrary degree and modular depth in~\cite{brown2017multiple}.

\subsubsection{\texorpdfstring{Equivariant iterated Eisenstein integrals for $\SLZ$ and zeta generators}{Equivariant iterated Eisenstein integrals for SL(2,Z) and zeta generators}}
\label{sec222}

The cocycles in (\ref{eq: I1T}) spoil the equivariance of the generating series (\ref{eq: I1}) of holomorphic iterated Eisenstein integrals. It was shown in \cite{Brown:2017qwo,Brown:2017qwo2,Dorigoni:2022npe,Dorigoni:2024oft, Schlotterer:2025qjv} that one can build an equivariant completion of that generating series, which shares the modular properties $\AN{\gamma\tau} = U_\gamma^{-1} \AN{\tau} U_\gamma$ of the connection (\ref{eq: N1conn}) under $\gamma \in \SLZ$,
\beq
\INeqv{\gamma \tau} = U_\gamma^{-1} \INeqv{\tau} U_\gamma\, ,
 \label{eq:sl2targ}
\eeq
which is the $\theta_\gamma \rightarrow 1$ counterpart of our target $\SLZ$ action (\ref{eq: equiv2}) for $\Gamma(2)$. An explicit solution to the equivariance property (\ref{eq:sl2targ}) is constructed by combining iterated Eisenstein integrals with their complex conjugates and $\mathbb Q$-linear combinations of single-valued MZVs~\cite{Dorigoni:2024oft}:
\beq
\INeqv{\tau} = \mathbb M^{\rm sv}_{\SLZ}(z)^{-1} \, \overline{\IN{\tau}^T} \, \mathbb M^{\rm sv}_{\SLZ}(\sigma) \, \IN{\tau}\,.
 \label{eq: eqvN1}
\eeq
The transposition operation of the complex conjugate path-ordered exponential acts via 
\beq
(\ldots \ep_{k_1}^{(j_1)} \ldots \ep_{k_2}^{(j_2)} \ldots)^T = (-1)^{j_1+j_2+\ldots}(\ldots \ep_{k_2}^{(j_2)} \ldots \ep_{k_1}^{(j_1)}\ldots)\, ,
\label{Taction}
\eeq
where the alternating signs can be understood from $\ep_{k}^{(j)} = {\rm ad}_{\ep_0}^j\ep_{k}$. The single-valued MZVs enter (\ref{eq: eqvN1}) through the group-like generating series denoted $\mathbb M^{\sv}_{\SLZ}(z)$ and $\mathbb M^{\sv}_{\SLZ}(\sigma)$. With the notation $\rho^{-1}$ for the inverse of an isomorphism $\rho$ into the $f$-alphabet of MZVs~\cite{Brown:2011ik, BrownTate}\footnote{We suppress the distinction between motivic MZVs $\zeta^\mot_{n_1,\ldots , n_r}$ (required for a well-defined $f$-alphabet isomorphism) and their real counterparts $\zeta_{n_1,\ldots , n_r}$ obtained from the period map \cite{Brown:2011ik, BrownTate}, see appendix \ref{app:a.1}. 
The isomorphism $\rho$ is non-canonical (although there exist arguments for a canonical choice, see \cite{Keilthy, Dorigoni:2024iyt}).} 
reviewed in appendix \ref{app:a.1}, the series $\mathbb M^{\sv}_{\SLZ}(\sigma)$ is built from
\beq
\mathbb M_{\SLZ}(\sigma) = 1 + \sum_{r=1}^\infty \sum_{i_1,\ldots,i_r \in 2\mathbb N +1} \rho^{-1} \big( f_{i_1}\ldots f_{i_r} \big) \sigma_{i_1}^{\SLZ}\ldots \sigma_{i_r}^{\SLZ}\,,
 \label{eq: zetagen}
\eeq
where the non-commuting derivations $\sigma_{w}^{\SLZ}$ are known as \emph{zeta generators}~\cite{EnriquezEllAss,tsunogai11,Brown:depth3,Schneps:2015mzv,hain_matsumoto_2020}.
The series composition $\mathbb{M}_{\SLZ}^\sv(\sigma)=\mathbb{M}_{\SLZ}(\sigma)^T\,\mathbb{M}_{\SLZ}(\sigma)$ with transposition $(\ldots \sigma_{i_1}\ldots \sigma_{i_2} \ldots)^T$ $= \ldots \sigma_{i_2}\ldots \sigma_{i_1} \ldots$ then implements the single-valued map (\ref{svfalp}) on the $f$-alphabet representation of~MZVs, with leading orders
\begin{align}
\mathbb M^{\rm sv}_{\SLZ}(\sigma) = 1 +  2\sum_{i_1 \in 2\mathbb N +1} \zeta_{i_1} \sigma_{i_1}^{\SLZ} + 2  \sum_{i_1,i_2 \in 2\mathbb N +1} 
\zeta_{i_1}  \zeta_{i_2} \sigma_{i_1}^{\SLZ}\sigma_{i_2}^{\SLZ} + \ldots\,,
\end{align}
and terms of order $(\sigma^{\SLZ})^3$ in the ellipsis (involving single-valued indecomposable MZVs ${\rm sv}\, \zeta_{n_1,\ldots ,n_r}$ of depth $r>1$). The zeta generators $\sigma_w^{\SLZ}$ at $w=3,5,\ldots$ generate a free algebra $\mathfrak{s}_{\SLZ} = {\rm Lie}[\sigma_3^{\SLZ},\sigma_5^{\SLZ},\dots]$ of derivations
acting on the aforementioned free Lie algebra $\mathrm{Lie}[a,b]$ which furthermore satisfies $[\mathfrak{s}_{\SLZ},\mathfrak{u}_{\SLZ}]\subset\mathfrak{u}_{\SLZ}$, i.e., also acts by derivation on the $\epsilon_k^{(j)}$ \cite{tsunogai11, EnriquezEllAss, Brown:depth3, Schneps:2015mzv, hain_matsumoto_2020}. Zeta generators decompose into two contributions of distinct origins,
\begin{align}
\sigma_w^{\SLZ}=z_w^{\SLZ}+\sigma_w^{\SLZ,\rm g}\,,
    \label{eq: zeta}
\end{align}
that consist of a single arithmetic term $z_w^{\SLZ}$ of total degree $2w$ (also known as the \emph{key degree} of $\sigma_w$) as well as an infinite series of terms of unbounded $b$-degree captured in $\sigma_w^{\SLZ,\rm g}$ that takes values in the algebra of geometric derivations $\mathfrak{u}_{\SLZ}$. Replacing zeta generators $\sigma_w^{\SLZ}$ by their arithmetic parts $z_w^{\SLZ}$ gives the leftmost series $\mathbb M^{\rm sv}_{\SLZ}(z) = \mathbb M^{\rm sv}_{\SLZ}(\sigma) |_{\sigma \rightarrow z}$ in (\ref{eq: eqvN1}). 
In view of $\sigma_w^{\SLZ,\rm g}=\sigma_w^{\SLZ}{-}z_w^{\SLZ} \in \mathfrak{u}_{\SLZ}$, the normalisation property $[\mathfrak{s}_{\SLZ},\mathfrak{u}_{\SLZ}]\subset\mathfrak{u}_{\SLZ}$ implies $[z_w^{\SLZ} , \mathfrak{u}_{\SLZ} ]\subset \mathfrak{u}_{\SLZ}$, 
which ensures that the right side of (\ref{eq: eqvN1}) takes values in the Lie group of $\mathfrak{u}_{\SLZ}$.
We note that the decomposition~\eqref{eq: zeta} into an arithmetic and a geometric part is not uniquely defined by the properties below, but we can fix it as in section 5.4 of~\cite{Dorigoni:2024iyt} by assigning all $\mathfrak{sl}_2$ singlets of $\sigma_w^{\SLZ}$ to $z_w^{\SLZ}$.

Extracting explicit equivariant iterated Eisenstein integrals of modular depth $r$ from the generating series \eqref{eq: eqvN1} requires the following ingredients:
\begin{itemize}
    \item Geometric contributions $\sigma_w^{\SLZ,\rm g}$ to $\sigma_w^{\SLZ}$ up to modular depth $r$.
    \item Commutators $[z_w^{\SLZ},\ep_k]$, which are expressible as Lie polynomials in $\ep_{k_1}^{(j_1)}\dots \ep_{k_r}^{(j_r)}$ of total degree $k_1{+}\dots{+}k_r=2w{+}k$.
\end{itemize}
There is no obstruction to determining the geometric terms of $\sigma_w^{\SLZ}$ and the resulting brackets with $\ep_k$ to arbitrary $b$-degrees from the action on ${\rm Lie}[a,b]$ as detailed in \cite{Dorigoni:2024iyt}. Alternatively, one can determine the action of zeta generators on $\mathfrak{u}_{\SLZ}$ by imposing equivariance of the generating series \eqref{eq: eqvN1} \cite{Dorigoni:2024oft, Dorigoni:2024iyt}:\footnote{This second approach represents some computational shortcuts compared to the first-principles construction in section 5 of~\cite{Dorigoni:2024iyt}. The first-principles construction, however,  
additionally yields the action of $\sigma_w$ on ${\rm Lie}[a,b]$, i.e., going beyond their action on $\mathfrak{u}_{\SLZ}$.}  Acting with the modular $S$ and $T$-transformations, we find two types of necessary conditions for equivariance of $\INeqv{\tau}$~\cite{Brown:2017qwo2, Dorigoni:2024oft}. Firstly
\begin{align}
\big[\ep_0,z_w^{\SLZ}\big]=0=\big[\ep_0^\vee,z_w^{\SLZ}\big]\,,
   \label{eq: conditions2}
\end{align}
equivalent to $U_\gamma^{-1}\mathbb{M}^\sv_{\SLZ}(z)^{-1}U_\gamma=\mathbb{M}^\sv_{\SLZ}(z)^{-1}$ for any $\gamma \in \SLZ$, and secondly
\begin{align}
   e^{-2\pi i \mathfrak{N}_{\SLZ}}\mathbb{M}^\sv_{\SLZ}(\sigma) e^{2\pi i \mathfrak{N}_{\SLZ}}&=\mathbb{M}^\sv_{\SLZ}(\sigma)\,,\notag\\
\overline{\mathbb{S}_{\SLZ}(\ep)^T}\mathbb{M}^\sv_{\SLZ}(\sigma) \mathbb{S}_{\SLZ}(\ep) &= U_S^{-1} \mathbb{M}^\sv_{\SLZ}(\sigma) U_S \, .\label{eq: conditions1}
\end{align}
The former conditions (\ref{eq: conditions2})
amount to $\mathfrak{sl}_2$-invariance of the arithmetic generators $z_w^{\SLZ}$ and are known to be met from \cite{hain_matsumoto_2020}.
The first line of \eqref{eq: conditions1} is most conveniently studied at an infinitesimal level in the following sense: Solely equating terms involving $\mathbb Q$-multiples of $2\pi i \rho^{-1}(f_w)$ for different values of $w$ while discarding higher orders in $2\pi i $ or $r\geq 2$ letters $f_{w_1}\ldots f_{w_r}$ in the $f$-alphabet (with $w_i\geq 3$ odd) results in~\cite{brown2017multiple} 
\begin{equation}
\big[\sigma_w^{\SLZ},\mathfrak{N}_{\SLZ}\big]= 0 \,,\ \ \ \ \text{for all}\,\ w\geq 3\, . \       \label{eq: sigmaN}
\end{equation}
Given that (\ref{eq: Tcoc1simple}) expresses $\mathfrak{N}_{\SLZ} = -\ep_0$ plus terms of modular depth one, (\ref{eq: sigmaN}) translates into the recursive procedure of \cite{Dorigoni:2024iyt} to construct both $\big[z_w^{\SLZ},\ep_k\big]$ and $\sigma^{\SLZ,\rm g}_w$ modular depth by modular depth. However, the new contributions to $\sigma^{\SLZ}_w$ when incrementing the modular depth by one are only determined up to highest-weight vectors of $\mathfrak{sl}_2$, i.e., terms in the kernel of ${\rm ad}_{\ep_0}$; see section 7.3 of \cite{Dorigoni:2024iyt}. 
Hence, it remains to supplement this recursive procedure with a separate method to reinstate these highest-weight vectors in the expansion of $\sigma^{\SLZ,\rm g}_w$. 

\subsubsection{\texorpdfstring{Zeta generators and the $S$-cocycle}{Zeta generators and the S-cocycle}}
\label{sec:zetaScoc}

The missing highest-weight-vector contributions to $\sigma^{\SLZ,\rm g}_w$ that remain undetermined by \eqref{eq: sigmaN} can in principle be extracted from the $S$-equivariance condition in the second line of~\eqref{eq: conditions1}. However, it is more convenient to use the information from $S$-equivariance through the alternative form of the $S$-cocycle
\cite{brown2017multiple,Kleinschmidt:2025dtk} 
\begin{align}
    \mathbb{S}_{\SLZ}(\ep)=\mathbb{M}_{\SLZ}(\sigma)^{-1}\mathbb{X}_{\SLZ}(\ep) U_S^{-1} \mathbb{M}_{\SLZ}(\sigma) U_S\,,
    \label{eq: Scocalt}
\end{align}
which in fact implies \eqref{eq: conditions1}, see appendix E.1 of \cite{Schlotterer:2025qjv}. This factorised form of the $S$-cocycle is conjectured to generalise to level 2,
see (\ref{eq:Scoc2}) below for the explicit formula, section \ref{sec:sequv} for its implications on $S$-equivariance and section \ref{sec: HW vectors} for its input on level-2 zeta generators.

The series $\mathbb{M}_{\SLZ}(\sigma)$ in zeta generators on the right-hand side of (\ref{eq: Scocalt}) was defined in (\ref{eq: zetagen}) and exposes the contributions from all MZVs associated with the odd-weight $f$-alphabet generators $f_{2k+1}$ in the preimage of the single-valued map (\ref{svfalp}). The second factor $\mathbb{X}_{\SLZ}(\ep)$ on the right-hand side gathers the remaining contributions from powers of $\zeta_2=\rho^{-1}(f_2)$. It is a group-like series valued in  $\exp(\mathfrak{u}_{\SLZ})$ with $\mathbb{Q}[\pi^2]$ coefficients,
and the derivation of (\ref{eq: conditions1}) from (\ref{eq: Scocalt}) relies on its property $\mathbb{X}_{\SLZ}(\ep)^T\mathbb{X}_{\SLZ}(\ep)=1$
\cite{Kleinschmidt:2025dtk,Schlotterer:2025qjv}.

Based on the representation \eqref{eq: Scocalt} of the $S$-cocycle, one can infer the highest-weight vectors of $\mathfrak{sl}_2$ that contribute to $\sigma_w^{\SLZ,\rm g}$ in a particularly streamlined way by comparing the left- and right-hand sides of the equation. On the left-hand side, we have access to numerical computations of the MMVs accompanying the $\ep_k^{(j)}$ in the expansion (\ref{eq: MMVgen}). On the right-hand side, the unknown highest-weight vectors are part of the zeta generators that make up the generating series $\mathbb{M}_{\SLZ}(\sigma)$. One can compare the terms linear in $\zeta_w$ on the right-hand side coming from $U_S^{-1}\sigma^{\SLZ,\rm g}_w U_S-\sigma^{\SLZ,\rm g}_w$ to the left-hand side at the level of individual words in $\ep_{k}^{(j)}$ which are independent under Pollack relations. 

The target contributions to $\sigma^{\SLZ,\rm g}_w$ are highest-weight vectors of $\geq 3$-dimensional representations of $\mathfrak{sl}_2$ (as the highest-weight vectors in $\mathfrak{sl}_2$-singlets were assigned to $z^{\SLZ}_w$). As a consequence, their images within $U_S^{-1}\sigma^{\SLZ,\rm g}_w U_S$ have a different Cartan eigenvalue under $h$ in (\ref{eq: sl2})\footnote{More precisely, the Cartan eigenvalue $2j{+}2{-}k$ of $\ep_k^{(j)}$ changes sign under the conjugation (\ref{eq: US}) by $U_S$ which maps $\ep_k^{(j)}$ to a multiple of $\ep_k^{(k-2-j)}$, and similarly for the Lie polynomials entering $\sigma^{\SLZ}_w$.} and can thus be cleanly separated from the preimages within $\sigma^{\SLZ}_w$.
Moreover, individual terms in $\mathbb{X}_{\SLZ}(\ep)$ are of even $a$- and $b$-degree, whereas the terms  of interest in $\mathbb{M}_{\SLZ}(\sigma)$ with a single $\sigma^{\SLZ}_w$ factor are of odd $a$- and $b$-degrees, and we recall the $(a,b)$-degrees $(j{+}1,k{-}j{-}1)$ of $\ep_k^{(j)}$. On these grounds, the contributions of $\pi^{2k}$ to the $S$-cocycle can be computed separately from the highest-weight vector contributions to zeta generators --- they necessarily occur along with different words in $\ep_{k}^{(j)}$.

The highest-weight-vector contributions to zeta generators which were undetermined by the $T$-equivariance condition (\ref{eq: sigmaN}) are therefore completely fixed by the form (\ref{eq: Scocalt}) of the $S$-cocycle. Hence, (\ref{eq: sigmaN}) combined with (\ref{eq: Scocalt}) to reinstate highest-weight vectors determines both the geometric parts $\sigma^{\SLZ,\mathrm{g}}_w$ and the brackets $[z^{\SLZ}_w,\ep_k]$ of the arithmetic parts recursively in modular depth. In section~\ref{sec: level2}, we will generalise this procedure to level 2.

By combining \eqref{eq: sigmaN} and (\ref{eq: Scocalt})  we can now determine the action of $\sigma_w^{\SLZ,\mathrm{g}}$ 
on $\mathfrak{u}_{\SLZ}$ to any desired modular depth and degree.
However, this does not yet fix the action of the arithmetic part $z^{\SLZ}_w$ on ${\rm Lie}[a,b]$ (see section \ref{sec:221ts} on Tsunogai's derivations), whose explicit form for instance enters the construction of single-valued elliptic multiple polylogarithms \cite{Schlotterer:2025qjv}. 
The  generalisation of (\ref{eq: sigmaN}) and (\ref{eq: Scocalt}) to level 2 in section~\ref{sec: level2} will face a similar limitation: It will determine the action of (cyclotomic) zeta generators $\sigma_w$ for level 2 on the algebra $\mathfrak{u}$ of derivations $\tilde{\ep}_{2|\mu} $, $ \ep_{k | \mu}^{(j)}$ appearing in (\ref{eq: N2conn}). However, the method of section~\ref{sec: level2} will not be able to fix the action of $\sigma_w$ on the level-2 analogue of ${\rm Lie}[a,b]$  to be introduced in section \ref{sec:fltcg}. The first-principles construction of level-1 zeta generators with the full information on $[z^{\SLZ}_w,{\rm Lie}[a,b]]$ is reviewed in appendix \ref{sec:furth.1}. In appendix \ref{sec:furth.12}, we shall outline a similar method based on conjectural bracket relations of $\sigma_w$ to determine the action of $z_w$ for level 2 on a more fundamental algebra.

\subsubsection{Explicit results for zeta generators}

The expansion procedure for zeta generators deduced from the $\SLZ$-equivariance property of (\ref{eq:sl2targ}) or (\ref{eq: conditions1}) leads to the following terms of modular depth one and two \cite{Dorigoni:2024iyt}
\begin{align}
\sigma_w^{\SLZ} &= z_w^{\SLZ} - \frac{\ep_{w+1}^{(w-1)} }{(w{-}1)!} -\sum_{d=5}^{w} \BF_{d-1} s_{d-1,w+1}^{d}  - \frac{1}{2}  \BF_{w+1} s_{w+1,w+1}^{w+2}  
   \label{eq: sigma}  \\
&\quad
-\frac{1}{2}\sum_{\substack{d=3 \\ {d \ {\rm odd}}}}^{w-2} \frac{ \BF_{d-1}}{\BF_{w-d+2}} \sum_{k=d+1}^{w-1}  \BF_{k-d+1} \BF_{w-k+1} s_{k,w-k+d}^{d} 
\notag \\
&\quad +
 \sum_{k=w+3}^\infty \BF_k \sum_{j=0}^{w-2} \frac{(-1)^j\left(\begin{smallmatrix} k-2\\ j\end{smallmatrix}\right)^{-1}}{j!(w{-}2{-}j)!} \left[\epsilon_{w+1}^{(w{-}2{-}j)},\epsilon_k^{(j)} \right]+\dots\,,
\notag
\end{align}
with a Lie series involving modular depths $3,4,\ldots,w$ in the ellipsis. Moreover, the terms of lowest modular depth in the brackets of the arithmetic zeta generators $z_w^{\SLZ}$ with $\epsilon_k$ are~\cite{hain_matsumoto_2020},
\begin{align}
    \label{eq: bracket}
\big[z_w^{\SLZ},\ep_k\big]=\frac{\BF_{w+k-1}}{\BF_k}t^{w+1}_{w+1,w+k-1}+\dots\,,
\end{align}
with a Lie polynomial in $\ep_k^{(j)}$ of modular depth $3,4,\ldots,w{+}1$ and total degree $2w{+}k$ in the ellipsis.
Both (\ref{eq: sigma}) and (\ref{eq: bracket}) are expressed in terms of the combinations of brackets:
\begin{align}
    \label{eq: HWst}
    &s_{p,q}^{d} = \frac{(d{-}2)!}{(p{-}2)!(q{-}2)!} \sum_{i=0}^{d-2} (-1)^{i} \left[\epsilon_p^{(p-2-i)}, \epsilon_{q}^{(q-d+i)}\right]\,,\\
    &t_{p,q}^{d} = \frac{(d{-}2)!}{(p{-}2)!(q{-}2)!} \sum_{i=0}^{d-2} (-1)^{i}\frac{(p{-}2{-}i)!(q{-}d{+}i)!}{i!(d{-}2{-}i)!}\left[\epsilon_p^{(i)}, \epsilon_{q}^{(d-2-i)}\right]\notag\,,
\end{align}
which furnish $\mathfrak{sl}_2$ highest- and lowest-weight vectors, respectively, within their standard range $2\leq d\leq \min(p,q)$, i.e., $[\ep_0^\vee,t^d_{p,q}] = 0 = [\ep_0,s^d_{p,q}]  $. The instances $s_{d-1,w+1}^{d} $, $s_{w+1,w+1}^{w+2}$ of (\ref{eq: HWst}) in the first line of (\ref{eq: sigma}), however, are outside this range, and thus they are not highest-weight vectors (brackets involving $\ep_{k}^{(j<0)}$ with
negative superscripts  are understood to vanish).

\subsubsection{Single-valued iterated Eisenstein integrals}
\label{sec:226}

Throughout the above discussions, we have appealed to the canonical choice of the arithmetic zeta generators which defines $z^{\SLZ}_w$ to be the $\mathfrak{sl}_2$-singlet component of the degree-$2w$ contributions to $\sigma^{\SLZ}_w$, see Theorem 5.4.1 of 
\cite{Dorigoni:2024iyt}. At the same time, one may redefine $z^{\SLZ}_w$ by $\mathfrak{sl}_2$-invariant Lie polynomials in $\ep_k^{(j)}$ of total degree $2w$ while preserving the $\SLZ$ property (\ref{eq: conditions2}). The associated ambiguity in defining the generating series of equivariant iterated Eisenstein integrals in \eqref{eq: eqvN1} was identified in \cite{Brown:2017qwo2}. Even though the canonical $z^{\SLZ}_w$ of \cite{Dorigoni:2024iyt} single out a preferred representative of $\INeqv{\tau} $, it is worth highlighting that the following result of replacing $z\mapsto \sigma$ in the left-most generating series of \eqref{eq: eqvN1} is invariant under redefinitions of the arithmetic zeta generators (see also~\cite{Brown:2017qwo2}):
\begin{align}
\mathbb{I}_{\SLZ}^\sv(\ep;\tau)=\mathbb{M}_{\SLZ}^\sv(\sigma)^{-1}\overline{\IN{\tau}^T} \mathbb{M}_{\SLZ}^\sv(\sigma)\IN{\tau}\,.
    \label{eq: svI1}
\end{align}
On the one hand, this generating series produces single-valued iterated Eisenstein integrals that are not $\SLZ$-equivariant, since the left-most series $\mathbb{M}_{\SLZ}^\sv(\sigma)^{-1}$ does not commute with $U_S^{-1}$ and $U_T^{-1}$. On the other hand, its form was exploited in \cite{Kleinschmidt:2025dtk} to propose a possible form of the motivic coaction acting on the path-ordered exponential \eqref{eq: I1}, assuming that its motivic version exists~\cite{Matthes:talk}.\footnote{The modular transformations of de Rham iterated Eisenstein integrals prescribed in section 5 of \cite{Kleinschmidt:2025dtk} should be corrected as explained in \cite{wip:coactionlong}.} The coaction of~\cite{Kleinschmidt:2025dtk} for $\IN{\tau}$ is inspired by the formal similarity of $\mathbb{I}_{\SLZ}^\sv(\ep;\tau)$ in (\ref{eq: svI1}) with the series in single-valued multiple polylogarithms on the Riemann sphere in (\ref{revgg.02}) below.
This idea will lead us to a similar proposal for the coaction of (motivic) iterated Eisenstein integrals for $\Gamma(2)$ (see section \ref{sec:svlv2} and \cite{wip:coactionlong}).

A change of the $f$-alphabet isomorphism $\rho$ alters both the MZVs $\rho^{-1}(f_a f_b\ldots f_c)$ in \eqref{eq: zetagen}
and the zeta generators $\sigma_w^{\SLZ}$, which are defined through the coefficient of $f_w$
in the $\rho$-image of the Drinfeld associator \cite{Dorigoni:2024iyt, Kleinschmidt:2025dtk}
based on~\cite{EnriquezEllAss, Brown:depth3, Schneps:2015mzv, hain_matsumoto_2020} (see appendix \ref{sec:furth.1}).
In the adjoint action of $\mathbb M^{\sv}_{\SLZ}(\sigma)$ these two changes cancel, so $\mathbb{I}_{\SLZ}^\sv(\ep;\tau)$ in \eqref{eq: svI1} does not
depend on~$\rho$. However, these changes do not cancel completely from the ratio  $\mathbb M^{\sv}_{\SLZ}(z)^{-1} \mathbb M^{\sv}_{\SLZ}(\sigma)$ between $\mathbb{I}_{\SLZ}^\sv(\ep;\tau)$ and the expression (\ref{eq: eqvN1}) for $\INeqv{\tau}$. A change in $\rho$ modifies $\sigma_w^{\SLZ}$ by nested commutators of lower-weight zeta generators. The resulting modification of $z_w^{\SLZ}$, again the $\mathfrak{sl}_2$ singlet of the modified $\sigma_w^{\SLZ}$, receives contributions from both the arithmetic terms and the geometric terms of the shifts by lower-weight zeta generators. As a result, the change in $z_w^{\SLZ}$ goes beyond the nested brackets of lower-weight arithmetic generators which could be compensated by the modified $\rho^{-1}(f_a f_b\ldots f_c)$ in $\mathbb M^{\sv}_{\SLZ}(z)^{-1}$.
Hence, our representative of $\INeqv{\tau}$ depends on $\rho$,
first through terms of degree 22.

\subsection{Equivariance of the level-2 generating series}
\label{tau level 2}

In this section, we outline our proposal for a generating series of equivariant iterated Eisenstein integrals for level 2.

\subsubsection{\texorpdfstring{Expanding the series in iterated Eisenstein integrals for $\Gamma(2)$}{Expanding the series in iterated Eisenstein integrals for Gamma(2)}}
\label{tau level 2.1}

We start by generalising the expansion \eqref{eq: I1} of the generating series of iterated Eisenstein integrals from $\SLZ$ to $\Gamma(2)$. The expansion of the required path-ordered exponential (\ref{prevpexp}) takes a particularly convenient form upon rewriting the degree-two part of the level-2 connection \eqref{eq: N2conn}~as
\beq
\sum_{\nu=3}^4\tilde \ep_{2|\nu} f^{(2)}(\omega_\nu,\tau)
= \sum_{\mu=2}^4  \ep_{2|\mu} f^{(2)}(\omega_\mu,\tau)\,,
\label{rewrd2}
\eeq
which can then be lined up with a $k=2$ extension of the infinite sum in \eqref{eq: N2conn}:
\beq
\ANN{\tau} = -  \dd \tau \sum_{k=2}^{\infty} (k{-}1) \sum_{j=0}^{k-2} \frac{(-1)^j}{j!} (2\pi i \tau)^{j} \sum_{\mu = 2}^4 \frac{ f^{(k)}(\omega_\mu,\tau) }{(2\pi i)^{k-1}} \ep_{k | \mu}^{(j)}\,.
\label{newlv2con}
\eeq
This connection takes values in the level-2 analogue of Tsunogai's derivation algebra (see section~\ref{sec:221ts}), which we shall denote by $\mathfrak{u}$. It is non-freely generated by elements $\tilde{\ep}_{2|\nu=3,4}$ and $\ep_{k|\mu=2,3,4}^{(j)}$ with $k\geq 4$ even and $j=0,1,\ldots,k{-}2$ and also carries an $\mathfrak{sl}_2$ structure that we shall explore more in section~\ref{sec:335}.
We assign \emph{(total) degree} $k$  to $\ep^{(j)}_{k|\mu}$, which will be seen in section \ref{sec:fltcg} to originate as a sum of $a$- and $b$-degrees just like in the level-1 case. 
The $a$- and $b$-degrees of $\ep_{k|\mu}^{(j)}$ will turn out to be $(j{+}1,k{-}j{-}1)$, independently of $\mu$, just as for $\SLZ$.

In view of the relation $\sum_{\mu=2}^4 f^{(2)}(\omega_\mu,\tau) = 0$, one is free to simultaneously shift the degree-two derivations $\ep_{2|\mu} = \ep^{(0)}_{2|\mu}$ by $ \ep_{2|\mu}\rightarrow \ep_{2|\mu}{+} \lambda$ for an arbitrary, $\mu$-independent generator $\lambda$. The latter drops out when recovering the $\tilde  \ep_{2|\nu}$ with the restricted range $\nu \in \{3,4\}$ via (see section \ref{subapp.e1} for more details)
\beq
\tilde \ep_{2|\nu} = \ep_{2|\nu} - \ep_{2|2} \,, \ \ \ \ \text{with}\, \ \ \nu=3,4 \, .
\label{backtotild}
\eeq
In many situations throughout this work, it will be convenient to fix the ambiguity of adding $\lambda$ to all $\ep_{2|\mu}$ ($\mu=2,3,4$) by imposing  $\ep_{2|2}+\ep_{2|3}+\ep_{2|4}=0$. 

With the unified formulation (\ref{newlv2con}) of the level-2 connection in place, the expansion of its path-ordered exponential (\ref{prevpexp}) takes the convenient form
\begin{align}
\INN{\tau} &=  1+ \sum_{k=2}^\infty (k{-}1) \sum_{j=0}^{k-2} \frac{(-1)^j}{j!} \sum_{\mu=2}^4
\eez{j }{k}{ \omega_\mu}{\tau} \ep_{k|\mu}^{(j)} \label{eq: I2} \\
&\quad
+ \sum_{k_1,k_2=2}^\infty (k_1{-}1)(k_2{-}1) \sum_{j_1=0}^{k_1-2} \sum_{j_2=0}^{k_2-2} \frac{(-1)^{j_1+j_2}}{j_1! j_2!} 
\sum_{\mu_1,\mu_2=2}^4
\eez{j_1 &j_2}{k_1 &k_2}{\omega_{\mu_1} &\omega_{\mu_2}}{\tau} \ep_{k_1| \mu_1}^{(j_1)} \ep_{k_2 | \mu_2}^{(j_2)}+ \ldots \notag\,.
\end{align}
Similar to the iterated Eisenstein integrals $\ee{j_1 &\ldots &j_r}{k_1 &\ldots &k_r}{\tau}$ for $\SLZ$ in the expansion of (\ref{eq: I1}), we denote iterated integrals of  Eisenstein series $f^{(k)}(\omega_\mu,\tau)$ for $\Gamma(2)$ by (see section 3.2.1 of~\cite{Hidding:2022vjf})
\beq
\eez{j_1 &\ldots &j_r}{k_1 &\ldots &k_r}{z_1 &\ldots &z_r}{\tau} = - (2\pi i)^{1+j_r - k_r} \int_\tau^{i\infty} \dd \tau_r \, \tau_r^{j_r} f^{(k_r)}(u_r \tau_r {+} v_r,\tau_r) \eez{j_1 &\ldots &j_{r-1}}{k_1 &\ldots &k_{r-1}}{z_1 &\ldots &z_{r-1}}{\tau_r} \,,
\label{eq: eez}
\eeq
with $k_i\geq 2$ and $0\leq j_i\leq k_i{-}2$ as well as $\eez{\emptyset}{\emptyset}{\emptyset}{\tau} = 1$. The integrals of (\ref{eq: eez}) are performed at fixed co-moving coordinates $u_i, v_i \in \mathbb R$ of $z_i = u_i \tau{+} v_i$, such that the kernels $f^{(k_i)}(u_i\tau{+}v_i,\tau)$ are holomorphic in $\tau$ and the iterated integrals are homotopy-invariant.
We again employ tangential base-point regularisation \cite{brown2017multiple} for the endpoint divergence at $\tau_r=i\infty$ from the zero modes of (\ref{eq: FKroneis}), and we assign \emph{modular depth} $r$ to both $\eez{j_1 &\ldots &j_r}{k_1 &\ldots &k_r}{z_1 &\ldots &z_r}{\tau}$ (according to the number of integrations (\ref{eq: eez})) and the accompanying
word $\ep_{k_1|\mu_1}^{(j_1)}\ldots \ep_{k_r|\mu_r}^{(j_r)}$ in (\ref{eq: I2}).
Once we assemble the degree-two derivations $\ep_{2|\mu=2,3,4}$ into their well-defined combinations (\ref{backtotild}), the series (\ref{eq: I2}) takes values in the (pro-unipotent) group with Lie algebra $\mathfrak{u}$.

As will be detailed in section \ref{sec: pollack}, there is a wealth of relations among (nested) brackets of $\tilde{\ep}_{2|\nu}$ and $\ep^{(j)}_{k|\mu}$ akin to the Pollack relations at level 1. Within the expansion (\ref{eq: I2}) of the generating series $\mathbb I(\ep;\tau)$, the level-2 Pollack relations project the iterated Eisenstein integrals (\ref{eq: eez}) to those linear combinations which can be realised as the twisted ($\mathfrak{A}$-cycle) elliptic MZVs of \cite{Broedel:2017jdo}. This will be proven in appendix \ref{app:eMZV}, where an ellipsitomic associator \cite{Calaque:2020ell} generating  twisted elliptic MZVs for level 2 will be expressed in terms of the series $\mathbb I(\ep;\tau)$ conjugating a known, $\tau$-independent initial value. 

\subsubsection{The main proposal for the equivariant series for level 2}\label{tau level 2.2}

Similarly to the level-1 case in section \ref{sec:lvl1}, the main goal of this section is to combine $\INN{\tau}$ into an $\SLZ$-equivariant generating series $\INNeqv{\tau}$, so that it transforms as in \eqref{eq: equiv2} and the cocycles $\mathbb C_\gamma(\ep)$ of $ \INN{\gamma\tau}$ in (\ref{trfpexp}) cancel. In order to ensure that $\INNeqv{\tau}$ obeys the same holomorphic differential equation as $\INN{\tau}$ and to keep as close as possible to the structure at level~1 of~\eqref{eq: eqvN1}, we propose the following main result of this work:
\begin{align}
\INNeqv{\tau} = \mathbb M^{\rm sv}(z)^{-1} \, \overline{\INN{\tau}^T} \, \mathbb M^{\rm sv}(\sigma) \, \INN{\tau}\,.
 \label{eq: eqv2proposal}
\end{align}
Besides the generating series $\INN{\tau}$ of iterated Eisenstein integrals for $\Gamma(2)$ and its complex conjugate transpose,\footnote{The alternating signs of (\ref{Taction}) generalise to $(\ldots \ep_{k_1|\mu_1}^{(j_1)} \ldots \ep_{k_2|\mu_2}^{(j_2)} \ldots)^T {=} (-1)^{j_1+j_2+\ldots}(\ldots \ep_{k_2|\mu_2}^{(j_2)} \ldots \ep_{k_1|\mu_1}^{(j_1)}\ldots)$.} we again have two additional generating series adapted to level 2, denoted $\mathbb{M}^\sv(\sigma)$ and $\mathbb{M}^\sv(z)$. They are group-like series in single-valued cyclotomic MZVs for level 2, i.e., special values of single-valued multiple polylogarithms built from $\dd\!\log(z)$ and $\dd\!\log(z{\pm}1)$, see appendix \ref{app:a.2} (in fact, these are harmonic polylogarithms~\cite{Remiddi:1999ew}). Cyclotomic MZVs for level 2 also admit an isomorphism $\rho$ (from their motivic versions) into an $f$-alphabet which exposes all their relations over $\mathbb Q$ \cite{Deligne2010,brown2017notes,GLANOIS2016334,glanois2016unramifiedeulersumshoffman,Jin2018MotivicMZ}.
Besides the non-commuting $f_{2k+1}=\rho(\zeta_{2k+1})$ and commuting $f_2 = \rho(\zeta_2)$ from the level-1 case, the $f$-alphabet for level 2 features a single additional non-commuting letter $f_1 = \rho(\log(2))$, see appendix \ref{app:a.3}. In direct analogy with the level-1 series (\ref{eq: zetagen}), the first and third factors on the right side of (\ref{eq: eqv2proposal}) are built from zeta generators (augmented by $\sigma_1$ associated with~$\rho^{-1}f_1 = \log(2)$),
\begin{align}
\mathbb M(\sigma) = 1 + \sum_{r=1}^\infty \sum_{i_1,\ldots,i_r \in 2\mathbb N -1} \rho^{-1} \big( f_{i_1}\ldots f_{i_r} \big) \sigma_{i_1}\ldots \sigma_{i_r}\,,
 \label{eq: zetagen2}
\end{align}
where the $f$-alphabet representation (\ref{svfalp}) of the single-valued map applies in identical form to level 2 and gives rise to leading orders
\begin{align}
\mathbb M^\sv(\sigma) &= 1 + 2 \log(2) \sigma_1  +  2\sum_{i_1 \in 2\mathbb N +1} \zeta_{i_1} \sigma_{i_1}  +2 \log(2)^2 \sigma_1 \sigma_1 \label{lv2msv} \\
&\quad 
+ 2\log(2) \sum_{i_1 \in 2\mathbb N +1} \zeta_{i_1} (\sigma_{i_1} \sigma_1 + \sigma_{1} \sigma_{i_1} )
+ 2  \sum_{i_1,i_2 \in 2\mathbb N +1} 
\zeta_{i_1}  \zeta_{i_2} \sigma_{i_1} \sigma_{i_2} 
+ \ldots \, .\notag
\end{align}
The terms in the ellipsis are of order $\sigma^3$ and involve single-valued cyclotomic MZVs of depth $>1$ starting from $\rho^{-1}({\rm sv}(f_1 f_1 f_3))$, which has weight five, see (\ref{falph.sv}). The algebra $\mathfrak{s} = {\rm Lie[\sigma_1,\sigma_3,\dots]}$ is free~\cite{Deligne2010,GLANOIS2016334} and normalises $\mathfrak{u}$, so that $[\mathfrak{s},\mathfrak{u}]\subset\mathfrak{u}$. The factor of $\mathbb M^\sv(\sigma)$ in (\ref{eq: eqv2proposal}) plays a crucial role to compensate for the cocycle factors $\mathbb C_\gamma(\ep)$ in the modular transformations (\ref{trfpexp}) of the series $\overline{\INN{\tau}^T}$ and $\INN{\tau}$ that obstruct the equivariant transformation, as we shall explain in more detail in section~\ref{sec: zetaconstraints}.

Similar to the decomposition (\ref{eq: zeta}) of zeta generators $\sigma_w^{\SLZ}$ at level 1, their counterparts $\sigma_w$ at level 2 are conjectured to decompose as
\begin{align}
    \sigma_w=z_w+\sigma_w^{\rm g}\,,\ \ \ \ \text{for all}\,\ w\geq 1 \ \, \text{odd}\,,
    \label{eq: sigma2}
\end{align}
with a single \emph{arithmetic} term $z_w$ of total degree $2w$ (which will again be referred to as the \emph{key degree} of $\sigma_w$). The \emph{geometric} part $\sigma_w^{\rm g}$ comprises an infinite Lie series in the generators  $\tilde{\ep}_{2|\mu}$ and $\ep_{k|\mu}^{(j)}$ of $\mathfrak{u}$.
Replacing the zeta generators $\sigma_w$ by their arithmetic parts $z_w$ gives the leftmost series $\mathbb M^{\rm sv}(z) = \mathbb M^{\rm sv}(\sigma) |_{\sigma \rightarrow z}$ in (\ref{eq: eqv2proposal}). Given that $\sigma_w^{\rm g} = \sigma_w{-}z_w$ lies in $\mathfrak{u}$ by definition, the above normalisation property $[\mathfrak{s},\mathfrak{u}]\subset \mathfrak{u}$ implies that $[z_w,\mathfrak{u}]\subset\mathfrak{u}$ for all odd $w\geq 1$. On these grounds, one easily verifies that the right-hand side of (\ref{eq: eqv2proposal}) takes values in $\exp(\mathfrak{u})$, i.e., furnishes a group-like series with an expansion solely in words in $\epsilon_{k|\mu}^{(j)}$.

As a necessary condition for the $\SLZ$ equivariance of $\INNeqv{\tau}$, the $\mathfrak{sl}_2$ invariance property (\ref{eq: conditions2}) of the arithmetic zeta generators $z_w^{\SLZ}$ at level 1 needs to carry over to level~2
\beq
[\ep_0 , z_w] = 0 = [\ep^\vee_0 , z_w]\, ,
\label{zwconj.1}
\eeq
(which is equivalent to $U^{-1}_\gamma z_w U_\gamma = z_w$ for any $\gamma \in \SLZ$) and to furthermore extend to the operations $\theta_\gamma$  in (\ref{eq: theta}) and (\ref{eq: thetatild}) that implement permutations of the torsion points,
\beq
\theta_\gamma^{-1} z_w  \theta_\gamma = z_w \,, \forall \ \gamma \in \SLZ \, ,
\label{zwconj.2}
\eeq
where only the quotient $\mathcal{S}_3$ acts non-trivially. In other words, the arithmetic zeta generators $z_w$ for level 2 are invariant under both of $\mathfrak{sl}_2$ and $\mathcal{S}_3$. The link between the $\mathfrak{sl}_2$ generators $\ep_0,\ep^\vee_0$ and the transformations (\ref{eq: US}) of $\ep_{k|\mu}^{(j)}$ under $U_T,U_S$ is discussed in
section \ref{sec:335}, and
a detailed derivation of (\ref{zwconj.1}) and (\ref{zwconj.2}) will be given in
section \ref{sec: zetaconstraints}. 

The decomposition~\eqref{eq: sigma2} of $\sigma_w$ is not unambiguously defined, since at key degree $2w$ there can be $\mathfrak{sl}_2$- and $\mathcal{S}_3$-invariant geometric pieces in $\mathfrak{u}$. We uniquely fix this ambiguity by the following convention: For a given $\sigma_w$,
the arithmetic parts $z_w$ in the decomposition~\eqref{eq: sigma2} are defined by the properties that $\sigma_w^{\rm g}\in \mathfrak{u}$ and that at key degree $2w$ all simultaneously $\mathcal{S}_3$- and $\mathfrak{sl}_2$-invariant pieces of this $\sigma_w$ belong to $z_w$, i.e.
\beq
z_w = \sigma_w \, \Big|^{{\rm degree} \ 2w}_{\mathfrak{sl}_2 \times {\cal S}_3 \ {\rm invariant}} \, .
\label{def:split}
\eeq

\subsubsection{Supporting evidence and open questions}\label{tau level 2.3}

Our proposal \eqref{eq: eqv2proposal} for the equivariant generating series $\INNeqv{\tau}$ of iterated Eisenstein integrals for $\Gamma(2)$ remains conjectural in this work. We shall here comment on both the supporting evidence and the open questions.

The equivariant transformation property \eqref{eq: equiv2} of the expression \eqref{eq: eqv2proposal} for $\INNeqv{\tau}$ has been exhaustively tested up to and including degree 10. These checks concern the iterated integrals of modular depth $\leq 5$ accompanying the words in the derivations $\tilde \epsilon_{2|\mu}$ and $\epsilon_{k| \mu}^{(j)}$ of total degrees 2 and $k$ 
which are independent under their level-2 Pollack relations in section~\ref{sec: pollack}.\footnote{When employing the unified form (\ref{eq: I2}) of the expansion of $\INN{\tau}$ in iterated integrals, it is convenient to reinstate the combinations $\tilde \ep_{2|\nu=3,4}$ of $\ep_{2|\mu =2,3,4}$ in (\ref{backtotild}) by setting $\ep_{2|2}= - \ep_{2|3} - \ep_{2|4}$ and then solving the resulting equations $\tilde \ep_{2|3} =2\ep_{2|3} + \ep_{2|4}$ and $3\leftrightarrow 4$.}

Moreover, the desired equivariance \eqref{eq: equiv2} can be established to arbitrary degrees and modular depths by assuming the properties of zeta generators and cocycles in section~\ref{sec: zetaconstraints}, which we have validated for all cases at degrees $\leq 10$. These assumed properties translate into a recursive procedure to determine the explicit form of the level-2 zeta generators $\sigma_w$ as well as their commutators with $\ep_{k|\mu}$, which will be explained in section \ref{sec: level2} and resembles the level-1 procedure outlined in section \ref{sec:lvl1}. This construction of level-2 zeta generators relies on the following central open conjecture, whose proof is under investigation:

\medskip
{\bf Conjecture:}
\textit{The $S$-cocycle $\mathbb S(\ep) = \mathbb C_S(\ep)$ of the level-2 connection (\ref{eq: N2conn}) defined by (\ref{defcocyc}) can be brought into the factorised form}
\begin{align}
    \mathbb S(\epsilon) = \mathbb{M}(\sigma)^{-1}\,\mathbb X(\epsilon)\,U_S^{-1}\,\theta_S^{-1}\mathbb M(\sigma)\theta_S\,U_S\,,
    \label{eq:Scoc2}
\end{align}
\textit{where $\mathbb M(\sigma)$ as defined in (\ref{eq: zetagen2}) carries all the contributions from letters $f_1,f_3,\ldots$ of odd weight. The action of $\theta_S$ and $U_S$ is given by (\ref{eq: theta}), (\ref{eq: thetatild}) and (\ref{eq: US}), respectively. Assigning $(a,b)$-degrees $(j{+}1,k{-}j{-}1)$ to $\ep_{k|\mu}^{(j)} \in \mathfrak{u}$, we make the following additional assumptions:
\begin{itemize}
\item[(i)] the second factor $\mathbb X(\epsilon)$ on the right side is a group-like series in $\exp(\mathfrak{u})$ involving only even $a$- and $b$-degrees, 
\item[(ii)] for a given term in $\mathbb{X}(\epsilon)$, the coefficient of level-2 derivations with combined $a$-degree $2w$ is a rational multiple of $\pi^{2w}$, and
\item[(iii)] the level-2 zeta generators $\sigma_w$ have uniform $a$-degree $w$ for any odd $w\geq 1$.
\end{itemize}}

\medskip
The level-1 analogue of this conjecture is well-established: First, the analogue of (\ref{eq:Scoc2}), which can be found in (\ref{eq: Scocalt}), was discussed in \cite{Kleinschmidt:2025dtk} and can be proven based on the motivic coaction of the $S$-cocycle for $\SLZ$ \cite{brown2017multiple, Schlotterer:2025qjv}. The references also confirm properties (i) and (ii) of the level-1 counterpart $\mathbb X_{\SLZ}(\epsilon)$ of $\mathbb X(\epsilon)$. Zeta generators $\sigma_w^{\SLZ}$ for level 1 are well-known to have uniform $a$-degree $w$ as one can for instance see from their defining bracket relations reviewed in (\ref{lv1r.02}).

The analogous proofs for level 2 will require conceptual groundwork to relate the construction of $\sigma_w$ in this work to the genus-1 generators of the motivic Galois group for level 2 in \cite{Hopper:2022ell, Hirose:2023}. The study of this link and the compatibility of the generators of the references with the decomposition (\ref{eq: sigma2}) as well as the properties (\ref{zwconj.1}) and (\ref{zwconj.2}) of the arithmetic parts $z_w$ is relegated to future work. A more concrete starting point into this direction is discussed in appendix \ref{sec:fp}.

\subsubsection{Single-valued iterated Eisenstein integrals at level 2}
\label{sec:svlv2}

We close this discussion of iterated Eisenstein integrals for $\Gamma(2)$ by looking at their single-valued rather than equivariant version, mimicking the single-valued iterated Eisenstein integrals for level 1 in section~\ref{sec:226}.

The definition of the equivariant integrals~\eqref{eq: eqv2proposal} rested on the decomposition (\ref{eq: sigma2}) of the zeta generator $\sigma_w$ that singles out an arithmetic part $z_w$ and that we have fixed in (\ref{def:split}). 
If one instead uses the full $\sigma_w$ everywhere, one obtains
\begin{align}
\INNsv{\tau}=\mathbb{M}^\sv(\sigma)^{-1}\overline{\INN{\tau}^T}\mathbb{M}^\sv(\sigma)\INN{\tau}\,,
    \label{eq: svI2}
\end{align}
as a proposed generating series of  single-valued iterated Eisenstein integrals at level 2, the direct generalisation of the level-1 formula~\eqref{eq: svI1}. 

Let us briefly compare the properties of $\INNeqv{\tau}$ and $\INNsv{\tau}$: First,
by the same arguments as in the level-1 case in section \ref{sec:226}, the generating series $\INNeqv{\tau}$ depends on the choice of the $f$-alphabet isomorphism $\rho$ starting from degree 10 whereas $\INNsv{\tau}$ is independent of $\rho$. Second, in contrast to $\lim_{\tau \rightarrow i \infty} \INNeqv{\tau} = \mathbb{M}^\sv(z)^{-1} \mathbb{M}^\sv(\sigma)$, the single-valued series $\INNsv{\tau}$ exhibits a trivial regularised boundary condition $\lim_{\tau \rightarrow i \infty}\INNsv{\tau} = 1$. Moreover, the single-valued series~\eqref{eq: svI2} makes the following links more transparent than its equivariant counterpart $\INNeqv{\tau}$ in (\ref{eq: eqv2proposal}):
\begin{itemize}
\item Once we project the Eisenstein series $f^{(k)}(\omega_\mu,\tau)$ of $\Gamma(2)$ in the integrand of
$\INN{\tau}$ to modular weight 2 (or equivalently the derivations $\epsilon_{k|\mu}$ to degree $k=2$), the single-valued series (\ref{eq: svI2}) will be shown in section \ref{sec:weight2} to reproduce the generating series of single-valued multiple polylogarithms on the genus-zero modular curve $\Gamma(2) \backslash \mathbb{H}$.
\item The right-hand side of (\ref{eq: svI2}) with all degrees $k\geq 2$ retained has the same structure as the generating series (\ref{mplapp.11}) of single-valued genus-0 multiple polylogarithms in one variable with singular points $\in \{ 0,1,-1\}$; in both generating series, the factor involving antiholomorphic iterated integrals is conjugated by zeta generators $\sigma_w$ with $w\geq 1$ odd via~$\mathbb{M}^\sv(\sigma)$.
\end{itemize}
Similar to the level-1 case \cite{Brown:2017qwo2, Dorigoni:2024oft}, a downside of the single-valued series (\ref{eq: svI2}) is its lack of equivariance (\ref{eq: eqvprop}) in the sense that 
\beq
\INNsv{\gamma\tau} = \mathbb{M}^\sv(\sigma)^{-1}U_\gamma^{-1}\theta^{-1}_\gamma  \mathbb{M}^\sv(\sigma) \INNsv{\tau} \theta_\gamma U_\gamma
\label{noneqv}
\eeq
One can combine the virtues of equivariant transformations and independence on the splitting of $\sigma_w = z_w{+}\sigma_w^{\rm g}$ by passing to the series
\beq
\mathbb{M}^\sv(\sigma)\INNsv{\tau} = \mathbb{M}^\sv(z)\INNeqv{\tau} = \overline{\INN{\tau}^T}\mathbb{M}^\sv(\sigma)\INN{\tau}
\label{bothvirt}
\eeq
with $\mathbb{M}^\sv(\sigma)$ as its regularised $\tau \rightarrow i\infty$ limit. In appendix B of \cite{wip:PRL}, we
exhibit a matrix representation of (\ref{bothvirt}) when expanding Gauss' hypergeometric function $_2F_1$ around half-odd integer values of its parameters.

\section{Derivations for level 2 from flat connections}
\label{sec: Connection}

The iterated Eisenstein integrals investigated in this work are constructed from the \mbox{level-2} connection $\ANN{\tau}$ introduced in (\ref{eq: N2conn}). The connection was built from modular forms for $\Gamma(2)$ and derivations $\epsilon_{k|\mu}^{(j)} \in \mathfrak{u}$ of even degree $k\geq 2$, with $j=0,1,\ldots,k{-}2$ and $\mu=2,3,4$, with a redundancy at $k=2$ discussed around (\ref{newlv2con}). This section will be dedicated to deriving the action of the derivations $\epsilon^{(j)}_{k|\mu}$ on a more fundamental Lie algebra that generalises the free ${\rm Lie}[a,b]$ of section \ref{sec:221ts} to the case of $\Gamma(2)$. In particular, we will present a method to derive relations in $\mathfrak{u}$ which generalise Pollack relations such as (\ref{eq: pollack}) from level 1 to level~2. The starting point for the discussions of this section is the level-2 instance of the flat ellipsitomic connection of Calaque-Gonzalez (CG) \cite{Calaque:2019ell,Calaque:2020ell}.

\subsection{Kronecker-Eisenstein series for level 2}
\label{defgtg}

Let us review the objects needed to describe the $z$-dependence of the CG connection.
In (\ref{eq: kronEis}) we introduced the doubly-periodic integration kernels $f^{(k)}$. 
Their generating series is the non-holomorphic Kronecker-Eisenstein series~\cite{BrownLev}
\begin{equation}
\Omega(z,\eta,\tau)=\sum_{m,n\in \mathbb Z}\frac{e^{2\pi i(m v-n u)}}{m\tau+n+\eta}=\sum_{k=0}^\infty \eta^{k-1}f^{(k)}(z,\tau)   \, ,
\end{equation}
with $f^{(0)}(z,\tau)=1$. 
The lattice-sum representation involves the co-moving coordinates $u,v\in~\mathbb R$ for $z=u\tau{+}v$ and exposes that the
Kronecker-Eisenstein series $\Omega(z,\eta,\tau)$ is doubly-periodic in its first argument $z\in \mathbb C$, i.e., $\Omega(z{+}\tau,\eta,\tau)=\Omega(z{+}1,\eta,\tau)=\Omega(z,\eta,\tau)$.

Let us furthermore introduce the meromorphic variant $F(z,\eta,\tau)$ in which the antiholomorphic $\bar z$-dependence of $\Omega(z,\eta,\tau)$ is removed at the cost of breaking the double periodicity,\footnote{This is the original Kronecker-Eisenstein series~\cite{Weil:1976}.}
\begin{equation}
F(z,\eta,\tau)=\exp\big({-}2\pi i\eta \tfrac{\Im z}{\Im \tau} \big)\Omega(z,\eta,\tau) = \frac{\vartheta'_1(0,\tau) \vartheta_1(z{+}\eta,\tau)}{ \vartheta_1(z,\tau) \vartheta_1(\eta,\tau) }\,,
\end{equation}
where $\vartheta_1(z,\tau)$ is the first Jacobi theta function.
Its expansion in the bookkeeping variable $\eta$ defines the meromorphic, but multi-valued, Kronecker-Eisenstein kernels $g^{(k)}(z,\tau)$,
\begin{equation}\label{KroneckerEisenstein}
F(z,\eta,\tau)=\sum_{k=0}^\infty \eta^{k-1} g^{(k)}(z,\tau) \, , \end{equation}
with $g^{(0)}(z,\tau)=1$. An essential
ingredient for the flatness of the CG connection is the mixed heat equation~\cite{BrownLev} 
\begin{equation}\label{mixedheat}
2\pi i\partial_\tau F(z,\eta,\tau)=\partial_z\partial_\eta F(z,\eta,\tau)\, ,
\end{equation}
see for instance section 2.1.1 of \cite{Gerken:2019cxz} for the modifications in passing to the mixed heat equation of the non-meromorphic counterpart $\Omega(z,\eta,\tau)$.

The CG connection adapted to level $2$ will be built from Kronecker-Eisenstein kernels $g^{(k)}(z,\tau)$ with shifts of $z \rightarrow z{+}\omega_\mu$ by torsion points $\omega_\mu$ in (\ref{eq: torsion}) with $\mu=2,3,4$. The naive first guess $F(z{+}\tfrac{\tau}{2},\eta,\tau)$ and $F(z{+}\tfrac{\tau}{2}{+}\tfrac{1}{2},\eta,\tau)$ for the generating series of the $\omega_4 = \frac{\tau}{2}$ and $\omega_3 = \frac{1{+}\tau}{2}$ cases will not satisfy the mixed heat equation (\ref{mixedheat}). This can be fixed by introducing the more general meromorphic generating series $\tilde{F}(z,  r ,\eta,\tau)$ with $r\in \mathbb R$ defined~by,\footnote{The definition of $\tilde{F}(z,r,\eta,\tau)$ in (\ref{g kernel tilde}) is
an alternative convention for the quantities denoted by $k_\alpha(x,z|\tau)$ in section 1.5 of \cite{Calaque:2019ell} and $h_\alpha(x,z|\tau)$ in section 7.2 of \cite{Hopper:2021ell}.} 
\begin{equation}
    \tilde{F}(z,r,\eta,\tau)=\exp(2\pi i r \eta)F(z{+}r\tau ,\eta,\tau) =\sum_{k=0}^\infty \eta^{k-1} \tilde g^{(k)}(z, r,\tau) \, ,
    \label{g kernel tilde}
\end{equation}
where the expansion coefficients $\tilde g^{(k)}(z, r,\tau)$ are recovered from those in (\ref{KroneckerEisenstein}) via 
\beq
\tilde{g}^{(k)}(z,r,\tau)=\sum_{n=0}^k \frac{(2\pi i r)^n}{n!} g^{(k-n)}(z{+} r\tau ,\tau) \,.
\label{moregtild}
\eeq
The exponential $e^{2 \pi i r \eta}$ in the series (\ref{g kernel tilde}) ensures that $\tilde{F}(z,0,\eta,\tau)=F(z,\eta,\tau)$ is extended to preserve the mixed heat equation~\eqref{mixedheat} also for $r\neq 0$:
\beq\label{mixed heat tilde}
2\pi i \partial_\tau\tilde F(z,r,\eta,\tau)=\partial_z\partial_\eta \tilde F(z,r,\eta,\tau)\, .
\eeq
Moreover, $\tilde F$ in (\ref{g kernel tilde}) obeys a direct analogue of the Fay relations of $F$ and $\Omega$, which can be thought of as the genus-1 uplift of the partial-fraction identity $\frac{1}{z_1z_2} = \frac{1}{z_1(z_2{-}z_1)}+\frac{1}{z_2(z_1{-}z_2)}$,
\begin{align}
\Omega(z_1,\eta_1,\tau)\Omega(z_2,\eta_2,\tau) &=\Omega(z_1,\eta_1{+}\eta_2,\tau)\Omega(z_2{-}z_1,\eta_2,\tau)+\Omega(z_2,\eta_1{+}\eta_2,\tau)\Omega(z_1{-}z_2,\eta_1,\tau) \,,\label{FayuntildeF}\\
F(z_1,\eta_1,\tau)F(z_2,\eta_2,\tau)&=F(z_1,\eta_1{+}\eta_2,\tau)F(z_2{-}z_1,\eta_2,\tau)+F(z_2,\eta_1{+}\eta_2,\tau)F(z_1{-}z_2,\eta_1,\tau) \, .\notag
\end{align}
Indeed, we have,
\begin{align}
\tilde F(z_1, r_1 ,\eta_1,\tau)\tilde F(z_2, r_2 ,\eta_2,\tau)&=\tilde F(z_1, r_1 ,\eta_1{+}\eta_2,\tau)\tilde F(z_2{-}z_1, r_2{-}r_1 ,\eta_2,\tau) \label{FaytildeF} \\
&\quad+\tilde F(z_2, r_2 ,\eta_1{+}\eta_2,\tau)\tilde F(z_1{-}z_2, r_1{-}r_2 ,\eta_1,\tau) 
\, .
\notag
\end{align}
The specific corollaries of (\ref{FayuntildeF}) and (\ref{FaytildeF}) that enter the flatness discussions of sections \ref{sec:flcee} and \ref{sec:fltcg} below can be found in appendix \ref{apc.sec}.

\subsection{Reviewing the flat CEE connection at level 1}
\label{sec:flcee}

As recalled in section~\ref{sec:lvl1}, iterated Eisenstein integrals for $\SLZ$ and their equivariant versions are generated by the level-1 connection $\mathbb A_{\SLZ}(\ep;\tau)$ in (\ref{eq: N1conn}) proportional to $\dd \tau$. The goal of this subsection is to review the derivation of the action (\ref{eq: derivations}) of the Tsunogai derivations $\ep_k$ from the flatness conditions of the meromorphic but multi-valued Calaque-Enriquez-Etingof (CEE) connection \cite{KZB},
\beq
\mathbb K_\text{CEE}(z,\tau) = \dd z \, \mathbb  K_\text{CEE}^{(z)}(z,\tau) +\dd \tau \, \mathbb  K_\text{CEE}^{(\tau)}(z,\tau)\, ,
\label{ceeconn}
\eeq
with components in both $\dd \tau$ and $\dd z$ directions.

\begin{figure}[t!]
\centering
\begin{tikzpicture}[
  x=1cm,y=1cm,line cap=round,line join=round,
  every node/.style={font=\large},
  cycle arrow/.style={line width=1.1pt,-{Stealth[length=2.7mm,width=1.9mm]}},
  puncture loop/.style={loopgreen,line width=1.05pt,
    preaction={draw=white,line width=3.5pt},
    -{Stealth[length=2.4mm,width=1.8mm]}}
]
  \coordinate (O) at (0,0);
  \coordinate (A) at (5.8,0);
  \coordinate (B) at (1.6,3.1);
  \coordinate (C) at ($(A)+(B)$);
  \coordinate (HalfA) at ($(O)!0.5!(A)$);
  \coordinate (HalfB) at ($(O)!0.5!(B)$);
  \coordinate (HalfC) at ($(O)!0.5!(C)$);

  \draw[black!65,line width=0.7pt]
    (O) -- (A) -- (C) -- (B) -- cycle;

  \draw[cycle arrow,basisblue]
    (0,-1.08) -- node[midway,below=2pt] {$a\leftrightarrow\mathfrak A$} (5.8,-1.08);
  \draw[cycle arrow,basisred]
    (-0.55,0) -- node[midway,left=3pt] {$b\leftrightarrow\mathfrak B$} (1.05,3.1);

  \foreach \P in {O} {
    \draw[puncture loop] ([shift=(35:0.27cm)]\P)
      arc[start angle=35,end angle=355,radius=0.27cm];
  }
  \foreach \P in {O,A,B,C}
    \fill (\P) circle (1.6pt);
  \node[anchor=center,inner sep=1pt] at ($(O)+(0,-0.62)$) {$0$};
  \node[anchor=center,inner sep=1pt] at ($(A)+(0,-0.62)$) {$1$};
  \node[above=4pt] at (B) {$\tau$};
  \node[above=4pt] at (C) {$1{+}\tau$};
  \node[loopgreen,anchor=west,inner sep=1pt] at (0.38,0.25) {$t_{2|1}$};
\end{tikzpicture}
\caption{\textit{The fundamental domain for the torus $\mathbb{C}/(\mathbb{Z}{+}\tau\mathbb{Z})$ with a puncture at~$z=0$.}}
\label{figlevel1}
\end{figure}
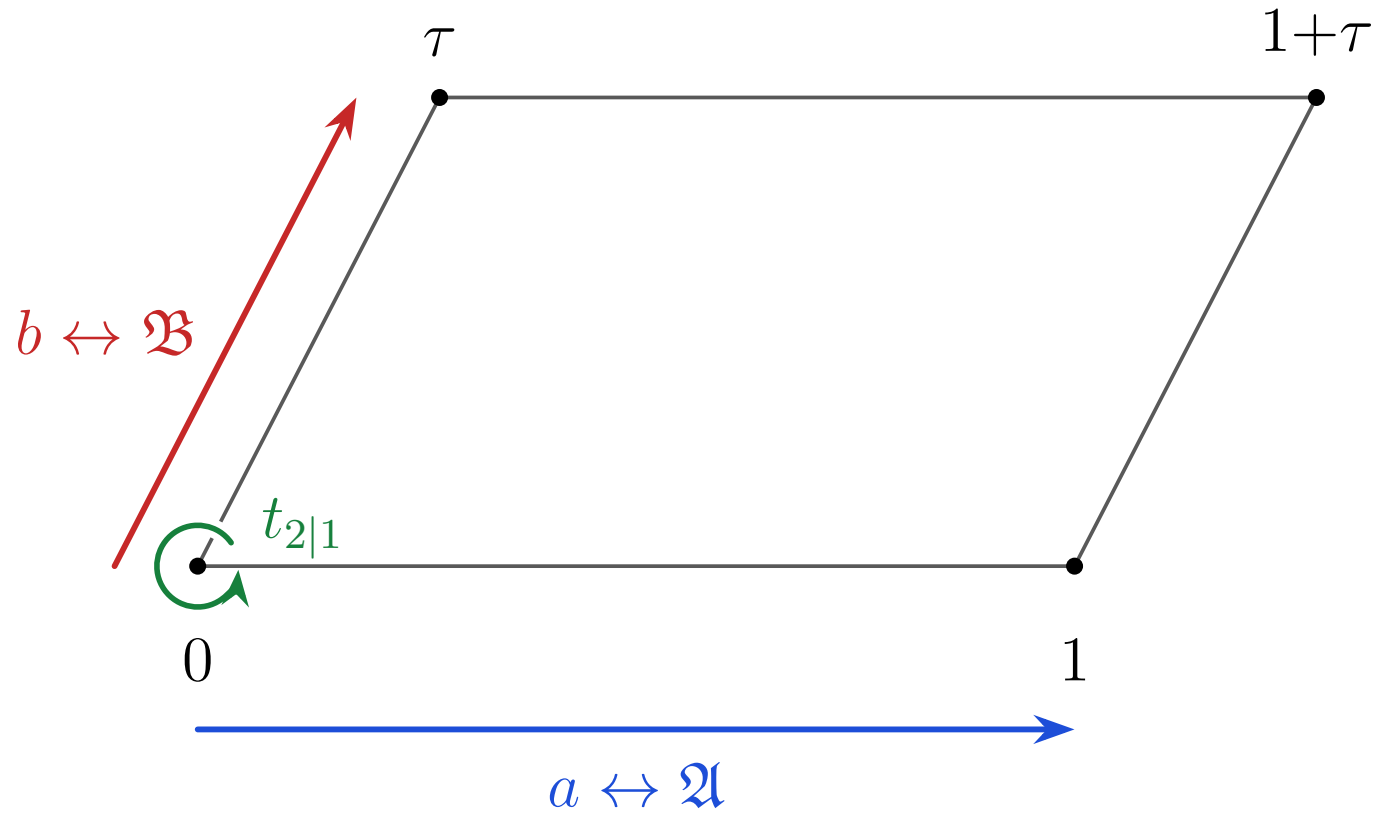

We start by adding a spurious generator $t_{2|1}$ to the free algebra Lie$[a,b]$ subject to the commutator relation
\begin{equation}\label{level1com}
    [\bgen,\agen]=t_{2|1} \, ,
\end{equation}
where the first subscript of $t_{2|1}$ tracks the degree in $a,b$ and the meaning of the second one will become clear through the generalisation to level 2 in section \ref{sec:fltcg}.
On the fundamental domain for the punctured torus depicted in figure \ref{figlevel1}, the generator $t_{2|1}$ corresponds to a local monodromy around the puncture at $z=0$, while the generators $\agen$ and $\bgen$ can be associated with translations along the $\mathfrak A$ and $\mathfrak B$ cycles of the torus, respectively.

The $\dd z$ component of the CEE connection (\ref{ceeconn}) is built from the meromorphic Kron\-ecker-Eisenstein kernels $g^{(k)}(z,\tau)$ in (\ref{KroneckerEisenstein})~\cite{KZB}
\begin{align}
\mathbb  K_\text{CEE}^{(z)}(z,\tau) &=      F(z,\ad_{\frac{b}{2\pi i}},\tau)t_{2|1}
= \sum_{k=0}^\infty (2\pi i)^{1-k}  g^{(k)}(z,\tau) {\rm ad}_b^{k}(a)
\label{CEEexpanded} \\
&= 2\pi i\agen+\sum^\infty_{k=1} (2\pi i)^{1-k} g^{(k)}(z,\tau)  t_{k+1|1}
\, , \notag
\end{align}
where the generating-series representation in the first line is based on the formal definition $\agen =\ad_\bgen^{-1}t_{2|1}$, and the shorthand notation
$t_{k|1}$ in the second line exposes the degree $k$ in $a,b$,
\beq
t_{k|1}=\ad_{\bgen}^{k-2} t_{2|1} = {\rm ad}_b^{k-1} (a)\, .
\label{deftk1}
\eeq 
Note that the generator $t_{2|1}$ is the residue of the pole of $g^{(1)}(z,\tau)=\frac{1}{z}+\dots$, and in fact of $\mathbb  K_\text{CEE}^{(z)}$, since $g^{(k)}(z,\tau)$ with $k\geq 2$ are regular as $z\rightarrow 0$.

\subsubsection{The flatness conditions}
\label{rev:3.2.1}

By meromorphicity in $z$, flatness of $\dd z\,\mathbb  K_\text{CEE}^{(z)}$ with respect to $\dd z \,\partial_z + \dd \bar z \,\partial_{\bar z}$ is automatic. However, flatness of (\ref{ceeconn}) on the (universal cover of the) moduli space of twice-punctured tori, i.e., with respect to $\dd = \dd z \,\partial_z + \dd \bar z\, \partial_{\bar z} +  \dd \tau\, \partial_\tau + \dd \bar \tau\, \partial_{\bar \tau}$,
\beq
\dd \mathbb K_\text{CEE}(z,\tau) = \mathbb K_\text{CEE}(z,\tau) \wedge \mathbb K_\text{CEE}(z,\tau) \,,
\label{fltns}
\eeq
requires an additional component $\dd \tau\, \mathbb K_\text{CEE}^{(\tau)}$ in a suitable Lie algebra. After stripping off $\dd z \wedge \dd \tau$ and imposing meromorphicity in $z$ and $\tau$,\footnote{Meromorphicity of $\mathbb K_\text{CEE}^{(z)}$ and $\mathbb K_\text{CEE}^{(\tau)}$ in both of $z$ and $\tau$ avoids flatness conditions involving $ \dd \bar z$ or $\dd \bar \tau$.} the flatness condition (\ref{fltns}) is equivalent to
\begin{equation}\label{flatnesslevel1}
\partial_z \mathbb K_\text{CEE}^{(\tau)}(z,\tau)-\partial_\tau \mathbb K_\text{CEE}^{(z)}(z,\tau)
= \big[\mathbb K_\text{CEE}^{(z)} (z,\tau),\mathbb K_\text{CEE}^{(\tau)}(z,\tau) \big]\, .
\end{equation}
Given that the $\dd z$ component $\mathbb K_\text{CEE}^{(z)}$ is of uniform $a$-degree 1, it is of particular interest if (\ref{flatnesslevel1}) admits $\dd \tau$ components $\mathbb K_\text{CEE}^{(\tau)}(z,\tau)$ of the same uniform $a$-degree 1. In this case, the two sides of the flatness condition (\ref{flatnesslevel1}) have different $a$-degrees and therefore need to vanish separately. As we will see, these preferred kinds of solutions indeed exist, and the vanishing two sides of (\ref{flatnesslevel1}) determine $\mathbb K_\text{CEE}^{(\tau)}$ up to $z$-independent terms which commute with $\mathbb K_\text{CEE}^{(z)}$.

Starting from the mixed heat equation (\ref{mixedheat}) of the Kronecker-Eisenstein series, we solve the $a$-degree 1 part of the flatness condition (\ref{flatnesslevel1}) by noting that
\begin{equation}
   2\pi i \partial_\tau \mathbb K^{(z)}_\text{CEE} (z,\tau)=\partial_z \partial_{\ad_{\frac{\bgen}{2\pi i}}} \mathbb K^{(z)}_\text{CEE}(z,\tau) \,,
   \label{heatcee}
\end{equation}
where the formal derivative with respect to ${\rm ad}_b$ is performed at constant $t_{2|1}$, i.e., we set $\partial_{\ad_{b}} t_{2|1}=0$. Therefore, we may set the $\dd \tau$ component of the full connection to be
\begin{align}
    \mathbb K_\text{CEE}^{(\tau)} (z,\tau) &=\partial_{\ad_{\bgen}} \big( \mathbb K^{(z)}_\text{CEE}(z,\tau) -2\pi i a \big)+\mathbb L_\text{CEE}(\tau) 
    \label{ktaulv1} \\
    &= \sum_{k=2}^\infty (k{-}1)(2\pi i)^{1-k} g^{(k)}(z,\tau) t_{k|1} +\mathbb L_\text{CEE}(\tau) \, ,
    \notag
\end{align}
where we used $\frac{1}{2\pi i}\partial_{\ad_{\frac{\bgen}{2\pi i}}}=\partial_{\ad_\bgen}$ and added an as-of-now undetermined integration constant $\mathbb L_\text{CEE}(\tau)$ independent of $z$. By construction, (\ref{ktaulv1}) solves $\partial_z \mathbb K_\text{CEE}^{(\tau)}-\partial_\tau \mathbb K_\text{CEE}^{(z)} = 0$, the flatness condition (\ref{flatnesslevel1}) at $a$-degree 1, and it remains to fix $\mathbb L_\text{CEE}$ by imposing the $a$-degree 2 condition $[\mathbb K_\text{CEE}^{(z)}, \mathbb K_\text{CEE}^{(\tau)}] = 0$ for $\mathbb K_\text{CEE}^{(\tau)}$ given by (\ref{ktaulv1}):
\begin{equation}\label{flatnesslevel1comms}
   \big[\mathbb K_\text{CEE}^{(z)}(z,\tau),\partial_{\ad_\bgen}(\mathbb K_\text{CEE}^{(z)}(z,\tau) - 2\pi i a) \big]=\big[\mathbb L_\text{CEE}(\tau),\mathbb K^{(z)}_\text{CEE}(z,\tau) \big]\, .
\end{equation}
This will be solved based on the ansatz 
\begin{equation}
    \mathbb L_\text{CEE}(\tau)=2\pi i\epsilon_0+\sum_{k=4}^\infty(k{-}1)(2\pi i)^{1-k} \est_k(\tau)(\epsilon_{k}+ t_{k|1}) \, ,
    \label{ansaL}
\end{equation}
where the non-commuting $\epsilon_{k}$ are introduced with a shift by $t_{k|1}$ for later convenience. The ansatz (\ref{ansaL}) cannot be made to obey
(\ref{flatnesslevel1comms}) for any choice of $\epsilon_{k}  \in {\rm Lie}[a,b]$, so instead the $\epsilon_{k}$ need to be outer derivations on $\agen$ and $\bgen$. Their action can be determined by simplifying the left-hand side of (\ref{flatnesslevel1comms})  through the corollaries of the Fay identities (\ref{FayuntildeF}) in appendix \ref{apc.1} which introduce holomorphic Eisenstein series via $g^{(k)}(0,\tau) =-{\rm G}_k(\tau)$. In particular, (\ref{app Fay 5}) together with the ansatz (\ref{ansaL}) translate into conditions for the brackets among $[\epsilon_{k_1},a]$ and $[\epsilon_{k_1},t_{k_2|1}]$ with $k_2\geq 2$. These conditions reproduce the bracket relations of Tsunogai's derivation algebra, namely the action of $\epsilon_0$ and $\epsilon_k$ on $\agen$ and $\bgen$ already given in (\ref{eq: derivations}).

\subsubsection{\texorpdfstring{Gauge equivalence to $\AN{\tau}$ at $z=0$}{Gauge equivalence to ASL2(\tau) at z=0}}

The shift of $\epsilon_{k}$ by $t_{k|1}$ in the conventions of (\ref{ansaL}) has the following two related benefits: First,
the residue $t_{2|1} =[b,a]$ of the pole of $\mathbb K_\text{CEE}^{(z)}$ at $z=0$ commutes not only with $\ep_0$ but also with all of $\ep_{k\geq 4}$,
\beq
[\ep_k , t_{2|1}] = 0 \,,\ \ \ {\rm for} \ \, k\geq 0 \, \ {\rm even} \, ,
\label{gaulv1.0}
\eeq
which would not be the case if the ansatz (\ref{ansaL}) had been written with $\ep_k$ in the place of $\ep_k{+}t_{k|1}$. The second
advantage becomes apparent when evaluating $\mathbb K_\text{CEE}^{(\tau)} (z,\tau)$ given by (\ref{ktaulv1}) and (\ref{ansaL}) at $z = 0$ via $g^{(k)}(0,\tau) =-{\rm G}_k(\tau)$ for $k\geq 2$: 
\beq
\mathbb K^{(\tau)}_\text{CEE}(0,\tau)=2\pi i \epsilon_0-\frac{ {\rm G}_2(\tau)}{2\pi i}t_{2|1}+\sum_{k=4}^{\infty}(k{-}1)(2\pi i)^{1-k}\est_k(\tau) \epsilon_k\,.
\label{gaulv1.1}
\eeq
The quasi-modular Eisenstein series ${\rm G}_2(\tau)$ is defined by the $q_1$ series in \eqref{eq: Eis}, since the lattice sum in \eqref{eq: Eis} is only conditionally convergent for $k=2$. This special value of the $\dd \tau$ component of the CEE connection plays a crucial role for elliptic associators \cite{KZB, EnriquezEllAss, Hain} and is related to the level-1 connection $\AN{\tau}$ of (\ref{eq: N1conn}) for iterated Eisenstein integrals via gauge transformation by the group-like element
\beq
{\cal V}(\tau) = \exp \bigg( {-} 2\pi i \ep_0 \tau  + \int^\tau_{i\infty} \frac{\dd \tau_1}{2\pi i} \, {\rm G}_2(\tau_1) t_{2|1}\bigg) \, ,
\label{gaulv1.3}
\eeq
namely 
\beq
\AN{\tau} = {\cal V}(\tau)
\dd \tau \, \mathbb K^{(\tau)}_\text{CEE}(0,\tau) {\cal V}(\tau)^{-1} + \big(\dd {\cal V}(\tau) \big) {\cal V}(\tau)^{-1}\, .
\label{gaulv1.2}
\eeq
One easily verifies via (\ref{gaulv1.0}) that the primitive of the quasi-modular form ${\rm G}_2(\tau)$ does not contribute to the first term on the right-hand side of (\ref{gaulv1.2}). Thus, the second term $(\dd {\cal V}(\tau)) {\cal V}(\tau)^{-1}$ simply cancels the degree $<4$ contributions to $\mathbb K^{(\tau)}_\text{CEE}(0,\tau)$ in (\ref{gaulv1.1}), and the exponential of $\ep_0 \tau$ in (\ref{gaulv1.2}) leads to the factors of $\tau^j$ in (\ref{eq: N1conn}). The bound $j\leq k{-}2$ within the degree-$k$ terms of (\ref{eq: N1conn}) arises from the nilpotency ${\rm ad}_{\ep_0}^{k-1}(\ep_k)=0$.

 Since the regularised $\tau \rightarrow i \infty$ limit of ${\cal V}(\tau)$ is trivial, the gauge transformation (\ref{gaulv1.2}) merely affects the path-ordered exponential (\ref{eq: I1}) of $\AN{\tau}$ by a right-multiplicative factor of (\ref{gaulv1.3}). Hence, its expansion in derivations generates a dictionary between the two flavors of iterated Eisenstein integrals for $\SLZ$ due to $\AN{\tau}$ and $\mathbb K^{(\tau)}_\text{CEE}(0,\tau)$ with polynomials in $\tau$ and $ \int^\tau_{i\infty}  \dd \tau_1\, {\rm G}_2(\tau_1)$ as coefficients.\footnote{See for instance section 3.3 and appendix D of \cite{Broedel:2018izr} as well as appendix G of \cite{Gerken:2020yii} for alternative methods of relating the flavors of iterated Eisenstein integrals produced by $\AN{\tau}$ and $\mathbb K^{(\tau)}_\text{CEE}(0,\tau)$.} In particular, the representations of elliptic associators involving path-ordered exponentials \cite{Broedel:2015hia, Lochak:2020} of $ \mathbb K^{(\tau)}_\text{CEE}(0,\tau)$ can be straightforwardly rewritten in terms of the iterated Eisenstein integrals of  $\AN{\tau}$.

\subsection{\texorpdfstring{The flat CG connection at level $2$}{The flat CG connection at level 2}}
\label{sec:fltcg}

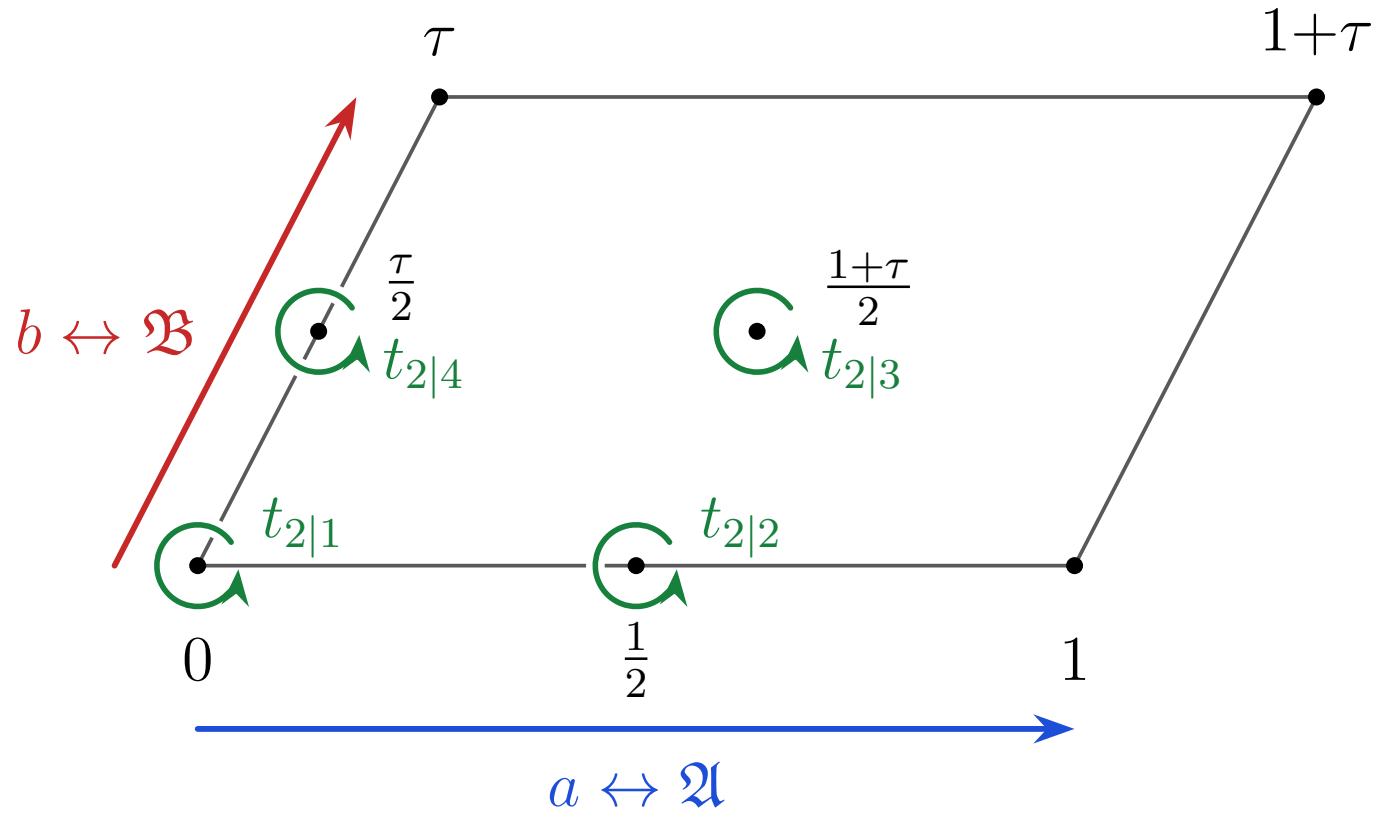
\begin{figure}[h!]
\centering
\begin{tikzpicture}[
  x=1cm,y=1cm,line cap=round,line join=round,
  every node/.style={font=\large},
  cycle arrow/.style={line width=1.1pt,-{Stealth[length=2.7mm,width=1.9mm]}},
  puncture loop/.style={loopgreen,line width=1.05pt,
    preaction={draw=white,line width=3.5pt},
    -{Stealth[length=2.4mm,width=1.8mm]}}
]
  \coordinate (O) at (0,0);
  \coordinate (A) at (5.8,0);
  \coordinate (B) at (1.6,3.1);
  \coordinate (C) at ($(A)+(B)$);
  \coordinate (HalfA) at ($(O)!0.5!(A)$);
  \coordinate (HalfB) at ($(O)!0.5!(B)$);
  \coordinate (HalfC) at ($(O)!0.5!(C)$);

  \draw[black!65,line width=0.7pt]
    (O) -- (A) -- (C) -- (B) -- cycle;

  \draw[cycle arrow,basisblue]
    (0,-1.08) -- node[midway,below=2pt] {$a\leftrightarrow\mathfrak A$} (5.8,-1.08);
  \draw[cycle arrow,basisred]
    (-0.55,0) -- node[midway,left=3pt] {$b\leftrightarrow\mathfrak B$} (1.05,3.1);

  \foreach \P in {O,HalfA,HalfB,HalfC} {
    \draw[puncture loop] ([shift=(35:0.27cm)]\P)
      arc[start angle=35,end angle=355,radius=0.27cm];
  }
  \foreach \P in {O,A,B,C}
    \fill (\P) circle (1.6pt);
  \node[anchor=center,inner sep=1pt] at ($(O)+(0,-0.62)$) {$0$};
  \node[anchor=center,inner sep=1pt] at ($(A)+(0,-0.62)$) {$1$};
  \node[above=4pt] at (B) {$\tau$};
  \node[above=4pt] at (C) {$1{+}\tau$};
  \node[loopgreen,anchor=west,inner sep=1pt] at (0.38,0.25) {$t_{2|1}$};

  \foreach \P in {HalfA,HalfB,HalfC}
    \fill (\P) circle (1.6pt);
  \node[anchor=center,inner sep=1pt] at ($(HalfA)+(0,-0.62)$) {$\frac{1}{2}$};
  \node[anchor=west,inner sep=1pt] at ($(HalfB)+(0.38,0.29)$) {$\frac{\tau}{2}$};
  \node[anchor=west,inner sep=1pt] at ($(HalfC)+(0.38,0.29)$) {$\frac{1+\tau}{2}$};
  \node[loopgreen,anchor=west,inner sep=1pt] at ($(HalfA)+(0.38,0.25)$) {$t_{2|2}$};
  \node[loopgreen,anchor=west,inner sep=1pt] at ($(HalfB)+(0.38,-0.25)$) {$t_{2|4}$};
  \node[loopgreen,anchor=west,inner sep=1pt] at ($(HalfC)+(0.38,-0.25)$) {$t_{2|3}$};
\end{tikzpicture}
\caption{\textit{The fundamental domain for the torus $\mathbb C/(\mathbb Z{+} \tau \mathbb Z)$ with four punctures $\frac{r\tau{+}s}{2}$ with $r,s=0,1$.}}
\label{figlevel2}
\end{figure}

We shall now introduce the level-2 instance of the CG connection \cite{Calaque:2019ell,Calaque:2020ell}
\beq
\mathbb K_\text{CG}(z,\tau) = \dd z \, \mathbb  K_\text{CG}^{(z)}(z,\tau) +\dd \tau \, \mathbb  K_\text{CG}^{(\tau)}(z,\tau)\,,
\label{cgconn}
\eeq
which takes values in a larger Lie algebra with generators $t_{2|1},t_{2|2},t_{2|3},t_{2|4},\agen,\bgen$ 
and one fundamental bracket relation 
\begin{equation}
    [\bgen,\agen]=t_{2|1}+t_{2|2}+t_{2|3}+t_{2|4}\,.
\label{brackab2}
\end{equation}
The generators $\agen$ and $\bgen$ correspond to translations around the homology cycles $\mathfrak{A}$, $\mathfrak{B}$ of the torus while the generators $t_{2|A}$ with $A=1,2,3,4$ correspond to the torsion points of order~2. Accordingly, the bracket relation (\ref{brackab2}) expresses that the path along the boundary of the parallelogram corresponding to $[b,a]$ is homotopic to the four infinitesimal loops around the punctures $(\omega_1,\omega_2,\omega_3,\omega_4)=(0,\frac{1}{2},\frac{1{+}\tau}{2},\frac{\tau}{2})$, see figure \ref{figlevel2}. 

In the remainder of this section, it will be shown how the flatness condition
\begin{align}
\dd \mathbb K_\text{CG}(z,\tau) = \mathbb K_\text{CG}(z,\tau) \wedge \mathbb K_\text{CG}(z,\tau) \,,
\label{flatCG}
\end{align}
leads to the emergence of the derivations $\tilde \ep_{2|\nu=3,4}$ and $\ep^{(j)}_{k| \mu=2,3,4}$ of section \ref{sec:motiv}, determines their action on the algebra generators $a$, $b$, $t_{2|A}$, introduces the $\mathfrak{sl}_2$ structure known from level~1 (see section \ref{sec:221ts}) and results in a gauge equivalent of the level-2 connection $\mathbb A(\ep;\tau)$ in~(\ref{eq: N2conn}).

\subsubsection{\texorpdfstring{The $\boldsymbol{\dd z}$ component}{The dz component}}

Similar to its level-1 counterpart in (\ref{CEEexpanded}), we shall express the $\dd z$
component $\mathbb K_\text{CG}^{(z)}$ of (\ref{cgconn}) through the  meromorphic Kronecker-Eisenstein series introduced in (\ref{KroneckerEisenstein}), but now acting on the more general Lie-algebra generators $t_{2|A}$ with $A=1,2,3,4$, 
\begin{align}
\mathbb K_\text{CG}^{(z)}(z,\tau)&= F(z,\ad_{\frac{\bgen}{2\pi i}},\tau)t_{2|1}+F(z{+}\tfrac{1}{2},\ad_{\frac{b}{2\pi i}},\tau)t_{2|2} 
    \label{dzexpanded}  
    \\
    &\quad  + \tilde{F}(z{+}\tfrac{1}{2},\tfrac{1}{2},\ad_{\frac{b}{2\pi i}},\tau) t_{2|3}+ \tilde F(z,\tfrac{1}{2},\ad_{\frac{b}{2\pi i}},\tau) t_{2|4} \notag \\ 
&=2\pi i\agen+\sum_{k=1}^\infty(2\pi i)^{1-k}\bigl( g^{(k)}(z,\tau) t_{k+1|1}+g^{(k)}(z{+}\tfrac{1}{2},\tau) t_{k+1|2} \notag\\
&\qquad\qquad\qquad\qquad\qquad
+\tilde{g}^{(k)}(z{+}\tfrac{1}{2},\tfrac{1}{2},\tau) t_{k+1|3}
+\tilde{g}^{(k)}(z,\tfrac{1}{2},\tau) t_{k+1|4}
\bigr)\,.\notag
\end{align}
In passing to the third line, we generalised the notation (\ref{deftk1}) to
\beq
t_{k|\mu}=\ad_{\bgen}^{k-2}t_{2|\mu}  \ \ \ \ {\rm for} \ \, k \geq 2\, ,
\label{deftkmu}
\eeq
and the generators $t_{k|3}$, $t_{k|4}$ in (\ref{dzexpanded}) are accompanied by the modified Kronecker-Eisenstein series $\tilde F$ and associated $\tilde g^{(k)}$ kernels defined by (\ref{g kernel tilde}) and (\ref{moregtild}). The degree-one term $2\pi i a$ of (\ref{dzexpanded}) arises from the formal division $a= ({\rm ad}_b)^{-1}(t_{2|1} + t_{2|2}+t_{2|3}+t_{2|4})$ of (\ref{brackab2}) by ${\rm ad}_b$.

The modified $\tilde F$ and associated $\tilde g^{(k)}$ in (\ref{dzexpanded}) are motivated by requiring the connection $\mathbb K_\text{CG}^{(z)}$ to satisfy the mixed heat equation just as in the level-1 case (\ref{heatcee}),
\begin{equation}
    2\pi i \partial_\tau\mathbb K_\text{CG}^{(z)}(z,\tau) =\partial_z\partial_{{\rm ad}_{ \frac{b }{2\pi i}} } \mathbb K_{\rm CG}^{(z)}(z,\tau)\, ,
    \label{heatcg}
\end{equation}
where we set $\partial_{{\rm ad}_b} t_{2|\mu}=0$.

\subsubsection{\texorpdfstring{The $\boldsymbol{\dd \tau}$ component}{The dtau component}}

We shall next determine the $\dd \tau$ component of the CG connection (\ref{cgconn}).
Just as in the level-$1$ case, one starts from the  equivalent
\beq\label{flatnessCG}
\partial_z \mathbb K_\text{CG}^{(\tau)}(z,\tau)-\partial_\tau \mathbb K_\text{CG}^{(z)}(z,\tau)= \big[\mathbb K^{(z)}_\text{CG}(z,\tau) ,\mathbb K^{(\tau)}_\text{CG}(z,\tau) \big]\,,
\eeq
of the flatness condition (\ref{flatCG}) and imposes
the as-of-yet unknown component $\mathbb K^{(\tau)}_\text{CG}$ to share the $a$-degree $1$ of $\mathbb K^{(z)}_\text{CG}$ in (\ref{dzexpanded}). As a result, the two sides of (\ref{flatnessCG}) have different $a$-degree and need to vanish separately, which in turn will determine $\mathbb K^{(\tau)}_\text{CG}$ up to $z$-independent terms that commute with $\mathbb K^{(z)}_\text{CG}$. Similar to (\ref{ktaulv1}) at level 1, the mixed heat equation (\ref{heatcg}) identifies the following solution to the flatness condition (\ref{flatnessCG}) of $a$-degree 1,
\begin{align}
\mathbb K_\text{CG}^{(\tau)}(z,\tau)&=\partial_{\ad_\bgen} \big(\mathbb K^{(z)}_\text{CG} (z,\tau) - 2\pi i a \big) +\mathbb L_\text{CG}(\tau) 
\label{solht2} \\
&=\sum_{k=2}^\infty(k{-}1) (2\pi i)^{1-k}\bigl(g^{(k)}(z,\tau)t_{k|1}+g^{(k)}(z{+}\tfrac{1}{2},\tau)t_{k|2} \notag \\
&\hspace{37mm}
+\tilde g^{(k)}(z{+}\tfrac{1}{2},\tfrac{1}{2},\tau)t_{k|3}+\tilde g^{(k)}(z,\tfrac{1}{2},\tau)t_{k|4}\bigr) + \mathbb L_\text{CG}(\tau) \, .\notag
\end{align}
The adjoint action of the $z$-independent integration constant $\mathbb L_\text{CG}(\tau)$  will once more be fixed from the flatness condition $[\mathbb K^{(z)}_\text{CG},\mathbb K^{(\tau)}_\text{CG}] = 0$ at $a$-degree two,
\beq
\big[\mathbb K_\text{CG}^{(z)}(z,\tau),\partial_{\ad_\bgen} (\mathbb K_\text{CG}^{(z)}(z,\tau) - 2\pi i a ) \big]=\big[\mathbb L_\text{CG}(\tau),\mathbb K_\text{CG}^{(z)}(z,\tau) \big] \,  .
\label{getlcg}
\eeq
More specifically, we shall generalise the ansatz (\ref{ansaL}) for the level-1 analogue $\mathbb L_\text{CEE}(\tau)$ to involve bases of Eisenstein series for $\Gamma(2)$ instead of those for $\SLZ$, 
\begin{align}
    \mathbb L_\text{CG}(\tau)&=2\pi i\epsilon_0-  (2\pi i)^{-1} \sum_{\nu=3}^4 f^{(2)}(\omega_\nu,\tau)(\tilde\epsilon_{2|\nu}+t_{2|\nu}-t_{2|2}) \label{anslv2}\\
    &\quad-\sum_{k=4}^\infty (k{-}1)(2\pi i)^{1-k}\sum_{\mu=2}^4 f^{(k)}(\omega_\mu,\tau)\Big(\epsilon_{k|\mu}+t_{k|\mu}{-}\tfrac{2^{k-2}}{2^{k-2}{-}1}t_{k|1} \Big) \, .\notag
\end{align}
Our conventions for shifting the non-commuting $\tilde\epsilon_{2|\nu}$, $\epsilon_{k|\mu}$ by combinations of $t_{k | \lambda}$ in (\ref{deftkmu}) are again motivated by the $z=0$ value $\mathbb K^{(\tau)}_\text{CG}(0,\tau)$ in (\ref{k0lv2}) below.
Moreover, we depart from the choice of \cite{Calaque:2019ell,Calaque:2020ell, Hopper:2021ell} (see also section \ref{sec: extensions}), where the connection features an overcomplete set of weight-$(k\geq 4)$ Eisenstein series $f^{(k)}(z,\tau)$ for $\Gamma(2)$ with all of $z \in \{ 0,\tfrac{1}{2},\tfrac{1+\tau}{2},\tfrac{\tau}{2} \}$, accompanied by four separate derivations of degree $k$.
Similar to the level-$1$ case, solutions to (\ref{getlcg}) can only be found when $\ep_0$, $\tilde\epsilon_{2|\nu}$, $\epsilon_{k|\mu}$ are allowed to be outer derivations as opposed to Lie polynomials in $a$, $b$, $t_{2|A}$.
As detailed in section \ref{sec:335}, the 
above $\tilde{\ep}_{2 | \nu}$ and $\ep_{k | \mu}$ with $k\geq 4$ are part of the derivations in $\mathfrak{u}$, where the level-2 connection $\mathbb A(\ep;\tau)$ of \eqref{newlv2con} takes values.

\subsubsection{\texorpdfstring{Action of the derivations in $\mathbb L_\text{CG}(\tau)$}{Action of the derivations in LCG(tau)}}
\label{sec:actderiv}

The next step is to determine the action of $\ep_0$, $\tilde\epsilon_{2|\nu}$, $\epsilon_{k|\mu}$ in the ansatz (\ref{anslv2}) for $\mathbb L_\text{CG}(\tau)$ on the algebra generators $a$, $b$, $t_{2|A}$ of $\mathbb K_\text{CG}^{(z)}(z,\tau)$. For this purpose, one simplifies the left-hand side of the flatness condition (\ref{getlcg}) through the corollaries of the Fay identities (\ref{FayuntildeF}) and (\ref{FaytildeF}) in appendix \ref{apc.2} which produce the $z\rightarrow 0$ limits $g^{(k)}(0,\tau) =-\est_k(\tau)$ and
\beq
\lim_{z\rightarrow 0} g^{(k)}(z{+}\tfrac{1}{2},\tau) = f^{(k)}(\tfrac{1}{2},\tau)\,,\quad \lim_{z\rightarrow 0}\tilde g^{(k)}(z{+}\tfrac{s}{2},\tfrac{1}{2},\tau)=f^{(k)}(\tfrac{s}{2}{+}\tfrac{\tau}{2},\tau)\,,\quad s\in \{0,1\} \, .
\label{spcgk}
\eeq
Moreover, we recall the shorthands $\omega_2=\tfrac{1}{2}$, $\omega_3=\tfrac{1{+}\tau}{2}$ and $\omega_4=\tfrac{\tau}{2}$ introduced in (\ref{eq: torsion}) as well as the relations (\ref{funrel.02}) and (\ref{funrel.01}) for the sums $\sum_{\mu=2}^4 f^{(k)}(\omega_\mu,\tau)$.
In order to compactly represent the resulting action of the derivations in (\ref{anslv2}), we parametrise the shift of doubly-periodic functions $f(z)$ by the torsion points $\omega_{\mu=2,3,4}$ via square matrices $\varphi_2$, $\varphi_3$, $\varphi_4$,
\beq
\varphi_2=\left(\begin{array}{cccc}0&1&0&0\\1&0&0&0\\0&0&0&1\\0&0&1&0\end{array}\right)\,,\quad \varphi_3=\left(\begin{array}{cccc}0&0&1&0\\0&0&0&1\\1&0&0&0\\0&1&0&0\end{array}\right)\,, \quad \varphi_4=\left(\begin{array}{cccc}0&0&0&1\\0&0&1&0\\0&1&0&0\\1&0&0&0\end{array}\right)\, .
\label{vphmat}
\eeq
These matrices permute the torsion points in the sense of 
\beq
{\bf f}(z{+}\omega_\mu)=\varphi_\mu\, {\bf f}(z)\,,\quad \text{for}\,\, {\bf f}(z)=\begin{pmatrix}f(z)\\f(z{+}\omega_2)\\f(z{+}\omega_3)\\f(z{+}\omega_4)\end{pmatrix}\, .
\eeq

\begin{itemize}
\item \textbf{The action of $\epsilon_0$:}
The derivation $\epsilon_0$ acts on $\agen$ and $\bgen$ exactly as in (\ref{eq: derivations}) for the level-$1$ case and annihilates the generators $t_{2|A}$,
\beq
[\epsilon_0,\agen]=0\,,\quad [\epsilon_0,\bgen]=\agen\,,\quad [\epsilon_0,t_{2|A}]=0\,,\quad A\in\{1,2,3,4\}\, .
\label{epac00}
\eeq
\item \textbf{The action of $\tilde\epsilon_{2|\mu}$:} 
The action of $\tilde \epsilon_{2|\mu}$ with $\mu\in\{3,4\}$ on $t_{2|A}$ with $A\in\{1,2,3,4\}$ is given by
\beq
[\tilde\epsilon_{2|\mu},t_{2|A}]=\sum_{B=1}^4\biggl((\varphi_2)_{AB} - (\varphi_\mu)_{AB}\biggr) [t_{2|A},t_{2|B}] +[ t_{2|2}{-}t_{2|\mu},t_{2|A}]\ , 
\label{epac01}
\eeq
and their action on $\agen$ and $\bgen$ reads
\beq
[\tilde\epsilon_{2|\mu},\agen]=[\agen,t_{2|\mu}-t_{2|2}]\,,\quad [\tilde\epsilon_{2|\mu},\bgen]=t_{3|\mu}-t_{3|2}\, .
\label{epac02}
\eeq
\item \textbf{The action of $\epsilon_{k|\mu}$:}
For the action of $\epsilon_{k|\mu}$ with $k\geq 4$, $\mu\in\{2,3,4\}$ on $t_{2|A}$ with $A\in\{1,2,3,4\}$, we find
\beq
[\epsilon_{k|\mu},t_{2|A}]  ={-}\sum_{B=1}^4\biggl((\varphi_{\mu})_{AB}[t_{2|A},  t_{k|B}]\biggr)+\tfrac{2^{k-2}}{2^{k-2}-1}[t_{2|A},t_{k|A}{-}t_{k|1}]+[t_{2|A},t_{k|\mu}]\, .
\label{eq: epsont}
\eeq
The analogous action on $\bgen$ and $\agen$ is given by
\beq
[\epsilon_{k|\mu},\bgen] =t_{k+1|\mu}-\tfrac{2^{k-2}}{2^{k-2}-1} t_{k+1|1}\,,
\label{epac03}
\eeq
as well as
\begin{align}
[\epsilon_{k|\mu},\agen]&=-%
\sum_{\ell=0}^{k/2-2}
({-}1)^\ell\sum_{A=1}^4\biggl(\sum_{B=1}^4(\varphi_{\mu})_{AB}[t_{2+\ell|A},  t_{k-1-\ell|B}]-\tfrac{2^{k-2}}{2^{k-2}{-}1}[ t_{2+\ell|A},t_{k-1-\ell|A}]\biggr)\notag\\
&\quad -[t_{k|\mu},\agen]+\tfrac{2^{k-2}}{2^{k-2}-1}[t_{k|1},\agen]\, .
\label{eq: epsona}
\end{align}
\item \textbf{$t_{2|1}$ commutes with $\ep_0$, $\tilde \epsilon_{2|\nu}$, $\epsilon_{k|\mu}$:} 
By inspecting the $A=1$ case of (\ref{epac00}), (\ref{epac01}), (\ref{eq: epsont}) and using the formula $(\varphi_\mu)_{1B} = \delta_{\mu,B}, \ \mu=2,3,4$ for the relevant entries of (\ref{vphmat}), we conclude that $t_{2|1}$ commutes with all the derivations in (\ref{anslv2}),
\beq
[\ep_0, t_{2|1}] = [ \tilde \epsilon_{2|\nu} , t_{2|1}] = [ \epsilon_{k|\mu} , t_{2|1}] = 0 \, , \ \ \ \ \nu \in \{ 3,4\} \, , \ \  \mu \in \{2,3,4\} \, , \ \ k\geq 4\, .
\label{t21cen}
\eeq
\end{itemize}

The above actions are compatible with the fundamental bracket relation \eqref{brackab2}. This can be shown by direct calculation -- summing the actions \eqref{epac01}, \eqref{eq: epsont} of the derivations on $t_{2|A}$ over $A=1,2,3,4$ and evaluating their action on $[b,a]$ through the Jacobi identity and \eqref{epac02}, \eqref{epac03}, \eqref{eq: epsona}. Compatibility of the $\ep_0$ action with \eqref{brackab2} is readily checked via (\ref{epac00}).

\subsubsection{Derivations as representations}
\label{sec:335}

In section~\ref{tau level 2.1} we introduced the algebra $\mathfrak{u}$ generated by elements $\ep_{k|\mu}^{(j)}$. We will now discuss the connection of these derivations to the $\ep_0$, $\tilde{\ep}_{2|\mu}$, and $\ep_{k|\mu}$ above.
The key identification is  
\begin{align}
\ep_{k|\mu}^{(j)}=\ad_{\ep_0}^j(\ep_{k|\mu})\,,
\label{idepkj}
\end{align} 
which defines an action of the derivations in $\mathfrak{u}$ on $a$, $b$ and $t_{2|A}$. That this is the right identification will become apparent when we relate the connection $\mathbb K_\text{CG}(z,\tau)$ at $z=0$ to $\ANN{\tau}$ in section~\ref{sec:3.3.x}.

In addition to $\ep_0$ one can also define a derivation $\ep_0^\vee$ that acts as a lowering operator by
\beq
[\epsilon^\vee_0,\bgen]=0\,,\quad [\epsilon^\vee_0,\agen]=\bgen\,,\quad [\epsilon^\vee_0,t_{2|A}]=0\,,\quad A\in\{1,2,3,4\}\, .
\label{defep0vee}
\eeq
On the generators $a$ and $b$ this is exactly the action familiar from level 1, see~\eqref{eq: derivations}. Together with the Cartan generator $h= [\ep_0,\ep_0^\vee]$, the derivations $\ep_0$, $\ep_0^\vee$ span the $\mathfrak{sl}_2$ algebra acting on~$\mathfrak{u}$. The identification (\ref{idepkj}) and representation theory of $\mathfrak{sl}_2$ (see section 5.1 of \cite{Dorigoni:2024iyt} for the level-1 analogue)
then lead to 
\begin{align}
\ad_{\ep_0}\ep_{k|\mu}^{(j)}=\ep_{k|\mu}^{(j+1)}\,,\ \ \ \ \ad_{\ep_0^\vee}\ep_{k|\mu}^{(j)}=j(k{-}1{-}j)\ep_{k|\mu}^{(j-1)}\,,\ \ \ \ {\rm for} \ \, k \geq 2\, ,
    \label{sl2epkmu}
\end{align}
identical to the level-1 case~\eqref{eq: sl2rep} (see (\ref{backtotild}) for relating $\tilde{\ep}_{2|\nu}$ to $\ep_{2|\mu}$) which again leads to Cartan eigenvalues $[h,\ep^{(j)}_{k|\mu}]=(2j{+}2{-}k) \ep^{(j)}_{k|\mu}$.

The fact that the level-2 connection (\ref{newlv2con}) only involves those $\ep_{k|\mu}^{(j)}$ in \eqref{idepkj} with $j=0,1,\ldots,k{-}2$ can be understood from the
nilpotency condition  for $ k\geq 2$,
\begin{align}    
\ep_{k|\mu}^{(k-1)}=0\,.
    \label{epmunilp}
\end{align}
This implies that $\ep_{k|\mu}=\ep^{(0)}_{k|\mu}$ is the lowest-weight vector of an irreducible $(k{-}1)$-dimensional $\mathfrak{sl}_2$ representation with $\ep_{k|\mu}^{(k-2)}$ as its highest-weight vector.

We shall now prove~\eqref{epmunilp}:
\begin{itemize}
\item The nilpotency property ${\rm ad}_{\ep_0}(\tilde \ep_{2|\mu})=0$ at degree $k=2$ can be derived from acting with $\ad_{\ep_0}$ on $\tilde \ep_{2|\mu}$ within equations \eqref{epac01} and \eqref{epac02}. Using the Jacobi identity on both sides of the three equations and using the action of $\ep_0$ from \eqref{epac00}, we find the vanishing of the brackets $[{\rm ad}_{\ep_0}(\tilde \ep_{2|\mu}),a]=[{\rm ad}_{\ep_0}(\tilde \ep_{2|\mu}),b]=[{\rm ad}_{\ep_0}(\tilde \ep_{2|\mu}),t_{2|A}]=0$ for $A=1,2,3,4$. Since the derivation ${\rm ad}_{\ep_0}(\tilde \ep_{2|\nu})$ yields zero when acting on all generators, it vanishes as a derivation, i.e., ${\rm ad}_{\ep_0}(\tilde \ep_{2|\nu})=0$. 

\item Similarly, at degree $k \geq 4$, one can show that acting with $\ad_{\ep_0}^{k-1}$ on the equations \eqref{eq: epsont}, \eqref{epac03}, and \eqref{eq: epsona} yields $[{\rm ad}_{\ep_0}^{k-1}(\ep_{k|\mu}),a]=[{\rm ad}_{\ep_0}^{k-1}(\ep_{k|\mu}),b]=[{\rm ad}_{\ep_0}^{k-1}(\ep_{k|\mu}),t_{2|A}]=0$. A quick way to see this is that, given a commutator involving several $a$, $t_{2|A}$ and $n$ number of $b$'s, one needs $n{+}1$ actions of $\ad_{\ep_0}$ to make the commutator vanish using \eqref{epac00} ($n$ of them are needed to turn every $b$ into an $a$ and one additional $\ad_{\ep_0}$ then makes the expression vanish). Hence, we have nilpotency of the derivations~\eqref{epmunilp}.
\end{itemize}

The $\SLZ$ group action $U_\gamma^{-1}\bullet U_\gamma$ for $\gamma=S,T$ from \eqref{eq: US} can be expressed in terms of the $\mathfrak{sl}_2$ algebra as
\begin{align}
    U_T=e^{2\pi i \ep_0}\,,\ \ \ \ \ \ U_S=e^{\ep_0^\vee/(2\pi i)}e^{-2\pi i \ep_0}e^{\ep_0^\vee/(2\pi i)}\,.
    \label{uSuT}
\end{align}

Beyond their $\mathfrak{sl}_2$ structure, the generators $\tilde{\ep}_{2|\mu}$, $\ep^{(j)}_{k|\mu}$ at fixed $k$, $j$ also carry an $\mathcal{S}_3$ representation as shown in section \ref{sec:2.1.3}. These two actions on $\mathfrak{u}$ commute: the $\mathcal{S}_3$ action in \eqref{eq: theta} and \eqref{eq: thetatild} permute the labels $\mu$ of $\tilde{\ep}_{2|\mu}$ and $\ep_{k|\mu}^{(j)}$, but keep $j$ (and $k$) inert, whereas the $\mathfrak{sl}_2$ action in \eqref{sl2epkmu} affects $j$ but leaves $\mu$ (and $k$) unchanged. This means that $\mathfrak{u}$ carries a representation of $\mathfrak{sl}_2\times\mathcal{S}_3$, and we shall frequently use decompositions into irreducible representations of this direct product.

For later reference, we define the $(a,b)$-degrees of the generators $a$, $b$ to be $(1,0)$ and $(0,1)$, respectively. The generators $t_{2|A}$ in turn are assigned $(a,b)$-degrees $(1,1)$, consistently with \eqref{brackab2}. Equation~\eqref{deftkmu} then implies that $t_{k|A}$ has $(a,b)$-degrees $(1,k{-}1)$. Similarly, we can infer the $(a,b)$ degrees of the derivations $\ep_0$, $\tilde{\ep}_{2|\mu}$, and $\ep_{k|\mu}^{(j)}$ (following the same definition for $a$ and $b$ degree of derivations as that for derivations in $\mathfrak{u}_{\SLZ}$ given in section \ref{sec:lvl1}) by considering their actions in \eqref{epac01} to \eqref{eq: epsona} and comparing their $(a,b)$ degrees on both sides of the equations. We find that $\ep_0$, $\tilde{\ep}_{2|\mu}$, and $\ep_{k|\mu}^{(j)}$ have $(a,b)$-degrees $(1,{-}1)$, $(1,1)$, and $(j{+}1,k{-}j{-}1)$, respectively, as anticipated below \eqref{eq: N2conn} and elsewhere. One similarly infers $(a,b)$-degrees $(-1,1)$ of $\ep_0^\vee$ from (\ref{defep0vee}) such that the $\mathfrak{sl}_2$ generators $\ep_0$, $\ep_0^\vee$ both have total degree $0$.

\subsubsection{\texorpdfstring{Gauge equivalence to $\mathbb A(\ep;\tau)$ at $z=0$}{Gauge equivalence to A(tau) at z=0}}
\label{sec:3.3.x}

We will again be interested in the special value $z=0$ of $\mathbb K^{(\tau)}_\text{CG}(z,\tau)$, which is related to the connection $\mathbb A(\ep;\tau)$ for $\Gamma(2)$ in (\ref{eq: N2conn}) by a simple gauge transformation. The special values (\ref{spcgk}) of the meromorphic Kronecker-Eisenstein kernels lead to the compact expression
\begin{align}
\mathbb K^{(\tau)}_\text{CG}(0,\tau) &=2\pi i \epsilon_0-\frac{\est_2(\tau)}{2\pi i} t_{2|1} -\sum_{\mu=3}^4\frac{f^{(2)}(\omega_\mu,\tau)}{2\pi i}\tilde\epsilon_{2|\mu}
\label{k0lv2} \\
&\quad 
-\sum_{k=4}^\infty(k{-}1)(2\pi i)^{1-k} \sum_{\mu=2}^4 f^{(k)}(\omega_\mu,\tau)\epsilon_{k|\mu}\,,
\notag
\end{align}
which justifies the conventions for the shifts of $\tilde \epsilon_{2|\nu}$, $\epsilon_{k|\mu}$ in (\ref{anslv2}) and resembles $\mathbb K^{(\tau)}_\text{CEE}(0,\tau)$ for level 1 in (\ref{gaulv1.1}). In fact, the gauge equivalence (\ref{gaulv1.2}) of $\AN{\tau}$ and  $\mathbb K^{(\tau)}_\text{CEE}(0,\tau)$ in the level-1 case straightforwardly generalises to level 2 via
\beq
\mathbb A(\ep;\tau) = {\cal V}(\tau)
\dd \tau \, \mathbb K^{(\tau)}_\text{CG}(0,\tau) {\cal V}(\tau)^{-1} + \big(\dd {\cal V}(\tau) \big) {\cal V}(\tau)^{-1}\, ,
\label{gaulv1.9}
\eeq
with the same group-like series ${\cal V}(\tau) $ in (\ref{gaulv1.3}) to cancel the terms proportional to $ \epsilon_0$ and $ \est_2(\tau)  t_{2|1}$ in (\ref{k0lv2}). The level-2 gauge equivalence (\ref{gaulv1.9}) relies on the fact that $t_{2|1}$ commutes with 
$\mathfrak{sl}_2$ and
$\mathfrak{u}$, see (\ref{t21cen}) and (\ref{defep0vee}). Moreover, it confirms the identification~\eqref{idepkj} by using the nilpotency properties~\eqref{epmunilp} of $\ep_{k|\mu} $ under ${\rm ad}^{k-1}_{\ep_0}$ that lead to the terms with powers of $\tau^j$ in (\ref{eq: N2conn}).

As explained below (\ref{gaulv1.3}), the gauge equivalence of (\ref{gaulv1.9}) generates relations between the two flavors of iterated Eisenstein integrals for $\Gamma(2)$ generated via $\mathbb K^{(\tau)}_\text{CG}(0,\tau)$ or $\mathbb A(\ep;\tau)$ with polynomials in $\tau$ and the primitive of ${\rm G}_2(\tau)$ as coefficients.

\section{\texorpdfstring{Relations among the level-2 generators $\ep_{k|\mu}$}{Relations among the level-2 generators epsilon(k|mu)}}
\label{sec: pollack}

Given the action of the level-2 Lie algebra $\mathfrak{u}$ of geometric derivations on the Lie algebra generated by $a,b,t_{2|A}$ via \eqref{epac00} to \eqref{eq: epsona}, we shall now discuss level 2 analogs of Pollack relations. Their existence implies that $\mathfrak{u}$ is not freely generated by the $\tilde{\ep}_{2|\mu}$ and $\ep^{(j)}_{k|\mu}$ for $k\geq 4$ and $j=0,1,\ldots,k{-}2$.
In analogy to level 1 (see section \ref{sec:221ts}), we expect that the existence of such relations at degree $k$ corresponds to the presence of cusp forms for $\Gamma(2)$ of weight $\leq k{-}2$ 
through the theory of period polynomials \cite{Pollack,Pa_ol_2013}. Generalisations of this correspondence beyond level 1 were discussed in \cite{Pollack:lv2}, and we shall find strong evidence  that our results line up with its incarnation for $\Gamma(2)$. However, we relegate a precise statement at level 2 that relates coefficients of period polynomials to coefficients of relations in $\mathfrak{u}$ to future work \cite{wip:cusp}. 

We note that the generators and relations of $\mathfrak{u}$ can be defined over the rational numbers $\mathbb{Q}$ and we study them with this coefficient field.

\subsection{Representation structure of relations}

Note that the existence of a relation in $\mathfrak{u}$ immediately implies the existence of an $\mathfrak{sl}_2$ multiplet of relations that can be obtained by the adjoint action (\ref{sl2epkmu}) of $\ep_0$ and $\ep_0^\vee$ on it. 
By analogy with known results at level 1~\cite{Pollack,brown2017multiple},
the dimensions of these multiplets are expected to be equal to the modular weight of the associated cusp form minus one. For example, the relations in \eqref{eq: pollack} define 11-plets associated to the (Ramanujan) cusp form of $\SLZ$ of modular weight 12 and coefficients in their modular-depth-2 terms are associated with the even and the odd part of the period polynomial, respectively~\cite{Pollack}. 

As discussed above in section~\ref{sec:2.1.3}, $\mathfrak{u}$ also carries an $\mathcal{S}_3$ structure beyond the $\mathfrak{sl}_2$ structure. The space of cusp forms $\mathcal{S}_k(\Gamma(2))$ carries the same $\mathcal{S}_3$ structure, so that the elements of $\mathcal{S}_k(\Gamma(2))$ can be organized into irreducible representations of $\mathcal{S}_3$, cf. appendix~\ref{app:mf2}. Accordingly, the associated ($\mathfrak{sl}_2$ multiplets of) level-2 Pollack relations are expected to decompose into the same irreducible $\mathcal{S}_3$ representations. 

Besides acting with $\mathfrak{sl}_2$ and $\mathcal{S}_3$ on Pollack relations, one can also consider the adjoint action of $\tilde{\ep}_{2|\mu}$, $\ep_{k|\mu}$, or $z_w$ on a given relation. 
These operations increase the (leading) modular depth and lead to derived relations (since  relations form an ideal) which we shall distinguish from \emph{fundamental} relations in $\mathfrak{u}$.
Fundamental relations are  defined as relations in $\mathfrak{u}$ that cannot be obtained from such adjoint actions on simpler relations and will therefore be in our main focus. Note that there are no derived relations of modular depth 2 since the elements $\tilde{\ep}_{2|\mu}$ and $\ep_{k|\mu}^{(j)}$ of modular depth one do not obey any Pollack relations. 

Based on the discussion of the space $\mathcal{S}_k(\Gamma(2))$   of cusp forms in appendix~\ref{app:mf2} and their irreducible $\mathcal{S}_3$ representations, we predict that the Pollack relations at degree $k$ and of maximal modular depth $r^{\rm max}$ decompose into the irreducible representations listed in Table \ref{Polltab}.\footnote{We give an explicit list of the relevant $\Gamma(2)$ cusp forms in appendix~\ref{app:mf2}.} The predicted pattern for the modular depths is in analogy with the structure known from level~1~\cite{Pollack}. We have confirmed it for all cases up to degree 12, as well as the cases of $\rmax=2$ and $\rmax=3$ up to degree 20 and 18, respectively. 
\begin{table}
\centering
\renewcommand{\arraystretch}{1.5}
\begin{tabular}{|c||c|c|c|c|c|c|c|c|c|}
\hline
\diagbox{$k$}{$\rmax$} & 2 & 3 & 4 & 5 \\
\hline\hline
8 & $\bep$ & $0$ & $0$ & $0$ \\
\hline
10 & $\mathbf{2}$ & $\bep$ & $0$ & $0$ \\
\hline
12 & $\bep\oplus \mathbf{2}$ & $\mathbf{2}$ & $\bep$ & $0$ \\
\hline
14 & $\mathbf{1}\oplus\bep\oplus\mathbf{2}$ & $\bep\oplus \mathbf{2}$ & $\mathbf{2}$ & $\bep$  \\
\hline
\end{tabular}
\caption{\textit{Collection of predicted decomposition of $\mathfrak{sl}_2$ multiplets of fundamental relations in $\mathfrak{u}$ into $\mathcal{S}_3$ irreducible representations, organized by total degree $k\leq 14$ and maximal modular depth $\rmax\leq 5$ based on the irreducible $\mathcal{S}_3$ representation of $\mathcal{S}_k(\Gamma(2))$.}}
\label{Polltab}
\end{table}

Let $n^{\rm Pol}(k,\rmax)$ denote the number of $\mathfrak{sl}_2$ multiplets of Pollack relations of degree $k$ and maximal modular depth $\rmax$. Using equation \eqref{dimsk} and Table \ref{dimsk}, we find 
\begin{align}
    n^{\rm Pol}(k,\rmax)=\dim\mathcal{S}_{k-2}\big(\Gamma(2) \big)+2-\rmax=\frac{k}{2}-\rmax-1\,.
    \label{numpoll}
\end{align}
In the rest of this section, we show that we find exact agreement with the prediction in Table~\ref{Polltab}, strongly suggesting a connection with the theory of period polynomials for $\mathcal{S}_k(\Gamma(2))$ that generalises the level 1 result in \cite{Pollack}. 

\subsection{Pollack relations at low degrees}
\label{sec:lowd}

In practice, the level-2 Pollack relations are determined as follows. In each $(a,b)$-degree, we form a general linear combination of a Lyndon basis of commutators using the letters $\tilde \ep_{2| \nu}$ and $\ep_{k|\mu}^{(j)}$ with $k\geq 4$, so that the commutators have compatible $(a,b)$-degree. We then require that this combination of derivations acts trivially on all of $a$, $b$ and $t_{2|A}$ with $A=1,2,3,4$. The solutions to these conditions can then be organized into irreducible representations of $\mathcal{S}_3$.

\subsubsection{Relations at degree 8}

The first cusp form for $\Gamma(2)$ has weight $6$ and sits in the antisymmetric $\bep$-representation of~$\mathcal{S}_3$. This is in agreement with the first relations we found in the derivation algebra $\mathfrak{u}$ at total degree $k=8$.
It has maximal modular depth $\rmax=2$ and spans a 5-plet of $\mathfrak{sl}_2$ with the following fundamental relation as its lowest-weight vector
\begin{align}
    \label{eq: pollack8}
    P_8^2=4[\epsilon_{4|2},\epsilon_{4|3}]{+}4[\epsilon_{4|3},\epsilon_{4|4}]{+}4[\epsilon_{4|4},\epsilon_{4|2}]{+}[\tilde\epsilon_{2|3},\epsilon_{6|2}]{-}[\tilde\epsilon_{2|3},\epsilon_{6|4}]{-}[\tilde\epsilon_{2|4},\epsilon_{6|2}]{+}[\tilde\epsilon_{2|4},\epsilon_{6|3}]
=0\,.
\end{align}
Furthermore, the relation transforms in the alternating representation $\bep$ of $\mathcal{S}_3$ under the action of $\theta_\gamma^{-1}\bullet \theta_\gamma$ for $\gamma=T,S$ in (\ref{eq: theta}) and (\ref{eq: thetatild}). This becomes manifest once we rewrite the $\tilde\epsilon_{2|\nu}$ of (\ref{eq: pollack8}) in terms of $\epsilon_{2|\mu}$ via (\ref{backtotild}) and regroup the expression as
\beq
P_8^2 = \sum_{\substack{(\mu,\nu)\in \{(2,3), \\  (3,4),(4,2)  \}  } } \Big( \big( 2 [\ep_{4|\mu},\ep_{4|\nu}]-[\ep_{2|\mu},\ep_{6|\nu}] \big) - (\mu \leftrightarrow \nu) \Big) = 0\, .
\label{poll:shrt}
\eeq
The existence of these relations and their transformation property under $\mathcal{S}_3$ agrees with the prediction in Table \ref{Polltab}. 

Finally, following the procedure discussed in~\eqref{eq:SL2limit} and in section~\ref{sec: extensions} for going from $\mathfrak{u}$ to $\mathfrak{u}_{\SLZ}$ by sending $\ep_{k|\mu}^{(j)}\mapsto \tfrac{2^{k-2}}{1-2^{k-2}}\ep_{k}^{(j)}$ and $\tilde \ep_{2|\nu} \mapsto 0$, this relation reduces to the trivial equality $0=0$.
This is consistent with $\dim\mathcal{S}_{6}(\SLZ)=0$ and the absence of any degree-8 relations in $\mathfrak{u}_{\SLZ}$. The fundamental relation~\eqref{eq: pollack8} and its $\mathfrak{sl}_2$ images under ${\rm ad}_{\ep_0}^j$ with $j=1,2,3,4$ exhaust all relations in $\mathfrak{u}$ at degree~8.

\subsubsection{Relations at degree 10}

Table~\ref{Polltab} predicts two $\mathfrak{sl}_2$-multiplets of fundamental relations at total degree~10. One of modular depth $r^{\text{max}}=2$ coming from the weight-$8$ cusp forms for $\Gamma(2)$, see appendix~\ref{app:mf2}, and another of modular depth $r^{\text{max}}=3$ that is expected to be related to the odd period polynomial of the weight-$6$ cusp form. Let us compare this with the explicit results we find from analysing the derivation algebra~$\mathfrak{u}$.

At total degree~10, we find both derived and fundamental $\mathfrak{sl}_2$ multiplets of relations. Let us first write all the derived relations: 
\begin{itemize}
\item Taking the lowest-weight vector of $\mathfrak{sl}_2$ at degree 8 from \eqref{eq: pollack8}, we can consider its images under the adjoint action of $\tilde{\epsilon}_{2\mid \mu}$ for $\mu=3,4$ to arrive at total degree~10. In this way we get two ($\mathfrak{sl}_2$ multiplets of) relations $[\tilde{\epsilon}_{2\mid \mu},P_8^2]=0$. As $P_8^2$ transforms in the alternating representation $\bep$ and the $\tilde{\epsilon}_{2\mid \mu}$ transform in the standard representation $\mathbf{2}$, the two derived relations transform in the $\bep \otimes \boldsymbol{2}\cong \boldsymbol{2}$ representation. 
\item Furthermore, we can consider the derived relation following from the adjoint action of $z_1$ on $P_8^2$, yielding one derived relation $[z_1,P_8^2]=0$ whose explicit form can be derived from the brackets $[z_1,\ep_{k | \mu}]$ stated in (\ref{newz1.02}) and (\ref{eq: zmatrix}) below. As $z_1$ is an ($\mathfrak{sl}_2$ and) $\mathcal{S}_3$-singlet, see (\ref{zwconj.2}), the derived relation transforms as $\boldsymbol{1}\otimes \bep\cong\bep$. 
\end{itemize}
This means that the derived relations together form a reducible $\mathcal{S}_3$ representation that decomposes as $\bep\oplus \boldsymbol{2}$. Note that all derived relations of degree 10 are of uniform modular depth~3 and therefore have $\rmax=3$. They moreover transform as 5-plets of $\mathfrak{sl}_2$.

We find a total of six relations at total degree $10$ and three of them can be recognised as the derived relations just discussed.
The three remaining fundamental relations feature cases of both $\rmax=2$ and $\rmax=3$:
\begin{itemize}
\item At modular depth $\rmax=2$, we find the expected two 7-plets of $\mathfrak{sl}_2$ of fundamental relations with lowest-weight vectors
\begin{align}
    P_{10(1)}^2&=7 [\tilde{\epsilon}_{2|3},\epsilon_{8|2}]{+}24 [\tilde{\epsilon}_{2|3},\epsilon_{8|3}]{-}10 [\tilde{\epsilon}_{2|3},\epsilon_{8|4}]{+}17 [\tilde{\epsilon}_{2|4},\epsilon_{8|2}]{-}17 [\tilde{\epsilon}_{2|4},\epsilon_{8|3}] \label{eq:p10a}\\
   &\quad{+}136 [\epsilon_{4|2},\epsilon_{6|2}]{+}17 [\epsilon_{4|2},\epsilon_{6|3}]{-}68 [\epsilon_{4|2},\epsilon_{6|4}]{-}17 [\epsilon_{4|3},\epsilon_{6|2}]{-}136 [\epsilon_{4|3},\epsilon_{6|3}]\notag\\
   &\quad{+}68 [\epsilon_{4|3},\epsilon_{6|4}]{-}85 [\epsilon_{4|4},\epsilon_{6|2}]{+}85 [\epsilon_{4|4},\epsilon_{6|3}]=0\,, \notag \\
    P_{10(2)}^2&=-10 [\tilde{\epsilon}_{2|3},\epsilon_{8|2}]{+}24 [\tilde{\epsilon}_{2|3},\epsilon_{8|3}]{+}7 [\tilde{\epsilon}_{2|3},\epsilon_{8|4}]{+}10 [\tilde{\epsilon}_{2|4},\epsilon_{8|2}]{-}7 [\tilde{\epsilon}_{2|4},\epsilon_{8|3}]\notag\\
   &\quad{-}24 [\tilde{\epsilon}_{2|4},\epsilon_{8|4}]{+}85 [\epsilon_{4|2},\epsilon_{6|3}]{-}85 [\epsilon_{4|2},\epsilon_{6|4}]{+}68 [\epsilon_{4|3},\epsilon_{6|2}]{-}136 [\epsilon_{4|3},\epsilon_{6|3}]\notag\\
   &\quad {-}17 [\epsilon_{4|3},\epsilon_{6|4}]{-}68 [\epsilon_{4|4},\epsilon_{6|2}]{+}17 [\epsilon_{4|4},\epsilon_{6|3}]{+}136 [\epsilon_{4|4},\epsilon_{6|4}]=0\,. \notag
\end{align}
The additional subscripts $(1)$ and $(2)$ on the left-hand sides label the two components of the standard representation $\mathbf{2}$ that transform in the same way as $\big( \smallmatrix \tilde{\ep}_{2|3} \\ \tilde{\ep}_{2|4} \endsmallmatrix \big)$ in~\eqref{eq: thetatild}, i.e.,
\begin{align}
   \!\!\! \theta_S^{-1} 
    \begin{pmatrix}
     P_{10(1)}^2\\P_{10(2)}^2
    \end{pmatrix}
    \theta_S
    = \begin{pmatrix}
        0 &1 \\ 1& 0
    \end{pmatrix}
    \begin{pmatrix}
     P_{10(1)}^2\\P_{10(2)}^2
    \end{pmatrix}
    \,,\quad \ \
    \theta_T^{-1} 
    \begin{pmatrix}
     P_{10(1)}^2\\P_{10(2)}^2
    \end{pmatrix}
    \theta_T
    = \begin{pmatrix}
        1 &-1 \\ 0& -1
    \end{pmatrix}
    \begin{pmatrix}
     P_{10(1)}^2\\P_{10(2)}^2
    \end{pmatrix}\,.
\end{align}
Since all derived relations have maximal modular depth $r^{\text{max}}=3$, the above fundamental relations can be identified uniquely.
\item At maximal modular depth $r^{\text{max}}=3$, one has to disentangle fundamental and derived relations.
The expected fundamental relation $P_{10}^3$ at total degree~10 
transforms in the alternating representation~$\bep$. The same applies to the derived relation $[z_1,P_8^2]$ which also carries the same $5$-dimensional $\mathfrak{sl}_2$ representation as $P_{10}^3$.
Therefore,
the fundamental relation $P_{10}^3$ is only well-defined up to the addition of 
rational multiples of $[z_1,P_8^2]$. We have verified the space of degree-10 relations transforming in the alternating representation~$\bep$ to have the expected dimension two from the span of $[z_1,P_8^2]$ and $P_{10}^3$. The following expression is a possible choice of representative for $P_{10}^3$, and its terms of modular depth~2 are well-defined since $[z_1,P_8^2]$ is of uniform modular depth 3:
\end{itemize}
\begin{align}
\label{p103rel}
    P_{10}^3&=-\frac{412}{55} \left[ \left[ \tilde{\epsilon}_{2\mid 3}, \epsilon_{4\mid 2} \right], \epsilon_{4\mid 2} \right]{+}\frac{683}{55} \left[ \left[ \tilde{\epsilon}_{2\mid 3}, \epsilon_{4\mid 3} \right], \epsilon_{4\mid 2} \right]{-}\frac{683}{55} \left[ \left[ \tilde{\epsilon}_{2\mid 3}, \epsilon_{4\mid 4} \right], \epsilon_{4\mid 3}\right]\\
    &+\frac{412}{55} \left[ \left[ \tilde{\epsilon}_{2\mid 3}, \epsilon_{4\mid 4} \right], \epsilon_{4\mid 4} \right]{-}\frac{10}{11} \left[ \left[ \tilde{\epsilon}_{2\mid 3}, \epsilon_{6\mid 2} \right], \tilde{\epsilon}_{2\mid 4} \right]{-}\frac{20}{11} \left[ \left[ \tilde{\epsilon}_{2\mid 3}, \epsilon_{6\mid 4} \right], \tilde{\epsilon}_{2\mid 4} \right]{+}\frac{412}{55} \left[ \left[ \tilde{\epsilon}_{2\mid 4}, \epsilon_{4\mid 2} \right], \epsilon_{4\mid 2} \right] \notag\\
    &- \frac{412}{55}\! \left[ \left[ \tilde{\epsilon}_{2\mid 4}, \epsilon_{4\mid 3} \right], \epsilon_{4\mid 3} \right]{-}\frac{683}{55} \left[ \left[ \tilde{\epsilon}_{2\mid 4}, \epsilon_{4\mid 4} \right], \epsilon_{4\mid 2} \right]{+}\frac{683}{55} \left[ \left[ \tilde{\epsilon}_{2\mid 4}, \epsilon_{4\mid 4} \right], \epsilon_{4\mid 3} \right]{+}\frac{10}{11}\! \left[ \tilde{\epsilon}_{2\mid 3}, \left[ \tilde{\epsilon}_{2\mid 3}, \epsilon_{6\mid 2} \right] \right] \notag\\
    &- \frac{10}{11} \left[ \tilde{\epsilon}_{2\mid 3}, \left[ \tilde{\epsilon}_{2\mid 3}, \epsilon_{6\mid 4} \right] \right]{-}\frac{10}{11} \left[ \tilde{\epsilon}_{2\mid 3}, \left[ \tilde{\epsilon}_{2\mid 4}, \epsilon_{6\mid 2} \right] \right]{-}\frac{20}{11} \left[ \tilde{\epsilon}_{2\mid 3}, \left[ \tilde{\epsilon}_{2\mid 4}, \epsilon_{6\mid 3} \right] \right]{+}\frac{62}{11} \left[ \tilde{\epsilon}_{2\mid 3}, \left[ \epsilon_{4\mid 2}, \epsilon_{4\mid 3} \right] \right] \notag\\
    &-\frac{63}{55} \left[ \tilde{\epsilon}_{2\mid 3}, \left[ \epsilon_{4\mid 2}, \epsilon_{4\mid 4} \right] \right]{-}\frac{373}{55} \left[ \tilde{\epsilon}_{2\mid 3}, \left[ \epsilon_{4\mid 3}, \epsilon_{4\mid 4} \right] \right]{-}\frac{10}{11} \left[ \tilde{\epsilon}_{2\mid 4}, \left[ \tilde{\epsilon}_{2\mid 4}, \epsilon_{6\mid 2} \right] \right]{+}\frac{10}{11} \left[ \tilde{\epsilon}_{2\mid 4}, \left[ \tilde{\epsilon}_{2\mid 4}, \epsilon_{6\mid 3} \right] \right] \notag\\
    &+\frac{63}{55} \left[ \tilde{\epsilon}_{2\mid 4}, \left[ \epsilon_{4\mid 2}, \epsilon_{4\mid 3} \right] \right]{-}\frac{62}{11} \left[ \tilde{\epsilon}_{2\mid 4}, \left[ \epsilon_{4\mid 2}, \epsilon_{4\mid 4} \right] \right]{+}\frac{62}{11} \left[ \tilde{\epsilon}_{2\mid 4}, \left[ \epsilon_{4\mid 3}, \epsilon_{4\mid 4} \right] \right]{-}\left[ \epsilon_{4\mid 2}, \epsilon_{6\mid 3}^{(1)} \right]{+} \left[ \epsilon_{4\mid 2}, \epsilon_{6\mid 4}^{(1)} \right] \notag\\
    &+2 \left[ \epsilon_{4\mid 2}^{(1)}, \epsilon_{6\mid 3} \right]{-}2 \left[ \epsilon_{4\mid 2}^{(1)}, \epsilon_{6\mid 4} \right]{+}\left[ \epsilon_{4\mid 3}, \epsilon_{6\mid 2}^{(1)} \right]{-}\left[ \epsilon_{4\mid 3}, \epsilon_{6\mid 4}^{(1)} \right]{-}2 \left[ \epsilon_{4\mid 3}^{(1)}, \epsilon_{6\mid 2} \right]{+}2 \left[ \epsilon_{4\mid 3}^{(1)}, \epsilon_{6\mid 4} \right] \notag\\
    &-\left[ \epsilon_{4\mid 4}, \epsilon_{6\mid 2}^{(1)} \right]{+}\left[ \epsilon_{4\mid 4}, \epsilon_{6\mid 3}^{(1)} \right]{+}2 \left[ \epsilon_{4\mid 4}^{(1)}, \epsilon_{6\mid 2} \right]{-}2 \left[ \epsilon_{4\mid 4}^{(1)}, \epsilon_{6\mid 3} \right]=0\,.\notag
\end{align}
Alternatively, we can write this relation as
\begin{align}
    P_{10}^3&= \sum_{(\mu,\nu)\in\{(2,3),(3,4),(4,2)\}}\bigg(\Big(-\tfrac{436}{55}\big[[\ep_{2|\mu},\ep_{4|\mu}],\ep_{4|\nu}\big]
-\tfrac{412}{55}\big[[\ep_{2|\nu},\ep_{4|\mu}],\ep_{4|\mu}\big]\\
&
-\tfrac{247}{55}\big[[\ep_{2|\mu},\ep_{4|\nu}],\ep_{4|\mu}\big]+\tfrac{30}{11}\big[[\ep_{2|\nu},\ep_{6|\mu}],\ep_{2|\mu}\big]
+2\big[\ep^{(1)}_{4|\mu},\ep_{6|\nu}\big]
-\big[\ep_{4|\mu},\ep^{(1)}_{6|\nu}\big]\Big)-(\mu\leftrightarrow \nu)\bigg)=0\notag\,,
\end{align}
making the alternating ${\cal S}_3$ representation manifest.

Both the derived and the fundamental relations at degree 10 reduce to the trivial equality $0=0$ under the specialisation~\eqref{eq:SL2limit} to level 1, which agrees with the absence of any relations in $\mathfrak{u}_{\SLZ}$ at this degree.

\subsubsection{Relations at degree 12}

The expected fundamental relations at total degree~12 are listed in Table~\ref{Polltab}. The three $\Gamma(2)$ cusp forms of weight $12$ lead to fundamental relations of maximal modular depth $r^{\text{max}}=2$ in the reducible  $\bep\oplus \mathbf{2}$-representation of $\mathcal{S}_3$. There are also echoes of the lower-weight cusp forms at higher modular depth as shown in the table. As the general discussion is similar to the one of the preceding section, we only give the salient features.

We again begin with the classification of derived relations at degree 12, along with a list of their irreducible representations of ${\cal S}_3$. At $\rmax=3$, we find the following derived relations, 
\begin{align}
    [\tilde\epsilon_{2\mid\mu},P_{10(1,2)}^2]&\longleftrightarrow\boldsymbol{2}\otimes\boldsymbol{2}\cong\boldsymbol{1}\oplus\bep\oplus\boldsymbol{2}\,,\notag\\
    [\epsilon_{4\mid \mu},P_8^2]&\longleftrightarrow(\boldsymbol{1}\oplus \boldsymbol{2})\otimes\bep\cong\bep\oplus\boldsymbol{2}\,,\notag\\
 [z_1,P_{10(1,2)}^2]&\longleftrightarrow\boldsymbol{1}\otimes\boldsymbol{2}\cong\boldsymbol{2}\,.
    \label{deg12pr}
\end{align}
Hence, the derived representations comprise a reducible $\mathcal{S}_3$ representation that decomposes as $\boldsymbol{1}\oplus(2\times \bep)\oplus(3\times\boldsymbol{2})$ with non-trivial multiplicities. This yields a total of 9 $\mathfrak{sl}_2$ multiplets of derived relations at degree 12 with $\rmax=3$. 

At $\rmax=4$, we have the following derived relations, again keeping track of the behaviour under the $\mathcal{S}_3$ action:
\begin{align}
    [\tilde{\epsilon}_{2\mid \mu}, P_{10}^3] &\longleftrightarrow\boldsymbol{2}\otimes\bep\cong \boldsymbol{2}\,,\notag\\
    [z_1,P_{10}^3]&\longleftrightarrow\boldsymbol{1}\otimes \bep\cong \bep\,,\notag\\
    [\tilde{\epsilon}_{2\mid \mu},[\tilde{\epsilon}_{2\mid \nu},P_8^2]]&\longleftrightarrow \boldsymbol{2}\otimes \boldsymbol{2}\otimes \bep\cong\boldsymbol{1}\oplus \bep\oplus \boldsymbol{2}\,,\notag\\
    [z_1,[\tilde{\epsilon}_{2\mid \mu},P_8^2]]&\longleftrightarrow \boldsymbol{1}\otimes \boldsymbol{2}\otimes \bep\cong\boldsymbol{2}\,,\notag\\
    [z_1,[z_1,P_8^2]]&\longleftrightarrow\boldsymbol{1}\otimes \boldsymbol{1}\otimes \bep\cong\bep\,,\notag\\
    t^3(\epsilon_{4|\mu},P_8^2)&\longleftrightarrow (\boldsymbol{1}\oplus\boldsymbol{2})\otimes \bep\cong \bep\oplus \boldsymbol{2}\,.
    \label{comprm4}
\end{align}
The projector $t^3$ appearing in the last line refers only to the $\mathfrak{sl}_2$-representation and works similarly to~\eqref{eq: HWst}. Its exact definition is given in~\eqref{eq:eqv:d3sdef} below.
Together the above relations comprise a reducible representation that decomposes as $\boldsymbol{1}\oplus (4\times \bep)\oplus (4\times \boldsymbol{2})$, giving a total of 13 $\mathfrak{sl}_2$ multiplets of derived relations at $\rmax=4$. 

From the action of $\mathfrak{u}$ as derivations, at $\rmax=2$ we find three $\mathfrak{sl}_2$ multiplets of relations that transform in $\bep\oplus \mathbf{2}$, which are all fundamental relations. Furthermore, at $\rmax=3$ we find 11 $\mathfrak{sl}_2$ multiplets of relations that transform as $\boldsymbol{1}\oplus(2\times \bep)\oplus(4\times\boldsymbol{2})$. This agrees with the derived relations of (\ref{deg12pr}) up to one standard representation $\mathbf{2}$. Hence, we find 2 $\mathfrak{sl}_2$ multiplets of fundamental relations transforming as $\mathbf{2}$. Finally, at $\rmax=4$, we find 14 $\mathfrak{sl}_2$ multiplets of relations that decompose as $\boldsymbol{1}\oplus (5\times \bep)\oplus (4\times \boldsymbol{2})$. This shows upon comparison with (\ref{comprm4}) that we find one $\mathfrak{sl}_2$ 5-plet of fundamental relation at depth 4 that transforms in the alternating representation of ${\cal S}_3$. 

Note that in the cases for which $\rmax=3,4$, the fundamental relations are not unique and well-defined only up to additions of derived relations that transform in the same $\mathcal{S}_3$ and $\mathfrak{sl}_2$ representation as the fundamental relation. Again, these results confirm the predictions made in Table \ref{Polltab}. For explicit formulae for the fundamental relations, we refer to the ancillary file. Again, the transcription \eqref{eq:SL2limit} shows that all derived and fundamental relations at degree 12 reduce to the trivial equality $0=0$ at level 1.

\subsubsection{Relations at degree 14}

Instead of repeating the analysis of finding all derived and fundamental relations at any modular depth of degree 14, we merely focus on the fundamental relations of maximal modular depth $\rmax= 2$. This result is particularly interesting because the prediction in Table \ref{Polltab} features a relation transforming in the trivial $\mathcal{S}_3$ representation $\mathbf{1}$. By its $\mathcal{S}_3$ representation and its $\rmax=2$, this relation is unambiguously defined. As this relation is expected to arise from the Ramanujan cusp form $\Delta$ of weight 12 for the full modular group, we also anticipate that this relation in $\mathfrak{u}$ degenerates to the relation in $\mathfrak{u}_{\SLZ}$ given in the first line of \eqref{eq: pollack}. At $\rmax=2$, we indeed find four $\mathfrak{sl}_2$ multiplets of fundamental relations that transform as $\mathbf{1}\oplus \bep\oplus \mathbf{2}$. We identify the lowest-weight vector in the trivial representation to be
\begin{align}
    P_{14}^{2}&=[\tilde\ep_{2|3},\ep_{12|2}] - 2[\tilde\ep_{2|3},\ep_{12|3}]+ [\tilde\ep_{2|3},\ep_{12|4}] + [\tilde\ep_{2|4},\ep_{12|2}] + [\tilde\ep_{2|4},\ep_{12|3}] - 2[\tilde\ep_{2|4},\ep_{12|4}]\notag\\
    &\quad + \frac{218}{9}[\ep_{4|2},\ep_{10|2}] - \frac{691}{36}[\ep_{4|2},\ep_{10|3}]- \frac{691}{36}[\ep_{4|2},\ep_{10|4}]- \frac{691}{36}[\ep_{4|3},\ep_{10|2}] + \frac{218}{9}[\ep_{4|3},\ep_{10|3}]
    \notag\\
    &\quad - \frac{691}{36}[\ep_{4|3},\ep_{10|4}]- \frac{691}{36}[\ep_{4|4},\ep_{10|2}] - \frac{691}{36}[\ep_{4|4},\ep_{10|3}]+ \frac{218}{9}[\ep_{4|4},\ep_{10|4}]- \frac{188}{3}[\ep_{6|2},\ep_{8|2}] \notag\\
    &\quad + \frac{691}{12}[\ep_{6|2},\ep_{8|3}]+ \frac{691}{12}[\ep_{6|2},\ep_{8|4}]+ \frac{691}{12}[\ep_{6|3},\ep_{8|2}] - \frac{188}{3}[\ep_{6|3},\ep_{8|3}]+ \frac{691}{12}[\ep_{6|3},\ep_{8|4}]
    \notag\\
    &\quad+ \frac{691}{12}[\ep_{6|4},\ep_{8|2}] + \frac{691}{12}[\ep_{6|4},\ep_{8|3}]- \frac{188}{3}[\ep_{6|4},\ep_{8|4}]=0\,,
    \label{pollack14}
\end{align}
or, by making the trivial representation of $\mathcal{S}_3$ manifest, 
\begin{align}
    P_{14}^2&=\sum_{\mu,\nu\in\{2,3,4\}}\Big(
(1-3\,\delta_{\mu\nu})[\ep_{2|\mu},\ep_{12|\nu}]
 \\
&\qquad 
+\big(\tfrac{691}{12}-\tfrac{481}{4}\,\delta_{\mu\nu}\big)[\ep_{6|\mu},\ep_{8|\nu}]
+\big(\tfrac{521}{12}\,\delta_{\mu\nu}-\tfrac{691}{36}\big)[\ep_{4|\mu},\ep_{10|\nu}]
\Big)=0\,.\notag
\end{align}
One can easily check that, following the prescription \eqref{eq:SL2limit}, this relation reduces to (a rational multiple of) the Pollack relation in $\mathfrak{u}_{\SLZ}$ in the first line of \eqref{eq: pollack}. The fundamental relations of maximal modular depth $\rmax=2$ and $3$ at degrees 14, 16 and 18, and of $\rmax=2$ at degree 20, can be found in the ancillary file along with those of lower degree.

\subsection{Implications of Pollack relations and extension to cusp forms}
\label{sec:4.csp}

The relations among the outer derivations in $\mathfrak{u}$ derived in section \ref{sec:lowd} lead to the dropout of the corresponding combination of iterated Eisenstein integrals in the expansion of their generating series \eqref{eq: I2}. These dropouts propagate directly to the equivariant iterated Eisenstein integrals for $\Gamma(2)$ generated by \eqref{eq: eqv2proposal}.  Furthermore, the presence of relations in $\mathfrak{u}$ creates an ambiguity in extracting the coefficients of a given  word in $\ep_{k|\mu}^{(j)}$ in both series. To alleviate this, one can for instance prescribe an ordering convention for words in $\ep_{k|\mu}^{(j)}$ and consistently eliminate those terms in all Pollack relations that are `smallest' within this ad-hoc convention.

The phenomenon of dropouts among iterated Eisenstein integrals is already familiar from level 1 \cite{Brown:2017qwo,Brown:2017qwo2,Dorigoni:2021ngn, Dorigoni:2022npe,Dorigoni:2024oft}, where it had several consequences. First, it implied the absence of periods beyond MZVs in the $S$-cocycle $\mathbb{S}_{\SLZ}(\ep)$ defined by \eqref{lv1coc}, which is in fact a necessary condition for its factorised form (\ref{eq: Scocalt}). Second, it implied that iterated Eisenstein integrals whose modular completion 
requires (iterated) integrals of cusp forms for $\SLZ$ drop out as well. These completions would still define non-holomorphic modular forms obtained from the following modifications to the level-1 connection and the series in zeta generators described in~\cite{Dorigoni:2024oft}:
\begin{itemize}
\item[(i)] lift the Tsunogai derivations $\ep_{k}^{(j)} $ in $\mathbb{I}_{\SLZ}(\ep;\tau)$ to a free Lie algebra of symbols ${\rm e}_{k}^{(j)}$ with $k\geq 4$ even and $j=0,1,\ldots,k{-}2$;
\item[(ii)]
augment the level-1 connection (\ref{eq: N1conn}) by $\mathfrak{sl}_2$ multiplets of weight-$k$ cusp forms $\Delta_k(\tau) \in \mathcal{S}_k(\SLZ)$ and associated letters (denoted by $\mathrm{e}_{\Delta^{\pm}_k}$ in \cite{Dorigoni:2024oft}) that are Lie series in the letters ${\rm e}_{k}^{(j)}$ (with ratios of L-values and more general periods as their coefficients) and vanish by Pollack relations upon reverting to ${\rm e}_{k}^{(j)} \mapsto \ep_{k}^{(j)}$;
\item[(iii)]augment the series $\mathbb M^{\rm sv}(\sigma^{\SLZ})$ by additional periods and associated Lie algebra generators (introduced in~\cite{Brown2019} and denoted by $\hat{\sigma}_\varpi$ in \cite{Dorigoni:2024oft}) which have $\mathfrak{sl}_2$-invariant arithmetic parts and are otherwise Lie series in ${\rm e}_{k}^{(j)}$ that again vanish upon reverting to~$\ep_{k}^{(j)}$;
\item[(iv)] also lift the $\ep_{k}^{(j)} $ in $\sigma_w^{\SLZ,{\rm g}}$ and $[z_w,\ep_k]$ to ${\rm e}_{k}^{(j)}$; the ambiguities of adding combinations that vanish upon reverting ${\rm e}_{k}^{(j)} \mapsto \ep_{k}^{(j)}$ can be either fixed by equivariance or absorbed into redefinitions of the new periods of (iii).
\end{itemize} 
In this setting, $\mathfrak{u}_{\SLZ}$ is realised as the quotient of the free Lie algebra ${\rm Lie}[\mathrm{e}_{k}^{(j)}]$ by the ideal
generated by the letters $\mathrm{e}_{\Delta_k^\pm}$ corresponding to the Pollack relations.

In all examples of equivariant iterated Eisenstein integrals that we constructed at level~2 (see section \ref{ex:sec44} and the ancillary file), we find an exact parallel of the implications at level~1. First, the relations in $\mathfrak{u}$ ensure that no periods beyond cyclotomic MZVs appear in the $S$-cocycle $\mathbb S(\ep)$ from \eqref{defcocyc} which is a necessary condition for its conjectural factorised form (\ref{eq:Scoc2}). 
Second, the relations lead to dropouts of equivariant iterated Eisenstein integrals that would 
require (iterated) integrals of cusp forms for $\Gamma(2)$ in their modular completion. 

Again, we could consider studying more general equivariant iterated integrals for $\Gamma(2)$  
by modifying the level-2 connection and zeta generators similar to the above steps (i) to (iv) of the level-1 case. This would for instance lead to a free Lie algebra on symbols $\tilde{\mathrm{e}}_{2|\nu}$ with $\nu=3,4$ and $\mathrm{e}_{k|\mu}^{(j)}$ with $k\geq 4$ even, $\mu=2,3,4$, and $j=0,1,\ldots,k{-}2$ as well as additional integration kernels in $\mathcal{S}_k(\Gamma(2))$. Then, $\mathfrak{u}$ can be recovered by quotienting the free Lie algebra ${\rm Lie}[\tilde{\mathrm{e}}_{2|\nu},\mathrm{e}_{k|\mu}^{(j)}]$by an ideal corresponding to the level-2 Pollack relations. We leave a detailed investigation of this to future work \cite{wip:cusp}.

\section{Zeta generators and equivariant iterated Eisenstein integrals for level 2}
\label{sec: level2}

In section \ref{sec:motiv}, we introduced the $\Gamma(2)$-connection (\ref{eq: N2conn}) that generates iterated Eisenstein integrals for $\Gamma(2)$ through its path-ordered exponential and originates from the flat CG connection as detailed in section \ref{sec:fltcg}. Furthermore, we proposed an $\SLZ$-equivariant generating series of iterated Eisenstein integrals for $\Gamma(2)$ in section \ref{tau level 2}, which we shall repeat here for the reader's convenience 
\begin{align}
\INNeqv{\tau} &=\mathbb{M}^\sv(z)^{-1}\overline{\INN{\tau}^T}\mathbb{M}^\sv(\sigma)\INN{\tau} \,,
  \label{eq: eqvprop} \\
\INNeqv{\gamma\tau} &=U_\gamma^{-1}\theta^{-1}_\gamma \INNeqv{\tau} \theta_\gamma U_\gamma\,, \hspace{20mm} \forall \ \gamma \in \SLZ
\,,
  \notag
\end{align}
in analogy with the construction of equivariant iterated Eisenstein integrals for $\SLZ$ reviewed in section \ref{sec:lvl1}. The goals of this section are 
\begin{itemize}
\item[(a)] to derive the conditions on the constituents of the generating series, so that the series furnishes 
equivariant iterated integrals, see section \ref{sec: zetaconstraints}, 
\item[(b)] to recursively determine the explicit form of the zeta generators $\sigma_w$ for level 2, including the action of their arithmetic parts $z_w$ on the $\epsilon^{(j)}_{k|\mu}$, see sections \ref{sec: recursion} and \ref{sec: HW vectors}, and
\item[(c)] to spell out representative examples of equivariant iterated Eisenstein integrals obtained from the expansion coefficients of (\ref{eq: eqvprop}) beyond modular depth 1, see section \ref{ex:sec44}.
\end{itemize}
We will in fact construct the expansion of $\INNeqv{\tau}$ and that of $\sigma_w^{\rm g}$ and $[z_w,\ep_{k|\mu}]$ in parallel.
The contributions to $\sigma_w^{\rm g}$ and $[z_w,\ep_{k|\mu}]$ will be derived from necessary conditions for the equivariance (\ref{eq: eqvprop}) of $\INNeqv{\tau}$, and the sufficient conditions hinge among other things on the conjectural factorised form (\ref{eq:Scoc2}) of the $S$-cocycle.

Appendix \ref{app: examples} and the ancillary file provide additional examples of (b) and (c) at higher degrees and modular depths. In particular, the ancillary file contains all equivariant iterated Eisenstein integrals through degree 16 at modular depth 1 and 2, and through degree 10 at modular depth 3, 4 and 5. We have checked their $T$-equivariance numerically at all these degrees and their $S$-equivariance modulo Pollack relations through degree 12 at modular depth 1 and 2 and through degree 10 at modular depth 3, 4 and 5.

\subsection{Constraining zeta generators}
\label{sec: zetaconstraints}

Based on the proposed form of the equivariant generating series for level 2 in the first line of \eqref{eq: eqvprop}, let us derive the constraints enforced on the generating series $\mathbb M^{\rm sv}(\sigma)$ and $\mathbb M^{\rm sv}(z)$ of zeta generators by its $\SLZ$ equivariance in the second line.

\subsubsection{\texorpdfstring{$T$-equivariance}{T-equivariance}}

We first investigate the $T$ transformation of the path-ordered exponential $\mathbb I(\ep;\tau)$ of the level-2 connection in \eqref{prevpexp}. By specialising the $\SLZ$ properties (\ref{trfpexp}) of path-ordered exponentials to $\gamma= T$ and denoting the associated cocycle by $\mathbb{C}_T(\ep) = \mathbb{T}(\ep)$,
we find 
\begin{align}
\label{eq: I2Tcoc}
\INN{\tau{+}1} &= 
 \mathbb{T}(\ep) U_T^{-1}\theta_T^{-1}   \INN{\tau}\theta_T U_T
\, .
\end{align}
The $\gamma = T$ instance of the general formula (\ref{defcocyc}) for cocycles leads to the following integrals, which are entirely determined by the behaviour of the level-2 connection at the cusp and most conveniently performed by taking residues in the variable $q_1 = e^{2\pi i \tau}$, 
\begin{align}
    \mathbb{T}(\ep)={\rm Pexp}\bigg(\int_{i\infty+1 }^{i\infty} \ANN{\tau}\bigg)= e^{2\pi i \mathfrak{N}}e^{2\pi i \ep_0} = e^{2\pi i \mathfrak{N}}U_T \,.
    \label{eq: Tcoc2}
\end{align}
Here, we have introduced the zero mode 
\begin{align}
    \mathfrak{N}&=-\ep_0 +\sum_{k=2}^\infty (k{-}1) \, \bigg\{\frac{B_k}{k!}\ep_{k|2}+\frac{B_k(1/2)}{k!}(\ep_{k|3}+\ep_{k|4})\bigg\}
    \label{eq: N2} \\
&=-\ep_0 +\sum_{k=2}^\infty (k{-}1) \,\BF_k \,\Big\{ \ep_{k|2}+r_k(\ep_{k|3}+\ep_{k|4})\Big\}
    \notag \\
    &= - \ep_0 - \frac{1}{24} (\tilde\ep_{2|3}+ \tilde\ep_{2|4})
- \frac{1}{240} \ep_{4| 2} + \frac{7 }{1920} ( \ep_{4| 3}{+} \ep_{4|4} )+ \ldots \, ,
    \notag
\end{align}
of a $T$-invariant auxiliary connection $\tilde{\mathbb A}(\ep;\tau)  = ( \mathbb K^{(\tau)}_\text{CG}(0,\tau) +\frac{\est_2(\tau)}{2\pi i} t_{2|1}  ) \dd \tau = -2\pi i \mathfrak{N} \dd \tau+{\cal O}(q_1)$, with $\mathbb K^{(\tau)}_\text{CG}(0,\tau)$ given by (\ref{k0lv2}). Since $\tilde{\mathbb A}(\ep;\tau)$ and $\mathbb A(\ep;\tau)$ are gauge equivalent in the sense that $\tilde{\mathbb A}(\ep;\tau) = e^{2\pi i \ep_0 \tau}\ANN{\tau} e^{-2\pi i \ep_0 \tau} + 2\pi i \ep_0 \dd \tau$, the path-ordered exponential of $\ANN{\tau}$ factorises into $\mathbb{I}(\ep;\tau) = {\rm Pexp}(\int_{\tau }^{i\infty} \tilde{\mathbb A}(\ep;\tau)) e^{2\pi i \ep_0 \tau} $, which explains the form $\mathbb{T}(\ep)= e^{2\pi i \mathfrak{N}}e^{2\pi i \ep_0}$ of the $T$-cocycle in (\ref{eq: Tcoc2}); see \cite{brown2017multiple, Dorigoni:2024oft} for the analogous discussion at level 1.

The ellipsis of (\ref{eq: N2}) refers to terms of degree $\geq 6$, and we have introduced the shorthand
\beq
r_n = 
\frac{1-2^{n-1}}{2^{n-1}}\,,
\,\qquad \! \! \text{for}
\ n\in \mathbb Z \ \ \ \ \Rightarrow \ \ \ \
r_k = \frac{B_k(1/2)}{B_k} \,,\qquad \! \! \text{for} \ k\geq 2 \ {\rm even}\,,
\label{defrk}
\eeq
and keep on using the definition $\BF_k=B_k/k!$ of (\ref{defbf}). The degree-two terms of the third line of (\ref{eq: N2}) are presented in terms of the  derivations $\tilde \ep_{2|\nu}$ in (\ref{backtotild}) and reduce to $ \frac{1}{8}\ep_{2|2}$ when imposing $\sum_{\mu=2}^4 \ep_{2|\mu}=0$; see the discussion in section \ref{tau level 2.1}.

The form (\ref{eq: Tcoc2}) of the $T$-cocycle together with $U_T=e^{2\pi i \ep_0}$ further simplifies the transformation \eqref{eq: I2Tcoc} to
\begin{align}
    \INN{\tau{+}1}=e^{2\pi i \mathfrak{N}}\theta_T^{-1}\INN{\tau}\theta_T U_T\,,
    \label{eq: I2Tcocsimple}
\end{align}
where $\mathfrak{N}$ in (\ref{eq: N2}) is evidently invariant under the $\ep^{(j)}_{k|3} \leftrightarrow \ep^{(j)}_{k|4}$ swap $\theta_{T}^{-1}\mathfrak{N}\theta_T=\mathfrak{N}$. With an analogous result for the transposed complex conjugate series, we find the modular $T$ transformation of \eqref{eq: eqvprop} to be
\begin{align}
    \label{eq: Ieqv2Tcoc}
    \INNeqv{\tau{+}1} &= 
    \mathbb M^{\rm sv}(z)^{-1}U_T^{-1} \,  \theta_T^{-1} \overline{\INN{\tau}^T} \theta_T \,
    e^{-2\pi i \mathfrak{N}} \,
    \mathbb M^{\rm sv}(\sigma) \,
     e^{2\pi i \mathfrak{N}}  
    \theta_T^{-1} \INN{\tau} \theta_T U_T\\
    &\overset{!}{=}U_T^{-1} \theta_T^{-1} \mathbb M^{\rm sv}(z)^{-1}\overline{\INN{\tau}^T}  \,\mathbb M^{\rm sv}(\sigma)  \INN{\tau} \theta_T U_T\,,\notag
\end{align}
where the second equality is the result required by $T$-equivariance. The equality needs to hold separately for the non-commuting series on the left and on the right of $\overline{\INN{\tau}^T}$, so we arrive at the following two implications of $T$-equivariance (see section \ref{sec222} for the analogous discussion at level 1):
\begin{itemize}
    \item In the first place, matching the terms in (\ref{eq: Ieqv2Tcoc}) on the \emph{left} of $\overline{\INN{\tau}^T}$ imposes the series in arithmetic zeta generators to obey $\mathbb M^{\rm sv}(z)^{-1}U_T^{-1} \theta_T^{-1} = U_T^{-1} \theta_T^{-1} \mathbb M^{\rm sv}(z)^{-1}$. Since non-trivial $U_T$ transformations in the second line of (\ref{eq: US}) always introduce powers of $2\pi i $ while $\theta_T$ in (\ref{eq: theta}) and (\ref{eq: thetatild}) does not, we conclude that $\mathbb M^{\rm sv}(z)$ is separately invariant under conjugation by $U_T$ and $\theta_T$. Moreover, this has to hold along with each $\mathbb Q$-linearly independent cyclotomic MZV for level 2, and thus for the coefficients $z_w$ of $\rho^{-1}(f_w)$ for all $w\geq 1$ odd, leading to the $\gamma=T$ instance of the conditions previewed in (\ref{zwconj.1}) and~(\ref{zwconj.2}):
\begin{align}
   \theta_T^{-1} \mathbb M^{\rm sv}(z) \theta_T
   &= \mathbb M^{\rm sv}(z)
   \ \ \ \Rightarrow \ \ \ \ \theta_T^{-1} {z}_w \theta_T = {z}_w\,,\notag\\
   U_T^{-1}\mathbb M^\sv(z) U_T&=\mathbb{M}^\sv(z) \ \ \  \Rightarrow \ \ \ \ \ \ \, [\ep_0,z_w]=0\,.
\label{eq: Tconditions} 
\end{align}
\item Matching the terms in (\ref{eq: Ieqv2Tcoc}) on the \emph{right} of $\overline{\INN{\tau}^T}$ imposes as a condition on the series in the full level-2 zeta generators that $\theta_T  e^{-2\pi i \mathfrak{N}}   
    \mathbb M^{\rm sv}(\sigma) 
    e^{2\pi i \mathfrak{N}}   \theta_T^{-1}
    = \mathbb M^{\rm sv}(\sigma)$. Again, this has to hold separately along with each power of $2\pi i$ and cyclotomic MZV $\rho^{-1}(f_{i_1}\ldots f_{i_r})$ with $r\geq 0$ and $i_1,\ldots,i_r\geq 1$ odd from $\mathbb M^{\rm sv}(\sigma)$, which refines the condition to
\begin{align}
    \sigma_{w} = \theta_T^{-1}\sigma_{w}\theta_T \, , \ \ \ \ 
    [\sigma_w , \mathfrak{N}] = 0 \, , \ \ \ \ \textrm{for} \ \, w\geq 1 \, \ \textrm{odd}\, .
\label{eq: moreTinf} 
\end{align}
\end{itemize}
As we will see in section \ref{sec: recursion} below, the properties (\ref{eq: Tconditions}) and (\ref{eq: moreTinf}) of level-2 zeta generators $\sigma_w$ and their arithmetic parts $z_w$ have rich input on the recursive construction of the geometric parts $\sigma_w-z_w$ and the brackets $[z_w, \epsilon_{k|\mu}]$. 

\subsubsection{\texorpdfstring{$S$-equivariance}{S-equivariance}}
\label{sec:sequv}

The next step is to explore the analogous equivariance properties for the generator $S$ of $\SLZ$. The modular transformation (\ref{trfpexp}) of path-ordered exponentials for $\gamma = S$ yields
\begin{align}
    \INN{-\tfrac{1}{\tau}}&=
    \mathbb{S}(\ep)U_S^{-1}\theta_S^{-1}\INN{\tau}\theta_S U_S\, ,
     \label{eq: IS}
\end{align}
where the $S$-cocycle $\mathbb{S}(\ep) = \mathbb{C}_S(\ep)$ is given by
\beq
\mathbb{S}(\ep) = {\rm Pexp}\bigg(\int_{0 }^{i\infty} \ANN{\tau}\bigg)\,,
\label{scoc.01}
\eeq
and is found to enjoy the following inversion property by equating the two expressions in (\ref{defcocyc}) for $\gamma=S$,
\beq
\mathbb{S}(\ep)^{-1}=\theta_SU_S\mathbb{S}(\ep)U_S^{-1}\theta_S^{-1}\, .
\label{scoc.02}
\eeq
While the $T$-cocycle follows the compact closed-form expression (\ref{eq: Tcoc2}), the expansion of $\mathbb{S}(\ep)$ in the level-2 derivations is more complicated and exhibits a richer number-theoretic structure. By the definition of the path-ordered exponential in (\ref{scoc.01}), the terms 
$\ep_{k_1|\mu_1}^{(j_1)} \ldots \ep_{k_r|\mu_r}^{(j_r)} $ of modular depth $r\leq 2$
in its expansion read
\begin{align}
    \label{eq: Scocgen}
    &\mathbb S(\ep) = 1 + \sum^\infty_{k=2} (k{-}1)\sum_{j=0}^{k-2}\,\sum_{\mu=2}^{4} \frac{ (-1)^j}{j!}\,(2\pi i)^{j+1-k}\,\MMVo{j}{k}{\omega_\mu}\,\ep_{k|\mu}^{(j)} \\
    &\quad \! \!  \! + \! \! \!  \sum_{k_1,k_2=2}^\infty \! \! \!  (k_1{-}1)  (k_2{-}1)\sum_{j_1=0}^{k_1-2}\,\sum_{j_2=0}^{k_2-2}\,\sum_{\mu,\nu=2}^{4}  \! \frac{(-1)^{j_1+j_2}}{j_1!\,j_2!}\,(2\pi i)^{j_1+j_2+2-k_1-k_2}\,\MMVo{j_1 & j_2}{k_1 & k_2}{\omega_\mu & \omega_\nu}\,\ep_{k_1|\mu}^{(j_1)}\ep_{k_2|\nu}^{(j_2)}+\ldots\notag\, ,
\end{align}
with the following regularised $\tau \rightarrow 0$ limits of the iterated Eisenstein integrals (\ref{eq: eez}) for $\Gamma(2)$: 
\begin{align}
	\MMVo{j_1 &\ldots &j_r}{k_1 &\ldots &k_r}{\omega_{\mu_1} &\ldots &\omega_{\mu_r}} &=(-1)^r  \int^{i \infty}_{0} \dd \tau_r \, \tau_r^{j_r} f^{(k_r)}(\omega_{\mu_r},\tau_r)\int^{i \infty}_{\tau_r} \dd \tau_{r-1} \, \tau_{r-1}^{j_{r-1}} f^{(k_{r-1})}(\omega_{\mu_{r-1}},\tau_{r-1})\notag\\ 
    &\quad\quad\quad \ldots \int^{i \infty}_{\tau_3} \dd \tau_2 \, \tau_2^{j_2} f^{(k_2)}(\omega_{\mu_2},\tau_2) \int^{i \infty}_{\tau_2} \dd \tau_1 \, \tau_1^{j_1} f^{(k_1)}(\omega_{\mu_1},\tau_1)  \, .
	\label{eq: mmvdef}
\end{align}
By analogy with their counterparts for $\SLZ$ in section~\ref{sec:lvl1coc}, these limits of iterated Eisenstein integrals are referred to as MMVs, and in fact necessitate tangential-base-point regularisation \cite{brown2017multiple} at both endpoints $\tau_1= i\infty$ and $\tau_r= 0$. The treatment of this two-fold regularisation in the numerical evaluation of MMVs is discussed in appendix \ref{sec:num.mmv}.

While MMVs for modular forms of $\SLZ$ were for instance studied in \cite{brown2017multiple, Brown2019, saad2020multiple, Dorigoni:2021ngn, Dorigoni:2024oft, Kleinschmidt:2025dtk}, generalisations to congruence subgroups and to iterated integrals of $f^{(k)}(u\tau{+}v,\tau)$ at fixed $u,v\in \mathbb R$ are discussed in \cite{Duhr_2020,Duhr:2020gdd, Drewitt:2023con, Duhr:2025lvr} and \cite{Zudilin:2023}, respectively. The regularisations of the MMVs in (\ref{eq: mmvdef}) preserve the reality properties          \begin{align}
	\MMVo{j_1 &\ldots &j_r}{k_1 &\ldots &k_r}{\omega_{\mu_1} &\ldots &\omega_{\mu_r}} = (-1)^{r+j_1+\ldots+j_r}\overline{\MMVo{j_1 &\ldots &j_r}{k_1 &\ldots &k_r}{\omega_{\mu_1} &\ldots &\omega_{\mu_r}}} \,,
	\label{eq: mmvconj}
\end{align}
due to $f^{(k)}(u\tau{+}v,\tau) \in \mathbb R$ for $u,v \in \mathbb Z/2$ and $\tau \in i \mathbb R$ (see e.g.\ (\ref{eq: FKroneis})) and the reflection~property
\begin{align}
		\MMVo{j_1 &\ldots &j_r}{k_1 &\ldots &k_r}{\omega_{\mu_1} &\ldots &\omega_{\mu_r}} = (-1)^{r+j_1+\ldots+j_r}\MMVo{k_r-j_r-2 &\ldots &k_1-j_1-2}{k_r &\ldots &k_1}{\omega_{6-\mu_r} &\ldots & \omega_{6-\mu_1}}
		\label{eq: mmvref} \, ,
\end{align}
which follows from the change of integration variables $\tau_i ' =-1/\tau_i$ in (\ref{eq: mmvdef}), and where $\omega_\mu \rightarrow \omega_{6-\mu}$ swaps $\omega_2 \leftrightarrow \omega_4$ according to the $S$ transformation of half-periods. While the generating-series form of (\ref{eq: mmvref}) is simply the inversion formula (\ref{scoc.02}), the complex-conjugation properties (\ref{eq: mmvconj}) together with the factor of $2\pi i$ in (\ref{eq: Scocgen}) translate into the reality of
\beq
\overline{\mathbb S(\epsilon)} = \mathbb S(\epsilon) \, .
\label{realscoc}
\eeq

With these properties in place, we shall now insert the $S$-transformation \eqref{eq: IS} of $\mathbb I(\ep;\tau)$ into the equivariance condition (\ref{eq: eqvprop}),
\begin{align}
 \INNeqv{-1/\tau}&=\mathbb{M}^\sv(z)^{-1} U_S^{-1} \theta_S^{-1} \overline{\INN{\tau}^T}\theta_S U_S \overline{\mathbb{S}(\ep)^T} \mathbb{M}^\sv(\sigma)\mathbb{S}(\ep) U_S^{-1}\theta_S^{-1}\INN{\tau}\theta_SU_S\notag\\
    &\overset{!}{=}U_S^{-1} \theta_S^{-1}\mathbb{M}^\sv(z)^{-1} \overline{\INN{\tau}^T} \mathbb{M}^\sv (\sigma)\INN{\tau} \theta_S U_S\, ,
    \label{eq: Seqvforced}
\end{align}
where the exclamation mark in the second line signals an imposed equality. The implications can again be divided into a matching of the series on the left and right of $\overline{\INN{\tau}^T}$ in (\ref{eq: Seqvforced}):
\begin{itemize}
\item The terms on the \emph{left} of $\overline{\INN{\tau}^T}$
impose $\mathbb M^{\rm sv}(z)^{-1}U_S^{-1} \theta_S^{-1} = U_S^{-1} \theta_S^{-1} \mathbb M^{\rm sv}(z)^{-1}$. Isolating the coefficient of $\rho^{-1}(f_w)$ results in 
\beq
z_w = U_S^{-1}\theta_S^{-1} z_w \theta_S U_S\,,
\label{intermed}
\eeq
which can in fact be used to demonstrate separate invariance of $z_w$ under both $U_S$ and $\theta_S$: Since we have already shown $[\ep_0,z_w] = 0$ in (\ref{eq: Tconditions}) and $[\ep_0, U_S^{-1} \delta U_S] \sim U_S^{-1}[\ep_0^\vee , \delta] U_S$ for arbitrary derivations $\delta$ by virtue of (\ref{eq: US}), applying ${\rm ad}_{\ep_0}$ to both sides of (\ref{intermed}) yields $0 = \theta_S^{-1} U_S^{-1}[\ep_0^\vee , z_w]U_S \theta_S$ and therefore $[\ep_0^\vee , z_w] = 0$, using that $\theta_S$ commutes with $\ep_0$ and $U_S$. Given that $z_w$ is annihilated by both of ${\rm ad}_{\ep_0}$ and ${\rm ad}_{\ep_0^\vee}$, (\ref{intermed}) then implies that
\begin{align}
   \theta_S^{-1}\mathbb M^{\rm sv}(z) \theta_S
   &= \mathbb M^{\rm sv}(z)
   \ \ \  \Leftrightarrow \ \ \ \ \ \,\theta_S^{-1} z_w \theta_S = z_w \, ,\notag\\
   U_S^{-1}\mathbb M^{\rm sv}(z) U_S
   &= \mathbb M^{\rm sv}(z)
   \ \ \ \Leftrightarrow \ \ \ \ U_S^{-1} z_w U_S = z_w\, .
\label{eq: Sconditions} 
\end{align}
This concludes our derivation of $\mathfrak{sl}_2$-invariance $[\ep_0,z_w]=[\ep_0^\vee,z_w]=0$ and ${\cal S}_3$ invariance $\theta_\gamma^{-1} z_w \theta_\gamma =z_w \ \forall \ \gamma \in \SLZ$ of the arithmetic generators,  as previewed in (\ref{zwconj.1}),~(\ref{zwconj.2}).
\item By the reality (\ref{realscoc}) of the $S$-cocycle, the terms on the \emph{right} of $\overline{\INN{\tau}^T}$ in (\ref{eq: Seqvforced}) result in the equivariance condition
\begin{align}
   \mathbb{S}(\ep)^T \mathbb{M}^{\rm sv}(\sigma) \mathbb{S}(\ep) &= U_S^{-1} \theta_S^{-1} \mathbb{M}^{\rm sv}(\sigma)\theta_S U_S\,.
    \label{eq: ScondSigma}
\end{align}
As we shall now demonstrate, this bilinear relation in the $S$-cocycle $\mathbb{S}(\ep)$ follows from its conjectural factorised form (\ref{eq:Scoc2}). The latter can be used to rewrite the left side of (\ref{eq: ScondSigma}) as follows upon introducing the shorthand $R_S = \theta_S U_S$ (such that $\mathbb{S}(\ep)=\mathbb M(\sigma)^{-1} \mathbb X(\ep) R_S^{-1} \mathbb M(\sigma) R_S$): 
\begin{align}
\mathbb{S}(\ep)^T \mathbb{M}^{\rm sv}(\sigma) \mathbb{S}(\ep) &= R_S^T \mathbb M(\sigma)^T (R_S^T)^{-1} \mathbb X(\ep)^T \big( \mathbb M(\sigma)^T \big)^{-1} \mathbb M^{\rm sv}(\sigma) \mathbb M(\sigma)^{-1} \mathbb X(\ep) R_S^{-1} \mathbb M(\sigma) R_S \notag \\
&= R_S^{-1} \mathbb M(\sigma)^T R_S \mathbb X(\ep)^T   \mathbb X(\ep) R_S^{-1} \mathbb M(\sigma) R_S 
\notag \\
&= R_S^{-1} \mathbb M(\sigma)^T \mathbb M(\sigma) R_S\notag \\
&= U_S^{-1} \theta_S^{-1} \mathbb{M}^{\rm sv}(\sigma)\theta_S U_S\, .
\label{prf.sequi}
\end{align}
In passing to the second line, we have used $\mathbb{M}^{\rm sv}(\sigma) = \mathbb{M}(\sigma)^T \mathbb{M}(\sigma)$ (equivalent to (\ref{svfalp}) for the single-valued map) and the transposition property $R^{T}_S \mathbb M(\sigma)^T (R_S^T)^{-1} = R_S^{-1} \mathbb M(\sigma)^T R_S$. The latter hinges on the odd dimensionality of the $\mathfrak{sl}_2$ representations in the expansion of $\mathbb M(\sigma)^T$ (to convert the conjugation by $U^{T}_S$ into that by $U^{-1}_S$), and the general property $(\theta_S^{-1} \mathbb Y \theta_S)^T = \theta_S^{-1} (\mathbb Y)^T \theta_S$ of the transposition for arbitrary group-like series $\mathbb Y$ in $z_w$ and $\ep_{k| \mu}^{(j)}$.\footnote{The property $(\theta_S^{-1} \mathbb Y \theta_S)^T = \theta_S^{-1} (\mathbb Y)^T \theta_S$ can be seen from the rewriting $(\mathbb Y)^T =  {\rm sgn}_a (\mathbb Y^{-1} )$, where ${\rm sgn}_a ( \ep_{k|\mu}^{(j)} ) =(-1)^{1+j}  \ep_{k|\mu}^{(j)}$ and ${\rm sgn}_a(z_w) = -z_w$ according to the parity of the $a$-degree. The inverse within ${\rm sgn}_a (\mathbb Y^{-1} )$ reverses the concatenation order of $\ep_{k|\mu}^{(j)}$, $z_w$ with alternating signs which then conspires with the eigenvalues of ${\rm sgn}_a$ to match $(\ldots \ep_{k_1|\mu_1}^{(j_1)} \ldots \ep_{k_2|\mu_2}^{(j_2)} \ldots)^T = (-1)^{j_1+j_2+\ldots}(\ldots \ep_{k_2|\mu_2}^{(j_2)} \ldots \ep_{k_1|\mu_1}^{(j_1)}\ldots)$. Since conjugation by $\theta_S$ commutes with the series inversion and does not modify the $a$-degree, it also commutes with the transposition $(\mathbb Y)^T =  {\rm sgn}_a (\mathbb Y^{-1} )$.} The third line of (\ref{prf.sequi}) is obtained from the consequence $\mathbb X(\ep)^T  = \mathbb X(\ep)^{-1}$ of  assuming even $a$-degrees for each term in the expansion of the group-like series $\mathbb X(\ep)$, see the discussion around (\ref{eq:Scoc2}). 
\end{itemize}

Note that (\ref{eq:Scoc2}) is easily checked to obey the reality property (\ref{realscoc}), and the inversion formula (\ref{scoc.02}) requires the contributions that are series with $\mathbb Q[\pi^2]$ coefficients to obey $\mathbb{X}(\ep)^{-1}=\theta_SU_S\mathbb{X}(\ep)U_S^{-1}\theta_S^{-1}$. Moreover, the $\mathfrak{sl}_2$- and ${\cal S}_3$-invariance of $z_w$, together with the property $[z_w, \mathfrak{u}] \subset \mathfrak{u}$, are crucial for (\ref{eq:Scoc2}) to be expressible in $\exp(\mathfrak{u})$, i.e., as a group-like series in $\ep_{k|\mu}^{(j)}$.

\subsubsection{Multiple modular values at depth 1 and beyond}

Here we shall spell out the MMVs (\ref{eq: mmvdef}) of modular depth $r=1$ and comment on their number-theoretic content at modular depth $r\geq 2$.

For the extremal values $j=0$ and $j=k{-}2$, which coincide at $k=2$, MMVs $\MMVo{j}{k}{\omega_\mu}$ of modular depth 1 evaluate~to
\begin{align}
	\MMVo{0}{2}{\omega_\mu} &= 2\pi i \log(2)\, \big(\delta_{\mu,4}-\delta_{\mu,2}\big) \, , \notag \\
	\MMVo{k-2}{k}{\omega_\mu} &= -\delta_{\mu,2}\, \frac{2\pi i}{k{-}1} \frac{2^{k-2}-1}{2^{k-2}} \zeta_{k-1} \, , \ \ \ \ k > 2\, , \notag \\
	\MMVo{0}{k}{\omega_\mu} &= \delta_{\mu,4}\, \frac{2\pi i}{k{-}1} \frac{2^{k-2}-1}{2^{k-2}} \zeta_{k-1} \, , \ \ \ \ k > 2 \, ,
	\label{eq: mmvclosed}
\end{align}
which are special cases of a more general formula in \cite{Duhr:2025lvr} in terms of Clausen values (see also~\cite{Heumann2014}).
The remaining cases of $1\leq j \leq k{-}3$ can be readily expressed for general $u,v\in[0,1)$ in terms of Bernoulli polynomials \cite{Zudilin:2023, Duhr:2025lvr}
\begin{align}
	\MMVo{j}{k}{u\tau+v} = (-1)^{j+1} \frac{(2\pi i)^{k}}{(k{-}1)!} \frac{B_{k-j-1}(u)}{k{-}j{-}1}\frac{B_{j+1}(v)}{j{+}1}\, , \ \ \ \ {\rm for} \ \, j=1,2,\ldots,k{-}3 \, .
	\label{eq: mmvbernoulli}
\end{align}
At higher modular depths, MMVs fall into much wider classes of periods. Already their level-1 instances are known to exhaust the $\mathbb Q$ algebra of MZVs \cite{saad2020multiple} as reflected in (\ref{eq: Scocalt}). At level 2, by
tracking the appearance of $\rho^{-1}(f_{i_1}\ldots f_{i_r})$ with $i_1,\ldots,i_r \geq 1$ odd in the conjectured form (\ref{eq:Scoc2}) of the $S$-cocycle for $\Gamma(2)$, we find that similarly, all $\mathbb Q$ linear combinations of (motivic) cyclotomic MZVs for level 2 are realised as MMVs.
Moreover, individual MMVs of modular depth $\geq 2$ and level $\geq 1$ generically go beyond (cyclotomic) MZVs, and for instance include non-critical $L$-values of holomorphic cusp forms, see for example \cite{brown2017multiple, Brown2019, Duhr:2020gdd,Dorigoni:2021ngn, Dorigoni:2024oft}. 

However, the MMVs entering the expansion (\ref{eq: Scocgen}) of the $S$-cocycle for $\Gamma(2)$ are accompanied by derivations $\ep_{k|\mu}^{(j)}$ subject to the
level-2 generalisations of Pollack relations discussed in section \ref{sec: pollack}. These relations among $\ep_{k|\mu}^{(j)}$ are expected to project out all the number-theoretic content of level-2 MMVs beyond cyclotomic MZVs. In upcoming work, we plan to elaborate on MMVs beyond cyclotomic MZVs and their significance for the iterated Eisenstein integrals from a more general level-2 connection where $\ep_{k|\mu}^{(j)} \rightarrow \mathrm{e}_{k|\mu}^{(j)}$ with $k\geq 2$ even and $j=0,1,\ldots,k{-}2$ are promoted to generate a free Lie algebra \cite{wip:cusp}.

\subsection{\texorpdfstring{Level-2 recursion for $\sigma^{\rm g}_w$ and $[z_w,\ep_{k|\mu}]$}{Level-2 recursion for sigma(g,w) and [zw,epsilon(k|mu)]}}
\label{sec: recursion}

Extracting the explicit form of equivariant iterated Eisenstein integrals of modular depth~$r$
from the generating series \eqref{eq: eqvprop} requires the following ingredients:
\begin{itemize}
    \item geometric contributions $\sigma_w^{\rm g}\in\mathfrak{u}$ to level-2 zeta generators up to modular depth $r$, which are Lie series in $\ep_{k|\mu}^{(j)}$ of  unbounded degree, and
    \item commutators $[z_w,\ep_{k|\mu}]\in\mathfrak{u}$ of their arithmetic parts up to modular depth $r$, which are expressible as Lie polynomials in $\ep_{k|\mu}^{(j)}$ of total degree $2w{+}k$.
\end{itemize}
This section presents the first out of two steps in constructing $\sigma^{\rm g}_w $ and $[z_w,\ep_{k|\mu}] $ recursively in the modular depth of the contributing Lie polynomials. The subsequent first step exploits the $T$-equivariance conditions (\ref{eq: Tconditions}) and (\ref{eq: moreTinf}) of $\INNeqv{\tau} $, but leaves highest-weight vectors of $\mathfrak{sl}_2$ (annihilated by ${\rm ad}_{\ep_0}$) undetermined. The second step of the recursion, to be described in section \ref{sec: HW vectors} below, will fix the missing highest-weight vectors at each modular depth from the $S$-equivariance conditions of section \ref{sec:sequv}, more specifically by imposing the factorised form (\ref{eq:Scoc2}) of the $S$-cocycle. The two-step construction of $\sigma^{\rm g}_w $ and $[z_w,\ep_{k|\mu}] $ for $\Gamma(2)$ closely follows the recursion for zeta generators at level 1 (see sections \ref{sec222} and \ref{sec:zetaScoc}) and determines the action of level-2 zeta generators on $\mathfrak{u}$ to any desired modular depth and degree.

Throughout this section, whenever we write $\ep_{k|\mu}^{(j)}$, it is understood to include the case $k=2$ where we fix the shift redundancy of degree-2 derivations $\ep_{2|\mu}$ in the unified form (\ref{newlv2con}) of the level-2 connection by imposing their sum to vanish,
\beq
\sum_{\mu=2}^4\ep_{2|\mu}= 0 \ \ \ \ \Rightarrow \ \ \  \ \ep_{2|3} = \frac{1}{3} (2 \tilde \ep_{2|3} - \tilde \ep_{2|4}) \, , \ \ \ \ 
\ep_{2|4} = \frac{1}{3} (2 \tilde \ep_{2|4} - \tilde \ep_{2|3}) \, .
\label{fixeps}
\eeq
This can be used to express the subsequent results in terms of the derivations $\tilde \ep_{2|\nu}$ accompanying the independent $f^{(2)}(\omega_\nu,\tau)$ with $\nu=3,4$ in (\ref{eq: N2conn}).

Let $X^{[r]}$ denote the contribution of modular depth $r$ to some Lie series $X$, i.e., from Lie polynomials in $\ep_{k_1|\mu_1}^{(j_1)}\ldots \ep_{k_r|\mu_r}^{(j_r)}$. Then we have 
\begin{align}
\sigma_w&=z_w+\sigma_w^{[1]}+\sigma_w^{[2]}+\dots+\sigma_w^{[w]}\,,\label{eq: sig[z,ep]}\\
    [z_w,\ep_{k|\mu}]&=[z_w,\ep_{k|\mu}]^{[2]}+[z_w,\ep_{k|\mu}]^{[3]}+\dots+[z_w,\ep_{k|\mu}]^{[w+1]}\,.
    \notag
\end{align}
Recall that generic Pollack relations in $\mathfrak{u}$ such as (\ref{p103rel}) relate commutator brackets of different modular depths. Consequently, the contributions to \eqref{eq: sig[z,ep]} are not well-defined until one fixes a scheme for applying those Pollack relations that mix modular depths. Nevertheless, the recursive relations derived below for the right-hand sides of \eqref{eq: sig[z,ep]} hold for any such scheme, provided the same choice is applied consistently at all modular depths $r \geq 2$. Moreover, we will only compute terms of modular depth $\geq 3$ up to degree 10. In this range, the only Pollack relations that mix modular depths are $P_{10}^3$ in (\ref{p103rel}) and its $\mathfrak{sl}_2$-descendants. Since these transform in the alternating representation of ${\cal S}_3$, they cannot contribute to the $\theta_T$-invariant $\sigma_w$ or to the commutators $[z_w,\ep_{k|\mu}]$, which transform like the $\ep_{k|\mu}$. Our results therefore do not depend on this choice.

The bounds on the modular depth $r\leq w$ on $\sigma_w^{[r]}$ and $r\leq w{+}1$ on $[z_w,\ep_{k|\mu}]^{[r]}$ in (\ref{eq: sig[z,ep]}) follow from the fact that $\ep_{k|\mu}^{(j)}$ have $a$-degree $j{+}1$, and the conjecture that each term of $\sigma_w$ including $z_w$ has $a$-degree $w$ (see (iii) below (\ref{eq:Scoc2})), leading to $a$-degree $w{+}1$ of $[z_w,\ep_{k|\mu}]$. Similarly, the vanishing of $[z_w,\ep_{k|\mu}]^{[1]}$ follows from the fact that the only expression in $\mathfrak{u}$ of modular depth 1 compatible with its $a$-degree $w{+}1$ and total degree $2w{+}k$ is $\ep_{2w+k|\nu}^{(w)}$ which is ruled out by its $\mathfrak{sl}_2$ properties.\footnote{By $w\geq 1$ and (\ref{sl2epkmu}), a putative term $[z_w,\ep_{k|\mu}]^{[1]} \sim \ep_{2w+k|\nu}^{(w)}$ is in conflict with the fact that $z_w$ and $\ep_{k|\mu}$, and thus also $[z_w,\ep_{k|\mu}]$, are lowest-weight vectors, i.e., they are annihilated by~${\rm ad}_{\ep_0^\vee}$.}

Employing $(a,b)$-degree counting arguments similar to those in \cite{Dorigoni:2024iyt}, one can furthermore infer that $\sigma_w$ can only exhibit highest-weight vectors up to and including degree $2w$ in the expansion of its geometric part $\sigma_w^{\rm g} \in\mathfrak{u}$. Moreover, the highest-weight vectors at the key degree $2w$ of $\sigma_w$ are at the same time $\mathfrak{sl}_2$ singlets. 
Extending the logic in Theorem 5.4.1 (vi) of \cite{Dorigoni:2024iyt} to level 2, we try to absorb as many highest-weight vectors of $\sigma_w$ at key degree $2w$ as possible into the $\mathfrak{sl}_2$-invariant arithmetic part $ z_w$. However, by the additional $\theta_S,\theta_T$-invariance of $z_w$ at level 2, this is only possible for highest-weight vectors at key degree $2w$, which are at the same time in the trivial $\mathbf{1}$ representation of $\mathcal{S}_3$.
Hence, the highest degree of the geometric part $\sigma_w^{\rm g}$ that is not yet determined by
the $T$-equivariance analysis of this section is $2w{-}2$ for $\mathcal{S}_3$-invariant highest-weight vectors and $2w$ for highest-weight vectors which transform non-trivially under $\theta_S$. By the first condition of (\ref{eq: moreTinf}), even those contributions to $\sigma_w$ in non-trivial representations of ${\cal S}_3$ are automatically $\theta_T$-invariant.

\subsubsection{\texorpdfstring{Organising the $T$-equivariance condition according to modular depth}{Organising the T-equivariance condition according to modular depth}}
\label{org:mdp}

The driving force of this section is to organize the $T$ equivariance condition (\ref{eq: moreTinf}) into
\beq
\big[\sigma_w , -\ep_0 + \mathfrak{N}^{[1]} \big] = 0 \, , \quad \quad \mathfrak{N}^{[1]} = \sum_{k=2}^\infty (k{-}1)\, \BF_k \, \Big\{ \ep_{k|2}+r_k(\ep_{k|3}+\ep_{k|4})\Big\}\, ,
\label{spltn}
\eeq
where $\mathfrak{N} = -\ep_0 + \mathfrak{N}^{[1]}$ given by (\ref{eq: N2}) is split into its terms of modular depth zero and one. The rational numbers $\BF_k$ and $r_k$ (not to be confused with our placeholder $r$ for modular depth)  are defined in (\ref{defbf}) and (\ref{defrk}), respectively. The vanishing commutator in (\ref{spltn}) then decomposes as follows according to the terms in (\ref{eq: sig[z,ep]}) of different modular depths (using $[\ep_0,z_w]=0$ and $[z_w,\ep_k]^{[1]}=0$),
\begin{align}
    \label{eq: fullrecursion}
    0=[\sigma_w,\mathfrak{N}]&=
  \big[  \sigma_w^{[1]}+\sigma_w^{[2]}+\dots+\sigma_w^{[w]} ,  \mathfrak{N}^{[1]} \big] 
 + \big[ \ep_0 , \sigma_w^{[1]}+\sigma_w^{[2]}+\dots+\sigma_w^{[w]}\big] \\
&\quad + \big[ z_w , \mathfrak{N}^{[1]} \big]^{[2]}
+ \big[ z_w , \mathfrak{N}^{[1]} \big]^{[3]}
+\ldots 
+  \big[ z_w , \mathfrak{N}^{[1]} \big]^{[w+1]}
\,.\notag
\end{align}
Isolating contributions to \eqref{eq: fullrecursion} at fixed modular depth $r=1,2,\dots,w{+}1$, we find
\begin{align}
    \label{eq: recursion}
\Big[\ep_0,\sigma_w^{[r]}\Big]+ \big[z_w,\mathfrak{N}^{[1]}  \big]^{[r]}  = \big[
\mathfrak{N}^{[1]} 
,\sigma_w^{[r-1]}\big]\,.
\end{align}
By $[z_w,\ep_k]^{[1]}=0$ and $\sigma_w^{[0]}=0$, the $r=1$ instance of \eqref{eq: recursion} turns out to yield $[\ep_0,\sigma_w^{[1]}]=0$. This implies that all the geometric contributions at modular depth 1 must be highest-weight vectors to be quoted in (\ref{eq: depth1}) below.

\subsubsection{\texorpdfstring{Isolating lowest-weight vectors of $\mathfrak{sl}_2$ and their complements}{Isolating lowest-weight vectors of sl2 and their complements}}
\label{org:sl2}

Following section 7.3 of \cite{Dorigoni:2024iyt}, each term in \eqref{eq: recursion} can be uniquely decomposed into $\mathfrak{sl}_2$ lowest-weight vectors annihilated by  $\ad_{\ep^\vee_0}$ and their complement, namely terms in the image of $\ad_{\ep_0}$.
This decomposition disentangles the two unknowns $\sigma_w^{[r]}$ and $\big[z_w,\mathfrak{N}^{[1]}  \big]^{[r]}$
on the left-hand side of (\ref{eq: recursion}) and relates both of them to the terms $\sigma_w^{[r-1]}$ of lower modular depth, which we consider as known.
We shall separately discuss the two types of refined equations obtained from projecting \eqref{eq: recursion} into ${\rm Im}(\ad_{\ep_0})$ and ${\rm ker}(\ad_{\ep^\vee_0})$ which involve only one of $\sigma_w^{[r]}$ and $\big[z_w,\mathfrak{N}^{[1]}  \big]^{[r]}$, respectively.
\begin{itemize}
\item \textbf{Image of $\ad_{\ep_0}$:}
 As we know that the commutators $[z_w,\ep_{k|\mu}]$ furnish lowest-weight vectors of $\mathfrak{sl}_2$, they cannot be in the image of $\ad_{\ep_0}$. This leads to the following projection of \eqref{eq: recursion} upon inserting the expression (\ref{spltn}) for $\mathfrak{N}^{[1]} $,
\begin{align}
\Big[\ep_0,\sigma_w^{[r]}\Big]=\sum_{k=2}^\infty (k{-}1) \, \BF_k\, \Big[ \ep_{k|2} + r_k(\ep_{k|3}+\ep_{k|4}),\sigma_w^{[r-1]}\Big] \, \Big\rvert_{{\rm Im}(\ad_{\ep_0})}\,,
    \label{eq: sigmarec}
\end{align}
where $\rvert_{{\rm Im}(\ad_{\ep_0})}$ on the right-hand side instructs us to isolate the terms that lie in the image of $\ad_{\ep_0}$. Note that this equation can equivalently be formulated by acting with $\ad_{\ep_0^\vee}$ on \eqref{eq: recursion}, yielding
\begin{align}
\Big[{\ep_0^\vee},\Big[{\ep_0},\sigma_w^{[r]}\Big]\Big]=\sum_{k=2}^\infty (k{-}1) \, \BF_k \Big[\ep_{k|2}
+ r_k (\ep_{k|3}+\ep_{k|4}),\big[\ep_0^\vee,\sigma_w^{[r-1]}\big]\Big]\,,
\label{eq: altsigmarec}
\end{align}
and thus resulting in the claimed recursion to determine the $\sigma_w^{[r]}$ in terms of $\sigma_w^{[r-1]}$ up to the kernel of $\ad_{\ep_0^\vee}\ad_{\ep_0}$. It was shown in \cite{Dorigoni:2024iyt} that $\ad_{\ep_0^\vee}$ is invertible on the image of $\ad_{\ep_0}$. This carries over to level 2, because its $\mathfrak{sl}_2$ structure is unchanged compared to level 1. This means that the $\sigma_w^{[r]}$ is completely determined up to the kernel of $\ad_{\ep_0}$ in terms of $\sigma_w^{[r-1]}$, i.e., up to $\mathfrak{sl}_2$ highest-weight vectors.
\item \textbf{Lowest-weight vectors:} As the first term on the left-hand side of \eqref{eq: recursion} is clearly not a lowest-weight vector, the projection $\vert_{\ker (\ad_{\ep_0^\vee})}$ to the kernel of $\ad_{\ep_0^\vee}$ gives rise to
\begin{align}
    \label{eq: kerade0c}
    \sum_{k=2}^\infty &(k{-}1)\, \BF_k\, \Big[z_w,\ep_{k|2} + r_k(\ep_{k|3}+\ep_{k|4})\Big]^{[r]}\\&=\sum_{k=2}^\infty (k{-}1)\,\BF_k\Big[  \ep_{k|2} + r_k (\ep_{k|3}+\ep_{k|4}),\sigma_w^{[r-1]}\Big] \, \Big\rvert_{\ker (\ad_{\ep_0^\vee})}\,.\notag
\end{align}
We will be interested in the summands on the left-hand side at fixed value of $k$, i.e., Lie polynomials $[z_w,\ep_{k|\mu} ]^{[r]}$ of degree $2w{+}k$. Hence,  isolating terms of degree $2w{+}k$ on the right-hand side of (\ref{eq: kerade0c}) to be indicated via $|^{{\rm deg} \ 2w{+}k}$ allows us to solve for
\begin{align}
    \label{eq:kerad2}
\Big[z_w,\ep_{k|2} + r_k(\ep_{k|3}{+}\ep_{k|4})\Big]^{[r]}
=  \sum_{\ell=2}^\infty  \frac{(\ell{-}1)\,\BF_\ell}{(k{-}1)\, \BF_k} \Big[  \ep_{\ell|2} + r_\ell (\ep_{\ell|3}{+}\ep_{\ell|4}),\sigma_w^{[r-1]}\Big] \, \Big\rvert^{{\rm deg} \ 2w{+}k}_{\ker (\ad_{\ep_0^\vee})}\,.
\end{align}
\end{itemize}
We reiterate that in order to solve either type of equations, (\ref{eq: altsigmarec}) or (\ref{eq:kerad2}), for $\sigma_w^{[r]}$ modulo highest-weight vectors or $[z_w,\ep_{k|2} + r_k(\ep_{k|3}{+}\ep_{k|4})]^{[r]}$,
one needs the complete contribution $\sigma_w^{[r-1]}$ of modular depth $r{-}1$  to $\sigma_w$, including highest-weight vectors. Hence, it remains to augment (\ref{eq: altsigmarec}) and (\ref{eq:kerad2}) by the method to be explained in section \ref{sec: HW vectors} to complete the recursion for $\sigma^{\rm g}_w$ and $[z_w,\ep_{k|\mu}]$ in modular depth.

\subsubsection{\texorpdfstring{Isolating $[z_w,\ep_{k|\mu}]^{[r]}$ at fixed $\mu$ via ${\cal S}_3$ action}{Isolating [zw,epsilon(k|mu)][r] at fixed mu via S3 action}}
\label{isozep}

Since (\ref{eq:kerad2}) only provides an explicit Lie polynomial in $\mathfrak{u}$ for the specific linear combination
$[z_w,\ep_{k|2} + r_k(\ep_{k|3}{+}\ep_{k|4})]^{[r]}$, it remains to find two additional independent equations at fixed values of $w,k,r$ in order to solve for the individual $[z_w,\ep_{k|\mu}]^{[r]}$ with $\mu = 2,3,4$. This can be accomplished by exploiting the ${\cal S}_3$ invariance of $z_w$ and the simple action of $\theta_S^{-1} \bullet \theta_S$ and $\theta_T^{-1} \bullet \theta_T$ on $\ep_{k|\mu}$ by permutation on the index $\mu$ as in (\ref{eq: theta}).

Referring to the right side of (\ref{eq:kerad2}) as
\beq
\dumz^r_{w,k} =  \sum_{\ell=2}^\infty  \frac{(\ell{-}1)\,\BF_\ell}{(k{-}1)\, \BF_k} \Big[  \ep_{\ell|2} + r_\ell (\ep_{\ell|3}{+}\ep_{\ell|4}),\sigma_w^{[r-1]}\Big] \, \Big\rvert^{{\rm deg} \ 2w{+}k}_{\ker (\ad_{\ep_0^\vee})}
\, ,
\label{zswk}
\eeq
we can complete the desired equation system for $[z_w,\ep_{k|\mu}]^{[r]} $ with $\mu = 2,3,4$ via
conjugations by $\theta_S$ and $\theta_S \theta_T$:
\begin{align}
\big[z_w,\ep_{k|2} + r_k(\ep_{k|3}{+}\ep_{k|4}) \big]^{[r]} &= \dumz^r_{w,k} \, , \label{zeqs} \\
\big[z_w,\ep_{k|3} + r_k(\ep_{k|2}{+}\ep_{k|4})\big]^{[r]} &= \theta_T^{-1}\theta^{-1}_S\dumz^r_{w,k} \theta_S \theta_T\, , \notag \\
\big[z_w,\ep_{k|4} + r_k(\ep_{k|2}{+}\ep_{k|3}) \big]^{[r]} &= \theta_S^{-1}\dumz^r_{w,k}\theta_S \, .\notag
\end{align}
The conjugations by $\theta_S$ and $ \theta_T$ on the right side are straightforward, because $\dumz^r_{w,k} $ is considered to be a known Lie polynomial in $\mathfrak{u}$ from (\ref{eq:kerad2}), assuming that $\sigma_w^{[r-1]}$ at lower modular depth has already been determined.

It is convenient to arrange the left sides of (\ref{zeqs}) into a 3-component column vector indexed by $\mu=2,3,4$ and to invert the $3\times 3$ coefficient matrix 
\begin{align}
    \left( \begin{matrix} 1 &r_k &r_k \\ r_k &1 &r_k \\ r_k &r_k &1 \end{matrix} \right)^{-1} = \frac{1}{(r_k{-}1)(2r_k{+}1)} \, 
    \left( \begin{matrix} {-}1{-}r_k &r_k &r_k \\ r_k &{-}1{-}r_k &r_k \\ r_k &r_k &{-}1{-}r_k \end{matrix} \right)\,.
    \label{zet.57}
\end{align}
However, this matrix inverse does not exist for $k= 0,2$ in view of the poles due to $r_0=1$ and $r_2 = -\frac{1}{2}$, so we need to restrict the following solution of (\ref{zeqs}) to $k\geq 4$:
\beq
 \left( \begin{matrix} [z_w,\ep_{k|2}]^{[r]} \\ [z_w,\ep_{k|3}]^{[r]}\\ [z_w,\ep_{k|4}]^{[r]}
    \end{matrix} \right) = \frac{1}{(r_k{-}1)(2r_k{+}1)}  
    \left( \begin{matrix} {-}1{-}r_k \! &r_k &r_k \\ r_k &{-}1{-}r_k &r_k \\ r_k &r_k &\! {-}1{-}r_k \end{matrix} \right) \! \left( \begin{matrix} \dumz^r_{w,k}    \\  \theta_T^{-1}\theta^{-1}_S\dumz^r_{w,k} \theta_S \theta_T  \\ \theta^{-1}_S\dumz^r_{w,k} \theta_S     \end{matrix} \right)\, , \ \ \ \, {\rm for}\ k\geq 4\, .
\label{zepk4}
\eeq
For $k=2$, one passes to the two independent degree-2 derivations $\tilde{\ep}_{2|3},\tilde{\ep}_{2|4}$ in the original expression (\ref{eq: N2conn}) for the level-2 connection, which are related to $\ep_{2|\mu}$ via (\ref{backtotild}). After simplifying the $k=2$ instance of (\ref{eq:kerad2}) via $\ep_{2|2} + r_2(\ep_{2|3}{+}\ep_{2|4})=-\frac{1}{2} (\tilde \ep_{2|3}{+} \tilde \ep_{2|4})$ and adjoining the $\theta_S$ image obtained from (\ref{eq: thetatild}), we arrive at
\beq
 \left( \begin{matrix}  [z_w,\tilde \ep_{2|3}]^{[r]}\\ [z_w,\tilde \ep_{2|4}]^{[r]}
    \end{matrix} \right) =  
  \frac{2}{3}\left( \begin{matrix} -2 &-1 \\ -1 &1 \end{matrix} \right) \! \left( \begin{matrix} 
  \dumz^r_{w,2} \\ \theta^{-1}_S\dumz^r_{w,2} \theta_S   \end{matrix} \right) \, ,
\label{zepk2}
\eeq
completing (\ref{zepk4}) to all positive even degrees. 
In the remainder of this section, we show how the recursive strategy based on $T$-equivariance provides us with the modular depth 2 contributions to both $\sigma_w$ and $[z_w,\ep_{k|\mu}]$. For results at modular depth 3, we refer to section \ref{s43.9} below; see also the ancillary file for higher modular depth.

Note that, by the $\mathfrak{sl}_2$ invariance of $z_w$, its brackets with $\ep^{(j)}_{k|\mu}$ for $j\neq 0$ are readily determined by (\ref{zepk4}) via $[z_w,\ep^{(j)}_{k|\mu}]^{[r]} = {\rm ad}^j_{\ep_0} [z_w,\ep_{k|\mu}]^{[r]}$, where the action of each ${\rm ad}_{\ep_0}$ on the Lie polynomials in $\mathfrak{u}$ from $[z_w,\ep_{k|\mu}]^{[r]}$ is evaluated through the Leibniz rule.

\subsubsection{\texorpdfstring{$\mathfrak{sl}_2$ lemmas for modular depth 2}{sl2 lemmas for modular depth 2}}
\label{sec:sl2lem}

In preparation for the computation of $\sigma_w^{[2]}$ and $\big[z_w,\mathfrak{N}^{[1]}  \big]^{[2]}$, we shall here introduce lemmas that project the expressions we encounter at modular depth 2 to ${\rm Im}(\ad_{\ep_0})$ and ${\rm ker}(\ad_{\ep^\vee_0})$. For this purpose, let us generalise the highest- and lowest-weight vectors \eqref{eq: HWst} for level-1 brackets of $\ep^{(j)}_k$ to those of level-2 derivations $\ep^{(j)}_{k| \mu}$:
\begin{align}
s^d(\ep_{p| \mu},\ep_{q| \nu}) &= \frac{(d{-}2)!}{(p{-}2)!(q{-}2)!} \sum_{i=0}^{d-2} (-1)^i \big[ \ep^{(p-2-i)}_{p|\mu},\ep^{(q-d+i)}_{q| \nu} \big]\,,
\label{eq: stvectors}
\\
t^d(\ep_{p| \mu},\ep_{q| \nu}) &= \frac{(d{-}2)!}{(p{-}2)!(q{-}2)!} \sum_{i=0}^{d-2} (-1)^i \frac{ (p{-}2{-}i)! (q{-}d{+}i)! }{i! (d{-}2{-}i)!}\big[ \ep^{(i)}_{p| \mu},\ep^{(d-2-i)}_{q| \nu} \big]\,.
\notag
\end{align}
In the range of $d=2,3,\ldots,{\rm min}(p,q)$, the combinations $s^d(\ep_{p| \mu},\ep_{q| \nu})$ and $t^d(\ep_{p| \mu},\ep_{q| \nu})$ are highest- and lowest-weight vectors of $\mathfrak{sl}_2$, respectively,
\begin{align}
\big[\ep_0,s^d(\ep_{p| \mu},\ep_{q| \nu})\big] = 0 \, , \ \ \ \ \ \
\big[\ep_0^\vee,
t^d(\ep_{p| \mu},\ep_{q| \nu}) \big] = 0 \, , \ \ \ \ \ \ {\rm for}\ 2\leq d \leq {\rm min}(p,q)\,.
\label{eq:hwconds}
\end{align}
Still, we will also make use of $s^d(\ep_{p| \mu},\ep_{q| \nu})$ with $d> {\rm min}(p,q)$ which are no longer highest-weight vectors and where contributions to the sums over $i$ in \eqref{eq: stvectors} with negative superscript $\ep_{k| \lambda}^{(j<0)}$ are understood to vanish. These combinations are instrumental to compactly formulate the following straightforward key lemmas for $\mathfrak{sl}_2$-decompositions:
\begin{align}
\frac{(k{-}1)}{(w{-}1)!} [ \ep_{k | \mu} , \ep_{w+1 | \nu}^{(w-1)}] &= \big[\ep_0 , s^{k+1}(\ep_{k | \mu},\ep_{w+1 | \nu}) \big] \, , \ \ \ \ \ \ 4 \leq k \leq w{-}1\,,
\label{eq: usefulids} \\
\frac{(k{-}1)}{(w{-}1)!} [   \ep_{w+1 | \nu}^{(w-1)} , \ep_{k | \mu}]&=   \bigg[ \ep_0,\sum_{j=0}^{w-2} \frac{(-1)^j \binom{k-2}{j}^{-1}}{j! (w{-}2{-}j)!} [ \ep_{w+1|\nu}^{(w-j-2)}, \ep_{k|\mu}^{(j)}  ]  \bigg]
\notag \\
&\quad + (k{-}w) t^{w+1}(\ep_{w+1 | \nu} , \ep_{k | \mu}) \, , \ \ \ \ \ \ k\geq w{+}3\,.
\notag
\end{align}
The second line can in fact be extended to $k=w{+}1$ if $\mu \neq \nu$. For $\mu = \nu$, however, the $k=w{+}1$ case requires the adapted version of the first line of \eqref{eq: usefulids} with an extra factor of 2:
\begin{align}
\frac{2w}{(w{-}1)!} [ \ep_{w+1 | \mu} , \ep_{w+1 | \mu}^{(w-1)}] &= \big[\ep_0 , s^{w+2}(\ep_{w+1 | \mu},\ep_{w+1 | \mu}) \big] \,.
\label{eq: adaptedid}
\end{align}

\subsubsection{Example: geometric contributions at modular depth 2}
\label{sec:sd2}

As explained above, in order to determine $\sigma^{[2]}_w$, we need access to all of the terms in $\sigma^{[1]}_w$, which were identified as highest-weight vectors below (\ref{eq: recursion}). These can be obtained by translating the equivariant iterated integrals of depth 1 from \cite{Duhr:2025lvr} into our conventions and matching them with the expansion of (\ref{eq: eqvprop}). We find  
\begin{align}
\sigma_1^{[1]} &= -\ep_{2|3}-\ep_{2|4} \label{eq: depth1} \,,\\
\sigma_w^{[1]} &= -r_w
\frac{ \ep_{w+1 |  2}^{(w-1) } }{(w{-}1)!}  \, , \ \ \ \ \ \ w\geq 3 \ {\rm odd}\,,
\notag
\end{align}
consistently with the MMVs (\ref{eq: mmvclosed}) and the conjectured form (\ref{eq:Scoc2}) of the $S$-cocycle. With the translation (\ref{fixeps}) between $\ep_{2|\mu}$ and $\tilde \ep_{2|\nu}$, the first line of (\ref{eq: depth1})~yields
\beq
\sigma_1 = z_1 - \frac{1}{3} ( \tilde \ep_{2|3} + \tilde \ep_{2|4})\,,
\label{exactz1}
\eeq
which is in fact an exact result since the $a$-degree 1 of $\sigma_1$ does not admit any terms of modular depth $\geq 2$.
It is amusing to compare the second line of (\ref{eq: depth1}) with the terms $\sigma_w^{\SLZ} = - \ep_{w+1}^{(w-1)} /(w{-}1)! + \ldots$ of modular depth 1 in the level-1 case, see (\ref{eq: sigma}):
The ratio $r_w$ of (\ref{defrk}) for odd $w$ turns out to be a natural conversion factor between zeta generators at level 1 and 2, in the same way as $r_w$ for even $w$ matches the ratio $B_w(1/2)/B_w$ seen in the expression (\ref{eq: N2}) for $\mathfrak{N}$.

Plugging \eqref{eq: depth1} into \eqref{eq: sigmarec} and projecting onto the image of $\ad_{\ep_0}$ using \eqref{eq: usefulids}, we find the following expression 
for $w\geq 3$ which generalises the level-1 case in \eqref{eq: sigma}: 
\begin{align}
\sigma_w^{[2]} &= r_w \bigg\{ \frac{1}{8(w{-}1)!} [\ep_{2|3}{+}\ep_{2|4} , \ep_{w+1|2}^{(w-2)}]  - \frac{1}{2} \BF_{w+1} s^{w+2}( \ep_{w+1|2} ,  \ep_{w+1|2}) \label{sysxi2.09} \\
&\quad - \sum_{k=4}^{w-1} \BF_k \big( s^{k+1}( \ep_{k|2}, \ep_{w+1|2}) + r_k s^{k+1}( \ep_{k|3}, \ep_{w+1|2}) + r_k s^{k+1}( \ep_{k|4}, \ep_{w+1|2}) \big) \notag \\
&\quad + \sum_{k=w+1}^\infty r_k \BF_k \sum_{j=0}^{w-2} \frac{(-1)^j \binom{k-2}{j}^{-1}}{j! (w{-}2{-}j)!} [ \ep_{w+1|2}^{(w-j-2)}, \ep_{k|3}^{(j)}{+}\ep_{k|4}^{(j)} ] \notag \\
&\quad + \sum_{k=w+3}^\infty \BF_k \sum_{j=0}^{w-2} \frac{(-1)^j \binom{k-2}{j}^{-1}}{j! (w{-}2{-}j)!} [ \ep_{w+1|2}^{(w-j-2)}, \ep_{k|2}^{(j)}  ] \bigg\} \ {\rm mod} \ {\rm ker}(\ep_0)\,.\notag
\end{align}
This illustrates that contributions $\sigma_w^{[r]}$ with $w\geq 3$ of modular depth $r\geq 2$ are Lie series of unbounded degree. Nevertheless, the missing highest-weight vectors in ${\rm ker}(\ep_0)$ only occur up to the key degree $2w$ of $\sigma_w$, as argued before section \ref{org:mdp}.

In combination with the modular depth 1 terms in (\ref{eq: depth1}), the simplest non-trivial examples of (\ref{sysxi2.09}) with $w=3,5,7$ yield
\begin{align}
\sigma_3 &= z_3 + \frac{3}{8} \ep_{4|2}^{(2)} + \frac{3}{64} [ \ep_{4|2}^{(1)} ,\ep_{2|3}{+}\ep_{2|4}] 
\label{zet.61} \\
&\quad + \frac{1}{640} [ \ep_{4|2}^{(1)} , \ep_{4|2} ]
- \frac{7}{7680}  [ \ep_{4|2}^{(1)} , \ep_{4|3}  {+} \ep_{4|4} ]
+\frac{7}{15360}  [ \ep_{4|2} , \ep_{4|3}^{(1)}  {+} \ep_{4|4}^{(1)} ] \notag \\
&\quad 
- \frac{ 31 }{ 5160960 } [\ep_{4| 2}, \ep^{(1)}_{6| 3} + \ep^{(1)}_{6| 4}] 
+ \frac{ 31 }{ 1290240 } [\ep^{(1)}_{4| 2}, \ep_{6|3} + \ep_{6| 4}]  \notag \\
&\quad + \frac{1}{161280} [\ep_{4| 2}, \ep^{(1)}_{6| 2}]
- \frac{1}{40320} [\ep^{(1)}_{4| 2}, \ep_{6|2}]
+ \ldots\,,\notag \\
\sigma_5 &= z_5 + \frac{5}{128} \ep_{6|2}^{(4)}
+ \frac{5}{1024} \big[ \ep^{(3)}_{6|2} , \ep_{2|3}{+} \ep_{2|4} \big] \notag \\
&\quad
 + \frac{7}{ 49152 }
\big( [\ep_{4|3} + \ep_{4| 4}, \ep^{(3)}_{6| 2}] 
  -  [\ep^{(1)}_{4| 3} + \ep^{(1)}_{4| 4}, \ep^{(2)}_{6| 2}] 
  +   [\ep^{(2)}_{4| 3} + \ep^{(2)}_{4| 4}, \ep^{(1)}_{6| 2}] \big) \notag \\
  &\quad + \frac{1}{6144} \big( {-} 
 [\ep^{(2)}_{4|2}, \ep^{(1)}_{6|2}]+ [\ep^{(1)}_{4| 2}, \ep^{(2)}_{6| 2}] - [\ep_{4| 2}, \ep^{(3)}_{6| 2}] \big)
\ {\rm mod} \ {\rm ker}(\ep_0)  +  \ldots \,,\notag\\
\sigma_7 &= z_7 + \frac{7}{5120} \ep_{8|2}^{(6)}
+ \frac{7}{ 40960 }  \big[ \ep^{(5)}_{8|2} , \ep_{2|3}{+} \ep_{2|4} \big]
\ {\rm mod} \ {\rm ker}(\ep_0)
+ \ldots\,,
\notag
\end{align}
with terms of degree $\geq 12$ and/or modular depth $\geq 3$ in the ellipses. The ancillary file contains these terms through degree 16 at modular depth 2 and through degree 10 at modular depth $\geq 3$.
The missing highest-weight vectors of $\sigma_5$ and $\sigma_7$ at modular depth 2 can be found in (\ref{eq: HWd2}) below.

\subsubsection{\texorpdfstring{Example: complete expressions for $[z_1,\ep_{k|\mu}]$}{Example: complete expressions for [z1,epsilon(k|mu)]}}
\label{sec:z1ep}

In preparation for contributions to $[z_w,\ep_{k|\mu}]$ of modular depth 2, we shall here consider the case of $w=1$ for all $k \geq 2$. By the $a$-degree 1 of $z_1$ and $\ep_{k|\mu}$, their bracket cannot produce any terms of modular depth $\geq 3$, and $[z_1,\ep_{k|\mu}]^{[2]} = [z_1,\ep_{k|\mu}]$ are already the complete expressions.
We start from the specialisation of (\ref{eq:kerad2}) to $w=1$ and $r=2$ using $\sigma_1^{[1]}$ of \eqref{eq: depth1},
\beq
\Big[z_1,\ep_{k|2} + r_k(\ep_{k|3}{+}\ep_{k|4})\Big]
=  - \Big[  \ep_{k|2} + r_k (\ep_{k|3}{+}\ep_{k|4}),\ep_{2|3}{+}\ep_{2|4}\Big]\, ,
\label{newz1.01}
\eeq
where the projection to $\ker (\ad_{\ep_0^\vee})$ is automatic with lowest-weight vectors in both entries of the commutator. 
\begin{itemize}
\item For $k=2$, the right side of (\ref{newz1.01}) vanishes (see the comment below (\ref{zepk4})) and we find $[z_1,\ep_{2|3}{+}\ep_{2|4}] = 0$. By taking combinations with its $\theta_S$ image, we deduce
\beq
[z_1,\ep_{2|\nu}] = 0\,,  \ \ \ \ {\rm for} \ \nu = 3,4\, .
\label{newz1.02}
\eeq
\item For $k\geq 4$, matching (\ref{newz1.01}) with the first line of (\ref{zeqs}) at $w=1$ and $r=2$ identifies $\dumz^2_{1,k}= [  \ep_{k|2} + r_k (\ep_{k|3}{+}\ep_{k|4}),\ep_{2|2}]$ after rewriting $\ep_{2|3}{+}\ep_{2|4}= - \ep_{2|2}$. The general result (\ref{zepk4}) then specialises to
\begin{align}
     \! \! \left( \begin{matrix} [z_1,\ep_{k|2}] \\ [z_1,\ep_{k|3}]\\ [z_1,\ep_{k|4}]
    \end{matrix} \right) = \frac{1}{(r_k{-}1)(2r_k{+}1)} \, 
    \left( \begin{matrix} {-}1{-}r_k &r_k &r_k \\ r_k &{-}1{-}r_k &r_k \\ r_k &r_k &{-}1{-}r_k \end{matrix} \right)  \left( \begin{matrix}  [ \ep_{k|2}  + r_k (\ep_{k|3}{+}\ep_{k|4}) , \ep_{2|2}] \\  [ \ep_{k|3}  + r_k (\ep_{k|2}{+}\ep_{k|4}) , \ep_{2|3}] \\  [ \ep_{k|4}  + r_k (\ep_{k|2}{+}\ep_{k|3}) , \ep_{2|4}] 
    \end{matrix} \right)\,.
    \label{eq: zmatrix}
\end{align}
At $k=4$ we find
\begin{align}
\label{eq:z1e4}
    [z_1, \ep_{4|2}] &= \frac{19}{30} [\ep_{2|2}, \ep_{4|2}]+\frac{7}{10} \big( [\ep_{2|3}, \ep_{4|3}]{+} [\ep_{2|4}, \ep_{4|4}] \big)-\frac{7}{15} \big( [\ep_{2|3}, \ep_{4|4}]{+} [ \ep_{2|4}, \ep_{4|3}] \big) \,,\notag \\
    [z_1, \ep_{4|3}] &= \frac{7}{6} [\ep_{2|4}, \ep_{4|4}{-} \ep_{4|2}] - \frac{7}{10}[\ep_{2|3}, \ep_{4|2}] +\frac{19}{30}  [\ep_{2|3}, \ep_{4|3}]
    +\frac{7}{15}  [\ep_{2|3}, \ep_{4|4}]\,, \notag \\
    [z_1, \ep_{4|4}] &= \frac{7}{6} [\ep_{2|3}, \ep_{4|3}{-} \ep_{4|2}] -\frac{7}{10}[\ep_{2|4}, \ep_{4|2}] +\frac{19}{30}  [\ep_{2|4}, \ep_{4|4}]+\frac{7}{15}  [\ep_{2|4}, \ep_{4|3}]\,, 
\end{align}
and the $k=6$ instance of \eqref{eq: zmatrix} reads
\begin{align}
    [z_1,\ep_{6|2}] &=  \frac{341}{630}\big( [\ep_{2|3}, \ep_{6|3}]{+}[\ep_{2|4}, \ep_{6|4}] \big) +\frac{331}{630} [\ep_{2|2}, \ep_{6|2}]- \frac{31}{63} \big( [\ep_{2|3}, \ep_{6|4}]  {+} [\ep_{2|4}, \ep_{6|3}] \big)\,,
\notag \\
    [z_1,\ep_{6|3}] &= \frac{331}{630} [\ep_{2|3}, \ep_{6|3}] -\frac{341}{630} [\ep_{2|3}, \ep_{6|2}]  +\frac{31}{63} [\ep_{2|3}, \ep_{6|4}] 
    +\frac{31}{30} [\ep_{2|4}, \ep_{6|4}{-}\ep_{6|2}]\,,\notag \\
    [z_1,\ep_{6|4}] &=
   \frac{331}{630} [\ep_{2|4}, \ep_{6|4}] 
   - \frac{341}{630} [\ep_{2|4}, \ep_{6|2}] 
   + \frac{31}{63} [\ep_{2|4}, \ep_{6|3}]
   + \frac{31}{30} [\ep_{2|3}, \ep_{6|3}{-}\ep_{6|2}] \,.
    \label{zet.59}
\end{align}
\end{itemize}

\subsubsection{\texorpdfstring{Example: $[z_w,\ep_{k|\mu}]$ with $w\geq 3$ at modular depth 2}{Example: [zw,epsilon(k|mu)] with w>=3 at modular depth 2}}
\label{sec:zwep}

We shall next apply the strategy of sections \ref{org:sl2} and \ref{isozep} to determine
$[z_w,\ep_{k|\mu}]^{[r]}$ for modular depth $r=2$ and general $w\geq 3$. In this case, (\ref{eq:kerad2}) with $\sigma_w^{[1]}$ given by the second line of (\ref{eq: depth1}) specialises to
\begin{align}
&\Big[z_w,\ep_{k|2} + r_k(\ep_{k|3}{+}\ep_{k|4})\Big]^{[2]}
=  - \frac{ r_w (w{+}k{-}2) \BF_{w+k-1}}{(w{-}1)!(k{-}1) \BF_k}  \label{wkd2.11} \\
&\quad \times \Big[  \ep_{w+k-1|2} + r_{w+k-1} (\ep_{w+k-1|3}{+}\ep_{w+k-1|4}), \ep^{(w-1)}_{w+1|2} \Big] \, \Big\rvert_{\ker (\ad_{\ep_0^\vee})}\,.
\notag
\end{align}
The projection to lowest-weight vectors proceeds differently for $k=2$ compared to $k\geq 4$:
\begin{itemize}
\item For $k=2$, (\ref{wkd2.11}) simplifies as follows with the relation (\ref{fixeps}) among $\ep_{2|\mu}$, 
\beq
\big[z_w,\ep_{2|2} \big]^{[2]}
= - \frac{8wr_w \BF_{w+1}}{(w{-}1)!} \, \Big[  \ep_{w+1|2} + r_{w+1} (\ep_{w+1|3}{+}\ep_{w+1|4}), \ep^{(w-1)}_{w+1|2} \Big] \, \Big\rvert_{\ker (\ad_{\ep_0^\vee})}\, .
\label{wkd2.12}
\eeq
The first bracket $[  \ep_{w+1|2}, \ep^{(w-1)}_{w+1|2} ]$ on the right side is in the image of $\ad_{\ep_0}$ by (\ref{eq: adaptedid}) and does not contribute to $\ker (\ad_{\ep_0^\vee})$. Hence, all lowest-weight vectors are obtained from the terms involving $\ep_{w+1|3}{+}\ep_{w+1|4}$ by the extension of (\ref{eq: usefulids}) to $k=w{+}1$ with $\mu\neq \nu$. By combining the result with its images under $\theta_S$ and $\theta_T \theta_S$, we arrive at
\begin{align}
\big[z_w,\ep_{2|2} \big]^{[2]}
&=  8 r_w r_{w+1} \BF_{w+1}t^{w+1}(\ep_{w+1|2}, \ep_{w+1|3}{+}\ep_{w+1|4})\, ,
\label{wkd2.13} \\
\big[z_w,\ep_{2|3} \big]^{[2]}
&=  8 r_w r_{w+1} \BF_{w+1}t^{w+1}(\ep_{w+1|3}, \ep_{w+1|2}{+}\ep_{w+1|4})\, ,
\notag \\
\big[z_w,\ep_{2|4} \big]^{[2]}
&=  8 r_w r_{w+1} \BF_{w+1}t^{w+1}(\ep_{w+1|4}, \ep_{w+1|2}{+}\ep_{w+1|3})\, ,
\notag
\end{align}
whose simplest examples are the following bracket and its images under $\theta_S$ and $\theta_T \theta_S$,
\begin{align}
\big[z_3,\ep_{2|3} \big]^{[2]} &= - \frac{7}{1920} \Big(  [ \ep_{4|3}^{(2)} , \ep_{4|2}{+}\ep_{4|4}  ]  - [ \ep_{4|3}^{(1)} , \ep_{4|2}^{(1)}{+}\ep_{4|4}^{(1)}  ] + [ \ep_{4|3} , \ep_{4|2}^{(2)}{+}\ep_{4|4}^{(2)}  ]  \Big) \, .
\end{align}
\item For $k\geq 4$, the brackets on the right side of (\ref{wkd2.11}) can all be projected to $\ker (\ad_{\ep_0^\vee})$ by using (\ref{eq: usefulids}) and only retaining the lowest-weight vector $t^{w+1}(\bullet,\bullet)$,
\begin{align}
&\Big[z_w,\ep_{k|2} + r_k(\ep_{k|3}{+}\ep_{k|4})\Big]^{[2]}
=   \frac{ r_w  \BF_{w+k-1}}{\BF_k}   \label{wkd2.16} \\
&\quad \times t^{w+1}\big(\ep_{w+1|2}, \ep_{w+k-1|2} + r_{w+k-1} (\ep_{w+k-1|3}{+}\ep_{w+k-1|4}) \big) \,.
\notag
\end{align}
The ratios of Bernoulli numbers and the brackets of schematic form $t^{w+1}(\ep_{w+1}, \ep_{w+k-1})$ are
familiar from the level-1 case.
In fact, the level-1 counterparts $[z_w^{\SLZ},\ep_k]^{[2]}$ in (\ref{eq: bracket}) are recovered by applying the transcription $\ep_{k|\mu}^{(j)} \mapsto \ep_k^{(j)}/r_{k-1}$ of (\ref{eq:SL2limit}) to (\ref{wkd2.16}) and noting that $(1{+}2r_k)/r_{k-1}=1$.
The individual $[z_w,\ep_{k|\mu} ]^{[2]}$ in \eqref{wkd2.16} at fixed $\mu=2,3,4$ are isolated as in 
section \ref{isozep}, resulting in
\begin{align}
    &\left( \begin{matrix} [z_w,\ep_{k|2}]^{[2]}\\ [z_w,\ep_{k|3}]^{[2]}\\ [z_w,\ep_{k|4}]^{[2]}
    \end{matrix} \right) = \frac{r_w \BF_{w+k-1}}{(r_k{-}1)(2r_k{+}1) \BF_k} \, 
    \left( \begin{matrix} {-}1{-}r_k &r_k &r_k \\ r_k &{-}1{-}r_k &r_k \\ r_k &r_k &{-}1{-}r_k \end{matrix} \right)  \label{sysxi2.17} \\
    &\qquad \times \left( \begin{matrix}  
    t^{w+1}\big(\ep_{w+1 | 2} , \ep_{w+k-1|2} + r_{w+k-1} (\ep_{w+k-1|3} {+}  \ep_{w+k-1|4} ) \big) \\
    t^{w+1}\big(\ep_{w+1 | 3} , \ep_{w+k-1|3} + r_{w+k-1}( \ep_{w+k-1|2} {+} \ep_{w+k-1|4})  \big) \\
    t^{w+1}\big(\ep_{w+1 | 4} , \ep_{w+k-1|4} + r_{w+k-1} (\ep_{w+k-1|2} {+}  \ep_{w+k-1|3})  \big) 
    \end{matrix} \right) \,,
    \notag
\end{align}
with the following simplest example:
\begin{align}
\big[z_3,\ep_{4|2} \big]^{[2]} &=
-\frac{1}{630} t^4( \ep_{4| 2}, \ep_{6| 2}) + 
 \frac{31 }{20160}  t^4( \ep_{4| 2}, \ep_{6| 3}{+}\ep_{6| 4})   + 
 \frac{31}{2880}   t^4( \ep_{4| 3}{+}\ep_{4| 4}, \ep_{6| 2})   \notag\\
&\quad   
 -  \frac{1}{90} \big(t^4( \ep_{4| 3}, \ep_{6| 3}) {+} t^4( \ep_{4| 4}, \ep_{6| 4}) \big) 
 + 
\frac{ 31}{2880} (t^4( \ep_{4| 3}, \ep_{6| 4}) {+} t^4( \ep_{4| 4}, \ep_{6| 3}) \big)\, .
\end{align}
\end{itemize}
With the $2\times 2$ cases of
$[z_w,\ep_{k|\mu}]^{[2]}$ with
$w=1$ vs.\ $w\geq 3$ and $k=2$ vs.\ $k\geq 4$ in (\ref{newz1.02}), (\ref{eq: zmatrix}), (\ref{wkd2.13}) and (\ref{sysxi2.17}), we have covered all cases of $[z_w,\ep_{k|\mu}]$ at modular depth 2. Examples at higher modular depths can be found in section \ref{s43.9} below and in the ancillary file.

\subsection{Highest-weight vectors of zeta generators}
\label{sec: HW vectors}

In the previous section, we outlined a recursive procedure to determine both $\sigma_w^{\rm g}$ and $[z_w,\ep_{k|\mu}]$ modular depth by modular depth. The partial recursive step of section \ref{org:sl2} requires the exact form of $\sigma_w^{[r]}$ including highest-weight vectors to determine $[z_w,\ep_{k|\mu}]^{[r+1]}$ and $\sigma_w^{[r+1]}$, but one cannot obtain highest-weight vectors of $\sigma_w^{[r]}$ from $\sigma_w^{[r-1]}$. This section provides the missing ingredient to complete the recursion, namely the computation of  $\mathfrak{sl}_2$ highest-weight vectors entering the terms~$\sigma_w^{[r]}$ of modular depth $r$.
More specifically, we generalise the 
method for level 1 in section \ref{sec:zetaScoc} which builds upon the factorised form (\ref{eq: Scocalt}) of the $S$-cocycle. Its generalisation to level 2 relies on the conjecture
\begin{align}
    \mathbb S(\epsilon) = \mathbb{M}(\sigma)^{-1}\,\mathbb X(\epsilon)\,U_S^{-1}\,\theta_S^{-1}\mathbb M(\sigma)\theta_S\,U_S\,,
    \label{ag:Scoc2}
\end{align}
which is a repetition of \eqref{eq:Scoc2} for the reader's convenience and was shown in \eqref{prf.sequi} to imply the cancellation of $S$-cocycles from the generating series~\eqref{eq: eqvprop}.
 The series $\mathbb{X}(\epsilon) \in \exp(\mathfrak{u})$ has coefficients in $\mathbb Q[\pi^2]$ along with each word in $\ep_{k|\mu}^{(j)}$, see section \ref{s43.2} below for details.

First of all, recall that, according to \eqref{eq: moreTinf} and (iii) below (\ref{eq:Scoc2}), only $\theta_T$-invariant combinations of brackets in~$\mathfrak{u}$ of $a$-degree $w$ are viable candidates for highest-weight vectors of~$\sigma_w^{\rm g}$. We furthermore reiterate that the highest degree of the geometric part $\sigma_w^{\rm g}$ that is not yet determined by $[\sigma_w,\mathfrak{N}] = 0$ is $2w{-}2$ for $\theta_S,\theta_T$-invariant highest-weight vectors and $2w$ for solely $\theta_T$-invariant highest-weight vectors. Hence, we only need to extract a finite number of terms from (\ref{ag:Scoc2}) for the completion of a given $\sigma_w^{[r]}$ (and by (\ref{eq: sig[z,ep]}) for the entire zeta generator $\sigma_w$).

The highest-weight vectors of the zeta generators can be inferred from comparing the left- and right-hand sides of the factorised $S$-cocycle \eqref{ag:Scoc2}. On the left-hand side, we have access to numerical values of the MMVs defined in (\ref{eq: mmvdef}) that accompany the words in $\ep_{k|\mu}^{(j)}$, see appendix \ref{sec:num.mmv} for details. On the right-hand side of \eqref{ag:Scoc2}, the unknown highest-weight vectors are part of the level-2 zeta generators that make up the generating series $\mathbb M(\sigma)^{\pm 1}$. The words in $\ep_{k|\mu}^{(j)}$ which occur in the expanded commutator brackets of the unknown highest-weight vectors may also arise from compositions of zeta generators $\sigma_a \sigma_b\ldots$ (contributing with lower modular depth) and derivations from the series $\mathbb X(\epsilon)$ in $\exp(\mathfrak{u})$ with $\mathbb Q[\pi^{2k}]$ coefficients. Hence, 
the words in $\ep_{k|\mu}^{(j)}$ from the unknown highest-weight vector in~$\sigma_w$ will not only be accompanied by $\zeta_w$, but also by known contributions $\pi^{2k}\rho^{-1}(f_af_b\ldots)$. Computing these extra contributions besides $\zeta_w$ from (\ref{ag:Scoc2}) reduces the number of rational coefficients that need to be numerically determined via PSLQ, cf.\
appendix \ref{sec:num.mmv}. The unknown terms on the right-hand side of \eqref{ag:Scoc2} that are linear in $\zeta_w$ and come from  $U_S^{-1}\theta_S^{-1}\sigma_w\theta_SU_S-\sigma_w$ can then be numerically matched with the left-hand side at the level of individual words in $\ep_{k|\mu}^{(j)}$ which are independent under Pollack relations.

There are three distinct cases to consider when solving for highest-weight vectors, depending on their $\mathfrak{sl}_2\times\mathcal{S}_3$ properties:
\begin{itemize}
\item[(i)] For singlets under both $\mathfrak{sl}_2$ and $\mathcal{S}_3$, the highest-weight vector drops out from the combination $U_S^{-1}\theta_S^{-1}\sigma_w\theta_SU_S-\sigma_w$ and can be absorbed into $z_w$ as discussed above section~\ref{org:mdp}.
\item[(ii)] Highest-weight vectors of $\sigma_w$ in non-singlet representations of $\mathfrak{sl}_2$ have positive Cartan eigenvalues under $h= [\ep_0,\ep_0^\vee]$ as can be seen from $[h,\ep^{(j)}_{k|\mu}]=(2j{+}2{-}k) \ep^{(j)}_{k|\mu}$. The images of unknown highest-weight vectors of $\sigma_w$ within $U_S^{-1}\theta_S^{-1}\sigma_w\theta_SU_S$ in turn have negative $h$-eigenvalues by (\ref{eq: US}). Since the search for highest-weight vectors in non-singlet representations of $\mathfrak{sl}_2$ will only require terms of positive $h$-eigenvalues in (\ref{ag:Scoc2}), their unknown images within $U_S^{-1}\theta_S^{-1}\sigma_w\theta_SU_S$ are not encountered.
\item[(iii)] If a highest-weight vector is a singlet of $\mathfrak{sl}_2$, but it is in a non-trivial representation of $\mathcal{S}_3$, its contributions to $\sigma_w$ and $U_S^{-1}\theta_S^{-1}\sigma_w\theta_SU_S$ are slightly harder to separate. First of all, $\mathfrak{sl}_2$-singlet contributions $v_w$ to $\sigma_w$ obey $U_S^{-1}\theta_S^{-1}v_w\theta_SU_S = \theta_S^{-1}v_w\theta_S$, so it remains to reconstruct $v_w$ from the combination $\theta_S^{-1}v_w\theta_S {-} v_w$ accompanying $\zeta_w$ on the right side of (\ref{ag:Scoc2}). By $\theta_T$ invariance of $\sigma_w$, and thus $v_w$, the latter cannot be in the alternating representation of ${\cal S}_3$, so its assumed non-trivial ${\cal S}_3$ property enforces $v_w$ to be in the standard representation $\mathbf{2}$ of ${\cal S}_3$. For $v_w \in \mathbf{2}$ (as opposed to elements of $\mathbf{1}$ or $\bep$) in turn, we have
\beq
v_w = \frac{1}{3} \Big\{
 v_w - \theta_S^{-1}v_w\theta_S + 
\theta_T^{-1}\big(  v_w - \theta_S^{-1}v_w\theta_S \big) \theta_T
\Big\}\,, \ \ \ \ {\rm for} \ v_w \in \mathbf{2}\,,
\label{getvw}
\eeq
as one can for instance verify for the $\theta_T$ invariant $\ep_{2|3}{+}\ep_{2|4}$ using  (\ref{eq: theta}) and the irreducibility condition $\ep_{2|2}{+}\ep_{2|3}{+}\ep_{2|4}=0$ of the $\mathbf{2}$ representation.
Based on (\ref{getvw}), we can isolate the $\mathfrak{sl}_2$-singlet highest-weight vector $v_w \in \mathbf{2}$,
assuming that the combination $v_w-\theta_S^{-1}v_w\theta_S$ on the right side has been determined from  (\ref{ag:Scoc2}). 
\end{itemize}

\subsubsection{\texorpdfstring{The series $\mathbb{X}(\ep)$}{The series X(epsilon)}}
\label{s43.2}

In order to facilitate the PSLQ procedure for the numerical computation of highest-weight vectors in $\sigma_w^{[r]}$, it will be useful to give more information on the series $\mathbb X(\epsilon) \in \exp(\mathfrak{u})$ in \eqref{ag:Scoc2} with $\mathbb Q[\pi^2]$-coefficients. One way of computing $\mathbb{X}$ order by order in $\ep_{k|\mu}^{(j)}$ is to solve \eqref{ag:Scoc2} for
\begin{align}
    \mathbb X(\epsilon)= \mathbb M(\sigma)\mathbb S(\ep) U_S^{-1}\,\theta_S^{-1}\mathbb{M}(\sigma)^{-1}\theta_S U_S\,.
\label{eq: X}
\end{align}
At modular depth 1, the MMVs \eqref{eq: mmvclosed} and \eqref{eq: mmvbernoulli} lead to the following contributions
\begin{align}
\label{os:Xdepth1}
    \mathbb X(\epsilon) = 1 + \sum_{k=4}^\infty\ \sum_{j=1}^{k-3} &\frac{2\zeta_{j+1}}{(k{-}2)!}\,\frac{B_{k-j-1}}{k{-}j{-}1}\,\Big(
    r_{j+1 }\ep_{k|2}^{(j)}
    +r_{k-j-1 }\ep_{k|4}^{(j)}
    +r_{j+1 } r_{k-j-1 }\ep_{k|3}^{(j)}
    \Big) +\ldots \, , 
\end{align}
with $r_n$ defined by (\ref{defrk}) and terms of modular depth $\geq 2$ in the ellipsis.
Given the equality $r_{j+1}r_{k-j-1}=r_{k-1}{-}r_{j+1}{-}r_{k-1-j}$, we recover the analogous level-1 series of \cite{Kleinschmidt:2025dtk} (also see section 15 
of \cite{brown2017multiple}) upon sending $\ep_{k\mid\mu}^{(j)}\mapsto \ep_{k}^{(j)}/r_{k-1}$, 
\begin{align}
    \mathbb{X}_{\SLZ}(\ep)=1+\sum_{k=4}^\infty\ \sum_{j=1 }^{k-3} \frac{2\zeta_{j+1}}{(k{-}2)!}\,\frac{B_{k-j-1}}{k{-}j{-}1}\ep_k^{(j)}
   +\ldots \, ,
    \label{Xlvl1}
\end{align}
again modulo terms of modular depth $\geq 2$ in the ellipsis, which serves as a consistency check of the generalised formula in \eqref{os:Xdepth1}. Note that only odd values of
$j$ contribute to \eqref{os:Xdepth1} and (\ref{Xlvl1}) by the factors of $B_{k-j-1}$ which  results in $\zeta_{j+1}\in\mathbb Q\,\pi^{j+1}$ and guarantees that the coefficients at modular depth 1 in $\mathbb{X}(\ep)$ are indeed $\mathbb Q[\pi^2]$-valued as conjectured. 

At modular depth $\geq 2$, we do not attempt to find a closed formula and content ourselves to spelling out examples at low degrees here and in appendix \ref{appd.pisq}, I rearranged the formula to present $\mathbb X(\ep)$ rather than $\rho(\mathbb X(\ep))$ (there were inconsistencies before), please check that conferted $f_{2k} \rightarrow \zeta_{2k}$ everywhere
\begin{align}
    \mathbb X(\ep) &= 1 + \mathbb X^{[1]}(\ep)  -\tfrac{1}{16} \zeta_2\,[\ep_{2|3},\ep_{2|4}]    \label{eq: rhoX}  \\
    &\quad + \zeta_2\Big({-}\tfrac{1}{192}[\ep_{2|3},\ep_{4|2}] + \tfrac{13}{1280}[\ep_{2|3},\ep_{4|3}] + \tfrac{1}{1920}[\ep_{2|3},\ep_{4|4}] \nn\\
    &\qquad \quad + \tfrac{1}{960}[\ep_{2|4},\ep_{4|2}] + \tfrac{7}{7680}[\ep_{2|4},\ep_{4|3}] + \tfrac{31}{2560}[\ep_{2|4},\ep_{4|4}]\Big) \nn\\
    &\quad + \zeta_4\Big({-}\tfrac{89}{256}[\ep_{2|3},\ep_{4|2}^{(2)}] + \tfrac{71}{256}[\ep_{2|3},\ep_{4|3}^{(2)}] - \tfrac{3}{16}[\ep_{2|3},\ep_{4|4}^{(2)}] \nn\\
    &\qquad \quad - \tfrac{93}{256}[\ep_{2|4},\ep_{4|2}^{(2)}] - \tfrac{7}{256}[\ep_{2|4},\ep_{4|3}^{(2)}] - \tfrac{1}{32}[\ep_{2|4},\ep_{4|4}^{(2)}]\Big)  
    + \ldots\, , \nn
\end{align}
where $\mathbb X^{[1]}(\ep)$ refers to the terms of modular depth 1 in (\ref{os:Xdepth1}), and the terms in the ellipsis have degree $\geq 8$. The terms of degree 8 and modular depth 2 can be found in appendix \ref{appd.pisq} and exemplify that the logarithm of $\mathbb X(\ep)$ is Lie-algebra valued, and the ancillary file contains all terms through degree 10 and those of modular depth 1 through degree 16.
Note that the expression (\ref{eq: X}) for $\mathbb{X}(\ep)$, and already the degree-6 terms in (\ref{eq: rhoX}), depend on the choice of the $f$-alphabet  isomorphism $\rho$ for level-2. This can  be seen from the fact that $\mathbb M(\sigma)$ in (\ref{eq: X}) changes if one chooses to modify the expression (\ref{falph.02}) for $\rho^{-1}(f_1 f_3)$ by a rational multiple of $\zeta_4^\mot$.
The coefficients in (\ref{eq: rhoX}) and the higher-degree terms in appendix \ref{appd.pisq} are those for the choice (\ref{falph.02}).

Given the expansion of $\mathbb{X}(\ep)$,  $\sigma^{\rm g}_w$ and $[z_w,\ep_{k|\mu}]$ to modular depth $r$, one can already predict all the appearances of $\pi^{2k}\rho^{-1}(f_{i_1}\ldots f_{i_r})$ with $k\neq 0$ and $r\geq 1$ in the terms of modular depth $r{+}1$ in $\mathbb S(\ep)$. This is a computational advantage in numerically determining the appearance of $\zeta_w$ in (\ref{ag:Scoc2}) at modular depth $r{+}1$, since only a single rational unknown needs to be numerically fixed per independent word in $\ep_{k|\mu}^{(j)}$. Since the entire expansion of $\mathbb{X}(\ep)$ only involves even $a$-degrees $2k$ (accompanied by rational multiples of $\pi^{2k}$), the new terms of $\mathbb{X}(\ep)$ at modular depth $r{+}1$ always involve distinct words in $\ep_{k|\mu}^{(j)}$ compared to the contributions from a single zeta generator $\sigma_w$, which necessarily have odd $a$-degree $w$. The analogous practical advantages of the factorised form (\ref{eq: Scocalt}) of the level-1 $S$-cocycle are discussed in section 4.4 of \cite{Kleinschmidt:2025dtk}.

\subsubsection{Example: highest-weight vectors at modular depth 2}
\label{s43.1}

Highest-weight-vector contributions to zeta generators of modular depth 1 have already been quoted in (\ref{eq: depth1}). Together with the terms (\ref{os:Xdepth1}) of modular depth 1 in the series $\mathbb{X}(\ep)$, this fixes all products $\zeta_{2k}\zeta_w$ and $\zeta_{2k}\log(2)$ with $k\neq 0$ and $w$ odd in the $S$-cocycle (\ref{ag:Scoc2}) at modular depth 2. We can then solve (\ref{ag:Scoc2}) for the as-of-now unknown highest-weight-vector terms $\sigma_w^{[2]}$ by numerically matching a single rational coefficient per word $\epsilon^{(j_1)}_{k_1|\mu_1}\epsilon^{(j_2)}_{k_2|\mu_2}$ with odd $j_1{+}j_2=w{-}2$ and $k_1{+}k_2 \leq 2w$ which is independent under Pollack relations. 

For modular depth 2, the $a$-degree 3 of $\sigma_3$ does not allow for a $T$-invariant highest-weight vector, and hence the first non-trivial highest-weight vector of $\sigma_w^{[2]}$ occurs for $w=5$. Using the shorthand notation
\beq
s^d_{p|\mu,q|\nu} = s^d(\ep_{p|\mu},\ep_{q|\nu})
\label{shsmunu}
\eeq
for the projectors \eqref{eq: stvectors} and $\sigma_w^{[2]}\big\rvert_{\rm HW}$ for the highest-weight vectors of $\sigma_w^{[2]}$, we find for example
\begin{align}
\sigma_5^{[2]}\Big\rvert_{\rm HW}&=\frac{3}{64} s_{4|2,\, 4|2}^{3} + \frac{5}{128} (s_{4|3,\, 4|2}^{3} {+} s_{4|4,\, 4|2}^{3}) - \frac{1}{16}( s_{4|3,\, 4|3}^{3} {+} s_{4|4,\, 4|4}^{3}) + \frac{15}{128} s_{4|4,\, 4|3}^{3}  \, ,\label{eq: HWd2}\\
    \sigma_7^{[2]}\Big\rvert_{\rm HW}&=\frac{97}{768} s_{4|2,\, 6|2}^{3} + \frac{7}{1024} (s_{4|2,\, 6|3}^{3} + s_{4|2,\, 6|4}^{3}) + \frac{35}{1024} (s_{4|3,\, 6|2}^{3}+s_{4|4,\, 6|2}^{3}) \notag\\
    &\quad
    - \frac{5}{48} (s_{4|3,\, 6|3}^{3}+ s_{4|4,\, 6|4}^{3})
    + \frac{105}{1024} (s_{4|3,\, 6|4}^{3}   + s_{4|4,\, 6|3}^{3})  - \frac{11}{15360} s_{6|2,\, 6|2}^{5}\notag\\
    &\quad - \frac{49}{40960} (s_{6|3,\, 6|2}^{5}+s_{6|4,\, 6|2}^{5}) + \frac{7}{3840} (s_{6|3,\, 6|3}^{5} +s_{6|4,\, 6|4}^{5}) - \frac{147}{40960} s_{6|4,\, 6|3}^{5}  \,,
   \notag
\end{align}
with the next example $\sigma_9^{[2]}$ in appendix \ref{appd.1}. We have grouped pairs of terms that are related by the $\theta_T$-invariance of $\sigma_w$ into parentheses. Note that $s^d_{p|\mu,q|\nu} = (-1)^{d+1}s^d_{q|\nu,p|\mu}$ implies that $s_{k|4,\, k|3}^{d}$ is by itself  $\theta_T$-invariant for odd $d$.

After numerically inferring all highest-weight vectors of $\sigma_w^{[2]}$ up to and including degree 16, we arrived at the following conjectural closed formula for modular depth 2,
\begin{align}
\sigma_w^{[2]}\Big\rvert_{\rm HW}= -\frac12\sum_{\substack{ d=3 \\{d \ {\rm odd}}}}^{w-2}
    \frac{\BF_{d-1}}{\BF_{w-d+2}}\sum_{k=d+1}^{w-1}\BF_{k-d+1}\,\BF_{w-k+1} \! \! \!
    \sum_{\mu,\nu\in\{2,3,4\}}  \! \! \! c_{\mu\nu}^{(w,d,k)}\,
  s^d_{k|\mu,\,(w-k+d)|\nu}\, ,
\label{hwlevel2}
\end{align}
where $c_{\mu\nu}$ defines a $3\times 3$ matrix of rational numbers 
\begin{align}
    c_{23}^{(w,d,k)}&=c_{24}^{(w,d,k)} = - \frac{r_{w - k + 2} r_{w} r_{d - 1} r_{k - d + 2}}{
    r_{k - w}  r_{w - d + 3}}\, ,&
    c_{33}^{(w,d,k)}&=c_{44}^{(w,d,k)} = -\,\frac{r_{d-1}\,r_{k-d+2}\,r_{w-k+2}}{r_{w-d+3}}\, , \nn\\
    c_{32}^{(w,d,k)}&=c_{42}^{(w,d,k)} = -  \frac{ r_{k - d + 2} r_w r_{d - 1} r_{w - k + 2} }{ r_{d - k}
    r_{w - d + 3}} \, ,&
    c_{34}^{(w,d,k)}&=c_{43}^{(w,d,k)}=r_w\,c_{33}^{(w,d,k)}\,,\nn\\
    c_{22}^{(w,d,k)} &= r_{k-1}\,r_{w-k+d-1}-\!\!\sum_{(\mu,\nu)\neq(2,2)}\!\! c_{\mu\nu}^{(w,d,k)}
    \,.
\label{hwdressing}
\end{align}
In view of $s^d_{p|\mu,q|\nu}=s^d_{q|\nu,p|\mu}$ for odd $d$, 
(\ref{hwlevel2}) automatically gathers the two orderings of $p|\mu$ and~$q|\nu$. Furthermore, using $\sum_{\mu,\nu=2}^4 c_{\mu\nu}^{(w,d,k)}=r_{k-1}r_{w-k+d-1}$, \eqref{hwlevel2} reduces to the closed formula for highest-weight vectors at level 1 in the second line of (\ref{eq: sigma}) under the transcription $\ep_{k|\mu}^{(j)}\mapsto \ep_{k}^{(j)}/r_{k-1}$. 
Together with the remaining terms $\sigma_w^{[2]} - \sigma_w^{[2]}\big\rvert_{\rm HW}$ in (\ref{sysxi2.09}), the highest-weight vectors of (\ref{hwlevel2}) complete our conjectural explicit form of level-2 zeta generators at modular depth 2. The resulting terms $\sigma_w^{[2]}$ through degree 16 are listed in the ancillary file.

Closed formulas for highest-weight vectors of $\sigma_w^{[r]}$ at modular depth $r\geq 3$ are currently out of reach, but we shall display several examples in section \ref{s43.3} and appendix \ref{appd.2} below.

\subsubsection{\texorpdfstring{Example: $[z_w,\ep_{k|\mu}]$ with $w\geq 3$ at modular depth 3}{Example: [zw,epsilon(k|mu)] with w>=3 at modular depth 3}}
\label{s43.9}

With the complete expression (\ref{sysxi2.09}) and (\ref{hwlevel2}) for $\sigma_w^{[2]}$ in place, we can obtain the terms $[z_w,\ep_{k|\mu}]^{[3]}$ of modular depth 3 from (\ref{zepk4}) and (\ref{zepk2}). It suffices to consider $w\geq 3$, since the brackets of $z_1$ do not go beyond modular depth 2 and were fully determined in section \ref{sec:z1ep}. The lowest-degree bracket $[z_w,\ep_{k|\mu}]^{[3]}$ with $w\geq 3$ vanishes,
\beq
[z_3, \ep_{2|\mu}]^{[3]} = 0 \, .
\eeq
In order to compactly represent the non-vanishing $[z_w,\ep_{k|\mu}]^{[3]}$ with $w\geq 3$, we generalise the projectors (\ref{eq: stvectors}) to
\begin{align}
s^{d}\big(\ep_{k|\mu},X_r\big) &=\frac{(d{-}2)!}{(k{-}2)!\,(r{-}2)!}\sum_{i=0}^{d-2}(-1)^i\,
\big[\ep_{k|\mu}^{(k-2-i)},\,\ad_{\ep_0}^{\,r-d+i}X_r\big]\,,
\label{eq:eqv:d3sdef} \\
t^{d}\big(\ep_{k|\mu},X_r\big) &=\frac{(d{-}2)!}{(k{-}2)!\,(r{-}2)!}\sum_{i=0}^{d-2}(-1)^i \, \frac{(k{-}2{-}i)!\,(r{-}d{+}i)!}{i! \, (d{-}2{-}i)!}\,
\big[\ep_{k|\mu}^{(i)},\,\ad_{\ep_0}^{\,d-2-i}X_r\big]\,,
\notag
\end{align}
where $X_r=t^{d_1}(\ep_{k_1|\mu_1},\ep_{k_2|\mu_2})$ is taken to be the  lowest-weight vector of an $r{-}1$-dimensional representation of $\mathfrak{sl}_2$ with $r=k_1{+}k_2{-}2d_1{+}2$.
This notation (\ref{eq:eqv:d3sdef}) for instance leads to  
\begin{align}
[z_3,\ep_{4|3}]^{[3]}=\ &\tfrac{7}{160}t^2\big(\ep_{2|3},t^3(\ep_{4|2},\ep_{4|2})\big)-\tfrac{91}{1920}t^2\big(\ep_{2|3},t^3(\ep_{4|2},\ep_{4|3})\big)+\tfrac{7}{240}t^3\big(\ep_{4|2},t^2(\ep_{2|3},\ep_{4|3})\big)\nn\\
&-\tfrac{49}{960}t^2\big(\ep_{2|3},t^3(\ep_{4|2},\ep_{4|4})\big)+\tfrac{49}{1920}t^3\big(\ep_{4|2},t^2(\ep_{2|3},\ep_{4|4})\big)-\tfrac{1}{160}t^2\big(\ep_{2|3},t^3(\ep_{4|3},\ep_{4|3})\big)\nn\\
&+\tfrac{7}{960}t^2\big(\ep_{2|3},t^3(\ep_{4|3},\ep_{4|4})\big)-\tfrac{7}{1920}t^3\big(\ep_{4|3},t^2(\ep_{2|3},\ep_{4|4})\big)+\tfrac{7}{160}t^2\big(\ep_{2|4},t^3(\ep_{4|2},\ep_{4|2})\big)\nn\\
&-\tfrac{49}{960}t^2\big(\ep_{2|4},t^3(\ep_{4|2},\ep_{4|3})\big)+\tfrac{49}{1920}t^3\big(\ep_{4|2},t^2(\ep_{2|4},\ep_{4|3})\big)-\tfrac{49}{1920}t^2\big(\ep_{2|4},t^3(\ep_{4|2},\ep_{4|4})\big)\nn\\
&+\tfrac{49}{960}t^3\big(\ep_{4|2},t^2(\ep_{2|4},\ep_{4|4})\big)+\tfrac{49}{1920}t^2\big(\ep_{2|4},t^3(\ep_{4|3},\ep_{4|4})\big)+\tfrac{49}{1920}t^3\big(\ep_{4|3},t^2(\ep_{2|4},\ep_{4|4})\big)\nn\\
&-\tfrac{7}{160}t^2\big(\ep_{2|4},t^3(\ep_{4|4},\ep_{4|4}) \big)\,,
\label{sysxi3.11}
\end{align}
as well as
\begin{align}
[z_5,\ep_{2|3}]^{[3]} = \ & \tfrac{31}{30720}t^4\big(\ep_{4|2},t^3(\ep_{4|2},\ep_{4|3})\big) - \tfrac{217}{92160}t^4\big(\ep_{4|2},t^3(\ep_{4|2},\ep_{4|4})\big) - \tfrac{31}{46080}t^4\big(\ep_{4|2},t^3(\ep_{4|3},\ep_{4|3})\big) \nn\\
& + \tfrac{31}{18432}t^4\big(\ep_{4|3},t^3(\ep_{4|2},\ep_{4|4})\big) + \tfrac{217}{184320}t^4\big(\ep_{4|2},t^3(\ep_{4|4},\ep_{4|4})\big) \nn\\
& + \tfrac{31}{23040}t^4\big(\ep_{4|3},t^3(\ep_{4|3},\ep_{4|4})\big) - \tfrac{31}{61440}t^4\big(\ep_{4|3},t^3(\ep_{4|4},\ep_{4|4}) \big)\,.
\label{sysxi3.08}
\end{align}
At modular depth 4, the brackets of $z_3$ with $\tilde\ep_{2|\nu}$ read, in terms of the sum-zero $\ep_{2|\mu}$ of (\ref{fixeps}),
\begin{align}
[z_3,\tilde\ep_{2|3}]^{[4]} &= \tfrac{1}{192}
\Big(2\,\ad^2_{\ep_{2|3}} + 2 \ad_{\ep_{2|3}} \ad_{\ep_{2|4}} - \ad^2_{\ep_{2|4}}\Big) [\ep_{2|3},\ep_{2|4}]\,, \label{eq:z3tail4}\\
[z_3,\tilde\ep_{2|4}]^{[4]} &= \tfrac{1}{192}
\Big(\ad^2_{\ep_{2|3}} - 2 \ad_{\ep_{2|3}} \ad_{\ep_{2|4}} - 2\,\ad^2_{\ep_{2|4}}\Big) [\ep_{2|3},\ep_{2|4}]\,. \notag
\end{align}
The complete expressions for all $[z_w,\ep_{k|\mu}]^{[r]}$ of degree $2w{+}k\leq 10$ (including all modular depths $r=2,3,\ldots,w{+}1$) can be found in the ancillary file. Their terms of modular depth $r=2$ are given there through degree $2w{+}k\leq 16$.

\subsubsection{Example: zeta generators at modular depth 3}
\label{s43.3}

The closed-form expressions (\ref{sysxi2.09}) and (\ref{hwlevel2}) for $\sigma_w^{[2]}$ also fix the geometric terms $\sigma_w^{[3]}$ of modular depth~3 up to highest-weight vectors via (\ref{eq: sigmarec}). We shall here gather examples of $\sigma_w^{[3]}$ including highest-weight vectors which are determined from the factorised $S$-cocycle as discussed above section \ref{s43.2}.

By the $a$-degree $j{+}1$ of $\ep_{k|\mu}^{(j)}$, terms of modular depth 3 in $\sigma_w^{\rm g}$ must be $[\ep_{k_1|\mu_1}^{(j_1)} , [\ep_{k_2|\mu_2}^{(j_2)} , \ep_{k_3|\mu_3}^{(j_3)}]]$ with $j_1{+}j_2{+}j_3 = w{-}3$. At the same time, the $\mathfrak{sl}_2$ condition $j_i\leq k_i{-}2$ bounds the superscripts to obey $j_1{+}j_2{+}j_3 \leq k_1{+}k_2{+}k_3{-}6$, such that the expansion of $\sigma_w^{[3]}$ can only involve degrees $k_1{+}k_2{+}k_3 \geq w{+}3$. The lowest-degree terms, which saturate this bound, are necessarily highest-weight vectors, since they are Lie polynomials in highest-weight vectors $\ep_{k_i|\mu_i}^{( k_i{-}2)}$. 

At $w=3$, for instance, the expansion of $\sigma_3^{[3]}$ is numerically found to start with the highest-weight vector $\tfrac{1}{192}[\ep_{2|3}{-}\ep_{2|4},[\ep_{2|3},\ep_{2|4}]]$ of degree $w{+}3=6$. This is the simplest instance of  highest-weight vectors $v_w$ in $\mathfrak{sl}_2$ singlets but non-trivial representations of ${\cal S}_3$, which are extracted from the $S$-cocycle condition (\ref{ag:Scoc2}) via (\ref{getvw}). The leftover degree $\geq 8$ parts of the Lie series $\sigma_3^{[3]}$, by contrast, no longer involve highest-weight vectors (the Cartan eigenvalues are negative term by term) and are thus determined by (\ref{eq: sigmarec}). In ordinary brackets, the expansion of $\sigma_3^{[3]}$ through degree twelve reads,
\begin{align}
\sigma_3^{[3]} &= 
\tfrac{1}{192}\big[\ep_{2|3}{-}\ep_{2|4},[\ep_{2|3},\ep_{2|4}]\big]
+\tfrac{3}{512}\,\big[\ep_{2|3}{+}\ep_{2|4},[\ep_{2|3}{+}\ep_{2|4},\ep_{4|2}]\big] \label{sysxi3.03}\\
&\quad+\tfrac{1}{10240}\,\big[\ep_{4|2},[\ep_{2|3}{+}\ep_{2|4},\ep_{4|2}]\big]+\tfrac{7}{61440}\,\big[\ep_{2|3}{+}\ep_{2|4},[\ep_{4|2},\ep_{4|3}{+}\ep_{4|4}]\big] \nn\\
&\quad-\tfrac{7}{81920}\,\big[\ep_{4|2},[\ep_{2|3}{+}\ep_{2|4},\ep_{4|3}{+}\ep_{4|4}]\big]+\tfrac{1}{322560}\,\big[\ep_{2|3}{+}\ep_{2|4},[\ep_{4|2},\ep_{6|2}]\big] \nn\\
&\quad-\tfrac{1}{387072}\,\big[\ep_{4|2},[\ep_{2|3}{+}\ep_{2|4},\ep_{6|2}]\big]-\tfrac{31}{10321920}\,\big[\ep_{2|3}{+}\ep_{2|4},[\ep_{4|2},\ep_{6|3}{+}\ep_{6|4}]\big] \nn\\
&\quad+\tfrac{31}{12386304}\,\big[\ep_{4|2},[\ep_{2|3}{+}\ep_{2|4},\ep_{6|3}{+}\ep_{6|4}]\big] \nn\\
&\quad+\tfrac{49}{88473600}\,\Big( \big[\ep_{4|2},[\ep_{4|3},\ep_{4|4}]\big]
-\big[\ep_{4|4},[\ep_{4|2},\ep_{4|4}]\big]
-
\big[\ep_{4|3},[\ep_{4|2},\ep_{4|3}]\big] \Big) 
\nn \\
&\quad+\tfrac{7}{11059200}\,\big[\ep_{4|2},[\ep_{4|2},\ep_{4|3}{+}\ep_{4|4}]\big]-\tfrac{49}{44236800}\,\big[\ep_{4|3},[\ep_{4|2},\ep_{4|4}]\big] + \ldots\,,
\notag
\end{align}
with terms of degree $\geq 14$ in the ellipsis.

The expansion of $\sigma_5^{[3]}$ in turn starts with the highest-weight vectors 
\begin{align}
\sigma_5^{[3]} &= -\tfrac{3487}{3072}\Big( \big[\ep_{2|3},[\ep_{2|3},\ep^{(2)}_{4|2}]\big] + \big[\ep_{2|4},[\ep_{2|4},\ep^{(2)}_{4|2}]\big]\Big)  -\tfrac{1757}{1920}\Big( \big[\ep_{2|3},[\ep_{2|4},\ep^{(2)}_{4|2}]\big] + \big[\ep_{2|4},[\ep_{2|3},\ep^{(2)}_{4|2}]\big]\Big) \nn\\
&\quad +\tfrac{2249}{3072}\Big( \big[\ep_{2|3},[\ep_{2|3},\ep^{(2)}_{4|3}]\big] + \big[\ep_{2|4},[\ep_{2|4},\ep^{(2)}_{4|4}]\big]\Big)  -\tfrac{121}{192}\Big( \big[\ep_{2|3},[\ep_{2|3},\ep^{(2)}_{4|4}]\big] + \big[\ep_{2|4},[\ep_{2|4},\ep^{(2)}_{4|3}]\big]\Big) \nn\\
&\quad -\tfrac{3383}{7680}\Big( \big[\ep_{2|3},[\ep_{2|4},\ep^{(2)}_{4|3}]\big] + \big[\ep_{2|4},[\ep_{2|3},\ep^{(2)}_{4|4}]\big]\Big) +\tfrac{2261}{3840}\Big( \big[\ep_{2|3},[\ep_{2|4},\ep^{(2)}_{4|4}]\big] + \big[\ep_{2|4},[\ep_{2|3},\ep^{(2)}_{4|3}]\big]\Big)
\notag \\
&\quad +\ldots \,,
\label{eq:eqv:lam58}
\end{align}
followed by degree $\geq 10$ terms in the ellipsis. The pairs of terms in parentheses are related by the $\theta_T$-invariance of $\sigma_5$.
The degree 10 contributions to $\sigma^{[3]}_5$ and $\sigma^{[3]}_7$ can be found in appendix \ref{appd.2}, and terms in $\sigma^{[r]}_w$ of higher degree and/or modular depth are given in the ancillary~file.

\subsection{Examples of equivariant Eisenstein integrals beyond modular depth 1}
\label{ex:sec44}

We define the equivariant iterated Eisenstein integrals as the expansion coefficients of the series $\INNeqv{\tau}$ in (\ref{eq: eqvprop}) with the normalisation factors of (\ref{eq: I2}). We expand in the degree-two derivations $\tilde\ep_{2|\nu}$ with $\nu=3,4$ of (\ref{rewrd2}), which accompany the independent $f^{(2)}(\omega_\nu,\tau)$ in the connection (\ref{eq: N2conn}), and in the $\ep_{k|\mu}^{(j)}$ with $k\geq 4$ and $\mu=2,3,4$:
\begin{align}
\INNeqv{\tau} &=  1+ \sum_{\nu=3}^4 \eezeqv{0}{2}{\omega_\nu}{\tau}\, \tilde\ep_{2|\nu} + \sum_{k=4}^\infty (k{-}1) \sum_{j=0}^{k-2} \frac{(-1)^j}{j!} \sum_{\mu=2}^4
\eezeqv{j }{k}{ \omega_\mu}{\tau}\, \ep_{k|\mu}^{(j)} \label{eq:Eeqvdef} \\
&\quad + \sum_{\nu_1,\nu_2=3}^4 \eezeqv{0&0}{2&2}{\omega_{\nu_1}&\omega_{\nu_2}}{\tau}\, \tilde\ep_{2|\nu_1}\tilde\ep_{2|\nu_2} \notag \\
&\quad +  \sum_{k=4}^\infty (k{-}1) \sum_{j=0}^{k-2} \frac{(-1)^j}{j!} \sum_{\mu=2}^4 \sum_{\nu=3}^4 \eezeqv{0&j}{2&k}{\omega_\nu&\omega_\mu}{\tau}\, \tilde\ep_{2|\nu}\ep_{k|\mu}^{(j)} \notag \\
&\quad + \sum_{k=4}^\infty (k{-}1) \sum_{j=0}^{k-2} \frac{(-1)^j}{j!} \sum_{\mu=2}^4 \sum_{\nu=3}^4 \eezeqv{j&0}{k&2}{\omega_\mu&\omega_\nu}{\tau}\, \ep_{k|\mu}^{(j)}\tilde\ep_{2|\nu} \notag \\
&\quad
+ \sum_{k_1,k_2=4}^\infty (k_1{-}1)(k_2{-}1) \sum_{j_1=0}^{k_1-2} \sum_{j_2=0}^{k_2-2} \frac{(-1)^{j_1+j_2}}{j_1! j_2!}
\sum_{\mu_1,\mu_2=2}^4
\eezeqv{j_1 &j_2}{k_1 &k_2}{\omega_{\mu_1} &\omega_{\mu_2}}{\tau}\, \ep_{k_1| \mu_1}^{(j_1)} \ep_{k_2 | \mu_2}^{(j_2)} \notag \\
&\quad + \ldots\,.\notag
\end{align}
Starting from degree 8 and modular depth 2, individual ${\cal E}^{\rm eqv}[\ldots]$ are no longer well-defined whenever the accompanying derivations obey Pollack relations, see (\ref{eq: pollack8}) for the lowest-degree example. The expansion is thus understood as defining those linear combinations of ${\cal E}^{\rm eqv}$ that accompany words in $\ep_{k|\mu}^{(j)}$ that are independent under Pollack relations.

In words of any length, a letter $\tilde\ep_{2|\nu}$ comes with the entries $0$, $2$, $\omega_\nu$ in its column of ${\cal E}^{\rm eqv}[\ldots]$ and the factor 1, and a letter $\ep^{(j)}_{k|\mu}$ with $k\geq 4$ comes with the entries $j$, $k$, $\omega_\mu$ and the factor $(k{-}1)(-1)^j/j!$. In particular, the label $\omega_2$ never occurs at $k_i=2$. The integrals ${\cal E}[\ldots]$ in the expansion (\ref{eq: I2}) of $\INN{\tau}$ in the same letters carry the same labels and factors.

The linear combinations of ${\cal E}^{\rm eqv}[\ldots]$ 
defined by the expansion (\ref{eq:Eeqvdef}) obey the differential equations of their holomorphic counterparts (\ref{eq: eez}),
\beq
\partial_\tau \eezeqv{j_1&\ldots&j_r}{k_1&\ldots&k_r}{\omega_{\mu_1}&\ldots&\omega_{\mu_r}}{\tau}
= (2\pi i)^{1+j_r-k_r}\, \tau^{j_r} f^{(k_r)}(\omega_{\mu_r},\tau)\, \eezeqv{j_1&\ldots&j_{r-1}}{k_1&\ldots&k_{r-1}}{\omega_{\mu_1}&\ldots&\omega_{\mu_{r-1}}}{\tau}\, ,
\label{eq:Eeqvdiff}
\eeq
At modular depth one, the terms (\ref{eq: depth1}) of $\sigma_w$ give
\begin{align}
\eezeqv{0}{2}{\omega_\nu}{\tau} &= \eez{0}{2}{\omega_\nu}{\tau} + \overline{\eez{0}{2}{\omega_\nu}{\tau}} - \tfrac{2}{3}\log(2)\, , \ \ \ \ \nu=3,4\, ,
 \label{eq:Eeqvd1} \\
 \eezeqv{j}{k}{\omega_\mu}{\tau} &= \eez{j}{k}{\omega_\mu}{\tau} + (-1)^j\, \overline{\eez{j}{k}{\omega_\mu}{\tau}} - \delta_{j,k-2}\,\delta_{\mu,2}\, \frac{2 r_{k-1}}{k{-}1}\, \zeta_{k-1}\, , \ \ \ \ k\geq 4\, , \ \ \ \ \mu = 2,3,4\, ,
 \notag
\end{align}
with $r_{k-1}$ given by (\ref{defrk}), in agreement with~\cite{Duhr:2025lvr}.

\subsubsection{Modular depth 2}

At modular depth two, the highest-weight vectors (\ref{eq: HWd2}) of $\sigma_5^{[2]}$ first enter at degree eight, for instance in
\begin{align}
\eezeqv{1&2}{4&4}{\omega_2&\omega_2}{\tau} &= \eez{1&2}{4&4}{\omega_2&\omega_2}{\tau} - \overline{\eez{2&1}{4&4}{\omega_2&\omega_2}{\tau}} - \eez{2}{4}{\omega_2}{\tau}\, \overline{\eez{1}{4}{\omega_2}{\tau}} \label{eq:Eeqvex1} \\
&\quad - \tfrac{1}{2}\,\zeta_3\, \overline{\eez{1}{4}{\omega_2}{\tau}} + \tfrac{1}{96}\,\zeta_5 \, . \notag
\end{align}
Brackets of arithmetic zeta generators already contribute at degree six, for instance in
\begin{align}
\eezeqv{0&0}{2&4}{\omega_3&\omega_2}{\tau} &= \eez{0&0}{2&4}{\omega_3&\omega_2}{\tau} + \overline{\eez{0&0}{4&2}{\omega_2&\omega_3}{\tau}} + \eez{0}{4}{\omega_2}{\tau}\, \overline{\eez{0}{2}{\omega_3}{\tau}} - \tfrac{2}{3}\log(2)\, \eez{0}{4}{\omega_2}{\tau} \label{eq:Eeqvex2} \\
&\quad + \tfrac{1}{45}\log(2) \Big( 19\, \overline{\eez{0}{4}{\omega_2}{\tau}} + 7\, \overline{\eez{0}{4}{\omega_3}{\tau}} + 49\, \overline{\eez{0}{4}{\omega_4}{\tau}} \Big) \, , \notag
\end{align}
where the second line can be traced back to $[z_1, \ep_{4|\mu}]$ in (\ref{eq:z1e4}).

\subsubsection{Modular depth 3}

At modular depth three and degree eight, the cubic terms of $\mathbb M^\sv(\sigma)$ in (\ref{lv2msv}) bring in the single-valued cyclotomic MZV $\rho^{-1}(\sv(f_1f_1f_3))$ of (\ref{falph.sv}), for instance in
\begin{align}
\eezeqv{0&0&2}{2&2&4}{\omega_3&\omega_3&\omega_2}{\tau} &= \eez{0&0&2}{2&2&4}{\omega_3&\omega_3&\omega_2}{\tau} + \overline{\eez{2&0&0}{4&2&2}{\omega_2&\omega_3&\omega_3}{\tau}} \label{eq:Eeqvex3} \\
&\quad + \eez{0&2}{2&4}{\omega_3&\omega_2}{\tau}\, \overline{\eez{0}{2}{\omega_3}{\tau}} + \eez{2}{4}{\omega_2}{\tau}\, \overline{\eez{0&0}{2&2}{\omega_3&\omega_3}{\tau}} \notag \\
&\quad - \tfrac{2}{3}\log(2) \Big( \eez{0&2}{2&4}{\omega_3&\omega_2}{\tau} + \eez{2}{4}{\omega_2}{\tau}\, \overline{\eez{0}{2}{\omega_3}{\tau}} \Big) + \tfrac{2}{9}\log^2(2)\, \eez{2}{4}{\omega_2}{\tau} \notag \\
&\quad + \tfrac{1}{45}\log(2) \Big( 19\, \overline{\eez{2&0}{4&2}{\omega_2&\omega_3}{\tau}} + 7\, \overline{\eez{2&0}{4&2}{\omega_3&\omega_3}{\tau}} + 49\, \overline{\eez{2&0}{4&2}{\omega_4&\omega_3}{\tau}} \Big) \notag \\
&\quad + \tfrac{1}{135}\log^2(2) \Big( 79\, \overline{\eez{2}{4}{\omega_2}{\tau}} + 7\, \overline{\eez{2}{4}{\omega_3}{\tau}} + 49\, \overline{\eez{2}{4}{\omega_4}{\tau}} \Big) \notag \\
&\quad + \tfrac{1}{2}\,\zeta_3\, \overline{\eez{0&0}{2&2}{\omega_3&\omega_3}{\tau}} - \tfrac{11}{180}\log(2)\,\zeta_3\, \overline{\eez{0}{2}{\omega_3}{\tau}} \notag \\
&\quad + \tfrac{49}{360}\, \rho^{-1}\big(\sv(f_1f_1f_3)\big) - \tfrac{19}{270}\log^2(2)\,\zeta_3 - \tfrac{3439}{11520}\,\zeta_5 \, . \notag
\end{align}
Examples at modular depth four and five are given in appendix \ref{appd.3}, and the ancillary file contains the series (\ref{eq:Eeqvdef}) through degree 16 at modular depth 1 and 2 and through degree 10 at modular depth 3, 4 and 5.

\subsection{Relation to non-holomorphic modular forms}
\label{sec:nonholo}

By the equivariant transformation (\ref{eq: equiv2}) of the series $\mathbb{I}^{\rm eqv}(\ep;\tau)$ in (\ref{eq:Eeqvdef}), the combinations of ${\cal E}^{\rm eqv}[\ldots]$ along with the independent words in $\tilde \ep_{2|\nu}$, $\ep^{(j)}_{k|\mu}$ under Pollack relations can be combined to non-holomorphic modular forms for $\Gamma(2)$. The required linear combinations are formed in direct analogy with the $\SLZ$ case \cite{Brown:2017qwo, Dorigoni:2022npe, Dorigoni:2024oft}, mixing different values of the entries $j_i$ and involving rational functions of $i\pi \tau$, $i\pi \bar \tau$ as coefficients, see (3.26) of \cite{Dorigoni:2024oft}. The non-holomorphic modular forms constructed from ${\cal E}^{\rm eqv}[\ldots]$ are expected to be elliptic modular graph forms \cite{DHoker:2018mys, DHoker:2020tcq, Basu:2020pey, Dhoker:2020gdz} evaluated at half-periods or the origin as one can see from their iterated-integral representations \cite{Hidding:2022vjf, Schlotterer:2025qjv}. In particular, special values of single-valued elliptic multiple polylogarithms in one variable \cite{Baune:2025ndr, Schlotterer:2025qjv} at half-periods or the origin are expressible in terms of these modular forms, but not expected to form a spanning set.

\section{From weight 2 Eisenstein series to multiple polylogarithms}
\label{sec:weight2}

One new feature of modular forms at level 2 compared to those at level 1 is the existence of holomorphic modular forms of weight 2. Primitives of weight-2 Eisenstein series for a principal congruence subgroup~$\Gamma$ correspond to logarithms on the modular curve $Y_\Gamma = \Gamma\backslash \mathbb{H} $, and, as shown in \cite{Duhr:2025lvr}, their single-valued versions correspond to the single-valued logarithms.
Here, we shall explain how this carries over to higher-depth iterated integrals in the case of $\Gamma(2)$, i.e., how single-valued iterated Eisenstein integrals over $f^{(2)}(\omega_\mu,\tau)$ lead to a generating series of single-valued multiple polylogarithms (svMPLs) on the modular curve. 
We recall that for $\Gamma(2)$ the modular curve $Y_{\Gamma(2)}$ has genus 0.

\subsection{A Knizhnik-Zamolodchikov connection on the modular curve}
\label{sec:lam.1}

Let $\mathfrak{i}= \langle\ep_0,
\ep_0^\vee,
\ep_{k\mid\mu} \,\vert\,  k\geq 4,\ \mu\in\{2,3,4\} \rangle$ 
 be the ideal in $\mathfrak{sl}_2\ltimes \mathfrak{u}$ generated by $\ep_0$
, its dual $\ep_0^\vee$, 
and all generators $\ep_{k\mid\mu}$ of degree $\geq 4$. Consider the map 
\begin{equation}
    \pi_2: \mathfrak{sl}_2\ltimes\mathfrak{u}\longrightarrow  (\mathfrak{sl}_2\ltimes \mathfrak u)/\mathfrak i\,,
    \label{eq:pi2}
\end{equation}
which acts via
\begin{equation}
    \pi_2(\ep_{0}) =   \pi_2(\ep^\vee_{0}) = 0
\, ,
    \ \ \ \ \ \ 
    \pi_2(\tilde \ep_{2|\nu}) = \tilde \ep_{2|\nu} \, ,
    \ \ \ \ \ \ 
    \pi_2(\ep^{(j)}_{k|\mu}) = 0   \ \ \ {\rm for}\ k\geq 4
  \, .
    \label{eq:pi2expl}
\end{equation}
It projects onto the Lie subalgebra ${\rm Lie}\!\left[ \tilde\ep_{2|3},\tilde\ep_{2|4} \right]$ generated by the generators of degree $2$ inside~$\mathfrak{sl}_2\ltimes \mathfrak{u}$, which we shall demonstrate to be free in section \ref{sec:lam.3} below. Since $\mathfrak i$ is an ideal, the map $\pi_2$ is a Lie-algebra homomorphism, and thus compatible with the Lie bracket, i.e., for all $w_1,w_2 \in \mathfrak{sl}_2\ltimes \mathfrak{u}$, we have
\begin{equation}
    \pi_2(\left[ w_1,w_2 \right]) = \left[\pi_2(w_1), \pi_2(w_2) \right].
\end{equation}
Under the map $\pi_2$, the connection $\ANN{\tau}$ from~\eqref{eq: N2conn} only keeps the contribution of Eisenstein series of weight 2,
\begin{equation}
     \pi_2\left(\ANN{\tau}\right) =  -\frac{\dd \tau}{2\pi i }  \Big(\tilde\epsilon_{2|3} f^{(2)}(\omega_3,\tau) + \tilde\epsilon_{2|4} f^{(2)}(\omega_4,\tau)  \Big)\,.
     \label{eq:pi2A.1}
\end{equation}
Let $\lambda:Y_{\Gamma(2)}\to \mathbb C$ be a Hauptmodul  for the modular curve. In the following, we choose 
\begin{equation}
    \lambda(\tau) = \frac{\vartheta_2(\tau)^4}{\vartheta_3(\tau)^4}\,,
    \label{eq:hauptmodul}
\end{equation}
where $\vartheta_i(\tau) = \vartheta_i(0,\tau)$ is the $i^{\textrm{th}}$ Jacobi theta function.
As worked out in \cite{Broedel:2018rwm}, 
the integration kernels $f^{(k)}(\omega_\mu,\tau)$ spanning the Eisenstein space $\mathcal{E}_k(\Gamma(2))$ can be written in the form
\begin{equation}
    f^{(k)}(\omega_\mu,\tau) = \pi^k\,\vartheta_3(\tau)^{2k} {\cal P}_{k,\mu}\big(\lambda(\tau) \big)\, ,
    \label{fktokla}
\end{equation}
for some polynomial ${\cal P}_{k,\mu}$ of degree $\leq \tfrac{k}{2}$ in $\lambda(\tau)$ with coefficients in $\mathbb{Q}$. 
Note that, locally close to $\tau = i\infty$, we may change variables from $\tau$ to the Hauptmodul $\lambda$, and we have

\begin{equation}
    \frac{\pi}{2}\,\vartheta_3(\tau)^2 = \operatorname{K}(\lambda)\,,
\end{equation}
where
\begin{equation}
    \operatorname{K}(\lambda) = \int_0^1 \frac{\dd t}{\sqrt{(1-t^2)(1-\lambda t^2)}}\,,
\label{klam}
\end{equation}
denotes the complete elliptic integral of the first kind. Explicitly, we can express the Kron\-ecker-Eisenstein kernels of modular weight 2 in (\ref{eq:pi2A.1}) in terms of the Hauptmodul as follows
\begin{align}
    f^{(2)}(\omega_3,\tau) &= \frac{\pi^2}{6} \,\vartheta_3(\tau)^4\, \big(1-2\lambda(\tau) \big) \, ,\label{eq:f.ellK}\\
    f^{(2)}(\omega_4,\tau) &= \frac{\pi^2}{6} \,\vartheta_3(\tau)^4 \,\big(1+\lambda(\tau) \big)\,.\notag
\end{align}
In order to identify iterated integrals of (\ref{eq:f.ellK}) as polylogarithms on the modular curve, we compute the Jacobian for changing variables from $\tau$ to $\lambda$
\begin{equation}
    \frac{\dd \lambda(\tau)}{\dd \tau} =i\pi \,\vartheta_3(\tau)^4\,\big(1-\lambda(\tau) \big)\lambda(\tau)\,,
    \label{eq:Jac.lambda}
\end{equation}
which only cancels the powers of $\vartheta_3(\tau)$ in (\ref{fktokla}) for $k=2$ and motivates the projection to degree 2 in (\ref{eq:pi2A.1}). By virtue of (\ref{eq:f.ellK}) and (\ref{eq:Jac.lambda}), the projected connection
takes the Knizhnik-Zamolodchikov (KZ) form
\begin{equation}
    \pi_2\left(\ANN{\lambda}\right) =  \frac{1}{12}\dd \lambda \bigg( \big( \tilde\ep_{2|3} + \tilde\ep_{2|4} \big) \frac{1}{\lambda} + \big( \tilde\ep_{2|3} -2 \tilde\ep_{2|4} \big) \frac{1}{\lambda{-}1} \bigg)
    = -\dd \lambda \, \bigg( \frac{x}{\lambda}+ \frac{y}{\lambda {-}1}\bigg) \,,
    \label{eq:pi2A.2}
\end{equation}
with non-commuting variables
\begin{equation}
    x = -\frac{1}{12} (\tilde\ep_{2|3} + \tilde\ep_{2|4}) \, , \ \ \ \ 
    y = -\frac{1}{12} (\tilde\ep_{2|3} -2 \tilde\ep_{2|4})\, .
    \label{eq:xy.ep}
\end{equation}

\subsubsection{\texorpdfstring{Freeness of the Lie algebra $\pi_2(\mathfrak{sl}_2\ltimes \mathfrak{u})$}{Freeness of the Lie algebra pi2(sl2 x u)}}
\label{sec:lam.3}

It will be crucial in the remainder of this section that the image ${\rm Lie}[x,y]$ of $\mathfrak{sl}_2\ltimes \mathfrak{u}$ under $\pi_2$ is free. To show that this is the case, we have to prove that there are no Pollack relations (see section \ref{sec: pollack}) containing a term only in $\tilde \ep_{2|\mu}$, i.e., that each term of Pollack relations features at least one factor $\ep_{k|\mu}^{(j)}$ with $k\geq 4$ in the kernel of~$\pi_2$. Since Lie polynomials in $\tilde \ep_{2|\mu}$ are singlets under $\mathfrak{sl}_2$, it suffices to rule out $\mathfrak{sl}_2$ singlets of Pollack relations with a non-trivial image under~$\pi_2$.

The first step is to rule out fundamental Pollack relations in $\mathfrak{sl}_2$ singlets. This follows from the association of fundamental Pollack relations in irreducible $d$-dimensional representations of $\mathfrak{sl}_2$ with holomorphic cusp forms for $\Gamma(2)$ of modular weight $d{+}1$. The modular weights $\geq 6$ of cusp forms for $\Gamma(2)$ then translate into dimensions $\geq 5$ for the $\mathfrak{sl}_2$ representations of fundamental Pollack relations.

As the second step of our freeness proof, we show that all derived Pollack relations in $\mathfrak{sl}_2$ singlets trivialise under $\pi_2$. For this purpose, we note that all derived Pollack relations can be obtained from fundamental ones by (possibly repeated) adjoint action of $z_w$ followed by that of one or more $\epsilon_{k| \mu}^{(j)}$ (using $[z_w,\mathfrak{u}] \subset \mathfrak{u}$). By the $\mathfrak{sl}_2$-singlet property of $z_w$ and $\epsilon_{2| \mu}$, the only way to generate derived Pollack relations in $\mathfrak{sl}_2$ singlets from the fundamental ones in $(d\geq 5)$-plets is via outer brackets with $\epsilon_{k_1| \mu_1}^{(j_1)} \ldots \epsilon_{k_r| \mu_r}^{(j_r)}$ involving at least one $k_i\geq 4$. However, the corresponding $\epsilon_{k_i| \mu_i}^{(j_i)}$ vanishes under $\pi_2$
and trivialises the $\pi_2$-image of Pollack relations in $\mathfrak{sl}_2$ singlets. On these grounds, Lie polynomials solely in $\epsilon_{2| \mu}$ do not occur in any fundamental or derived Pollack relation. We conclude that the derivation algebra $\pi_2(\mathfrak{sl}_2\ltimes \mathfrak{u})$ is free.

\subsection{From the connection to single-valued MPLs}
\label{sec:lam.2}

In this section, we will demonstrate that the series (\ref{eq: svI2}) in single-valued iterated Eisenstein integrals for $\Gamma(2)$ projects to a generating series in svMPLs under $\pi_2$ in (\ref{eq:pi2}), up to the exponential of $x \log(2)$ in our final result (\ref{eq:pi2.IIsv}) below.

\subsubsection{The target series in single-valued MPLs}
\label{sec:tmpl}

The series in svMPLs will be identified through its representation of \cite{Frost:2023stm, Frost:2025lre} in terms of zeta generators. The starting point is the series
\beq
\mathbb G(x,y;\lambda) =  {\rm Pexp}\bigg( {-} \int_\lambda^{\vec{1}_0} \bigg( \frac{x}{\lambda'} + \frac{y}{\lambda'-1} \bigg) \dd\lambda'\bigg)\,,
\label{revgg.01}
\eeq
in MPLs in one variable $\lambda$, which is here taken to live on the modular curve $Y_{\Gamma(2)}=\mathbb{P}^1\setminus\{0,1,\infty\}$ of genus 0. The tangent vector $\vec{1}_0$ at the origin
shuffle-regularises the MPLs in the expansion of (\ref{revgg.01}) and assigns $\log(\lambda)$ as the coefficient of the one-letter word $x$. 

The single-valued version of the MPL series
(\ref{revgg.01}) is given by~\cite{Frost:2023stm,Frost:2025lre} 
\beq
\mathbb G^{\rm sv}(x,y;\lambda) = \mathbb M^{\rm sv}_{0}(M)^{-1} \,
\overline{ \mathbb G(x,y;\lambda)^T } \, \mathbb M^{\rm sv}_{0}(M) \, \mathbb G(x,y;\lambda)\, ,
\label{revgg.02}
\eeq
where the complex-conjugate series enters with reversed letters $(\ldots x \ldots y \ldots )^T = \ldots y \ldots x \ldots$ and is conjugated by the single-valued version of the following series in MZVs (excluding the letter $f_1$)
\beq
\mathbb M_{0}(M) = 1 +\sum_{r=1}^\infty \sum_{i_1,\ldots,i_r \in 2\mathbb N +1} \rho^{-1} \big( f_{i_1}\ldots f_{i_r} \big) M_{i_1} \ldots M_{i_r} \, .
\label{revgg.03}
\eeq
The genus-0 incarnation $M_{w}$ of zeta generators with $w \geq 3$ odd \cite{Drinfeld2, Ihara:stable, Furusho2000TheMZ, DG:2005} 
acts on the Lie-algebra generators $x$, $y$ in (\ref{revgg.01}) as Ihara derivations
\beq
[M_w,x] = 0 \, , \ \ \ \ \ \ [M_w,y] = \big[ y,P_w(x,y)\big] \, .
\label{revgg.04}
\eeq
The Lie polynomials $P_w(x,y)$ of degree $w$ on the right side can be defined through the motivic Drinfeld associator (the regularised $\lambda \rightarrow 1$ limit of (\ref{revgg.01})) \cite{Drinfeld2} 
\beq
\Phi(x,y) =
{\rm Pexp}\bigg[ {-} \int_{- \vec{1}_1}^{\vec{1}_0} \bigg( \frac{x}{\lambda'} + \frac{y}{\lambda'-1} \bigg) \dd\lambda'\bigg]\,,
\label{revgg.05}
\eeq
by extracting the coefficient of the primitive $f_w$ in the expansion of $\rho(\Phi^{\mot}(x,y) ) $  for a given choice of the $f$-alphabet isomorphism $\rho$ (see appendix \ref{app:a.1}),
\beq
P_w(x,y) = \rho \big(\Phi^{\mot}(x,y)  \big) \, \big|_{f_w}\, .
\label{revgg.06}
\eeq
We can expand~\eqref{revgg.02} into words in $x_0=x$ and $x_1=y$,
\beq
\mathbb{G}^{\sv}(x,y;\lambda) = 1+\sum_{w=1}^{\infty} \sum_{\substack{a_1,\ldots,a_w \\ \in \{0,1\} } } x_{a_1} \ldots x_{a_w} G^{\rm sv}(a_w,\ldots,a_1;\lambda) \, .
\label{GGsvdef}
\eeq
The coefficients are svMPL~\cite{svpolylog}, i.e., non-holomorphic analogues of MPLs that are free of branch cuts. They form a shuffle algebra and satisfy the same holomorphic differential equation as their holomorphic, multi-valued analogues defined from the generating series in~\eqref{revgg.01}.

\subsubsection{Converting the regularising tangent vectors}
\label{sec:reg}

In preparation for retrieving the MPLs in (\ref{revgg.01}) from the KZ connection (\ref{eq:pi2A.2}), we shall compare their regularisation with that of the iterated Eisenstein integrals in this work. For this purpose, we write out Brown's tangential base-point prescription \cite{brown2017multiple} that leads to the regularisation of the $\tau$-integral according to
\begin{align}
        \int_\tau^{\vec{1}_{i\infty}} \pi_2\left(\ANN{\tau^\prime}\right) &= \frac{\pi^2}{6} \int_0^\tau  \frac{\dd \tau^\prime}{2\pi i} (\tilde\epsilon_{2|3}+\tilde\epsilon_{2|4}) \\ 
        &\quad  -\int_\tau^{i\infty}  \frac{\dd \tau^\prime}{2\pi i} \bigg(\tilde\epsilon_{2|3} f^{(2)}(\omega_3,\tau^\prime) + \tilde\epsilon_{2|4} f^{(2)}(\omega_4,\tau^\prime) -\frac{\pi^2}{6}(\tilde\epsilon_{2|3}+\tilde\epsilon_{2|4})\bigg)\, ,
        \notag
    \end{align}
where the tangent vector $\vec{1}_{i\infty}$ on the left-hand side instructs us to integrate the zero mode $f^{(2)}(\omega_\mu,\tau)= \frac{\pi^2}{6}+{\cal O}(e^{i\pi \tau})$ for $\mu=3,4$ between $0$ and $\tau$ instead of $i\infty$ and $\tau$. This regularised expression transforms as follows under the change of variables (\ref{eq:Jac.lambda}) from $\tau'$ to $\lambda'$: 
\begin{align}
        \int_\tau^{\vec{1}_{i\infty}} \pi_2\left(\ANN{\tau^\prime}\right) & =  \pi i \, x\int_1^\lambda \frac{\dd \tau' }{\dd\lambda'} \dd\lambda'
    - \int_\lambda^{0} \bigg( \frac{x}{\lambda'} + \frac{y}{\lambda'-1} - \pi i \, x\, \frac{\dd \tau'}{\dd\lambda'} \bigg) \dd\lambda'
 \label{eq:pi2A.3}  \\
        &= \int_1^0 \bigg(  \pi i \, x\,\frac{\dd \tau'}{\dd\lambda'} - \frac{x}{\lambda'} \bigg) \dd\lambda'  - \int^{\vec{1}_0}_\lambda \bigg( \frac{x}{\lambda'} + \frac{y}{\lambda'-1} \bigg) \dd\lambda' \, .
   \notag
\end{align}
This equation allows us to write an integral along a path in $\mathbb{H}$ ending at the tangential-base-point $\vec{1}_{i\infty}$ as an integral along the image of this path in $\mathbb{P}^1\setminus\{0,1,\infty\}$, but ending at $\vec{1}_0$. The second term in the last line can readily be seen to define the series $\mathbb{G}$ in MPLs, cf.~\eqref{revgg.01}. The first term can be evaluated using $-\int^{\vec{1}_0}_\lambda \frac{\dd \lambda'}{\lambda'} = \log(\lambda)$. We can expand $\lambda$ in the variable $q_2 = \exp(\pi i\tau)$ according to $\lambda(q_2) = 16 q_2-128q_2^2+{\cal O}(q_2^3)$.
Changing variables from $\lambda$ to $q_2$, using  that $\lambda(q_2)|_{q_2=0}=0$ and $\lambda(q_2)|_{q_2=1}=1$ for the boundaries, the first term can then be evaluated to
\begin{equation}
    \int_1^0 \bigg(  \pi i \, x\, \frac{\dd \tau '}{\dd\lambda'} - \frac{x}{\lambda'} \bigg) \dd\lambda' = x\,\int_1^0 \big( \mathrm{dlog}(q_2)-\mathrm{dlog}(\lambda(q_2)) \big) = - x\, \log\bigg( \frac{\lambda(q_2)}{q_2} \bigg)\bigg\vert_{q_2 = 1}^{q_2=0} \, .
\end{equation}
By the $q_2$-expansion of $\lambda$, the upper boundary evaluates to $-x\log(16)$ and $\lambda(q_2)|_{q_2=1}=1$, so that the contribution from the lower boundary vanishes.
At the same time, it is the change of tangential base-point
\begin{equation}
    -\int_{\vec{1}_0}^{\vec{16}_0} \bigg( \frac{x}{\lambda} + \frac{y}{\lambda-1} \bigg) \dd\lambda = - x\,\int_{\vec{1}_0}^{\vec{16}_0} \frac{\dd\lambda}{\lambda} = -4x\,\log(2)\,,
\end{equation}
by definition of tangential base-point regularisation. The number 16 comes from the coefficient of $q_2$ in the Fourier expansion of $\lambda$
quoted above. We can thus treat (\ref{eq:pi2A.3}) as a composition of paths
\begin{equation}
    \int_\tau^{\vec{1}_{i\infty}} \pi_2\left(\ANN{\tau^\prime}\right) = - \int^{\vec{16}_0}_{\vec{1}_0} \bigg( \frac{x}{\lambda'} + \frac{y}{\lambda'-1} \bigg) \dd\lambda'\, 
    -\int^{\vec{1}_0}_\lambda \bigg( \frac{x}{\lambda'} + \frac{y}{\lambda'-1} \bigg) \dd\lambda'\, .
    \label{eq:pi2A.4}
\end{equation}
In passing to the path-ordered exponential
\begin{align}
    \pi_2\big(\INN{\tau} \big) &= {\rm Pexp}\bigg(\int_\tau^{\vec{1}_{i\infty}} \pi_2\left(\ANN{\tau^\prime}\right)  \bigg)\, ,
     \label{deftreg} 
\end{align} 
the relation (\ref{eq:pi2A.4}) translates into the concatenation product 
\begin{align}
    \pi_2\big(\INN{\lambda} \big) 
    &=  {\rm Pexp}\bigg({-} \int^{\vec{16}_0}_{\vec{1}_0} \bigg( \frac{x}{\lambda'} + \frac{y}{\lambda'-1} \bigg) \dd\lambda'\bigg) {\rm Pexp}\bigg({-}\int^{\vec{1}_0}_\lambda \bigg( \frac{x}{\lambda'} + \frac{y}{\lambda'-1} \bigg) \dd\lambda' \bigg) \notag\\
    &%
    = 2^{-4x}\, \mathbb G(x,y;\lambda)\, ,
    \label{eq:pi2.II}
\end{align}
according to the 
composition of paths formula.
Its left factor involves the tangent vector $\vec{16}_0$ inherited from $\vec{1}_{i\infty}$ in (\ref{deftreg}), and reduces to an exponential of $-x\log(16)$. Its right factor matches the MPL series (\ref{revgg.01}) with the tangent vector $\vec{1}_0$, and through its position ensures that 
$\lambda$-derivatives produce a right-multiplicative KZ connection (\ref{eq:pi2A.2}).

\subsubsection{Towards a translation of zeta generators}
\label{sec:lam.2.1}

In the remainder of this section, we shall extend the translation (\ref{eq:pi2.II}) of generating series in $\exp(\pi_2(\mathfrak{u}))$ to the respective zeta generators and relate the projected $\pi_2(\sigma_w)$ to the $M_w$ with defining Lie brackets in (\ref{revgg.04}). In order to extend (\ref{eq:pi2.II}) to the single-valued versions of $\pi_2(\INN{\lambda})$ and $\mathbb G(x,y;\lambda)$, it remains to show that $\pi_2(\sigma_w)$ and $M_w$ have the same action on $x$, $y$, and therefore on $\exp({\rm Lie}[x,y])$.

The genus-1 zeta generator $\sigma_w$ is of constant $a$-degree $w$. Under the projection $\pi_2$, its geometric part can only contain terms of modular depth $w$, since the $\tilde\ep_{2|\mu}$ have $a$-degree 1.
For example, the first term on the right side of (\ref{sysxi3.03}) translates into
\begin{equation}
\pi_2(\sigma_3^{\rm g}) = \tfrac{1}{576}\big[\tilde\ep_{2|3}-\tilde\ep_{2|4},[\tilde\ep_{2|3},\tilde\ep_{2|4}]\big]\,,
\label{pisigm3}
\end{equation}
where we write $\pi_2(\sigma_w)$ for $\pi_2\circ\sigma_w$.
The expansion of $\sigma_5^{\rm g}$ through the methods of sections \ref{sec: recursion} and \ref{sec: HW vectors} leads to
\begin{align}
\pi_2(\sigma_5^{\rm g}) &= \tfrac{1}{27648}[\tilde\ep_{2|3},[\tilde\ep_{2|3},[\tilde\ep_{2|3},[\tilde\ep_{2|3},\tilde\ep_{2|4}]]]] -\tfrac{1}{41472}[\tilde\ep_{2|3},[\tilde\ep_{2|3},[\tilde\ep_{2|4},[\tilde\ep_{2|3},\tilde\ep_{2|4}]]]] \notag\\
 &\quad -\tfrac{1}{41472}[\tilde\ep_{2|3},[\tilde\ep_{2|4},[\tilde\ep_{2|3},[\tilde\ep_{2|3},\tilde\ep_{2|4}]]]] +\tfrac{1}{41472}[\tilde\ep_{2|3},[\tilde\ep_{2|4},[\tilde\ep_{2|4},[\tilde\ep_{2|3},\tilde\ep_{2|4}]]]] \notag\\
&\quad -\tfrac{1}{41472}[\tilde\ep_{2|4},[\tilde\ep_{2|3},[\tilde\ep_{2|3},[\tilde\ep_{2|3},\tilde\ep_{2|4}]]]] +\tfrac{1}{41472}[\tilde\ep_{2|4},[\tilde\ep_{2|3},[\tilde\ep_{2|4},[\tilde\ep_{2|3},\tilde\ep_{2|4}]]]]\label{pisigm5}\\
&\quad +\tfrac{1}{41472}[\tilde\ep_{2|4},[\tilde\ep_{2|4},[\tilde\ep_{2|3},[\tilde\ep_{2|3},\tilde\ep_{2|4}]]]] -\tfrac{1}{27648}[\tilde\ep_{2|4},[\tilde\ep_{2|4},[\tilde\ep_{2|4},[\tilde\ep_{2|3},\tilde\ep_{2|4}]]]] \notag\\
&\quad +\tfrac{5}{165888}[[\tilde\ep_{2|3},\tilde\ep_{2|4}],[\tilde\ep_{2|3},[\tilde\ep_{2|3},\tilde\ep_{2|4}]]] +\tfrac{5}{165888}[[\tilde\ep_{2|3},\tilde\ep_{2|4}],[\tilde\ep_{2|4},[\tilde\ep_{2|3},\tilde\ep_{2|4}]]]\, . \notag
\end{align}
By applying $\pi_2$ to the definition (\ref{eq: N2}) of $\mathfrak{N}$, the vanishing bracket of a zeta generator $\sigma_w$ for $w\geq 1$ with $\mathfrak{N}$ projects under the map $\pi_2$ to 
\begin{equation}
    0 = \pi_2([\sigma_w,\mathfrak{N}]) = -\frac{1}{24}[\pi_2(\sigma_w),
      \tilde\ep_{2| 3}{+} \tilde\ep_{2|4} ]\,,
    \label{eq:pi2.com.1}
\end{equation}
where the equality follows from $[\sigma_w,\mathfrak{i}]\subset \mathfrak{i}$, i.e., $\sigma_w$ preserves the ideal. While $[\sigma_w,\mathfrak{i}]\subset \mathfrak{i}$ is obvious for the geometric part of $\sigma_w$, the
analogous identity $[z_w,\mathfrak{i}]\subset \mathfrak{i}$ for the arithmetic part requires further justification: By $\mathfrak{sl}_2$ invariance of $z_w$, its brackets with $\ep_0,\ep_0^\vee \in \mathfrak{i}$ vanish and $[z_w, \ep_{k | \mu}]$ with $k\geq 4$ are lowest-weight vectors of $\geq 3$-dimensional representations of $\mathfrak{sl}_2$. Hence, $[z_w, \ep_{k | \mu}]$ with $k\geq 4$ cannot produce any nested brackets of $\tilde \ep_{2 | \mu}$, which are $\mathfrak{sl}_2$ singlets and thus belong to $\mathfrak{i}$.

We therefore conclude that
\begin{equation}
    [\pi_2(\sigma_w), x ] = 0\,,
    \label{eq:pi2.com.2}
\end{equation}
for all $w\geq 1$ odd, which matches the relation $[M_w,x]= 0$ of (\ref{revgg.04}) for $w\geq 3$. 

Before turning to the action of $\sigma_w$ on $y$ under $\pi_2$ and comparing it with $[M_w,y]$, we like to point out that the generator $\sigma_1$ decouples from the other zeta generators. We have an exact expression for $\sigma_1$, given by (\ref{exactz1}). Its arithmetic part commutes with $x,y$  by (\ref{newz1.02}) and obeys $[z_1,\mathfrak{i}]\subset \mathfrak{i}$ by \eqref{eq: zmatrix}, which leads to the image under $\pi_2$
\begin{equation}
    \pi_2(\sigma_1) = 4 x\,,
    \label{eq:pi2.sigma1.2}
\end{equation}
and so it follows from (\ref{eq:pi2.com.2}), that
\begin{equation}
\pi_2(\left[\sigma_1,\sigma_w\right])= 0\,,
\end{equation}
for all zeta generators $\sigma_w$. On these grounds, the generating series of zeta generators factorises under the image of $\pi_2$,
\begin{equation}
    \pi_2 \big(\mathbb{M}(\sigma) \big) = 2^{4x}\, \mathbb{M}_{Y_{\Gamma(2)}}(\sigma) = \mathbb{M}_{Y_{\Gamma(2)}}(\sigma) \, 2^{4x}\, ,
    \label{eq:pi2.MM.1}
\end{equation}
where we introduced the following notation for the $\pi_2(\sigma_{w}) \leftrightarrow M_w$ counterpart of (\ref{revgg.03})
\begin{equation}
    \mathbb{M}_{Y_{\Gamma(2)}}(\sigma) = \pi_2\big(\mathbb{M}(\sigma)\vert_{\sigma_1\to 0} \big)
 = 1 +\sum_{r=1}^\infty \sum_{i_1,\ldots,i_r \in 2\mathbb N +1} \rho^{-1} \big( f_{i_1}\ldots f_{i_r} \big) \pi_2(\sigma_{i_1}) \ldots \pi_2(\sigma_{i_r}) \, .
 \label{myga2}
\end{equation}
Analogously to (\ref{eq:pi2.MM.1}), the single-valued version of the generating series of zeta generators factorises as well,
\begin{equation}
    \pi_2\big(\mathbb{M}^{\sv}(\sigma) \big) = 2^{8x} \, \mathbb{M}^{\sv}_{Y_{\Gamma(2)}}(\sigma) = \mathbb{M}^{\sv}_{Y_{\Gamma(2)}}(\sigma)\, 2^{8x}\,.
    \label{eq:pi2.MM.2}
\end{equation}
We used that the coefficients of $\sigma_1$ inside $\mathbb{M}(\sigma)$ and $\mathbb{M}^{\sv}(\sigma)$ are given by $\log(2) = \rho^{-1}(f_1)$ and $2\log(2) = \rho^{-1}(\sv f_1)$, respectively, see appendix~\ref{app:a.3}. 

\subsubsection{\texorpdfstring{The missing action on $y$ from the Drinfeld associator}{The missing action on y from the Drinfeld associator}}
\label{sec:lam.2.b}

We shall next investigate $\left[\pi_2(\sigma_w), y\right]$, prove it to match $[M_w,y]$ of (\ref{revgg.04}), and thereby derive the single-valued version of the link (\ref{eq:pi2.II}) between $\pi_2(\INN{\lambda})$ and $\mathbb G(x,y;\lambda)$. The first step is to apply $\theta_S$ to (\ref{eq:pi2.com.2}) by using the consequence $\theta_S^{-1} x \theta_S = y$ of (\ref{eq: thetatild}),
\begin{equation}
    \left[\theta_S^{-1}\pi_2(\sigma_w)\theta_S, y \right] = 0\,.
    \label{eq:pi2.com.3}
\end{equation}
Since $\theta_S$ preserves the degree of the derivation it acts on, it commutes with $\pi_2$. Let us consider the image of our equivariant iterated integrals and their $S$-transformations in (\ref{eq: eqvprop}) under $\pi_2$,
\begin{equation}
    \pi_2\big(\INNeqv{- \tfrac{1}{\tau}} \big) = \pi_2 \big(U_S^{-1} \theta_S^{-1}\, \INNeqv{\tau}\, \theta_S\, U_S \big)\,.
    \label{eq:pi2.IIeqv.1}
\end{equation}
By (\ref{eq: US}), $U_S$ operates trivially on $\ep_{2\vert \mu}$, so that we can drop it from the above equation. We shall next perform the projection $\pi_2$ of the series $\mathbb{I}^{\eqv}$ at the level of their expressions in (\ref{eq: eqvprop}), apply (\ref{eq:pi2.II}) and (\ref{eq:pi2.MM.2}) to their constituents and use the consequence $\lambda \rightarrow 1{-}\lambda$ of the  $\tau \rightarrow -1/\tau$ transformation to find that
\begin{align}
  \pi_2\big(\INNeqv{\tau} \big) &= \pi_2\big( \mathbb{M}^\sv(z) \big)^{-1}\, \overline{\mathbb{G}(x,y;\lambda)^{T}} \, \mathbb{M}^{\sv}_{Y_{\Gamma(2)}}(\sigma)\, \mathbb{G}(x,y;\lambda)\,,
    \label{eq:pi2.IIeqv.3} \\
    \pi_2 \big(\INNeqv{-\tfrac{1}{\tau}} \big) &= \pi_2\big( \mathbb{M}^\sv(z) \big)^{-1} \, \overline{\mathbb{G}(x,y;1-\lambda)^{T}} \, \mathbb{M}^{\sv}_{Y_{\Gamma(2)}}(\sigma)\, \mathbb{G}(x,y;1{-}\lambda)\,.\notag
\end{align}
The generating series of MPLs evaluated at $1{-}\lambda$ can be related to the one evaluated at $\lambda$ by the Drinfeld associator (\ref{revgg.05}) \cite{Drinfeld:1989st, Drinfeld2},
\begin{equation}
    \mathbb{G}(x,y;1{-}\lambda) = \Phi(x,y)\,\mathbb{G}(y,x; \lambda)\,.
    \label{eq:Drinfeld}
\end{equation}
Plugging (\ref{eq:pi2.IIeqv.3}) and (\ref{eq:Drinfeld}) into (\ref{eq:pi2.IIeqv.1})
and using the consequence $\theta_S^{-1} \mathbb{G}(x,y;\lambda)  \theta_S = \mathbb{G}(y,x;\lambda)$ of (\ref{eq: thetatild})
yields
\begin{equation}
    \overline{\Phi(x,y)^T} \,  \mathbb{M}^{\sv}_{Y_{\Gamma(2)}}(\sigma)\, \Phi(x,y) = \theta_S^{-1} \,  \mathbb{M}^{\sv}_{Y_{\Gamma(2)}}(\sigma)\, \theta_S\,,
    \label{eq:pi2.IIeqv.4}
\end{equation}
where the left-multiplicative series $\mathbb{M}^\sv(z)$ canceled because $\theta_S$ acts trivially on them.
By passing to the motivic version of (\ref{eq:pi2.IIeqv.4}) and equating the coefficient of $\rho^{-1}(f_w)$ with $w\geq 3$ in the $f$-alphabet images of both sides, we find
\begin{equation}
    \overline{P_w(x,y)^{T}} + \pi_2(2\sigma_w) + P_w(x,y) = \theta_S^{-1} \pi_2(2\sigma_w) \theta_S\,.
\end{equation}
The Lie polynomials $P_w(x,y)$ arise from their definition (\ref{revgg.06}) and obey
$\overline{P_w(x,y)^{T}} = P_w(x,y)$ by their reality and odd degree.
We therefore conclude
\begin{equation}
    P_w(x,y) = \theta_S^{-1} \pi_2(\sigma_w) \theta_S - \pi_2(\sigma_w)\,.
\end{equation}
Finally, taking the bracket with $y$ and using the vanishing commutator (\ref{eq:pi2.com.3}) 
yields the desired relation
\begin{equation}
    \left[\pi_2(\sigma_w), y \right] = \left[ y, P_w(x,y)\right]\,.
    \label{eq:pi2.com.6}
\end{equation}
We can now read off that the action of $\pi_2(\sigma_w)$ on $x$ in (\ref{eq:pi2.com.2}) and $y$ in (\ref{eq:pi2.com.6}) matches that of the zeta generator $M_w$ for the MPL series in (\ref{revgg.04}).\footnote{For $w=3$, this can be explicitly tested by combining the geometric part (\ref{pisigm3}) of $\pi_2(\sigma_3)$ with the only contributions $\pi_2([z_3, \tilde \ep_{2|\nu}]) = [z_3, \tilde \ep_{2|\nu}]^{[4]}$ given by (\ref{eq:z3tail4}).} As a consequence, the conjugations 
\begin{equation}
     \mathbb{M}_{Y_{\Gamma(2)}}(\sigma)^{-1} W(x,y) \mathbb{M}_{Y_{\Gamma(2)}}(\sigma) = \mathbb{M}_0(M)^{-1} W(x,y) \mathbb{M}_0(M)\,,
    \label{eq:pi2.MM.3}
\end{equation}
of the respective generating series (\ref{myga2}) and (\ref{revgg.03}) agree for arbitrary words $W(x,y)$ in the generators $x$, $y$. 

When applied to the single-valued series $\INNsv{\lambda}$ in (\ref{eq: svI2}), the expressions (\ref{eq:pi2.II}) and (\ref{eq:pi2.MM.2}) for the projected ingredients yield
\beq
 \pi_2\big(\INNsv{\tau} \big) =
2^{-8x}\, \mathbb M_{Y_{\Gamma(2)}}^\sv (\sigma)^{-1}\,\overline{\mathbb{G}(x,y;\lambda)^{T}}\,  \mathbb M_{Y_{\Gamma(2)}}^\sv (\sigma) \,  \mathbb{G}(x,y;\lambda) \,.
\label{pi2isv}
\eeq
Finally, the adjoint action of
$\mathbb M_{Y_{\Gamma(2)}}^\sv (\sigma)$ on the right side can be converted into that of $\mathbb M_{0}^\sv (M)$ by virtue of (\ref{eq:pi2.MM.3}), producing the representation (\ref{revgg.02}) of the svMPL series.
In this way, (\ref{pi2isv}) gives rise to the following main result of this section
\begin{align}
    \pi_2\big(\INNsv{\tau} \big) 
    &= 2^{-8 x} \,\mathbb{G}^\sv(x,y;\lambda)\,,
    \label{eq:pi2.IIsv}
\end{align}
namely the identification of single-valued iterated Eisenstein integrals over weight-two kernels $f^{(2)}(\omega_\mu,\tau)$ with svMPLs on the modular curve $Y_{\Gamma(2)}$. By the freeness of ${\rm Lie}[x,y]$ established in section \ref{sec:lam.3}, the single-valued version of each MPL in one variable with singular points $0$ and $1$ is realised as an iterated Eisenstein integral for $\Gamma(2)$. 

\subsection{Equivariant MPLs}
\label{sec:eqvmpl}

In the previous section we have seen in~\eqref{eq:pi2.IIsv} how we can recover the generating series of svMPLs from the one for the single-valued iterated Eisenstein integrals for $\Gamma(2)$ by restricting to Eisenstein series of weight 2. We conclude this section by briefly commenting what happens if we apply the projection $\pi_2$ to the generating series $\mathbb{I}^{\textrm{eqv}}$ of equivariant iterated Eisenstein integrals~\eqref{eq: eqv2proposal} instead of its single-valued counterpart $\mathbb{I}^{\textrm{sv}}$.

We define a series $\mathbb{G}^\eqv$ of equivariant MPLs through the image of $\mathbb{I}^{\textrm{eqv}}$ under $\pi_2$: 
\begin{equation}
    \mathbb{G}^\eqv(x,y;\lambda) = \pi_2 \big(\INNeqv{\tau} \big) = \pi_2 \big(\mathbb{M}^\sv(z) \big)^{-1} \, \overline{\mathbb{G}(x,y;\lambda)^{T}} \, \mathbb{M}^{\sv}_{Y_{\Gamma(2)}}(\sigma)\, \mathbb{G}(x,y;\lambda)\,,
    \label{defgeqv}
\end{equation}
which involves the constant series
\begin{equation}
    \Psi^{\sv}(x,y) = \pi_2 \big(\mathbb{M}^\sv(z) \big)^{-1} \mathbb M_{Y_{\Gamma(2)}}^\sv (\sigma)\, ,
\end{equation}
whose coefficients in depth 1 are the images of the geometric parts under $\pi_2$. By (\ref{pisigm3}) and (\ref{pisigm5}), we have
\begin{align}
\Psi^{\sv}(x,y) &= 1 + 2\zeta_3\, \pi_2(\sigma_3^{\rm g}) + 2\zeta_5\, \pi_2(\sigma_5^{\rm g}) + \ldots
\label{eq:Psisv.exp} \\
&= 1 + \tfrac{2}{3}\zeta_3 \big[x{+}2y,[x,y]\big] \notag \\
&\quad + \tfrac{1}{3}\zeta_5 \Big( 2[x,[x,[x,[x,y]]]] + 6[y,[x,[x,[x,y]]]] + 6[y,[y,[x,[x,y]]]] \notag \\
&\quad \phantom{+ \tfrac{1}{3}\zeta_5 \Big(} + 4[y,[y,[y,[x,y]]]] - [[x,y],[x,[x,y]]] + 2[[x,y],[y,[x,y]]] \Big) + \ldots \, . \notag
\end{align}
Note that $\Psi^{\sv}(x,y)$ is group-like, since it is the product of two group-like series. 
We may then cast~\eqref{defgeqv} into the form
\begin{equation}
    \mathbb{G}^\eqv(x,y;\lambda) = \Psi^\sv(x,y)\mathbb{G}^\sv(x,y;\lambda)\, .
\end{equation}
Note that also $\mathbb{G}^\eqv(x,y;\lambda)$ is group-like.

The quotient group $\mathcal{S}_3\cong \SLZ/\Gamma(2)$ acts on $Y_{\Gamma(2)} = \mathbb{P}^1\setminus\{0,1,\infty\}$ by permuting the three punctures.
From the transformation behaviour of $\INNeqv{\tau}$ under $\SLZ$, we deduce that $\mathbb{G}^\eqv(x,y;\lambda)$ transforms under this $\mathcal{S}_3$ action as
\begin{align} 
    \mathbb{G}^\eqv(x,y;1{-}\lambda) &= \theta_S^{-1} \,\mathbb{G}^\eqv(x,y;\lambda)\, \theta_S \, ,     \label{eq:Geqv.1}\\
    \mathbb{G}^\eqv(x,y;\tfrac{\lambda}{\lambda-1}) &= \theta_T^{-1} \,\mathbb{G}^\eqv(x,y;\lambda)\, \theta_T \,,
\notag
 \end{align}
or equivalently, by using (\ref{eq: thetatild}),
\begin{align} 
    \mathbb{G}^\eqv(x,y;1{-}\lambda) &= \mathbb{G}^\eqv(y,x;\lambda) \,,\\
    \mathbb{G}^\eqv(x,y;\tfrac{\lambda}{\lambda-1})& = \mathbb{G}^\eqv(x,-x-y;\lambda) \, . \notag
 \end{align}
The first examples for equivariant MPLs in low weight which depart from their single-valued counterparts are
\begin{align}
G^{\eqv}(1,1,1;\lambda) &= G^{\sv}(1,1,1;\lambda)\, , &
G^{\eqv}(0,0,0;\lambda) &= G^{\sv}(0,0,0;\lambda)\,,  \label{eq:Geqv.w3} \\
G^{\eqv}(0,0,1;\lambda) &= G^{\sv}(0,0,1;\lambda) + \tfrac{2}{3}\zeta_3 \, , &
G^{\eqv}(1,0,0;\lambda) &= G^{\sv}(1,0,0;\lambda) + \tfrac{2}{3}\zeta_3 \, ,  \notag \\
G^{\eqv}(0,1,0;\lambda) &= G^{\sv}(0,1,0;\lambda) - \tfrac{4}{3}\zeta_3 \, , &
G^{\eqv}(1,0,1;\lambda) &= G^{\sv}(1,0,1;\lambda) + \tfrac{8}{3}\zeta_3 \, , \notag \\
G^{\eqv}(0,1,1;\lambda) &= G^{\sv}(0,1,1;\lambda) - \tfrac{4}{3}\zeta_3 \, , &
G^{\eqv}(1,1,0;\lambda) &= G^{\sv}(1,1,0;\lambda) - \tfrac{4}{3}\zeta_3 \, , \notag
\end{align}
where the expansion of $\mathbb{G}^\eqv$ is defined in complete analogy with~\eqref{GGsvdef}. Its coefficients inherit the usual shuffle product, and they satisfy the same holomorphic differential equation as the $G^{\sv}$.
The equivariant combinations have the property~\eqref{eq:Geqv.1} that can be restated as saying that the $\mathcal{S}_3$ action on the letters $x,y$ intertwines equivariantly with the $\mathcal{S}_3$ action on $\lambda$ over~$\mathbb{Q}$.
Said differently, the $G^{\eqv}$ provide a basis of the space of svMPLs. They share numerous properties of svMPLs (e.g., shuffle product and holomorphic differentiation) but differ in their (regularized) limit $\lambda \rightarrow 0$, complex-conjugation properties and functional relations. As a noteworthy difference to the $G^{\sv}$, the equivariant MPLs $G^{\rm eqv}$ satisfy functional equations coming from the $\mathcal{S}_3$ action on its argument that are free of MZVs. This is very reminiscent of the clean svMPL defined in~\cite{Charlton:2021uhu}, which satisfy product-free functional equations, but include MZVs. It would be interesting to combine these two constructions to obtain a variant of svMPLs that satisfy functional equations coming from the $\mathcal{S}_3$ action that are free of both MZVs and product terms.

\section{Extensions to larger congruence subgroups of level 2}
\label{sec: extensions}

Until now, we have been concerned with the construction of equivariant iterated Eisenstein integrals for the principal congruence subgroup $\Gamma(N)$ for level $N=2$. However, the principal congruence subgroups are not the only congruence subgroups of $\SLZ$. Two important families of congruence subgroups are
\begin{align}
    \label{eq: gamma01}
    &\Gamma_0(N)=\bigg\{\gamma\in\SLZ \Bigm| \gamma\equiv\begin{pmatrix}
        \bullet&\bullet\\0& \bullet
    \end{pmatrix}\bmod N\bigg\}\,,\\
    &\Gamma_1(N)=\bigg\{\gamma\in\SLZ\Bigm|\gamma\equiv\begin{pmatrix}
        1&\bullet\\0&1
    \end{pmatrix}\bmod N\bigg\}\,,
\notag
\end{align}
with the inclusions $\SLZ\supset\Gamma_0(N)\supset\Gamma_1(N)\supset\Gamma(N)$. In fact, congruence subgroups comprise a hierarchy of subgroups that includes the full modular group. This follows from the inclusions above and the property that $M|N$ implies that $\Gamma(N)\subset\Gamma(M)$ (and similarly $\Gamma_0(N)\subset\Gamma_0(M)$ and $\Gamma_1(N)\subset\Gamma_1(M)$). This fact allows us to chain such inclusions together to create (semi-infinite) Hasse diagrams. For example, we have
\begin{equation}
\label{eq: hasse}
\begin{tikzpicture}[
    node distance=0.6cm,
    every node/.style={font=\large}
,baseline=(current bounding box.center)]

\node (SL2) {$\SLZ$};

\node (G02) [right=of SL2] {$\Gamma_0(2)=\Gamma_1(2)$};

\node (G04) [right=of G02, yshift=1.2cm] {$\Gamma_0(4)$};
\node (G2)  [right=of G02, yshift=-1.2cm] {$\Gamma(2)$};

\node (G14) [right=of G04] {$\Gamma_1(4)$};

\node (G4) [right=of G2,xshift=0.3cm] {$\Gamma(4)$};

\node (G8)  [right=of G4] {$\Gamma(8)$};
\node (G16) [right=of G8] {$\Gamma(16)$};
\node (dots) [right=of G16] {$\cdots$};

\node at ($(SL2)!0.4!(G02)$) {$\supset$};

\node[rotate=45] at ($(G02)!0.6!(G04)$) {$\supset$};
\node[rotate=-45] at ($(G02)!0.6!(G2)$) {$\supset$};

\node at ($(G04)!0.5!(G14)$) {$\supset$};

\node at ($(G2)!0.5!(G4)$) {$\supset$};
\node[rotate=-90] at ($(G14)!0.5!(G4)$) {$\supset$};

\node at ($(G4)!0.5!(G8)$) {$\supset$};
\node at ($(G8)!0.5!(G16)$) {$\supset$};
\node at ($(G16)!0.5!(dots)$) {$\supset$};
\end{tikzpicture}\,.
\end{equation}
The level of a congruence subgroup $\Gamma\subset \SLZ$ is defined as the smallest integer $N$ such that $\Gamma(N)\subset\Gamma$.

Passing instead to the spaces of modular forms (or of Eisenstein series or cusp forms) attached to these groups, the containments are reversed: $\Gamma\supset\Gamma'$ implies $\mathcal{E}_k(\Gamma)\subset \mathcal{E}_k(\Gamma')$. Consequently, the CG connection for a given congruence subgroup $\Gamma'$, built from the spaces $\mathcal{E}_k(\Gamma')$, already contains the data needed to construct the CG connection for any larger group $\Gamma$ with $\mathrm{SL}_2(\mathbb{Z})\supset\Gamma\supset\Gamma'$. Hopper \cite{Hopper:2021ell} exploited this principle to obtain \emph{a} flat  connection of CG type for every congruence subgroup $\Gamma$ of level $N$  from the CG connection for~$\Gamma(N)$. 
The connection resulting from the construction in~\cite[\S12]{Hopper:2021ell}, however, is not guaranteed to exhaust the subspace of all Eisenstein series for $\Gamma$ in the limit $z\to 0$. This problem becomes only pronounced when looking at level $N>2$.\footnote{One example arises for $\Gamma_0(3)$, where the construction of~\cite{Hopper:2021ell} leads to a connection that only uses a single derivation associated with the unique $\Gamma_0(3)$-fixed 3-torsion point, the origin. At the same time there are two cusps, and we therefore expect two Eisenstein series of weight $k>2$ along with their independent derivations in the appropriate CG connection.} 
For level $N=2$, Hopper's method is sufficient (see table~\ref{tab:lvl2grps} below), and in the present section we will explain this for all level-2 congruence subgroups as well as $\SLZ$, and we present full details for $\Gamma_0(2)$ and $\SLZ$. We leave the full analysis for $N>2$ to future~work.

A crucial ingredient in the construction is a set of  outer derivations $\epdem{k|A}$ that treat all torsion points $A$ of a congruence subgroup $\Gamma$ on equal footing and that we call here ``democratic''. For $\Gamma(2)$, there are the four torsion points $\omega_A$ in \eqref{eq: torsion}, and we were free to choose a preferred subset of  $\omega_A$ to span ${\cal E}_k(\Gamma(2))$ via $f^{(k)}(\omega_A,\tau)$ and label the associated outer derivations. Our choice~\eqref{eq: N2conn} selected a particular set of $\Gamma(2)$ torsion points, leaving out the origin $\omega_1 = 0$.
The main reason for picking a favored set of torsion points was to prevent ambiguities (up to Pollack relations) when extracting coefficients of words in outer derivations in the generating series $\mathbb{I}(\ep;\tau)$ and $\mathbb{I}^{\rm eqv}(\ep;\tau)$ of iterated Eisenstein~integrals.

Such a choice of preferred torsion points need not necessarily be consistent when going from $\Gamma(2)$ to a larger level-2 subgroup $\Gamma\supset \Gamma(2)$, where the choice of excluded torsion points could remove derivations that are necessary for the larger group $\Gamma$. 
It is advantageous to start with an overcomplete set of derivations $\epdem{k|A}$ for all $\Gamma(2)$ torsion points, rather than restricting to the $\ep_{k|\mu}$ with $\mu=2,3,4$.
This is in direct correspondence  to initially working with all $f^{(k)}(\omega_A,\tau)$ for $A=1,2,3,4$, which form a spanning set of ${\cal E}_k(\Gamma(2))$ for $k\geq 4$, but not a basis (such as $f^{(k)}(\omega_\mu,\tau)$ for $\mu=2,3,4$).

In this democratic approach, the specialisation 
$\mathfrak{u}\to\mathfrak{u}_{\SLZ}$ of \eqref{eq:SL2limit} becomes natural. We shall review the democratic derivations in section \ref{subapp.e1} and apply Hopper's procedure for level 2 to equivariant iterated Eisenstein integrals in section \ref{subapp.e2}. 

\begin{table}[t]
\centering
\renewcommand{\arraystretch}{1.4}
\begin{tabular}{c c c c c}
\toprule
Group $\Gamma$ & Generators & $\Gamma/\Gamma(2)$ & 
\begin{tabular}{c}set of $\Gamma$-invariant\\[-2mm] torsion points $\mathcal{T}_\Gamma$ \end{tabular}&  \begin{tabular}{c}\#\! cusps $=\dim\mathcal{E}_k(\Gamma)$ \\[-2mm]
for $k\geq 4$ even \end{tabular}%
\\
\midrule
$\Gamma(2)$        &$\langle T^2,\; ST^2S^{-1},-\mathds{1}\rangle$ & $\{e\}$        & all four          & 3
\\
$\Gamma_0(2)$      &$\langle T,\; ST^{2}S^{-1}\rangle$ & $\mathbb{Z}_2$ & $\{0,\,\tfrac{1}{2}\}$  & 2
\\
$\Gamma^0(2)$      &$\langle T^2,\; STS^{-1}\rangle$ & $\mathbb{Z}_2$ & $\{0,\,\tfrac{\tau}{2}\}$  & 2
\\
$\Gamma_\theta$    &$\langle S,\; T^2\rangle$ & $\mathbb{Z}_2$ & $\{0,\,\tfrac{1+\tau}{2}\}$  & 2%
\\
$\Gamma_{\text{even}}$ & $\langle ST,\; TS,\; T^2\rangle$ &$\mathbb{Z}_3$ & $\{0\}$      & 1
\\
$\mathrm{SL}_2(\mathbb{Z})$ &$\langle S,\; T\rangle$ & ${\cal S}_3$ & $\{0\}$         & 1
\\
\bottomrule
\end{tabular}
\caption{\textit{Collection of all congruence subgroups of level 1 and 2. 
For $N=2$, we note $\Gamma_1(2)=\Gamma_0(2)$ and $\Gamma^1(2)=\Gamma^0(2)$.
The space of Eisenstein series is $\mathcal{E}_k(\Gamma)= \langle f^{(k)}(z,\tau) \,|\, z\in \mathcal{T}_\Gamma\rangle$  (for $k\geq 4$). In all cases except for $\Gamma(2)$, the number of $\Gamma$-invariant torsion points is equal to the dimension of the space of Eisenstein series.}}
\label{tab:lvl2grps}
\end{table}

\subsection{Choice of sets of outer derivations}
\label{subapp.e1}

Outer derivations are, by definition, equivalence classes of derivations, where two representatives of the same class may differ by an inner derivation. In \cite{Calaque:2019ell}, a preferred set of representatives of outer derivations was introduced that we denote here as $\epdem{k|A}$, where $A=1,2,3,4$ labels a torsion point $\omega_A$ of the elliptic curve for $\Gamma(2)$.

The fact that the Kronecker-Eisenstein coefficients $f^{(k)}(z,\tau)$ evaluated at all torsion points $z\in\{0,\tfrac{1}{2},\tfrac{1+\tau}{2},\tfrac{\tau}{2}\}$ in the fundamental cell of the torus form an overcomplete spanning set of $\mathcal{E}_k(\Gamma(2))$ was already observed in section \ref{sec:N2conn}.
At weight 2 this was exemplified by the relation \eqref{funrel.02}, that we repeat here for convenience,
\beq
f^{(2)}(\omega_2 ,\tau)
+ f^{(2)}(\omega_3 ,\tau)
+ f^{(2)}(\omega_4 ,\tau)
= 0\,,
\label{e1}
\eeq
and the reasons for the exclusion of $f^{(2)}(0,\tau)$ were explained below \eqref{eq: Eis}. This leaves us with two possibilities for constructing the $\dd \tau$-part for the weight $k=2$ part of the CG connection: include Kronecker-Eisenstein coefficients for all torsion points and attach to each a derivation $\epdem{2|\mu}$ as done in \cite{Calaque:2019ell}, or pick a basis and attach only to the basis vectors a derivation $\tilde \ep_{2|\nu}$ with $\nu=3,4$. Using \eqref{e1}, one can compare the resulting connections by relating the derivations as explained in section \ref{tau level 2.1},
\begin{align}
    \sum_{\nu=3}^4\tilde \ep_{2|\nu} f^{(2)}(\omega_\nu,\tau)
    = \sum_{\mu=2}^4  \epdem{2|\mu} f^{(2)}(\omega_\mu,\tau) \implies \tilde\ep_{2|\nu} =\epdem{2|\nu}-\epdem{2|2} \ \ \ {\rm for} \ \nu\in \{3,4\} \,.
\label{e2}
\end{align}
The $\epdem{2|\mu}$ are equivalent to the $\epdem{2|\mu}=\ep_{2|\mu}$ from section~\ref{tau level 2.1} (without imposing the condition $\sum_{\mu=2}^4 \ep_{2|\mu}=0$) and therefore coincide with the preferred derivations in \cite{Calaque:2019ell}. However, in view of the implied relation, the derivations $\epdem{2|\mu}$ can be shifted simultaneously by an arbitrary derivation independent of $\mu$ that cancels when translating to $\tilde \ep_{2|\nu}$, see the discussion around~\eqref{backtotild}.

Although not explicitly discussed before, at higher even weights $k\geq 4$ we are in exactly the same situation. One could consider the Kronecker-Eisenstein coefficients evaluated at all four torsion points of order 2 (this time including the torsion point at $z=0$). The corresponding $f^{(k)}(z,\tau)$ span the space of holomorphic Eisenstein series, but they again form an overcomplete spanning set due to the relation \eqref{funrel.01}, which can be rewritten as (see (\ref{defrk}) for the rational numbers $r_{k-1}$)
\beq
f^{(k)}(\omega_2 ,\tau)
+ f^{(k)}(\omega_3 ,\tau)
+ f^{(k)}(\omega_4 ,\tau)
=r_{k-1} f^{(k)}(\omega_1,\tau)\,,
\label{e3}
\eeq
where we recall $\omega_1=0$. Once again, this means that we have the same possibilities when constructing the $\dd \tau$-part of the CG connection as before. We can either construct a connection with all $f^{(k)}(z,\tau)$ evaluated at all torsion points and attach to each a derivation $\epdem{k|A}$ (as done in \cite{Calaque:2019ell}), or we can single out a basis and attach a derivation $\ep_{k|\mu}$ only to those basis vectors (as done in our level-2 connection \eqref{eq: N2conn}). In the connection, these choices are related~via 
\begin{align}
    \sum_{\mu=2}^4f^{(k)}(\omega_\mu,\tau){\ep}_{k|\mu}=\sum_{A=1}^4f^{(k)}(\omega_A,\tau)\epdem{k|A} \implies \ep_{k|\mu}=\epdem{k|\mu}+r_{k-1}^{-1} \epdem{k|1} \ \ \ {\rm for} \ \mu\in \{2,3,4\} \,.
    \label{e4}
\end{align}
Similarly to weight $k=2$, we are again free to simultaneously shift 
$(\epdem{k|1},\epdem{k|2},\epdem{k|3},\epdem{k|4})$ by $ (-r_{k-1},1,1,1)\lambda$ which is engineered to cancel from 
\eqref{e4} for arbitrary derivations $\lambda$.\footnote{Our notation for the derivations chosen to appear in the connection~\eqref{eq: N2conn} is not uniform between $k=2$ and $k \geq 4$ in that, by analogy with \eqref{e2} and \eqref{e4}, one could have labelled the $k\geq 4$ derivations by $\tilde{\ep}_{k|\mu}$ instead of $\ep_{k|\mu}$.}

With the translations (\ref{e2}) and (\ref{e4}) in place, the representation (\ref{eq: N2conn}) of the level-2 connection $\mathbb{A}(\ep;\tau)$ is readily recovered from
its democratic formulation
\beq
\ANN{\tau} =-\dd \tau \sum_{\mu=2}^4 \frac{f^{(2)}(\omega_\mu,\tau)}{2\pi i} \epdem{2|\mu}  -  \dd \tau \sum_{k=4}^{\infty} (k{-}1) \sum_{j=0}^{k-2} \frac{(-1)^j}{j!} (2\pi i \tau)^{j} \sum_{A = 1}^4 \frac{ f^{(k)}(\omega_A,\tau) }{(2\pi i)^{k-1}} {\rm ad}^j_{\ep_0}\epdem{k | A}\,.
\label{eq:dema}
\eeq

Given the democratic formulation of the outer derivations, their actions on $a,b$ and $t_{2|A}$ from \eqref{epac01}--\eqref{eq: epsona} can be unified as follows. The action of $ \epdem{k|C}$ with $C\in\{1,2,3,4\}$ on $t_{2|A}$ with $A\in\{1,2,3,4\}$ is given by
\begin{align}
    [\epdem{k|C},t_{2|A}]=-\sum_{B=1}^4 (\varphi_C)_{AB} [t_{2|A},t_{2|B}] -[t_{2|C},t_{2|A}]\ , \ \ \ \text{for} \ \ k\geq 2\,,
\label{epac01dem}
\end{align}
with $\varphi_2,\varphi_3,\varphi_4$ in (\ref{vphmat}) and $\varphi_1=\mathds{1}_{4\times 4}$.
Their action on $\agen$ and $\bgen$ reads for all~$k\geq 2$
\begin{align}
[\epdem{k|C},b]&=t_{k+1|C}
\, ,  \label{epsonbdem}\\
[\epdem{k|C},\agen]&=-\sum_{\ell=0}^{k/2-2}({-}1)^\ell\sum_{A=1}^4\biggl(\sum_{B=1}^4(\varphi_C)_{AB}[t_{2+\ell|A},  t_{k-1-\ell|B}]\biggr) -[t_{k|C},\agen]
\,, \notag
\end{align}
and resembles the analogous derivation actions (\ref{eq: derivations}) on $a,b$ for level 1.

\subsection{Specialising connections and generating series}
\label{subapp.e2}

In this section, we restrict the procedure of~\cite{Hopper:2021ell} outlined in the beginning of the section to level 2. We start from the (degree-completed) level-2 Lie algebra introduced in section \ref{sec:fltcg}, that we here denote~by
\begin{align}
    \mathfrak{p} = {\rm Lie}[a,b,t_{2|1},t_{2|2},t_{2|3},t_{2|4}]\bigg/\bigg([b,a]=\sum_{A=1}^4t_{2|A} \bigg)\,.
    \label{pgammaN}
\end{align}
The generators $t_{2|\mu}$ carry the same action $\theta_\gamma^{-1}\bullet \theta_\gamma$ of $\SLZ$ (or equivalently ${\cal S}_3$) as the outer derivations in \eqref{eq: theta} for $\mu=2,3,4$, while $t_{2|1}$ and $a$, $b$ are invariant. Following \cite{Hopper:2021ell}, when moving to a larger congruence subgroup $\Gamma\supset \Gamma(2)$, we  only keep the $t_{2|A}$ that are fixed by $\Gamma$ and set the other generators to zero. As $\Gamma(2)$ has finite index in $\Gamma$, it suffices to consider the action of some coset representative of each coset of $\Gamma/\Gamma(2)$.  The resulting algebra defines a new Lie algebra $\mathfrak{p}_\Gamma$. We realise this procedure by defining a surjective map of Lie algebras, denoted by $\pi_\Gamma$, so that  
\begin{align}
    \pi_\Gamma: \mathfrak{p}\rightarrow \mathfrak{p}_{\Gamma}\,,\ \ \ \ \ker\pi_\Gamma= \big\langle t_{2|A}\in\mathfrak{p}\mid\ \exists\,\gamma\in\Gamma/\Gamma(2)\ \text{such that}\ \theta_\gamma^{-1}t_{2|A}\,\theta_\gamma\neq t_{2|A} \big\rangle\,,
\end{align}
where the angled brackets denote the Lie algebra ideal in $\mathfrak{p}$ generated by the elements inside.
Consequently, the actions of the democratic outer derivations $\epdem{k|A}$ that act on $\mathfrak{p}_\Gamma$ for a torsion point indexed by $A$ which is not fixed by $\Gamma$ trivialise in view of equations \eqref{epac01dem} and \eqref{epsonbdem}. Hence, one can infer that these derivations themselves vanish, and we define a new map $\tilde{\pi}_\Gamma$ that acts on the algebra of geometric derivations to implement this as
\begin{align}
\tilde{\pi}_\Gamma:\mathfrak{u}\rightarrow\mathfrak{u}_{\Gamma}\,,\ \ \ \ \ker \tilde \pi_\Gamma= \big\langle \epdem{2|A}\in\mathfrak{u}\mid\ \exists\,\gamma\in\Gamma/\Gamma(2)\ \text{such that}\ \theta_\gamma^{-1}\epdem{2|A}\,\theta_\gamma\neq \epdem{2|A} \big\rangle\, ,
\label{pitrule}
\end{align}
now as an ideal in the derivation algebra $\mathfrak{u}$.
Here, $\mathfrak{u}_\Gamma$ is the Lie algebra of geometric derivations associated with the congruence subgroup $\Gamma$.  The definition of $\tilde{\pi}_\Gamma$ is engineered such that for all generators $p \in \mathfrak{p}$ and all derivations 
$\delta \in \mathfrak{u}$, we have 
\beq
\pi_\Gamma\big( \delta(p) \big) = \tilde{\pi}_\Gamma (\delta)\big(\pi_\Gamma(p) \big) \, .
\label{manypi}
\eeq
The previous discussion has the important consequence that Pollack relations in $\mathfrak{u}$ propagate consistently to those of $\mathfrak{u}_\Gamma$. More precisely, the ideal $\mathfrak{i}_{\Gamma(2)}$ of Pollack relations for $\Gamma(2)$ projects precisely to the ideal $\mathfrak{i}_{\Gamma}$ of Pollack relations for $\Gamma$,
\beq
\tilde{\pi}_\Gamma (  \mathfrak{i}_{\Gamma(2)} ) = \mathfrak{i}_{\Gamma} \, .
\label{idealeq}
\eeq
In order to see this, note that every Pollack relation for $\Gamma$ is also a Pollack relation for $\Gamma(2)$, i.e.\ $\tilde{\pi}_\Gamma (  \mathfrak{i}_{\Gamma(2)} ) \supset \mathfrak{i}_{\Gamma}$ by the analogous inclusion of the spaces of Eisenstein series. 
The converse inclusion $\tilde{\pi}_\Gamma (  \mathfrak{i}_{\Gamma(2)} ) \subset \mathfrak{i}_{\Gamma}$ 
in turn follows from the fact that, for $p \in \mathfrak{p}$ and $\delta \in \mathfrak{i}_{\Gamma(2)}$, we have $ \delta(p)=0$, and so
(\ref{manypi}) translates into the vanishing of $ \tilde \pi_\Gamma (\delta)( \pi_\Gamma(p) )$. Since the possible $\pi_\Gamma(p)$ exhaust the generators of $\mathfrak{p}_\Gamma$, we have established that $\tilde \pi_\Gamma (\delta) \in \mathfrak{i}_{\Gamma}$.
As a consequence of (\ref{idealeq}), $\tilde{\pi}_\Gamma$ in (\ref{pitrule}) is a Lie-algebra homomorphism.

Applying the map (\ref{pitrule}) to the CG connection in the democratic formulation (\ref{eq:dema}) then results in the CG connection for the group $\Gamma$.
In particular, we arrive at the conclusion that for $\Gamma=\SLZ$ and every congruence subgroup $\Gamma$ of level 2 (cf. Table~\ref{tab:lvl2grps}), we have
\begin{align}
    \label{eq:piA}\tilde{\pi}_{\Gamma}\big(\ANN{\tau}\big)=\mathbb{A}_{\Gamma}(\ep;\tau)\,,\qquad 
\tilde{\pi}_{\Gamma}\big(\INN{\tau}\big)=\mathbb{I}_{\Gamma}(\ep;\tau)\,.
\end{align}
The map $\tilde{\pi}_{\Gamma}$ also extends to the zeta generators, and even to the full generating series of equivariant iterated Eisenstein integrals.
We will illustrate this explicitly in the remainder of this section on the examples of the subgroups $\SLZ$ and $\Gamma_0(2)$. The treatment of all other subgroups listed in Table~\ref{tab:lvl2grps} follows in a straightforward manner.

\subsubsection{\texorpdfstring{Specialising to $\SLZ$}{Specialising to SL(2,Z)}}
\label{specsl2}

We now apply the construction described in the previous subsection to the CG connection for $\Gamma(2)$ to recover the CEE connection for level 1. Under $\SLZ/\Gamma(2)\cong\mathcal{S}_3$, the only fixed torsion point is the point at $z=0$. This means that the maps $\pi_{\SLZ}$ and $\tilde\pi_{\SLZ}$ amount to setting $t_{k|\mu}$ and $\epdem{k|\mu}$ to zero for all $\mu=2,3,4$ and $k\geq 2$. We find that the Lie algebra $\mathfrak{p}$ specialises to the Lie algebra $\mathfrak{p}_{\SLZ}$ defined by
\begin{align}
    \mathfrak{p}_{\SLZ} = \pi_{\SLZ}(\mathfrak{p})={\rm Lie}[a,b,t_{2|1}]\big/\big([b,a]=t_{2|1}\big) \cong \mathrm{Lie}[a,b]
    \,,
\end{align}
which coincides with the algebra relevant to the level-1 CEE connection in section \ref{sec:flcee}. Furthermore, the map $\tilde\pi_{\SLZ}$ yields 
\begin{align}
\tilde{\pi}_{\SLZ}: \mathfrak{u} \rightarrow \mathfrak{u}_{\SLZ} \,,\ \ \ \text{with}\,\ \ \  \ker\tilde{\pi}_{\SLZ} = \big\langle\epdem{k|\mu}\mid\mu=2,3,4 \,\ \text{for all}\,\ k\geq 2 \big\rangle\,, 
\label{pimap}
\end{align}
which gives exactly $\mathfrak{u}_{\SLZ}$, provided that we identify $\epdem{k|1}=\ep_k$, and hence $\mathfrak{u}_{\SLZ}\subset \mathfrak{u}$. In order to apply both maps to the CG connection of section \ref{sec:fltcg},
we apply the maps $\pi_{\SLZ}$ and $\tilde \pi_{\SLZ}$ to the connection (\ref{eq:dema}) in democratic form. We find that the resulting connection precisely reproduces the CEE connection (\ref{ceeconn}), again provided that we identify $\epdem{k|1}=\ep_k$. 

The identification $\epdem{k|1}=\ep_k$ explains two phenomena observed in earlier sections. First, due to the inclusion $\mathfrak{u}_{\SLZ}\subset \mathfrak{u}$, it clarifies the appearance of Pollack relations of level 1 among those of level 2 in section \ref{sec: pollack}. Second, the implied relation in \eqref{e4} elucidates our proposal to recover the level-1 connection in \eqref{eq:SL2limit}. Indeed, to recover the level-1 connection \eqref{eq: N1conn} from that of level 2 \eqref{eq: N2conn}, we proposed to set $\tilde\ep_{2|\nu}\mapsto0$ for $\nu=3,4$ and $\ep_{k|\mu}^{(j)}\mapsto\ep_k^{(j)}/r_{k-1}$ for $\mu=2,3,4$. In terms of the democratic outer derivations $\epdem{k|\mu}$, this can be achieved by the (more uniform) map $\epdem{k|\mu}\mapsto 0$ for all $k\geq 2$ and $\mu=2,3,4$, which coincides with Hopper's prescription by specialising the Lie algebra $\mathfrak{u}$ to the subalgebra $\mathfrak{u}_{\SLZ}$.

Now that the connection correctly reduces to the CEE connection, we can use the gauge transformation of section \ref{sec:3.3.x} to transform the $\dd\tau$-part of the CEE connection into the form in \eqref{eq: N1conn} that we used to construct iterated Eisenstein integrals for full $\SLZ$. Hence we~find (cf.~\eqref{eq:piA})
\begin{align}
    \tilde{\pi}_{\SLZ}\big(\ANN{\tau}\big)=\AN{\tau}\,.
\end{align}
This extends to the generating series of iterated Eisenstein integrals and its complex conjugate transpose by simply acting on the connection 
\begin{align}
    \tilde{\pi}_{\SLZ}\big(\INN{\tau}\big)=\IN{\tau}\ \ \ \ \ \text{and}\ \ \ \ \ \ \tilde{\pi}_{\SLZ}\Big(\overline{\INN{\tau}^T}\Big)=\overline{\IN{\tau}^T}\,.
\end{align}
In order to further extend this to the full construction of the equivariant iterated Eisenstein integrals, we must investigate what happens to the adjoint actions of the arithmetic parts $z_w$ of the zeta generators $\sigma_w=z_w+\sigma_w^{\rm g}$ on $\mathfrak{u}$ when we specialise the latter to $\mathfrak{u}_{\SLZ}$. Recall that the equivariant iterated Eisenstein integrals for $\Gamma(2)$ are constructed via \eqref{eq: eqv2proposal}, which we repeat here,
\begin{align}
    \INNeqv{\tau}=\mathbb M^\sv(z)^{-1}\overline{\INN{\tau}^T}\mathbb M^\sv(\sigma)\INN{\tau}\,.
\end{align}
Since both sides of the equation are $\mathfrak{u}$-valued, we can apply the map $\tilde{\pi}_{\SLZ}$, which translates the $T$-invariance condition (\ref{eq: moreTinf}) to
\begin{align}
\tilde\pi_{\SLZ}\Big(\big[\sigma_w,\mathfrak{N}\big]\Big)=\big[\tilde\pi_{\SLZ}(\sigma_w),\tilde\pi_{\SLZ}(\mathfrak{N})\big]
\overset{!}{=}\big[\sigma_w^{\SLZ},\mathfrak{N}_{\SLZ}\big]=0\,,
    \label{projnrel}
\end{align}
where we defined $\tilde\pi_{\Gamma}(\sigma_w)=\sigma_w\rvert_{\mathfrak{u}_\Gamma}$ and use notation similar to~\eqref{eq:pi2.com.1}.
As $\tilde{\pi}_{\SLZ}\big(\mathfrak{N}\big)=\mathfrak{N}_{\SLZ}$, one necessary requirement to meet the condition is to impose 
\begin{equation}
   \sigma_w\rvert_{\mathfrak{u}_{\SLZ}}\overset{!}{=}\sigma_w^{\SLZ} \quad\text{for}\quad w\geq 3.
\end{equation}
In view of the decomposition $\sigma_w=z_w+\sigma^{\rm g}_w$ (with $\tilde \pi_{\SLZ}(\sigma^{\rm g}_w)=\sigma_w^{\SLZ,{\rm g}}$), this implies that we have to impose $z_w|_{\mathfrak{u}_{\SLZ}} \overset{!}{=}z_w^{\SLZ}$ for all $w\geq 3$. For $w=1$, we require instead $[\tilde\pi_{\SLZ}(\sigma_1),\mathfrak{N}_{\SLZ}]= 0$. Since $\sigma_1^{\rm g}\mapsto 0$ under $\tilde \pi_{\SLZ}$ by~\eqref{exactz1}, the
only condition that remains to be fulfilled is $[\tilde \pi_{\SLZ}(z_1),\mathfrak{N}_{\SLZ}]=0$. This follows directly from equation \eqref{eq: zmatrix} upon specialisation to $\mathfrak{u}_{\SLZ}$ which implies that $z_1|_{\mathfrak{u}_{\SLZ}}=0$, in agreement with the absence of an arithmetic generator for $w=1$ for $\SLZ$.

With all ingredients of the generating series of equivariant iterated Eisenstein integrals correctly mapping to their level-1 counterparts under the map $\tilde{\pi}_{\SLZ}$, we note that by (\ref{idealeq}) the Pollack relations propagate correctly to those of level 1 to complete the reduction. Indeed, applying the maps $\pi_{\SLZ}$ and $\tilde \pi_{\SLZ}$ to the actions of the derivations $\epdem{k|A}$ in \eqref{epac01dem} and \eqref{epsonbdem}, we confirmed explicitly that the actions correctly reduce to the actions of the Tsunogai derivations at level 1 in \eqref{eq: derivations}.
We thus conclude that
\begin{align}
    \tilde{\pi}_{\SLZ} \Big(\INNeqv{\tau}\Big)=\INeqv{\tau}\,.
    \label{concllv1}
\end{align}

\subsubsection{\texorpdfstring{Projecting to $\Gamma_0(2)$}{Projecting to Gamma0(2)}}
\label{specg02}

As the $\Gamma_0(2)$ connection has not yet been explicitly written down in this work, we will give more intermediate steps than in the previous example. Since $\Gamma(2)$ has index 2 in $\Gamma_0(2)$, we find one non-trivial coset of $\mathbb{Z}_2\simeq\Gamma_0(2)/\Gamma(2)$, whose representative we can take to be the generator $T$.
As $T$ exchanges $\tfrac{\tau}{2}\longleftrightarrow\tfrac{1+\tau}{2}$ (see~\eqref{eq: theta}), while keeping the other torsion points $\{0,\tfrac{1}{2}\}$ fixed, we set to zero all $t_{k|A}$ and $\epdem{k|A}$ with $A=3,4$ for $k\geq 2$. Hence, we have
\begin{align}
     \pi_{\Gamma_0(2)}(\mathfrak{p})={\rm Lie}[a,b,t_{2|1},t_{2|2}]\big/\big([b,a]=t_{2|1}+t_{2|2}\big)\,,
\end{align}
and we call this Lie algebra $\mathfrak{p}_{\Gamma_0(2)}$. This results in the $\dd z$-part of the CG connection starting from (\ref{dzexpanded}) 
\begin{align}
    \mathbb K_{\Gamma_0(2)}^{(z)}(z,\tau)&=\pi_{\Gamma_0(2)}\Big(\mathbb K_{\rm CG}^{(z)}(z,\tau)\Big)= F(z,\ad_{\frac{\bgen}{2\pi i}},\tau)t_{2|1}+F(z{+}\tfrac{1}{2},\ad_{\frac{b}{2\pi i}},\tau)t_{2|2} \label{dzgam02}\\
    &=2\pi i\agen+\sum_{k=1}^\infty(2\pi i)^{1-k}\Bigl( g^{(k)}(z,\tau) t_{k+1|1}+g^{(k)}(z{+}\tfrac{1}{2},\tau) t_{k+1|2}\Bigr) \notag\,.
\end{align}
Similarly, for the $\dd\tau$ part, (\ref{solht2}) projects to
\begin{align}
    \mathbb K_{\Gamma_0(2)}^{(\tau)}(z,\tau)&=\pi_{\Gamma_0(2)}\Big(\mathbb K_{\rm CG}^{(\tau)}(z,\tau)\Big)=\partial_{\ad_\bgen} \big(\mathbb K^{(z)}_{\Gamma_0(2)} (z,\tau) - 2\pi i a \big) +\mathbb L_{\Gamma_0(2)}(\tau) 
\label{dtgam02} \\
&=\sum_{k=2}^\infty (2\pi i)^{1-k}(k{-}1)\Bigl(g^{(k)}(z,\tau)t_{k|1}+g^{(k)}(z{+}\tfrac{1}{2},\tau)t_{k|2} \Bigr)  +\mathbb L_{\Gamma_0(2)}(\tau) \, .\notag
\end{align}
The integration constant $\mathbb L_{\Gamma_0(2)}(\tau)$  independent of $z$
is obtained from applying the maps $\pi_{\Gamma_0(2)}$ and $\tilde \pi_{\Gamma_0(2)}:\mathfrak{u}\rightarrow \mathfrak{u}_{\Gamma_0(2)}$ to the expression \eqref{anslv2} for $\mathbb L_{\rm CG}(\tau)$, such that 
\begin{align}
    \mathbb L_{\Gamma_0(2)}(\tau)&=2\pi i\epsilon_0-  (2\pi i)^{-1}f^{(2)}(\omega_2,\tau)(\epdem{2|2}+t_{2|2}) \label{anslv2_2}\\
    &\quad +\sum_{k=4}^\infty(k{-}1)(2\pi i)^{1-k} \Big\{ \GG_k(\tau)\big(\epdem{k|1}+t_{k|1} \big) -
    f^{(k)}(\omega_2,\tau)\big(\epdem{k|2}+t_{k|2}\big) \Big\} \, .\notag
\end{align}
The actions of the derivations $\epdem{k|C}$ on $a,b,t_{2|A}\in\mathfrak{p}_{\Gamma_0(2)}$ for $A,C\in \{1,2\}$ follow from flatness and coincide with those in \eqref{epac01dem} and \eqref{epsonbdem} without imposing additional restrictions.

Now that we have a CG connection for $\Gamma_0(2)$, we can use the gauge transformation from section \ref{sec:3.3.x} to obtain the $\dd\tau$-part of the CG connection that we build iterated Eisenstein integrals from. We find
\begin{align}
    \mathbb A_{\Gamma_0(2)}(\ep;\tau)&= - \dd \tau \,  (2\pi i)^{-1} f^{(2)}(\omega_2,\tau)\epdem{2|2} \label{anslv2_2b} \\
&\quad  +\dd \tau \sum_{k=4}^\infty \frac{(k{-}1)}{(2\pi i)^{k-1}} \sum_{j=0}^{k-2} \frac{(-1)^j}{j!} (2\pi i \tau {\rm ad}_{\ep_0})^j\Big\{ \GG_k(\tau)\epdem{k|1} -
    f^{(k)}(\omega_2,\tau)\epdem{k|2} \Big\} \, ,\notag
\end{align}
which coincides with $\tilde \pi_{\Gamma_0(2)}\big(\ANN{\tau}\big)$. Following the same arguments as in section \ref{specsl2}, we can construct equivariant iterated Eisenstein integrals from this connection by applying the map $\tilde \pi_{\Gamma_0(2)}$ to the equivariant iterated Eisenstein integrals for $\Gamma(2)$. To ensure that the procedure is consistent, we have to extend the definition of the map $\tilde \pi_{\Gamma_0(2)}$ to act on the zeta generators $\sigma_w$ for level 2. Repeating the argument from section \ref{specsl2}, we find the condition
\begin{align}
    \sigma_w\rvert_{\mathfrak{u}_{\Gamma_0(2)}}\overset{!}{=}\sigma_w^{\Gamma_0(2)}\ \ \ \text{for all}\ \ \ w\geq 1\,.
\end{align}
As we have not constructed zeta generators for $\Gamma_0(2)$ ab initio, we view this as a definition rather than a condition: $\sigma_w^{\Gamma_0(2)} = \sigma_w\rvert_{\mathfrak{u}_{\Gamma_0(2)}}$. Since $\tilde\pi_{\Gamma_0(2)}$ is already defined on the geometric part $\sigma_w^{\rm g}=\sigma_w{-}z_w\in\mathfrak{u}$, this definition implicitly defines the arithmetic generators $z_w^{\Gamma_0(2)} = z_w\rvert_{\mathfrak{u}_{\SLZ}}$. To find how these arithmetic generators act on $\epdem{k|A}$ for $A=1,2$, we rewrite the first rows of \eqref{newz1.01}, \eqref{wkd2.13}, and \eqref{sysxi2.17} in democratic form and afterwards apply the map $\tilde\pi_{\Gamma_0(2)}$. This gives formulae for $[z_w^{\Gamma_0(2)},\epdem{k|2}{+}r_{k-1}^{-1}\epdem{k|1}]$. We can isolate the adjoint actions on $\epdem{k|1}$ and $\epdem{k|2}$ by invoking $\epdem{k|1}=\ep_k$ and imposing $z_w^{\Gamma_0(2)}\rvert_{\mathfrak{u}_{\SLZ}}=z_w^{\SLZ}$, so that we can identify $[z_w^{\Gamma_0(2)},\epdem{k|1}]=[z_w^{\SLZ},\ep_k]$ from level 1 (see \eqref{eq: bracket}) and use it to isolate the adjoint action on $\epdem{k|2}$. We find the adjoint action of $z_1$ to all modular depths
\begin{align}
[z_1^{\Gamma_0(2)},\epdem{2|2}]&= 0\,, \\
    [z_1^{\Gamma_0(2)},\epdem{k|2}]&=-\frac{1{+}r_k}{(r_k{-}1)(1{+}2r_k)}[\epdem{k|2},\epdem{2|2}]\,, \ \ \ \ \ \text{for $k\geq 4$}\,. \notag
\end{align}
For the modular depth 2 contributions to $[z_w^{\Gamma_0(2)},\epdem{k|2}]$ for $w\geq 3$, we find 
\begin{align}
    &[z_w^{\Gamma_0(2)},\epdem{2|2}]^{[2]}=16r_{w+1}\BF_{w+1}t^{w+1}(\epdem{w+1|2},\ep_{w+1})\,,\\
    &[z_w^{\Gamma_0(2)},\epdem{k|2}]^{[2]}=\frac{\BF_{w+k-1}}{\BF_k}\bigg\{t^{w+1}(r_w\epdem{w+1|2}+\ep_{w+1},\epdem{w+k-1|2}) \notag \\&\qquad\qquad\qquad\qquad+r_w(1+2r_{w+k-1})r_{w+k-2}^{-1}t^{w+1}(\epdem{w+1|2},\epdem{w+k-1})\bigg\}\,,\ \ \ \text{for $k\geq 4$}\notag\,.
\end{align}
In the end, we find 
\begin{align}
   \label{eq:I_eqv_Gamma0(2)} \tilde\pi_{\Gamma_0(2)}\Big(\INNeqv{\tau}\Big)=\mathbb I^{\eqv}_{\Gamma_0(2)}(\ep;\tau)\,,
\end{align}
again recalling that the Pollack relations of $\Gamma(2)$ propagate consistently to those of $\Gamma_0(2)$ by (\ref{idealeq}).
We should make an important remark on the equivariance of this series, which is a downgraded version of the equivariance of the $\Gamma(2)$ series. That is, the series $\II^\eqv_{\Gamma_0(2)}(\ep;{\tau})$ is equivariant only under transformations $\gamma\in \Gamma_0(2)$, i.e.,
\beq
\mathbb I_{\Gamma_0(2)}^{\rm eqv}(\ep;\gamma  \tau) = U_\gamma^{-1}
\mathbb I_{\Gamma_0(2)}^{\rm eqv}(\ep;\tau) U_\gamma \ \ \ \ {\rm for} \, \ \gamma \in \Gamma_0(2) \, ,
\label{eqga02}
\eeq
but not for $\gamma \in \SLZ \setminus \Gamma_0(2)$.
The reason why the equivariance does not lift to full $\SLZ$ as it did for $\INN{\tau}$, is that $\Gamma_0(2)$ is not a normal subgroup of $\SLZ$. For instance, for $\gamma=S\notin \Gamma_0(2)$ we know that it does not preserve the space of  Eisenstein series for $\Gamma_0(2)$, see footnote~\ref{foot:normal}, and the right-hand side of~\eqref{eqga02} does not reassemble into iterated integrals of  Eisenstein series for $\Gamma_0(2)$.

The Pollack relations relevant to (\ref{eq:I_eqv_Gamma0(2)}) follow from specialising the actions \eqref{epac01dem} and \eqref{epsonbdem} to $\Gamma_0(2)$. 
The dimension of the space of cusp forms for $\Gamma_0(2)$ at weight $w=k{-}2$ can be checked to agree with the number of ($\mathfrak{sl}_2$ multiplets of) Pollack relations at degree $k$ and maximal modular depth $\rmax=2$. This generalises 
the level-1 result of \cite{Pollack} and the observations in section~\ref{sec: pollack}.
Indeed, taking the Pollack relations in $\mathfrak{u}$ of $\rmax=2$ from section \ref{sec: pollack} and applying the map $\tilde\pi_{\Gamma_0(2)}$, we find an exact match for all Pollack relations we have access to. For example, the Pollack relation for $\Gamma(2)$ of lowest degree in \eqref{eq: pollack8} trivialises upon degeneration, which is in line with $\dim \mathcal{S}_6(\Gamma_0(2))=0$. The first time a non-trivial Pollack relation arises upon degeneration to $\Gamma_0(2)$ is at degree 10. This is exactly when the associated cusp form of the original relation in $\mathfrak{u}$ was a newform for $\Gamma_0(2)$ (see appendix \ref{app:mf2} and Table \ref{dimnewf}). Explicitly, we find 
\begin{align}
    0&=8[\epdem{2|2},\ep_{8}]-9[\epdem{2|2},\epdem{8|2}]-17[\ep_4,\epdem{6|2}]-34[\epdem{4|2},\ep_6]+51[\epdem{4|2},\epdem{6|2}]\,,
\label{eq: pollgamm02}
\end{align}
where we used $\epdem{k|1}=\ep_k$. Note that this relation specialises to $0=0$ under the map $\tilde{\pi}_{\SLZ}$ as expected. 
Whether this simple connection between Pollack-type relations and cusp forms holds for any congruence subgroup is not known. See, however,~\cite{Pollack:lv2} for results on $\Gamma_1(N)$.

\section{Conclusion and outlook}
\label{sec:furth}

In this work, we have constructed equivariant iterated integrals of Eisenstein series for congruence subgroups of level 2, generalising their level-1 analogue in the mathematics \cite{brown2017multiple, Brown:2017qwo, Brown:2017qwo2} and string-theory literature \cite{Dorigoni:2024oft}.
We have analysed in detail the 
case of the principal subgroup $\Gamma(2)$ at the level of a generating series $\mathbb I^{\rm eqv}(\ep;\tau)$ of equivariant iterated integrals.
The Lie-algebraic backbone of $\mathbb I^{\rm eqv}(\ep;\tau)$ consists of derivations dual to Eisenstein series and subject to infinite families of bracket relations as well as zeta generators for level 2, which organize the appearance of single-valued cyclotomic MZVs. The series $\mathbb I^{\rm eqv}(\ep;\tau)$ is constructed in tandem with the expansion of zeta generators as a Lie series in the derivations and the bracket relations of their remaining, arithmetic parts. Our expansion method relies on a conjectural factorised form~\eqref{ag:Scoc2} of the cocycle series encoding the modular $S$-transformation of $\mathbb I^{\rm eqv}(\ep;\tau)$, which we confirm by a wealth of order-by-order checks of the equivariant $\SLZ$ transformations.

Our results unlock a variety of applications ranging from particle physics to number theory and raise numerous interdisciplinary follow-up questions.

A first task for future work is to generalise our construction to equivariant iterated Eisenstein integrals for the level-$N$ congruence subgroups  such as $\Gamma(N)$, $\Gamma_0(N)$ and $\Gamma_1(N)$ for arbitrary $N\geq 2$. The mathematical prerequisites of our construction are available for any $N$: both the flat CG  connection \cite{Calaque:2019ell, Calaque:2020ell}, which underpins the relations in the derivation algebra, and Hopper's procedure \cite{Hopper:2021ell} to recover certain connections and derivations of $\Gamma_0(N)$ and $\Gamma_1(N)$ from those of $\Gamma(N)$ although it is not clear to us that these connections are general enough for physics applications. Still, it remains to accommodate the new features of $N\geq 3$, such as Eisenstein series of odd modular weight,
cyclotomic MZVs with imaginary letters in their $f$-alphabet description and relations among the incarnations of zeta generators in the equivariant generating series for higher levels.

Another important follow-up problem is to elucidate the role of holomorphic cusp forms for congruence subgroups. Similar to the case of level $N=1$ \cite{Pollack}, the period polynomials of level $N\geq 2$ cusp forms are expected to determine certain terms of the relations in the derivation algebra  $\mathfrak{u}_{\Gamma(N)}$ \cite{Pollack:lv2}.
It would be interesting to use iterated integrals involving cusp forms for the equivariant completions of those iterated Eisenstein integrals which are projected out from their generating series through the relations in $\mathfrak{u}_{\Gamma(N)}$.
The level 1 prototypes of these additional non-holomorphic modular forms were studied in \cite{brown2017multiple, Brown:2017qwo, Dorigoni:2021ngn, Dorigoni:2024oft}, see section \ref{sec:4.csp}, and are expected, by analogy,   to lead to various generalisations of L-values of cusp forms in their Fourier expansions for level $N\geq 2$.

Finally, a major motivation for the study of equivariant iterated Eisenstein integrals comes from their significance for a coaction calculus beyond multiple polylogarithms. In future work \cite{wip:coactionlong}, we plan to test possible level $\geq 2$ extensions of the proposed coaction formula  for iterated Eisenstein integrals for $\SLZ$ \cite{Kleinschmidt:2025dtk}. Equivariant versions of iterated Eisenstein integrals serve as concrete representatives of their de Rham uplifts, assuming that the latter exist. In the companion paper \cite{wip:PRL}, Feynman integrals in dimensional regularisation are argued to furnish matrix representations of both the derivations (through their canonical differential equations) and of zeta generators (through their initial conditions). An important open question concerns the link between the de Rham entries of the tentative coaction of elliptic Feynman integrals and suitable notions of cut integrals, as discussed in~\cite{wip:PRL}. It would also be interesting to elucidate if there is any link of our candidate coaction and the theory of Mellin motives currently being developed~\cite{Mellin:progress}.

We hope that our results
feed into the symbiosis between 
scattering amplitudes in string theory and QFT on the one hand 
and mathematical developments in algebraic geometry and modular forms on the other
-- both at a conceptual level and with a concrete path towards hands-on computations.

\appendix
\section{\texorpdfstring{$f$-alphabet for cyclotomic MZVs at level 2}{f-alphabet for cyclotomic MZVs at level 2}}
\label{sec: falph}

We briefly recall the motivic $f$-alphabet associated to motivic incarnations of multiple zeta values (MZVs) and then describe its analogue at level~$2$.  The reviews are based on \cite{Brown:2011ik,BrownTate} for level 1 and \cite{brown2017notes,GLANOIS2016334} for level 2, also see \cite{Blumlein:2009cf} for a datamine of relations for both levels.

\subsection{\texorpdfstring{The MZV $f$-alphabet at level 1}{The MZV f-alphabet at level 1}}
\label{app:a.1}

Recall that MZVs are  real numbers defined by nested sums of the form 
\begin{align}
\zeta_{n_1,n_2,\dots,n_r}=
\sum_{1\leq m_1<m_2<\dots<m_r}\frac{1}{m_1^{n_1}m_2^{n_2}\dots m_r^{n_r}}\,,\ \ \ \ n_r\geq 2\,,
    \label{defmzvs}
\end{align}
which are assigned depth $r$ and weight $n_1{+}\ldots{+}n_r$. MZVs
arise as special values of iterated integrals between $0$ and $1$ of the differential forms
\begin{align}
    \omega_0=\frac{\dd t}{t}\,,\ \ \ \ \omega_1=\frac{\dd t}{t-1}\,,
    \label{cohbas1}
\end{align}
generating the first de Rham cohomology of $X = \mathbb P^1\backslash\{0,1,\infty\}$. 
Although not much is known about the arithmetic nature of MZVs, as for example irrationality of most of the odd zeta values could not be proven up to now, 
the $\mathbb{Q}$-linear relations of their motivic lifts are well understood. 
These motivic MZVs $\zeta_{n_1,\ldots,n_r}^\mot$ arise by considering the motive $[H^w_{\text{rel}},\omega,\gamma]$, with $w=\sum_i n_i$ and $H^w_{\text{rel}}$ a relative cohomology associated with the hyperplane arrangement fixed by the singularities of $\omega$, constructed out of the data underlying the integral representation of the MZV such that the period map applied to the motive yields the numerical value $\zeta_{n_1,\ldots,n_r}$ (up to $(-1)^r$).
In order to understand the structure of the $\mathbb{Q}$-algebra $\mathcal{MZ}$ of motivic MZVs, and later on of motivic cyclotomic MZVs, denoted $\mathcal{CMZ}_N$ for level $N$, it is helpful to introduce some facts about mixed Tate motives. 

The first de Rham cohomology group of $\mathbb{P}^1\setminus\{0,\infty\}$ is generated by only one element $\tfrac{\dd t}{t}$, its Betti homology is generated by the loop $\gamma$ winding once around 0. The corresponding period matrix is simply 
\begin{equation}
    \Big(\int_{\gamma} \frac{\dd t}{t} \Big) = (2\pi i)\, .
\end{equation}
This data is called the Lefschetz motive, often denoted by $\mathbb{Q}(-1)$. One can construct tensor powers $\mathbb{Q}(-1)^{\otimes n}$ of $\mathbb{Q}(-1)$, which are called simple Tate motives and denoted by $\mathbb{Q}(-n)$. Their periods are given by $(2\pi i)^{n}$, and the periods of their dual objects $\mathbb{Q}(n)$ are $(2\pi i)^{-n}$, so simple Tate motives only give powers of $2\pi i$. 

This changes when looking at {mixed} Tate motives $M$, that are extensions of simple/pure Tate motives built from short exact sequences
\begin{equation}
    0 \to \mathbb{Q}(n) \to M \to \mathbb{Q}(0) \to 0 \, .
    \label{eq:SES}
\end{equation}
At the level of period matrices, we are extending the $1\times 1$-matrix $(2\pi i)^0$ coming from $\mathbb{Q}(0)$ by the $1\times 1$-matrix $(2\pi i)^{-n}$ from $\mathbb{Q}(n)$. The result is a $2\times 2$-matrix 
\begin{equation}
    \left( \begin{smallmatrix}
    (2\pi i)^{-n} & \per \\ 0 & 1
    \end{smallmatrix} \right)\, ,
    \label{eq:PM}
\end{equation}
where $\per$ is a period of the mixed Tate motive $M$ represented by (\ref{eq:SES}). 

Two isomorphism classes of short exact sequences such as (\ref{eq:SES}) can be combined to a new short exact sequence (by the so-called Baer sum, see, e.g.~\cite{MacLane:1995}). In fact, they define a group called the first extension group of $\mathbb{Q}(0)$ by $\mathbb{Q}(n)$, denoted by $\textrm{Ext}^{1} (\mathbb{Q}(0),\mathbb{Q}(n))$. It has been shown that MZVs are periods of mixed Tate motives over $\mathbb{Z}$ \cite{DG:2005}\footnote{Roughly speaking, a motive is defined over $\mathbb{Z}$ if in the underlying geometry, written with integer coefficients, no special points collide modulo $p$ for any prime $p$. For example, $\mathbb{P}^1\setminus\{0,1,\infty\}$ has this property, whereas $\mathbb{P}^1\setminus \{0,\pm1,\infty\}$ degenerates modulo $2$ since $1\equiv-1$.} and in fact, any period of a mixed Tate motive over $\mathbb{Z}$, as the number $\per$ in (\ref{eq:PM}), is a $\mathbb{Q}[\tfrac{1}{2\pi i}]$-linear combination of MZVs. This is because the motivic fundamental group of $X$ at the tangential base-point $\vec{1}_0$ generates the category of mixed Tate motives $\mathcal{MT}(\mathbb{Z})$ \cite{BrownTate}, whose structure is determined by the first extension groups
\begin{equation}
    \textrm{Ext}^{1}_{\mathcal{MT}(\mathbb{Z})} \big(\mathbb{Q}(0),\mathbb{Q}(n) \big) \cong \begin{cases}
        \mathbb Q&\text{if }n\geq 3\text{ odd },\\
        0&\text{otherwise },
    \end{cases}
    \label{Ext.l1}
\end{equation}
together with the fact that all second Ext-groups vanish.\footnote{In an abelian category $\mathcal C$ (e.g.\ the category $\mathcal{MT}(\mathbb Z)$), the group $\textrm{Ext}^{2}_{\mathcal{C}} (A,B)$ for two objects $A$ and $B$ in $\mathcal C$ consists of equivalence classes of exact sequences $0 \to B \to M_1 \to M_2 \to A \to 0$ with $M_1$ and $M_2$ objects in~$\mathcal C$. This generalises the short exact sequence (\ref{eq:SES}) associated with $\textrm{Ext}^{1} (\mathbb{Q}(0),\mathbb{Q}(n))$.}
The category $\mathcal{MT}(\mathbb{Z})$ of mixed Tate motives over $\mathbb{Z}$ is equivalent to the category of representations of a group scheme $\mathcal{G}_{\mathcal{MT}}$ over $\mathbb{Q}$ \cite{Deligne1990}. There is a split short exact sequence 
\begin{equation}
  1 \to \mathcal{G}_{\mathcal U} \to \mathcal{G}_{\mathcal{MT}} \to \mathbb{G}_m \to 1\,,
\end{equation}
so that $\mathcal{G}_{\mathcal{MT}}$ can be written as the semidirect product $\mathcal{G}_{\mathcal{MT}} \cong \mathcal{G}_{\mathcal U} \rtimes \mathbb{G}_m $, where $\mathbb{G}_m$ is the multiplicative group and $\mathcal{G}_{\mathcal U}$ is a prounipotent group whose Lie algebra is freely generated by one element $\sigma_w$ in each odd degree $w\geq 3$. A choice of these generators is the zeta generators. The ring of affine functions on $\mathcal{G}_{\mathcal U}$ is a commutative graded Hopf algebra with one cogenerator $f_w$ in each odd degree $w\geq 3$. This follows from the structure (\ref{Ext.l1}) of $\mathcal{MT}(\mathbb{Z})$. The ring of affine functions $ \mathbb{Q}\left[f_2\right]$ of the multiplicative group $\mathbb G_m$ is generated by a single element $f_2$. The letter $f_2$ corresponds to the pure Tate part, i.e.\ to powers of the above mentioned Lefschetz motive. We conclude that the space $\mathcal{MZ}$ 
of motivic MZVs $\zeta_{n_1,\ldots,n_r}^\mot$ and the space
\begin{align}
    \mathcal{F} = \mathbb{Q}\langle f_3,f_5,f_7,\dots \rangle \otimes_\mathbb{Q} \mathbb{Q}[f_2]\,,
\end{align}
are isomorphic as graded $\mathbb{Q}$-algebra comodules. There is a coaction
\begin{equation}
    \Delta: \mathcal{F} \to \mathcal F \otimes_\mathbb{Q} \bar{\mathcal F}\,,
\end{equation}
with $\bar{\mathcal F} = \mathbb{Q}\langle f_3,f_5,f_7,\dots \rangle$, which is given by $\Delta(f_2) = f_2\otimes 1$ on the letter $f_2$, and by the deconcatenation coproduct on~$\bar{\mathcal F}$
\begin{equation}
    \Delta(f_{i_1}\ldots f_{i_r}) = \sum_{j=0}^r f_{i_1}\ldots f_{i_{j}} \otimes f_{i_{j+1}}\ldots f_{i_r}\,.
    \label{deconcop}
\end{equation}
Words $f_2^n f_{i_1} f_{i_2} \ldots f_{i_r}$ in the $f$-alphabet with $i_{1},\ldots,i_r \geq 3$ odd are said to have coradical degree $r$.

A given normalised choice of isomorphism $\rho: \mathcal{MZ} \to \mathcal F$ maps motivic (single) zeta values to the corresponding letter in the same degree
\begin{align}
    \rho(\zeta_{2k+1}^\mm)=f_{2k+1}\,,\ \ \ \ \rho(\zeta_2^\mm)=f_2\,,
\end{align}
and is compatible with the multiplication structure on both sides, i.e.
\begin{align}
    \rho(\zeta^\mm_{2k+1}\zeta^\mm_{2l+1})=f_{2k+1}\shuffle f_{2l+1}=f_{2k+1}f_{2l+1}+f_{2l+1}f_{2k+1}\,.
\end{align}
It is, moreover, compatible with the coaction structure determined by~\eqref{deconcop} for the $f$-alphabet and the Goncharov-Brown coaction on motivic MZVs~\cite{Goncharov:2001iea, Goncharov:2005sla, Brown:2011ik, BrownTate}.

A canonical choice of the isomorphism $\rho$ for level 1 is discussed in \cite{Keilthy, Dorigoni:2024iyt} and from this point onwards and everywhere else in the main body of the paper $\rho$ denotes this preferred isomorphism. In particular, this choice leads to the following $f$-alphabet images of the simplest indecomposable motivic MZVs
\begin{align}
\rho(\zeta^\mm_{3,5}) &=-5f_3f_5+\frac{100471}{35568}f_8\,, \label{lvrhoex}\\
\rho(\zeta^\mm_{3,7}) &=-14f_3f_7 -6f_5f_5+\frac{408872741707}{40214998720}f_{10}\,,\notag\\
\rho(\zeta^\mm_{3,3,5}) &= - 5 f_3 f_3 f_5 - 45 f_2 f_9 - \frac{6}{5} f_2^2 f_7 + \frac{4}{7} f_2^3 f_5+\frac{1119631493}{14735232}f_{11} \, , \notag
\end{align}
where $f_{2n}$ is a power of $f_2$ defined by $f_{2n} = \tfrac{\zeta_{2n}}{\zeta_2^n} f_2^n$.

There is a single-valued map on the $f$-alphabet given as follows for $i_1,\ldots,i_r$ odd:
\begin{align}
    \sv(f_2^n f_{i_1}\dots f_{i_r})=\delta_{n,0}\sum_{j=0}^rf_{i_j}\dots f_{i_1}\shuffle f_{i_{j+1}}\dots f_{i_r}\,.
    \label{svfalp}
\end{align}
It is defined in such a way that any chosen isomorphism $\rho$ commutes with the single-valued map on both sides. The single-valued map $\sv_{\mathcal{MZ}}$ on motivic MZVs was introduced in \cite{Schnetz:2013hqa, Brown:2013gia}. Hence, the following diagram commutes 
(we informally write $\sv$ for both of ${\rm sv}_{\mathcal{MZ}}$ and ${\rm sv}_{\mathcal{F} }$ outside the~diagram)
\begin{center}
\begin{tikzcd}
    \mathcal{MZ} \arrow[d, "\sv_{\mathcal{MZ}}"'] \arrow[rr, "\rho"] &  & \mathcal{F} \arrow[d, "\sv_{\mathcal{F}}"] \\
    \mathcal{MZ} \arrow[rr, "\rho"]            &  & \mathcal{F} 
\end{tikzcd}\,.
\end{center}
Note in particular that $\sv(f_2) = 0$ and $\sv(f_{2k+1}) = 2f_{2k+1}$ by (\ref{svfalp}) for $r\leq 1$, corresponding to ${\rm sv}(\zeta_{2k})=0$ and ${\rm sv}(\zeta_{2k+1})=2 \zeta_{2k+1}$ for Riemann zeta values. Moreover, we have $\sv(f_{2k+1}f_{2\ell+1}) = 2 f_{2k+1} \shuffle f_{2\ell+1} = 2 \rho( \zeta^\mot_{2k+1} \zeta^\mot_{2\ell+1} )$ by the $r=2$ case of (\ref{svfalp}). Starting from $r=3$, (\ref{svfalp}) produces combinations of $f_{i_1}\dots f_{i_r}$ that cannot be reduced to shuffles of depth 1, reflecting indecomposable single-valued motivic MZVs at weight $\geq 11$ such as 
\begin{equation}
    {\rm sv}(\zeta_{3,3,5}) = 2\zeta_{3,3,5} - 5 \zeta_3^2 \zeta_5 + 90 \zeta_2 \zeta_9 +\frac{12}{5} \zeta_2^2 \zeta_7 - \frac{8}{7}\zeta_2^3 \zeta_5\, .
    \label{eq:sv335}
\end{equation}
We furthermore note that the single-valued map is compatible with (shuffle-)multiplication, which in the $f$-alphabet means that (for $f_{i_1}\dots f_{i_r}$ and $ f_{j_1}\dots f_{j_r}$ in $ \bar {\cal F}$)
\beq
\sv( f_{i_1}\dots f_{i_r} \shuffle f_{j_1}\dots f_{j_r})
= \sv( f_{i_1}\dots f_{i_r}) \shuffle 
\sv( f_{j_1}\dots f_{j_r})\, .
\label{svshuff}
\eeq

\subsection{Cyclotomic MZVs at level 2}
\label{app:a.2}

In the previous section, we considered the $f$-alphabet at level 1, related to the motivic fundamental group of $\mathbb P^1\backslash\{0,1,\infty\}$. The MZVs appearing as periods of iterated integrals of Eisenstein series at level 2 are special values of iterated integrals of the differential forms 
\begin{align}
    \omega_0=\frac{\dd t}{t}\,,\ \ \ \ \omega_1=\frac{\dd t}{t-1}\,,\ \ \ \ \omega_{-1}=\frac{\dd t}{t+1}\,,
    \label{cohbas2}
\end{align}
which are the generators of the first de Rham cohomology group of the space $\mathbb{P}^1\setminus\{ 0,1,-1,\infty \}$. The new generator $\omega_{-1}$ contributes at weight and coradical degree 1 with
\begin{equation}
    \int_0^1 \frac{\dd t}{t+1} = \log(2)\,.
\end{equation}
It can as well be represented as the (conditionally convergent) alternating harmonic series
\begin{equation}
\label{eq:log2}
    \log(2) = \sum_{m = 1}^\infty \frac{(-1)^{m+1}}{m}\,.
\end{equation}
Similarly, all iterated integrals of (\ref{cohbas2}) evaluate to alternating versions of (\ref{defmzvs}), e.g.,
\begin{align}
    \zeta_{n_1,\dots\,, \bar{n}_s,\dots,n_r}=
    \sum_{1\leq m_1<\dots <m_r}\frac{(-1)^{m_s}}{m_1^{n_1}\dots m_s^{n_s}\dots m_r^{n_r}}\,,\ \ \ \ n_r\geq 2\,,
    \label{defcmzvs}
\end{align}
where $n_r=1$ is allowed in presence of $(-1)^{m_r}$, as in the case of $\log(2) = -\zeta_{\bar 1}$, and multiple indices can carry a bar. Following the terminology for level 1, the alternating sum (\ref{defcmzvs}) is said to have depth $r$ and weight $n_{1}{+}\ldots {+}n_r$.

\subsection{\texorpdfstring{Cyclotomic $f$-alphabet at level 2}{Cyclotomic f-alphabet at level 2}}
\label{app:a.3}

Because of the presence of $\log(2)$ in the evaluations of iterated integrals over (\ref{cohbas2}), the $f$-alphabet at level 2 has to contain an additional letter in degree 1. Indeed, Deligne showed \cite{Deligne2010} that for $N=2,3,4,6$ or $8$, the fundamental group of $\mathbb{P}^1\setminus\{0,\mu_N,\infty\}$, where $\mu_N$ is the group of $N^{\rm th}$ roots of unity, generates the category of mixed Tate motives over $\mathbb{Z}[\mu_N][\tfrac{1}{N}]$\footnote{The ring $\mathbb{Z}[\mu_N]$ is the ring of polynomials in elements of $\mu_N$ with integer coefficients. We localise this ring at $N$ to obtain $N^{-1}\mathbb{Z}[\mu_N]= \mathbb{Z}[\mu_N][\tfrac{1}{N}]$.} (resp.\ $\mathbb{Z}[\mu_6]$ in the case of $N=6$). For $N=2$, we have $\mathbb{Z}[\mu_2][\tfrac{1}{2}] = \mathbb{Z}[\tfrac{1}{2}]$, and the structure of $\mathcal {MT}(\mathbb{Z}[\tfrac{1}{2}])$ is determined by the groups
\begin{equation}
    \textrm{Ext}^{1}_{\mathcal{MT}(\mathbb{Z}[1/2])} (\mathbb{Q}(0),\mathbb{Q}(n)) \cong \begin{cases}
        \mathbb Q\,,&\text{if }n\geq 1\text{ odd ,}\\
        0\,,&\text{otherwise ,}
    \end{cases}
    \label{Ext.l2}
\end{equation}
and the fact that again all second Ext-groups vanish. We derive the existence of a letter $f_1$ in the $f$-alphabet from the fact that the first extension group of $\mathbb{Q}(0)$ by $\mathbb{Q}(1)$ is generated by one element. Analogously to level 1, there is a (non-canonical) isomorphism of graded $\mathbb{Q}$-algebra comodules
\begin{equation}
    \rho: \mathcal{CMZ}_{2} \to \mathcal{F}_{2}\,,
\end{equation}
from the space of motivic cyclotomic MZVs at level 2 to the comodule
\begin{align}
    \mathcal{F}_2 = \mathbb{Q}\langle f_1,f_3,f_5,\dots \rangle \otimes_\mathbb{Q} \mathbb{Q}[f_2]\,,
    \label{f2alph}
\end{align}
where similarly to level 1 the letter $f_2$ corresponds to the multiplicative group $\mathbb{G}_m$.
In coradical degree 1, a normalised choice of isomorphisms is given by
\begin{align}
    \rho(\log^\mm(2))=f_{1}\,,\ \ \ \ \rho(\zeta_{2k+1}^\mm)=f_{2k+1}\,,\ \ \ \ \rho(\zeta_2^\mm)=f_2\,.
\end{align}
Since the $f$-alphabet at level 2 only differs by the additional letter $f_1$ from the $f$-alphabet at level 1, we can define $\rho$ in such a way that the restriction of its inverse to words in $f_w$ with $w\geq 3$ is identical to our preferred choice of $\rho$ at level 1 presented in \cite{Dorigoni:2024iyt}. 
That is why we use the same notation $\rho: \mathcal{CMZ}_{2} \to \mathcal{F}_{2}$ for the $f$-alphabet isomorphism for level 2 as for its restriction to the space $\mathcal{MZ}$. Similar to the level-1 case, $\rho$ for level 2 is compatible with the Goncharov-Brown coaction on motivic cyclotomic MZVs and the extension of the $f$-alphabet coaction (\ref{deconcop}) to $i_1,\ldots,i_r\geq 1$ odd.

The preceding conditions do not determine $\rho$ uniquely. We specify our low-weight convention by choosing cyclotomic-MZV basis elements and giving their $f$-alphabet images; the images of products then follow by shuffle multiplication. The ancillary file lists these images through weight 8.

We use the following $f$-alphabet images of indecomposable cyclotomic MZVs of weight 4, 5 and 6 (see (\ref{lvrhoex}) for their level-1 analogues),
\begin{align}
\rho(\zeta^\mot_{\bar 1,\bar 3}) &= \tfrac74 f_3 f_1\,, \label{falph.rho}\\
\rho(\zeta^\mot_{\bar 1, 3,\bar 1}) &= -\tfrac72 f_3 f_1 f_1 - \tfrac{11}{40} f_2^2 f_1 + \tfrac{27}{8} f_5\,, \nonumber\\
\rho(\zeta^\mot_{\bar 1,\bar 5}) &= \tfrac34 f_3 f_3 + \tfrac{31}{16} f_5 f_1\,. \nonumber
\end{align}
With this choice of $\rho$, the following expressions for the preimages of the simplest shuffle-independent words of coradical degree~2, $f_1f_3$ and $f_1f_5$, contain no separate terms proportional to $\pi^4$ and $\pi^6$, respectively,
\begin{align}
\rho^{-1}(f_1 f_3) &= \log^\mot(2)\zeta^\mot_3 - \tfrac47 \zeta^\mot_{\bar 1,\bar 3}\, ,\label{falph.02}\\
\rho^{-1}(f_1 f_5) &= -\tfrac{16}{31} \,\zeta^\mot_{\bar 1,\bar 5} + \tfrac{6}{31} (\zeta_3^\mot)^2 + \log^\mot(2) \,\zeta^\mot_5 \nonumber\, ,
\end{align}
and in coradical degree 3,
\begin{align}
\rho^{-1}(f_1 f_1 f_3) &= -\tfrac27 \zeta^\mot_{\bar 1, 3,\bar 1} -\tfrac{11}{140} (\zeta^\mot_2)^2\log^\mot(2) -\tfrac47\zeta^\mot_{\bar 1,\bar 3} \log^\mot(2) +\tfrac12(\log^\mot(2))^2\zeta^\mot_3 + \tfrac{27}{28}\zeta^\mot_5\, ,\nonumber\\
\rho^{-1}(f_1 f_3 f_3) &= -\tfrac{16}{49} \zeta^\mot_{1 ,1,\bar 5} -\tfrac{16}{49} \zeta^\mot_{\bar 1,\bar 1, 5} - \tfrac{222}{1715} (\zeta^\mot_2)^3\log^\mot(2) + \tfrac{16}{49} \zeta^\mot_{\bar 1,\bar 5}\log^\mot(2) \nonumber\\
&\quad +\tfrac{8}{49}(\zeta^\mot_2)^2\zeta^\mot_3 +\tfrac9{98} \log^\mot(2)(\zeta^\mot_3)^2 -\tfrac9{14} \zeta^\mot_2\zeta^\mot_5 + \tfrac{769}{196}\zeta^\mot_7\, , \label{falph.d3}\\
\rho^{-1}(f_1 f_1 f_5) &= -\tfrac{48}{217} \zeta^\mot_{1, 1, \bar 5} + \tfrac{64}{217} \zeta^\mot_{\bar 1,\bar 1, 5} + \tfrac{888}{7595} (\zeta^\mot_2)^3\log^\mot(2) -\tfrac{64}{217} \zeta^\mot_{\bar 1,\bar 5}\log^\mot(2) \nonumber\\
&\quad + \tfrac{344}{1085} (\zeta^\mot_2)^2\zeta^\mot_3 -\tfrac{18}{217} \log^\mot(2)(\zeta^\mot_3)^2 + \tfrac{34}{31} \zeta^\mot_2\zeta^\mot_5 - \tfrac{7881}{1736}\zeta^\mot_7\nonumber\,.
\end{align}
It would be interesting to extend the level-1 construction of a canonical $f$-alphabet isomorphism in \cite{Keilthy, Dorigoni:2024iyt} to cyclotomic MZVs of level 2 and beyond.

The single-valued map (\ref{svfalp}) straightforwardly extends to $i_1,\ldots,i_r\geq 1$ odd, for instance $\sv(f_1) = 2f_1$ or equivalently $\sv(\log(2))=2\log(2)$.
The first indecomposable single-valued cyclotomic MZV at level~2 beyond depth 1 already appears at weight~5 (at level 1 it appeared at weight 11, (\ref{eq:sv335})), namely in the preimage of $ \sv(f_1 f_1 f_3) $. Together with two additional indecomposable instances at weight 7, the explicit expressions are
\begin{align}
\rho^{-1}(\sv (f_1 f_1 f_3)) &= -\tfrac47 \zeta^\mot_{\bar 1, 3,\bar 1} -\tfrac{11}{70} (\zeta^\mot_2)^2\log^\mot(2) -\tfrac87\zeta^\mot_{\bar 1,\bar 3} \log^\mot(2) +2(\log^\mot(2))^2\zeta^\mot_3 + \tfrac{27}{14}\zeta^\mot_5\,, \nonumber\\
\rho^{-1}(\sv (f_1 f_3 f_3)) &= -\tfrac{32}{49} \zeta^\mot_{1 ,1,\bar 5} -\tfrac{32}{49} \zeta^\mot_{\bar 1,\bar 1, 5} - \tfrac{444}{1715} (\zeta^\mot_2)^3\log^\mot(2) + \tfrac{32}{49} \zeta^\mot_{\bar 1,\bar 5}\log^\mot(2) \nonumber\\
&\quad +\tfrac{16}{49}(\zeta^\mot_2)^2\zeta^\mot_3 + \tfrac87 \zeta^\mot_{\bar 1,\bar 3}\zeta^\mot_3 +\tfrac9{49} \log^\mot(2)(\zeta^\mot_3)^2 -\tfrac97 \zeta^\mot_2\zeta^\mot_5 + \tfrac{769}{98}\zeta^\mot_7\,, \label{falph.sv}\\
\rho^{-1}(\sv (f_1 f_1 f_5)) &= -\tfrac{96}{217} \zeta^\mot_{1, 1,\bar 5} + \tfrac{128}{217} \zeta^\mot_{\bar 1,\bar 1 ,5} + \tfrac{1776}{7595} (\zeta^\mot_2)^3\log^\mot(2) -\tfrac{128}{217} \zeta^\mot_{\bar 1,\bar 5}\log^\mot(2) \nonumber\\
&\quad + \tfrac{688}{1085} (\zeta^\mot_2)^2\zeta^\mot_3 -\tfrac{36}{217} \log^\mot(2)(\zeta^\mot_3)^2 + \tfrac{68}{31} \zeta^\mot_2\zeta^\mot_5 + (\log^\mot(2))^2\zeta^\mot_5 - \tfrac{7881}{868}\zeta^\mot_7\,. \nonumber
\end{align}
Together with the permutations in the $f_a$ letters obtained from shuffle products, these preimages determine all contributions to the series $\mathbb M^{\rm sv}$ which enter the terms of degree $\leq 10$ of the equivariant series $\mathbb{I}^{\rm eqv}(\ep;\tau)$ in (\ref{eq: eqvprop}).
The ancillary file lists the preimages of all words in the $f$-alphabet through weight 8, and those of $\sv(f_{i_1}f_{i_2}f_{i_3})$ for all permutations of $f_1f_1f_3$, $f_1f_3f_3$ and $f_1f_1f_5$.

\section{\texorpdfstring{Modular forms for $\Gamma(2)$}{Modular forms for Gamma(2)}}
\label{app:mf2}

In this appendix, we summarise a few well-known facts about modular forms on $\Gamma(2)$. Standard references for this material are~\cite{diamond,Miyake:2006}.

The ring of modular forms for $\Gamma(2)$ is generated freely by two elements, which can be taken as any two of the linearly independent Eisenstein series at weight two, or alternatively as powers of Jacobi theta functions:
\begin{align}
\label{eq:etagens}
    F_1(\tau) &= \vartheta_2(\tau)^4 = 16 \frac{\eta(2\tau)^8}{\eta(\tau)^4} = 16q_1^{1/2} +64 q^{3/2}_1 + \ldots \,,\nn\\
    F_2(\tau) &= \vartheta_4(\tau)^4 = \frac{\eta(\tau/2)^8}{\eta(\tau)^4} = 1- 8 q_1^{1/2} +24 q_1 -32 q_1^{3/2}+ \ldots \,.
\end{align}
We record the relation of these generators to the $f^{(2)}$ kernels of~\eqref{eq: FKroneis}:
\begin{align}
    F_1(\tau) &= -\frac{2}{\pi^2} \left( f^{(2)}(\omega_3,\tau) -f^{(2)}(\omega_4,\tau)\right)\,,\nn\\
    F_2(\tau) &= \phantom{-}\frac{2}{\pi^2} \left( 2f^{(2)}(\omega_3,\tau) +f^{(2)}(\omega_4,\tau)\right)\,.
\end{align}
Using the relation \eqref{eq:f.ellK} between the $f^{(2)}$ and the Hauptmodul of the modular curve, we could alternatively express it in that basis.
Any modular form for $\Gamma(2)$ and weight $k=2n$ is of the form $\sum_{i=0}^n a_i F_1^i F_2^{n-i}$. There are exactly $n{+}1$ modular forms of weight $2n$ for $n\geq 1$. For $k=2n\geq 4$ there are always three Eisenstein series (see~\eqref{eq: dims}) and the cusp forms are in the complementary space of dimension $n{-}2$ for weight $k=2n$ and $n\geq 3$. In particular, the first cusp form (with trivial character) occurs at weight $k=6$, i.e., 
\begin{align}
    \dim \mathcal{S}_k(\Gamma(2)) = \left\{ \begin{array}{cl}
    \frac{k}2-2, &\quad  k\geq 6 \text{ even}\,,\\
    0,&\quad  \text{otherwise}\,.
    \end{array}
    \right.
    \label{dimsk}
\end{align}

In order to describe the space of cusp forms along with the action of the quotient group $\mathcal{S}_3$ on it (see section~\ref{sec:G2}), 
we first recall that cusp forms $f(\tau)$ for $\Gamma(2)$  can be viewed as modular forms of the same weight for $\Gamma_1(4)$ by considering $f(2\tau)$, which is invariant under $\tau\to\tau{+}1$. The space of $\Gamma_1(4)$ modular forms in turn decomposes into modular forms of $\Gamma_0(4)$, transforming potentially with a Dirichlet character $\chi$. In the present case, only trivial characters arise, so we will not need the full theory.

Modular forms on $\Gamma_0(4)$ can be either \emph{newforms} or \emph{oldforms}, where a newform $f(\tau)$ is characterised by being an eigenform of all the Hecke operators $T_p$~\cite{Atkin:1970} and not being a modular form on a bigger group $\Gamma_0(2)$ or $\Gamma_0(1)=\SLZ$ from which it can be constructed by multiplying the argument (and therefore is `old'). We also note that weight $k$ newforms $g$ for $\Gamma_0(N)$ have a so-called Fricke sign $\varepsilon_g$ which appears as the eigenvalue under the Fricke involution $g(-1/(N\tau))=\varepsilon_g N^{k/2} \tau^k  g(\tau)$.

In other words, we can classify cusp forms for $\Gamma(2)$ according to whether they come from newforms on $\Gamma_0(4)$, $\Gamma_0(2)$ or $\SLZ$. The properties of such newforms then also determine the $\mathcal{S}_3$ behaviour.
\begin{itemize}
    \item 
    Suppose there is a Hecke normalised newform $g(\tau)$ of $\Gamma_0(4)$ and weight $k$. Since the weight is even, it necessarily has trivial Dirichlet character. Then $f(\tau)=g(\tau/2)$ is a cusp form for $\Gamma(2)$. Moreover, Hecke theory for $\Gamma_0(4)$ tells us that $g(\tau/2)$ is annihilated by the Hecke operator $T_2$. This, together with the negative Fricke sign $\varepsilon_g=-1$ for such newforms, can be used to prove that
    \begin{align}
        f(\tau{+}1) = - f(\tau) \,,\hspace{10mm}
        f(-1/\tau) = -\tau^k f(\tau)\,,
    \end{align}
    so that the $\Gamma_0(4)$ newform induces a cusp form of $\Gamma(2)$ that is in the alternating representation $\bep$ of $\mathcal{S}_3$.
\item 
    Suppose there is a Hecke normalised newform $g(\tau)$ of $\Gamma_0(2)$ of weight $k$, again with trivial Dirichlet character. Then define
    \begin{align}
        f_1(\tau) = g(\tau) \,,\hspace{10mm} f_2(\tau) = \varepsilon_g 2^{-k/2} g(\tau/2)\,,
    \end{align}
    where $\varepsilon_g$ is the Fricke sign of $g$.
    Then one can show, using the second $\Gamma_0(2)$ Hecke eigenvalue $(T_2g)(\tau) = - \varepsilon_g 2^{k/2 -1} g(\tau)$, that
    \begin{align}
    \begin{pmatrix}
        f_1(\tau{+}1) \\ f_2(\tau{+}1)
    \end{pmatrix}
    = \begin{pmatrix}
    1 & 0 \\ -1 & -1   
    \end{pmatrix}
    \begin{pmatrix}
        f_1(\tau) \\ f_2(\tau)
    \end{pmatrix}\,,\hspace{10mm}
    \begin{pmatrix}
        f_1(-1/\tau) \\ f_2(-1/\tau)
    \end{pmatrix} 
    = \tau^k \begin{pmatrix}
    0 & 1 \\ 1 & 0   
    \end{pmatrix}
    \begin{pmatrix}
        f_1(\tau) \\ f_2(\tau)
    \end{pmatrix}\,.
    \end{align}
    Therefore, newforms on $\Gamma_0(2)$ give rise to cusp forms on $\Gamma(2)$ in the two-dimensional standard representations of $\mathcal{S}_3$.
\item 
    Suppose there is a Hecke normalised cusp form $g(\tau)$ of weight $k$ for $\SLZ$. Denote the second Hecke eigenvalue by $\lambda_2$. Define\footnote{One could simply look at $g(\tau), g(\tau/2), g(2\tau)$, but this definition also disentangles the $\mathcal{S}_3$ transformation.}
    \begin{align}
    \label{eq:sl2orig}
        f_1(\tau) = g(\tau) \,,\quad
        f_2(\tau) = \frac23 \lambda_2 g(\tau) -g(\tau/2)\,,\quad
        f_3(\tau) = \frac23 \lambda_2 g(\tau) -2^k g(2\tau)\,.
    \end{align}
    Then one can show that 
    \begin{align}
    \begin{pmatrix}
        f_1(\tau{+}1) \\ f_2(\tau{+}1) \\ f_3(\tau{+}1)
    \end{pmatrix} 
    = \begin{pmatrix}
    1 & 0  &0 \\ 0 & 1 & 0 \\ 0 &-1 & -1   
    \end{pmatrix}
    \begin{pmatrix}
        f_1(\tau) \\ f_2(\tau) \\ f_3(\tau)
    \end{pmatrix}\,,\hspace{10mm}
    \begin{pmatrix}
        f_1(-1/\tau) \\ f_2(-1/\tau) \\ f_3(-1/\tau)
    \end{pmatrix} 
    = \tau^k \begin{pmatrix}
    1 & 0 & 0 \\ 0 & 0 & 1 \\ 0 & 1 & 0   
    \end{pmatrix}
    \begin{pmatrix}
        f_1(\tau) \\ f_2(\tau) \\ f_3(\tau)
    \end{pmatrix}\,.
    \end{align}
    Therefore, cusp forms on $\SLZ$ give rise to the direct sum of a trivial (spanned by $f_1$) and a standard (spanned by $f_2$, $f_3$) representation of $\mathcal{S}_3$. The presence of the trivial representation follows since $f_1(\tau)=g(\tau)$ is invariant under all of $\SLZ$, and therefore under its $\mathcal{S}_3$ quotient. 
\end{itemize}

We list the dimensions of the weight-$k$ newforms for low $k$ and the associated $\mathcal{S}_3$ representations of cusp forms for $\Gamma(2)$ in Table~\ref{dimnewf}. In particular, the first $\SLZ$ cusp form appearing at weight $k=12$ is the Ramanujan discriminant.

\begin{table}
    \centering
    \begin{tabular}{c||c|c|c||c}
    $k$ & $\dim  \mathcal{S}_k^{\text{new}} (\Gamma_0(4)) $ & $\dim  \mathcal{S}_k^{\text{new}} (\Gamma_0(2))$ & $\dim  \mathcal{S}_k (\SLZ)$ & $\mathcal{S}_3\text{ representation}$\\\hline\hline
    $6$ & $1$ & -- & -- & $\bep$ \\
    $8$ & -- & $1$ & -- & $\mathbf{2}$ \\
    $10$ & $1$ & $1$ & -- & $\bep \oplus \mathbf{2}$ \\
    $12$ & $1$ & -- & $1$ & $\bep \oplus  \mathbf{1} \oplus \mathbf{2}$
    \end{tabular}
    \caption{\label{dimnewf}\textit{Dimensions of the newform spaces for $\Gamma_0(4)$, $\Gamma_0(2)$ and $\SLZ=\Gamma_0(1)$. These all give rise to cusp forms for $\Gamma(2)$ and the last column records the resulting $\mathcal{S}_3$ representations. This information can be obtained directly from the LMFDB website~\cite{LMFDB}.}}
\end{table}

{}From~\eqref{eq:etagens} and the ring structure one can find explicit expressions of the first few cusp forms in terms of eta products that we list for completeness.
\begin{itemize}
    \item[\underline{$k=6$:}] One cusp form coming from a $\Gamma_0(4)$ newform and given by $f(\tau)=\eta(\tau)^{12}$.
    \item[\underline{$k=8$:}] Two cusp forms coming from a $\Gamma_0(2)$ newform and given by
    \begin{align}
     f_1(\tau) = \eta(\tau)^{8} \eta(2\tau)^8\,,\hspace{10mm}       
     f_2(\tau) = \frac1{16}\eta(\tau)^{8} \eta(\tau/2)^8\,.
    \end{align}
\item[\underline{$k=10$:}] One cusp form coming from a $\Gamma_0(4)$ newform given by
\begin{align}
    f(\tau) = \eta(\tau)^4\eta(\tau/2)^{16} + 16 \eta(\tau)^4\eta(\tau/2)^8\eta(2\tau)^8+256 \eta(\tau)^4\eta(2\tau)^{16}\,.
\end{align}
Two cusp forms coming from a $\Gamma_0(2)$ newform that can be chosen as 
\begin{align}
    f_1(\tau) &= \eta(2\tau)^4 \eta(\tau)^{16} + 32  \eta(2\tau)^4\eta(\tau)^8\eta(4\tau)^8\nn\\
    &= \eta(\tau)^4\eta(\tau/2)^8\eta(2\tau)^8+8 \eta(\tau)^4\eta(2\tau)^{16}\,,\nn\\
    f_2(\tau) &= -\frac1{32}\left(\eta(\tau)^4 \eta(\tau/2)^{16} + 32 \eta(\tau)^4 \eta(\tau/2)^8\eta(2\tau)^8 \right)\,.
\end{align}
\item[\underline{$k=12$:}]
Three cusp forms coming from one for the full modular group (Ramanujan)
\begin{align}
f_1(\tau) &= \frac1{256} \left( F_1(\tau)^4 F_2(\tau)^2 + 2 F_1(\tau)^3F_2^3 + F_1(\tau)^2 F_2(\tau)^4\right)\nn\\
&= \eta(\tau)^{24}\,,\nn\\
f_2(\tau) &= \eta(\tau/2)^{24}\,,\nn\\
f_3(\tau) &= \eta(2\tau)^{24}\,.
\end{align}
Note that this basis is not adapted to the $\mathcal{S}_3$ transformation, see~\eqref{eq:sl2orig} for the appropriate choice where $\lambda=-24$ for Ramanujan's discriminant.

One cusp form coming from a $\Gamma_0(4)$ newform, leading to the  expression
\begin{align}
    f(\tau) &= -\frac{1}{16}F_1(\tau)^5F_2(\tau)-\frac{5}{32}F_1(\tau)^4F_2(\tau)^2+\frac{5}{32}F_1(\tau)^2F_2(\tau)^4+\frac{1}{16}F_1(\tau)F_2(\tau)^5\nn\\
    &= \frac{\eta(\tau/2)^{40}\eta(2\tau)^8}{\eta(\tau)^{24}}+40\frac{\eta(\tau/2)^{32}\eta(2\tau)^{16}}{\eta(\tau)^{24}}-10240 \frac{\eta(\tau/2)^{16} \eta(2\tau)^{32}}{\eta(\tau)^{24}} \nn\\
    &\quad -65536 \frac{\eta(\tau/2)^{8} \eta(2\tau)^{40}}{\eta(\tau)^{24}}\,.
\end{align}

\end{itemize}

\section{Fay relations and the flatness condition}
\label{apc.sec}

This appendix provides the corollaries of the Fay identities (\ref{FayuntildeF}) and (\ref{FaytildeF}) needed to simplify the flatness conditions (\ref{flatnesslevel1comms}) and (\ref{getlcg}) for the CEE and CG connections.

\subsection{\texorpdfstring{Level $1$}{Level 1}}
\label{apc.1}

In order to derive the action of the Tsunogai derivations $\epsilon_{k}$ on $\agen$ and $\bgen$ from the flatness condition (\ref{flatnesslevel1comms}), we
need to use the Fay relations (\ref{FayuntildeF}) to simplify its left-hand side, which we may expand as
\begin{align}\label{app Fay 1}
\big[\mathbb K^{(z)}_\text{CEE}(z,\tau),\partial_{\ad_\bgen}(\mathbb K^{(z)}_\text{CEE}(z,\tau){-}2\pi i\agen) \big]&= \frac{1}{2}\sum_{k,\ell =0}^{\infty} (2\pi i)^{1-k-\ell} [t_{k+1|1},t_{\ell+1|1}] \\
&\quad \times \left(\ell g^{(k)}(z,\tau)g^{(\ell+1)}(z,\tau)-kg^{(k+1)}(z,\tau)g^{(\ell)}(z,\tau)\right)\,, \notag
\end{align}
with the formal definition $t_{1|1}=a$. 
Here, we notice that we find products of the form $g^{(n_1)}(z,\tau)g^{(n_2)}(z,\tau)$, which we need to express in terms of $g^{(k_1)}(z,\tau)\est_{k_2}(\tau)$ with $k_1{+}k_2 = n_1{+} n_2$ via the (coincident limit $z_1\rightarrow z_2$ of the) Fay relations (\ref{FayuntildeF}).
To this end, let us use a corollary of the Fay identities (also explicitly derived in (E.2) of \cite{Broedel:2017jdo})
\begin{align}\label{app Fay 2}
&(\partial_{\alpha_2}{-}\partial_{\alpha_1})\big(F(z_2,\alpha_2,\tau)F(z_1,\alpha_1,\tau)\big)\\
&\hspace{10mm}=F(z_1,\alpha_1{+}\alpha_2,\tau)\partial_{\alpha_2}F(z_1{-}z_2,\alpha_2,\tau)-F(z_2,\alpha_2{+}\alpha_1,\tau)\partial_{\alpha_1}F(z_2{-}z_1,\alpha_1,\tau)\,.\notag
\end{align}
In the coincident limit $z_1=z_2=z$ this leads to 
\beq\label{app Fay 3}
(\partial_{\alpha_2}{-}\partial_{\alpha_1})\big(F(z,\alpha_1,\tau)F(z,\alpha_2,\tau)\big)
\!=\!
F(z,\alpha_1{+}\alpha_2,\tau)\Bigl(
\frac{1}{\alpha_1^2}-\frac{1}{\alpha_2^2}
-\sum_{k=4}^{\infty}(k{-}1)\est_k(\tau)\big(\alpha_2^{k-2}{-}\alpha_1^{k-2}\big)\!\Bigr)\,.
\eeq
We shall now use this relation to simplify the following generating function of the Kronecker-Eisenstein kernels in the second line of (\ref{app Fay 1}) with commuting bookkeeping variables $\alpha_1,\alpha_2$:
\begin{align}\label{app Fay 4}
&\sum_{k,\ell = 0}^{\infty}\left(\ell g^{(k)}(z,\tau)g^{(\ell+1)}(z,\tau){-}kg^{(k+1)}(z,\tau)g^{(\ell)}(z,\tau)\right)
\alpha_1^{\ell-1}\alpha_2^{k-1}\\
&\hspace{10mm}=-(\partial_{\alpha_2}{-}\partial_{\alpha_1})\big(F(z,\alpha_1,\tau)F(z,\alpha_2,\tau)\big)+\frac{1}{\alpha_1^2}F(z,\alpha_2,\tau)-\frac{1}{\alpha_2^2}F(z,\alpha_1,\tau)\notag\\
&\hspace{10mm}=\sum_{k=4}^{\infty}(k{-}1)\est_k(\tau) (\alpha_2^{k-2}{-}\alpha_1^{k-2})F(z,\alpha_1{+}\alpha_2,\tau)\notag\\
&\hspace{15mm}+\left(\frac{1}{\alpha_2^2}{-}\frac{1}{\alpha_1^2}\right)F(z,\alpha_1{+}\alpha_2,\tau)+\frac{1}{\alpha_1^2}F(z,\alpha_2,\tau)-\frac{1}{\alpha_2^2}F(z,\alpha_1,\tau)\,.\notag
\end{align}
We note that the double and triple poles in $\alpha_1$ and $\alpha_2$ all cancel upon expanding.
Translating the antisymmetric combinations $\alpha_1^{\ell-1} \alpha_2^{k-1}-(\alpha_1 \leftrightarrow \alpha_2)$ in the expansion of (\ref{app Fay 4}) into brackets $[t_{k+1|1},t_{\ell+1|1}]$ of (\ref{app Fay 1}), we conclude
\begin{align}\label{app Fay 5}
&\big[\mathbb K^{(z)}_\text{CEE},\partial_{\ad_\bgen}(\mathbb K^{(z)}_\text{CEE}{-}2\pi i\agen)\big]=
\sum_{k=2}^\infty (2\pi i)^{2-k}g^{(k)}(z,\tau)\sum_{j=0}^{k-2}\ad_\bgen^j [\agen,t_{k-j|1}]\\
&\hspace{40mm}+\sum_{k=4}^\infty
\sum_{\ell=0}^\infty (2\pi i)^{2-k-\ell} 
\est_k(\tau)g^{(\ell)}(z,\tau)\ad_\bgen^\ell\left(\sum_{j=1}^{k/2{-}1}({-}1)^j[t_{j+1|1},t_{k-j|1}]\right)\,.\notag
\end{align}
After inserting (\ref{app Fay 5})
into the left-hand side of (\ref{flatnesslevel1comms}) and the ansatz (\ref{ansaL}) for $\mathbb{L}_{\rm CEE}(\tau)$ into the right-hand side, we conclude that the flatness condition $[\mathbb K_\text{CEE}^{(z)}, \mathbb K_\text{CEE}^{(\tau)}] = 0$ implies the brackets of the Tsunogai derivations in section \ref{sec:221ts}.

\subsection{\texorpdfstring{Level $2$}{Level 2}}
\label{apc.2}

In (\ref{getlcg}) we need to use the Fay relations (\ref{FaytildeF}) in order to extract the action of the level-2 derivations $\epsilon_0$, $\tilde \epsilon_{2|\nu}$, $\epsilon_{k|\mu}$ on $\agen$ and $\bgen$. Similar to (\ref{app Fay 1}), we may expand the left side of (\ref{getlcg})~as
\begin{align}\label{app Fay 6}
&\big[\mathbb K_\text{CG}^{(z)}(z,\tau),\partial_{\ad_\bgen}(\mathbb K_\text{CG}^{(z)}(z,\tau)-2\pi i \agen) \big]
=\sum_{\mu=1}^4\sum_{k=2}^\infty (k{-}1)(2\pi i)^{2-k}\tilde g^{(k)}(z_\mu, r_\mu ,\tau)[\agen,t_{k|\mu}]\\
&\hspace{55mm}+\frac{1}{2}\sum_{\mu,\nu=1}^4\sum_{k,\ell = 1}^{\infty}(2\pi i)^{1-k-\ell}\biggl(\ell \tilde g^{(k)}(z_\mu, r_\mu ,\tau)\tilde g^{(\ell+1)}(z_\nu, r_\nu ,\tau)\notag\\
&\hspace{68mm}-k\tilde g^{(k+1)}(z_\mu, r_\mu ,\tau)\tilde g^{(\ell)}(z_\nu, r_\nu,\tau)\biggr)[t_{k+1|\mu},t_{\ell+1|\nu}]\,,\notag
\end{align}
where we introduce the shorthands $(z_\mu,r_\mu)$ corresponding to the torsion points
\beq\label{app Fay 7}
(z_1,r_1)=(z,0)\,,\quad  (z_2,r_2)=(z{+}\tfrac{1}{2},0)\,,\quad(z_3,r_3)=(z{+}\tfrac{1}{2},\tfrac{1}{2})\,,\quad(z_4,r_4)=(z,\tfrac{1}{2})\, .
\eeq
We notice that the products of the form $\tilde g^{(n_1)}(z_\mu,r_\mu,\tau)\tilde g^{(n_2)}(z_\nu,r_\nu,\tau)$ in (\ref{app Fay 6})
can be expressed in terms of $\tilde g^{(k_1)}(z_\mu,r_\mu,\tau)f^{(k_2)}(\omega_\nu,\tau)$ with $n_1{+}n_2=k_1{+}k_2$ via the (coincident limit $z_1\rightarrow z_2$ of the) Fay relations (\ref{FaytildeF}).
Let us again use a corollary of the Fay identities similar to (\ref{app Fay 2}), which in this specific case reduce to
\begin{align}\label{app Fay 8}
(\partial_{\alpha_2}{-}\partial_{\alpha_1})&\big(\tilde F(z_\mu,r_\mu,\alpha_2,\tau)\tilde F(z_\nu,r_\nu,\alpha_1,\tau) \big)
\\
&\notag=
\tilde F(z_\nu,r_\nu,\alpha_1{+}\alpha_2,\tau)\partial_{\alpha_2}\tilde F(z_\mu{-}z_\nu,r_\mu{-}r_\nu,\alpha_2,\tau)\\
&\notag \hspace{18mm} -\tilde F(z_\mu,r_\mu,\alpha_1{+}\alpha_2,\tau)\partial_{\alpha_1}\tilde F(z_\nu{-}z_\mu, r_\nu{-}r_\mu ,\alpha_1,\tau)\,.
\end{align}
For distinct labels, periodicity and parity at the half-periods give
\beq\label{app Fay 9}
\partial_\alpha \tilde F(z_\mu{-}z_\nu, r_\mu{-}r_\nu ,\alpha,\tau)
=-\frac{1}{\alpha^2}+\sum_{k= 2}^{\infty}(k{-}1)f^{(k)}(\omega_\mu{-}\omega_\nu,\tau)\alpha^{k-2}\,.
\eeq
For equal labels $\mu=\nu$, we use the coincident limit (\ref{app Fay 3}), with $f^{(k)}(0,\tau)=-\est_k(\tau)$.
Let us now use this relation to bring the following generating function of the bilinears in $\tilde{g}^{(n)}$ of (\ref{app Fay 6}) into the desired form
\begin{align}\label{app Fay 10}
&\sum_{k,\ell= 1}^\infty \biggl(\ell \tilde g^{(k)}(z_\mu, r_\mu ,\tau)\tilde g^{(\ell+1)}(z_\nu, r_\nu ,\tau)-k\tilde g^{(k+1)}(z_\mu, r_\mu ,\tau)\tilde g^{(\ell)}(z_\nu, r_\nu ,\tau)\biggr)
\alpha_2^{k-1}\alpha_1^{\ell-1}\\
&\hspace{10mm}={-}(\partial_{\alpha_2}{-}\partial_{\alpha_1})\biggl(\left(\tilde F(z_\mu,r_\mu,\alpha_2,\tau){-}\tfrac{1}{\alpha_2}\right)\left(\tilde F(z_\nu,r_\nu,\alpha_1,\tau){-}\tfrac{1}{\alpha_1}\right)\biggr)\notag\\
&\hspace{10mm}=\frac{\tilde F(z_\mu,r_\mu,\alpha_2,\tau){-}\tilde F(z_\mu,r_\mu,\alpha_1{+}\alpha_2,\tau)}{\alpha_1^2}-\frac{\tilde F(z_\nu,r_\nu,\alpha_1,\tau){-}\tilde F(z_\nu,r_\nu,\alpha_1{+}\alpha_2,\tau)}{\alpha_2^2}\notag\\
&\hspace{15mm}+\frac{1}{\alpha_1}\partial_{\alpha_2}\tilde F(z_\mu,r_\mu,\alpha_2,\tau)-\frac{1}{\alpha_2}\partial_{\alpha_1}\tilde F(z_\nu,r_\nu,\alpha_1,\tau)+\frac{1}{\alpha_2^2\alpha_1}-\frac{1}{\alpha_2\alpha_1^2}\notag\\
&\hspace{15mm}+\sum_{k = 2}^{\infty}(k{-}1)f^{(k)}(\omega_\mu{-}\omega_\nu,\tau)\biggl(\alpha_1^{k-2}\tilde F(z_\mu,r_\mu,\alpha_1{+}\alpha_2,\tau)-\alpha_2^{k-2}\tilde F(z_\nu,r_\nu,\alpha_1{+}\alpha_2,\tau)\biggr)\notag\,.
\end{align}
The double and triple poles in $\alpha_1$ and $\alpha_2$ cancel upon expanding.
Plugging (\ref{app Fay 9}) into (\ref{app Fay 6}) and using $\sum_{\mu=1}^4t_{2|\mu}=[\bgen,\agen]$ lets us conclude that
\begin{align}
    &\big[\mathbb K_\text{CG}^{(z)},\partial_{\ad_\bgen}(\mathbb K_\text{CG}^{(z)}{-}2\pi i \agen) \big]=\sum_{k= 2}^\infty (2\pi i)^{2-k} \sum_{\mu=1}^4 \tilde g^{(k)}(z_\mu,r_\mu,\tau)\sum_{j=0}^{k-2}\ad_\bgen^j[\agen,t_{k-j|\mu}]\\
    &\hspace{15mm}+\frac{1}{2}\sum_{k= 2}^{\infty}(k{-}1)(2\pi i)^{2-k}\sum_{\mu,\nu=1}^4f^{(k)}(\omega_\mu{-}\omega_\nu,\tau)\sum_{j=0}^{k-3}({-}1)^j[t_{j+2|\mu},t_{k-1-j|\nu}]\notag\\
    &\hspace{15mm}+\sum_{\ell=1}^\infty(2\pi i)^{-\ell}\sum_{1\leq \mu<\nu}^4f^{(2)}(\omega_\mu{-}\omega_\nu,\tau)\left(\tilde g^{(\ell)}(z_\mu,r_\mu,\tau){-}\tilde g^{(\ell)}(z_\nu,r_\nu,\tau)\right)\ad_\bgen^{\ell-1}[t_{2|\mu},t_{2|\nu}]\notag\\
    &\hspace{15mm}+\sum_{k= 4}^\infty \sum_{\ell=1}^\infty (k{-}1)(2\pi i)^{2-k-\ell}\sum_{\mu,\nu=1}^4f^{(k)}(\omega_\mu{-}\omega_\nu,\tau)\tilde g^{(\ell)}(z_\mu,r_\mu,\tau)\ad_\bgen^{\ell-1}[t_{2|\mu},t_{k|\nu}]\notag\,.
\end{align}
Upon insertion
into the left-hand side of (\ref{getlcg}) and by the ansatz (\ref{anslv2}) for $\mathbb{L}_{\rm CG}(\tau)$ on the right-hand side, we conclude that the flatness condition $[\mathbb K_\text{CG}^{(z)}, \mathbb K_\text{CG}^{(\tau)}] = 0$ implies the bracket relations of the level-2 derivations in section \ref{sec:actderiv}.

\section{\texorpdfstring{Single-valued genus-0 polylogarithms with singular points $0$, $1$, $-1$}{Single-valued genus-0 polylogarithms with singular points 0, 1, -1}}
\label{app:mpl}

This appendix is dedicated to multiple polylogarithms (MPLs) on the Riemann sphere with singular points at $\{0,1,-1\}$ (known as harmonic polylogarithms in the physics literature~\cite{Remiddi:1999ew}) which yield cyclotomic MZVs for level 2 (see appendix \ref{app:a.2}) as their special values. We will construct the single-valued versions of these level 2 MPLs from zeta generators (for alternative ways for their construction, see~\cite{Brown:unpubl,brown2017notes,DelDuca:2016lad,Broedel:2016kls}). Their generating series in (\ref{mplapp.11}) below exhibits the same structure as that of single-valued iterated Eisenstein integrals for level 2 in (\ref{eq: svI2}) and can therefore be viewed as its genus-0 prototype. We determine the action (\ref{lv2.28}) of genus-0 zeta generators for level 2 in the construction of single-valued MPLs whose implications for genus 1 will be discussed in section \ref{sec:furth.12}.

\subsection{MPL series and associators for level 2}
\label{ass.lv2}

In order to make contact with the degeneration $\tau \rightarrow i \infty$ of the CG connection in section~\ref{app:mpl.1}, we shall consider level 2 MPLs with base-point $w=1$ instead of the more conventional choice of $w=0$ in section~\ref{sec:tmpl} (also see appendix \ref{sec:furth.1}). The singular points $\{0,1,-1 \}$ of the integration kernels are associated with the free-Lie-algebra generators $\egen_{0}$, $\egen_{1}$, $\egen_{-1}$, and the integration base-point is reflected by the subscript of the generating series 
\beq
\Gcyc(\egen;w)=\text{Pexp}\biggl({-}\int_w^{-\vec{1}_1} \dd w_1\biggl(\frac{\egen_0}{w_1}+\frac{\egen_1}{w_1{-}1}+\frac{\egen_{-1}}{w_1{+}1}\biggr)\biggr)\,,
\label{mplapp.04}
\eeq
in $\exp({\rm Lie}[\egen_0,\egen_1,\egen_{-1}])$.
The tangent vector $-\vec{1}_1$ at $w_1=1$ shuffle-regularises the MPLs in the expansion in the generators and assigns $\log(1{-}w)$ as the series coefficient of the one-letter word~$\egen_{1}$.
We will also be interested in the regularised limits $w\rightarrow 0$ and $w\rightarrow -1$ of (\ref{mplapp.04}),
\begin{align}
Z_0(\egen_0,\egen_1,\egen_{-1}) &= 
 {\rm Pexp}\biggl({-}\int_{\vec{1}_0}^{-\vec{1}_1} {\rm d} w_1\biggl(\frac{\egen_0}{w_1}+\frac{\egen_1}{w_1{-}1}+\frac{\egen_{-1}}{w_1{+}1}\biggr)\biggr)\, ,
\label{mplapp.05} \\
Z_{-1}(\egen_0,\egen_1,\egen_{-1}) &= 
 {\rm Pexp}\biggl({-}\int_{\vec{1}_{-1}}^{-\vec{1}_1} {\rm d} w_1\biggl(\frac{\egen_0}{w_1}+\frac{\egen_1}{w_1{-}1}+\frac{\egen_{-1}}{w_1{+}1}\biggr)\biggr)\, ,
\notag
\end{align}
which will be referred to as associators. The tangent vectors $\vec{1}_{0}$ and $\vec{1}_{-1}$ at the points $w_1=0$ and $w_1=-1$ lead to shuffle-regularisations compatible with the assignments $\int^1_0 \frac{\dd w_1}{w_1} = 0$ and $\int^1_{-1} \frac{\dd w_1}{w_1{+}1} = \log(2)$. $Z_0$ and $Z_{-1}$ are group-like series in $\egen_0,\egen_1,\egen_{-1}$ with $\mathbb Q$-linear combinations of cyclotomic MZVs for level 2 as coefficients \cite{Vermaseren:1998uu, Remiddi:1999ew, Blumlein:2009cf, Jianqiang}.

\subsection{Single-valued MPLs for level 2}
\label{app:mpl.7}

The main result of this appendix is the following single-valued version of the MPL series (\ref{mplapp.04}) in terms of zeta generators for level 2:
\beq
\Gcyc^{\rm sv}({\rm e};w)=(\mathbb M^{\rm sv}(\sigma))^{-1} \overline{\Gcyc({\rm e};w)^T} \mathbb M^{\rm sv}(\sigma) \Gcyc({\rm e};w)\, .
\label{mplapp.11}
\eeq
The rightmost series ensures that $\Gcyc^{\rm sv}({\rm e};w)$ obeys the
same holomorphic differential equation as $\Gcyc({\rm e};w)$, and the
transposition 
acts via $(\ldots {\rm e}_{a_1}\ldots {\rm e}_{a_2} \ldots)^T = \ldots {\rm e}_{a_2}\ldots {\rm e}_{a_1} \ldots$ for $a_i \in \{0,1,-1\}$. 
The series $\mathbb M^{\rm sv}(\sigma)$ in cyclotomic MZVs for level 2 is the single-valued image of (\ref{eq: zetagen2}), and the bracket relations of the accompanying zeta generators~$\sigma_w$ are given by
\beq
[\sigma_w,{\rm e}_1] = 0 \, , \ \ \ \ \ \ \big[ \sigma_w, {\rm e}_0\big] = \big[ {\rm e}_0, Q_w({\rm e}) \big]
\, , \ \ \ \ \ \ \big[ \sigma_w, {\rm e}_{-1}\big] = \big[ {\rm e}_{-1}, R_w({\rm e}) \big]\, ,
  \label{lv2.28}
\eeq
involving Lie polynomials $Q_w$, $R_w$ in ${\rm e}_0,{\rm e}_1,{\rm e}_{-1}$ of odd degree $w\geq 1$. 
These Lie polynomials are defined by isolating the primitive $f_w$ in the $f$-alphabet decomposition of the motivic versions of the associators (\ref{mplapp.05}),
\beq
Q_w({\rm e}) = \rho\big( Z^\mot_0({\rm e}_0,{\rm e}_1,{\rm e}_{-1})\big) \, \big|_{f_w}
\, ,  \ \ \ \ \ \ 
R_w({\rm e}) = \rho\big( Z^\mot_{-1}({\rm e}_0,{\rm e}_1,{\rm e}_{-1})\big) \, \big|_{f_w} \, ,
  \label{lv2.27}
\eeq
in direct analogy with the definition (\ref{revgg.06}) of the Lie polynomials $P_w(x,y)$ for the level-1 case from the Drinfeld associator.

The simplest examples of the Lie polynomials (\ref{lv2.27}) at degree $w=1,3$ are given by
\beq
Q_1 =-\egen_{-1}\,,\quad R_1 =\egen_{1}-\egen_{-1}\,,
\eeq
and
\begin{align}
    Q_3&={}\frac{3}{4}[\egen_{0},[\egen_{-1},\egen_{0}]]-[\egen_{0},[\egen_{1},\egen_{0}]]-\frac{1}{8}[\egen_{-1},[\egen_{-1},\egen_{0}]]-[\egen_{-1},[\egen_{1},\egen_{0}]] \\
    &\quad-\frac{1}{8}[\egen_{-1},[\egen_{1},\egen_{-1}]]+\frac{13}{8}[\egen_{1},[\egen_{-1},\egen_{0}]]-[\egen_{1},[\egen_{1},\egen_{0}]]-\frac{7}{8}[\egen_{1},[\egen_{1},\egen_{-1}]] \, ,\nonumber\\
    R_3&=\frac{7}{8}[\egen_{-1},[\egen_{-1},\egen_{0}]]-[\egen_{-1},[\egen_{1},\egen_{-1}]]-\frac{21}{8} [\egen_{-1},[\egen_{1},\egen_{0}]]+\frac{7}{4} [\egen_{0},[\egen_{-1},\egen_{0}]] \nonumber\\
    &\quad-\frac{7}{4}[\egen_{0},[\egen_{1},\egen_{0}]]+\frac{21}{8}[\egen_{1},[\egen_{-1},\egen_{0}]]-[\egen_{1},[\egen_{1},\egen_{-1}]]-\frac{7}{8}[\egen_{1},[\egen_{1},\egen_{0}]]\, .
    \nonumber
\end{align}

While the individual Lie polynomials $Q_w({\rm e})$ and $R_w({\rm e})$ depend on the choice of the $f$-alphabet isomorphism $\rho$ for level 2, the right side of (\ref{mplapp.11}) is independent of $\rho$, as will be shown in section \ref{app:indrho} below. The use of the same notation $\sigma_w$ for the genus-0 zeta generators (\ref{mplapp.11}) and their genus-1 counterparts introduced in section~\ref{tau level 2.2} will be justified in appendix~\ref{sec:furth.1}.

The proof of (\ref{mplapp.11}) in sections \ref{sec:infco} and \ref{sec:incin} below will proceed by expressing the monodromies $\text{Mon}_{w=a}$ of (\ref{mplapp.04}) as the endpoint $w$ winds around the singular points $a \in \{0,1,-1\}$ in terms of zeta generators. A first expression for these monodromies follows from the composition-of-paths formula, 
\begin{align}
\text{Mon}_{w=1} \Gcyc(\egen;w)&= e^{2\pi i \egen_1}\Gcyc(\egen;w)\, ,
\label{mplapp.06}
\\
\text{Mon}_{w=0} \Gcyc (\egen;w) &=Z_0(\egen_0,\egen_1,\egen_{-1}) e^{2\pi i \egen_0} Z_0(\egen_0,\egen_1,\egen_{-1})^{-1} \Gcyc(\egen;w) \, , \notag \\
\text{Mon}_{w=-1} \Gcyc(\egen;w) &=Z_{-1}(\egen_0,\egen_1,\egen_{-1}) e^{2\pi i \egen_{-1}} Z_{-1}(\egen_0,\egen_1,\egen_{-1})^{-1} \Gcyc(\egen ;w) \, ,\notag
\end{align}
but it will be advantageous to rewrite the adjoint action of the associators $Z_a(\egen_0,\egen_1,\egen_{-1})$ in terms of $\mathbb M(\sigma)$ in order to see the cancellation of monodromies from (\ref{mplapp.11}). These rewritings of the monodromy factors in (\ref{mplapp.06}) will rely on their motivic coactions.

\subsection{Coactions of level 2 associator and monodromy factors}
\label{sec:infco}

In the first place, the
motivic coaction of the associators (\ref{mplapp.05}) is obtained from the Ihara formula \cite{Ihara1989TheGR, Brown:2019jng}
\begin{align}
\Delta Z^\mot_a ({\rm e}_0,{\rm e}_1,{\rm e}_{-1}) =  Z^\mot_a\Big(Z^\dr_0 {\rm e}_0 (Z^\dr_0)^{-1},{\rm e}_1, Z^\dr_{-1}{\rm e}_{-1} (Z^\dr_{-1})^{-1} \Big)  \, Z^\dr_a({\rm e}_0,{\rm e}_1,{\rm e}_{-1}) \, ,
  \label{lv2.25} 
\end{align}
which is adapted to the case that the integration path starts at the point $w=1$ associated with
the generator ${\rm e}_1$.
The superscripts $^\mot$ and $^\dr$ distinguish the motivic and de Rham periods produced by the motivic coaction \cite{brown2017notes}, and we will skip the $\otimes$ symbol between the motivic and de Rham entries.
The arguments ${\rm e}_0,{\rm e}_1,{\rm e}_{-1}$ of the de Rham associators $Z^\dr_0$, $Z^\dr_{-1}$ in the twisted letters of $Z^\mot_a$ on the right side of (\ref{lv2.25}) are suppressed for ease of notation. 

In order to pinpoint the implication of (\ref{lv2.25}) for the monodromy factors of (\ref{mplapp.06}), it will be convenient to pass to the infinitesimal version of the coaction in terms of non-commutative differential operators $\partial_{f_w} Z^\mot_a$ for odd $w\geq 1$ \cite{Brown:2011ik, BrownTate}. The operator $\partial_{f_w}$ isolates those terms in the $f$-alphabet image of $\Delta Z^\mot_a$ which only have the primitive $f_w$ in the de Rham entry,
\beq
\rho(\Delta Z^\mot_a) =  \sum_{w\in 2\mathbb N+1} Y_w^\mot f^\dr_w + \ldots  \ \ \ \Rightarrow \ \ \ \partial_{f_w} Z^\mot_a = \rho^{-1}(Y_w^\mot) \ \ \ {\rm for} \ \, w\geq 1 \ \, {\rm odd}\, ,
\label{mplapp.07}
\eeq
see appendix \ref{app:a.3} for the $f$-alphabet isomorphism $\rho$ for level 2.
In other words, the terms in the ellipsis of (\ref{mplapp.07}) are taken to have either 0 or $\geq 2$ letters $f_{w}$ with $w\geq 1$ odd in their de Rham entries. At the level of the $f$-alphabet, the deconcatenation form (\ref{deconcop}) of the coaction leads to the simple realisation of $\partial_{f_w}$ where it clips off the rightmost letter of $f_{i_1}\ldots f_{i_r}$ if $i_r=w$ and annihilates $f_{i_1}\ldots f_{i_r}$ if $i_r\neq w$.

With these prerequisites in place, the infinitesimal version of the coaction (\ref{lv2.25}) can be represented in terms of the Lie polynomials $Q_w({\rm e}) $ and $R_w({\rm e})$ defined by (\ref{lv2.27}).
According to the two possibilities of extracting $f_w^\dr$ from the rightmost de Rham associator of (\ref{lv2.25}) or the twisted letters $Z_b^\dr {\rm e}_b (Z_b^\dr)^{-1}$ in the expansion of $Z_a^\mot$, its infinitesimal version is given by
\begin{align}
\partial_{f_w} Z^\mot_0({\rm e}_0,{\rm e}_1,{\rm e}_{-1}) &=  Z^\mot_0({\rm e}_0,{\rm e}_1,{\rm e}_{-1}) \, Q_w({\rm e}) - \big[\sigma_w , Z^\mot_0({\rm e}_0,{\rm e}_1,{\rm e}_{-1}) \big] \, ,
  \label{lv2.26} \\
  \partial_{f_w}  Z^\mot_{-1}({\rm e}_0,{\rm e}_1,{\rm e}_{-1}) &=  Z^\mot_{-1}({\rm e}_0,{\rm e}_1,{\rm e}_{-1}) \, R_w({\rm e}) - \big[\sigma_w , Z^\mot_{-1}({\rm e}_0,{\rm e}_1,{\rm e}_{-1}) \big]\, .
  \notag
  \end{align}
The zeta generators $\sigma_w$ in the brackets on the right-hand side take the role of Ihara derivations of ${\rm Lie}[{\rm e}_0,{\rm e}_1,{\rm e}_{-1}]$ that implement the effect of the twisted arguments
$Z^\dr_0 {\rm e}_0 (Z^\dr_0)^{-1}$ and $ Z^\dr_{-1}{\rm e}_{-1} (Z^\dr_{-1})^{-1}$ of $Z^\mot_a$ in (\ref{lv2.25}). More specifically, the bracket relations (\ref{lv2.28}) of $\sigma_w$ ensure at an infinitesimal level that ${\rm e}_{1}$ is invariant and that ${\rm e}_0$ and ${\rm e}_{-1}$ are shifted by the contribution of $f_w^{\dr}$ to the conjugation by $Z^\dr_0$ and $Z^\dr_1$ which are given by a commutator with the Lie polynomial $Q_w$ and $R_w$, respectively.

In view of the infinitesimal coactions (\ref{lv2.26}), the inverse associators obey
\begin{align}
\partial_{f_w} Z^\mot_0({\rm e}_0,{\rm e}_1,{\rm e}_{-1})^{-1} &=  {-} Q_w({\rm e}) \, Z^\mot_0({\rm e}_0,{\rm e}_1,{\rm e}_{-1})^{-1} - \big[\sigma_w , Z^\mot_0({\rm e}_0,{\rm e}_1,{\rm e}_{-1})^{-1} \big] \, ,
  \label{lv2.29} \\
  \partial_{f_w}  Z^\mot_{-1}({\rm e}_0,{\rm e}_1,{\rm e}_{-1})^{-1} &= {-} R_w({\rm e})\,  Z^\mot_{-1}({\rm e}_0,{\rm e}_1,{\rm e}_{-1})^{-1}  - \big[\sigma_w , Z^\mot_{-1}({\rm e}_0,{\rm e}_1,{\rm e}_{-1})^{-1} \big]\, .
  \notag
  \end{align}
Given that $[Q_w, e^{2\pi i {\rm e}_0}] = - [\sigma_w,e^{2\pi i {\rm e}_0}]$ and $[R_w, e^{2\pi i {\rm e}_{-1}}] = - [\sigma_w,e^{2\pi i {\rm e}_{-1}}]$ as a consequence of (\ref{lv2.28}), we can combine (\ref{lv2.26}) and (\ref{lv2.29}) to find the following infinitesimal coactions of the monodromy factors in (\ref{mplapp.06}) with $a\in \{0,-1 \}$,
\beq
\partial_{f_w} \big( Z^\mot_a  e^{(2\pi  i)^\mot \egen_a} (Z^\mot_a)^{-1} \big) = - \big[\sigma_w ,Z^\mot_a  e^{(2\pi i)^\mot \egen_a} (Z^\mot_a)^{-1} \big] \, .
  \label{lv2.30}
\eeq
We have used that $\partial_{f_w}$ obeys the Leibniz property when acting on a product of series in (cyclotomic) MZVs, i.e., on a shuffle product of the non-commuting letters in the $f$-alphabet.

\subsection{Monodromies and their cancellation from zeta generators}
\label{sec:incin}

The infinitesimal coaction (\ref{lv2.30}) of the monodromy factors 
$Z_a  e^{2\pi  i  \egen_a} (Z_a)^{-1}$ will now be viewed as a ``differential equation'' in $f_w$ with $w\geq 1$ odd which will help us to find a convenient rewriting. The idea is to present an alternative expression for $Z_a  e^{2\pi  i  \egen_a} (Z_a)^{-1}$ which obeys the same ``differential equation'' and can only differ by a choice of ``initial condition'' which is a series with coefficients in $\mathbb Q[i\pi]$ in our case: The intersection of the kernels of all $\partial_{f_w}$ with $w\geq 1$ odd reduces to series in $(i\pi)^\mot$ when acting on the series $Z_a^\mot$, $e^{2\pi  i  \egen_a}$ in $i\pi$ and cyclotomic MZVs for level 2.

The alternative solution to (\ref{lv2.30}) will be constructed from the series $\mathbb M(\sigma)$ in zeta generators \cite{Goncharov:1998kja, GLANOIS2016334, Glanois:2016} and cyclotomic MZVs for level 2. Its motivic coaction $\Delta \mathbb M^\mot(\sigma)= \mathbb M^\mot(\sigma) \mathbb M^\dr(\sigma)$ is an immediate consequence of (\ref{eq: zetagen2}) and at an infinitesimal level yields $\partial_{f_w} \mathbb M^\mot(\sigma) = \mathbb M^\mot(\sigma) \sigma_w$ and 
$\partial_{f_w} \mathbb M^\mot(\sigma)^{-1} = -\sigma_w \,\mathbb M^\mot(\sigma)^{-1}$.
 As a consequence, any series $\mathbb X_{a}(\egen)$ with a trivial coaction in the sense of $\partial_{f_w} \mathbb X^\mot_{a}(\egen) = 0$ yields an alternative solution to (\ref{lv2.30}), 
\beq
\partial_{f_w} \big( \mathbb M^\mot(\sigma)^{-1}\, \mathbb X^\mot_{a}(\egen)\, \mathbb M^\mot(\sigma) \big) = - \big[\sigma_w ,\mathbb M^\mot(\sigma)^{-1}\, \mathbb X^\mot_{a}(\egen)\, \mathbb M^\mot(\sigma) \big] \, .
  \label{lv2.alt}
\eeq
By the matching of their ``differential equations'' (\ref{lv2.30}) and (\ref{lv2.alt}), there exists a unique choice of ``initial conditions'' $\mathbb X_{a}(\egen)$ for $a\in \{0,-1\}$ -- group-like series in $i\pi {\rm e}_{b}$ with 
rational coefficients and $b\in\{0,1,-1\}$ -- such that 
\beq
Z_a  e^{2\pi  i  \egen_a} (Z_a)^{-1} = \mathbb M(\sigma)^{-1}\, \mathbb X_{a}(\egen)\, \mathbb M(\sigma) \, .
  \label{lv2.31}
\eeq
Upon insertion into the monodromies (\ref{mplapp.06})
of the MPL series (\ref{mplapp.04}), we conclude that
\beq
\text{Mon}_{w=a} \Gcyc (\egen;w) = \mathbb M(\sigma)^{-1}\, \mathbb X_{a}(\egen)\, \mathbb M(\sigma)  \Gcyc(\egen;w)\,, \ \ \ {\rm for} \, \ a \in \{0,1,-1 \}\, .
\label{lv2.33}
\eeq
The innermost series $\mathbb X_{a}(\egen)$ with $\mathbb Q[i\pi]$ coefficients 
depends on $a \in \{0,1,-1 \}$. In the
case $a=1$, which is not covered by (\ref{lv2.31}), we have $ \mathbb X_{1} (\egen) = e^{2\pi i {\rm e}_1}$ which commutes with $\mathbb M(\sigma)$ by (\ref{lv2.28}), and thus matches the first line of (\ref{mplapp.06}) by
$\mathbb M(\sigma)^{-1}e^{2\pi i {\rm e}_1} \mathbb M(\sigma) = e^{2\pi i {\rm e}_1}$.

Before applying the new expression (\ref{lv2.33}) for the monodromies to demonstrate single-valuedness of $\Gcyc^{\rm sv}(\egen;w)$ in (\ref{mplapp.11}), we note the property
\beq
\overline{ \mathbb X_{a}(\egen)^{T} } \, \mathbb X_a(\egen) = 1\,, \ \ \ \ {\rm for} \ a \in \{0,1,-1\} \,,
\label{lv2.32}
\eeq
of the series $\mathbb X_a(\egen)$ in (\ref{lv2.31}). This follows from the facts that the series $\mathbb X_{a}(\egen)$ are group-like, such that $\mathbb X_{a}(\egen)^{-1} = \mathbb X_{a}(-\egen)^T$, and that complex conjugation acts on $\mathbb X_{a}(\egen)$ via ${\rm e}_b\rightarrow - {\rm e}_b$, since it expands in words in $i\pi {\rm e}_b$ with rational coefficients.

Based on (\ref{lv2.33}), (\ref{lv2.32}) and $\mathbb M^\sv(\sigma) = \mathbb M(\sigma)^T \mathbb M(\sigma)$, we can now show the cancellation of monodromies from (\ref{mplapp.11}) for all of $a\in \{0,1,-1\}$ as follows,
\begin{align}
\text{Mon}_{w=a}\Gcyc^\text{sv}(\egen;w)&=\mathbb M^\text{sv}(\sigma)^{-1} \overline{\Gcyc(\egen;w)^T} \overline{ \big( \mathbb M(\sigma)^{-1} \mathbb X_{a}(\egen) \mathbb M(\sigma)  \big)^T } \label{mplapp.17} \\
&\quad \times 
\mathbb M^\text{sv}(\sigma)  \mathbb M(\sigma)^{-1} \mathbb X_{a}(\egen) \mathbb M(\sigma)
\Gcyc(\egen;w)
\notag \\
&= \mathbb M^\text{sv}(\sigma)^{-1} \overline{\Gcyc(\egen;w)^T} \mathbb M(\sigma)^T  \overline{ \mathbb X_{a}(\egen)^{T} }  \mathbb X_a(\egen) 
 \mathbb M(\sigma)
\Gcyc(\egen;w) \notag \\
&= \mathbb M^\text{sv}(\sigma)^{-1} \overline{\Gcyc(\egen;w)^T} 
 \mathbb M^\sv(\sigma)
\Gcyc(\egen;w) 
 \notag \\
&=\Gcyc^\sv(\egen;w) \, ,  \notag
\end{align}
confirming single-valuedness under all monodromies as the integration path for $\Gcyc$, $\overline{\Gcyc^{T}}$ winds around the singular points $0$, $1$, $-1$ of its integrand.

\subsection{\texorpdfstring{Independence of the choice of isomorphism to the $f$-alphabet}{Independence of the choice of isomorphism to the f-alphabet}}
\label{app:indrho}

We shall here demonstrate that the expression (\ref{mplapp.11}) for $\Gcyc^\sv(\egen;w)$ does not depend on the choice of the $f$-alphabet isomorphism $\rho$. We will show by indirect means that, within (\ref{mplapp.11}), the dependence of the brackets $[\sigma_w,{\rm e}_a]$ on $\rho$ through the Lie
polynomials $Q_w$, $R_w$ in (\ref{lv2.28}) and (\ref{lv2.27}) is compensated by that of the MZVs $\rho^{-1}(f_{i_1}\ldots f_{i_r})$ accompanying $\sigma_{i_1}\ldots \sigma_{i_r}$ in the expression (\ref{eq: zetagen2}) for $\mathbb M(\sigma)$.

Our proof relies on the following variant of (\ref{lv2.31}),
\beq
Z^\dr_a  {\rm e}_a (Z^\dr_a)^{-1} = \mathbb M^\dr(\sigma)^{-1} {\rm e}_a \mathbb M^\dr(\sigma) \,, \ \ \ {\rm for} \ a\in \{0,-1\} \, ,
\label{mplapp.18} 
\eeq
where the de Rham superscripts on both sides instruct to work modulo $i\pi$. This again follows from the fact that both sides obey the same ``differential equation'' $\partial_{f_w}\bullet = -[\sigma_w, \bullet]$, see (\ref{lv2.26}), (\ref{lv2.29}) together with (\ref{lv2.28}) for the left side and the discussion above (\ref{lv2.alt}) for the right side. In the de Rham projected setting, the operator $\partial_{f_w}$ implements the infinitesimal coproduct of de Rham MPLs as opposed to the infinitesimal coaction of motivic MPLs \cite{Goncharov:2001iea, Goncharov:2005sla, Brown:2011ik, BrownTate}. Since both sides of (\ref{mplapp.18}) reduce to ${\rm e}_a$ as their ``initial condition'' upon discarding all $f^\dr_w$, the matching of their ``differential equation'' under $\partial_{f_w}$ already implies equality -- the de Rham projection prevents deviations by a series in $i\pi$ analogous to $\mathbb X_a(\egen)$ in (\ref{lv2.31}).

The image of (\ref{mplapp.18}) under the single-valued map,
\beq
Z^\sv_a  {\rm e}_a (Z^\sv_a)^{-1} = \mathbb M^\sv(\sigma)^{-1} {\rm e}_a \mathbb M^\sv(\sigma)\,, \ \ \ {\rm for} \ a\in \{0,-1\} \, ,
\label{mplapp.19} 
\eeq
now implies that $\mathbb M^\sv(\sigma)^{-1} {\rm e}_a \mathbb M^\sv(\sigma)$ is independent of the $f$-alphabet isomorphism $\rho$  since this is clearly the case for the left side. 
In order to deduce that the right side of (\ref{mplapp.11}) is $\rho$-independent as well, we pull the following conjugation by $\mathbb M^\sv$ into the individual letters by repeated insertion of $1=\mathbb M^\sv (\mathbb M^\sv)^{-1}$:
\beq
\mathbb M^\sv(\sigma)^{-1} \overline{\Gcyc(\egen;w)^T} \mathbb M^\sv(\sigma) = \overline{\Gcyc( \mathbb M^\sv(\sigma)^{-1}\egen \mathbb M^\sv(\sigma) ;w)^T}\, .
\label{mplapp.20} 
\eeq
The only reference to $\rho$ on the right side of (\ref{mplapp.20}) is now through the twisted letters, but the latter are independent of $\rho$ by virtue of (\ref{mplapp.19}) and $\mathbb M^\sv(\sigma)^{-1}\egen_1 \mathbb M^\sv(\sigma) = \egen_1$. Hence, the expression (\ref{mplapp.11}) for $\Gcyc^\sv(\egen;w)$ is independent of $\rho$.

\section{Towards a first-principles definition of level-2 zeta generators}
\label{sec:fp}

The recursive procedure of sections \ref{sec: recursion} and \ref{sec: HW vectors} determines all contributions of zeta generators to equivariant iterated Eisenstein integrals for $\Gamma(2)$: both the expansion of
the geometric zeta generators $\sigma_w-z_w$ and the brackets of the arithmetic parts $[z_w,\ep_{k|\mu}]$ in terms of level-2 derivations. However, by analogy with level 1 \cite{Dorigoni:2024oft}, the information from iterated Eisenstein integrals and their cocycles does not provide a first-principles definition of zeta generators. 

Instead, genus-1 zeta generators for level 1 are in the first place defined as derivations on the free Lie algebra ${\rm Lie}[a,b]$ of section \ref{sec:flcee} through the $\tau \rightarrow i\infty$ degeneration of the CEE connection  \cite{EnriquezEllAss, Brown:depth3, Schneps:2015mzv, hain_matsumoto_2020, Dorigoni:2024iyt} as reviewed in section \ref{sec:furth.1}.
This appendix generalises the defining equations of zeta generators for level 1 to a proposal for level 2, where the $\sigma_w$ are viewed as derivations on  the Lie algebra of section \ref{sec:fltcg} with generators $a,b$ and $t_{2|A}$ with $A=1,2,3,4$.  However, we will see that the equations of our proposal do not yet fully determine the action of $\sigma_w$ on the generators, and we leave it for future work to reinstate the missing information. Still, we will see that, if the geometric parts $\sigma^{\rm g}_w$ are known as an explicit Lie series in derivations of $\mathfrak{u}$, then our proposal in section \ref{sec:furth.12} fixes the adjoint action of $z_w$ on $a,b$ and $t_{2|A}$ with $A=1,2,3,4$.

\subsection{A level-2 KZ connection from degenerating the CG connection}
\label{app:mpl.1}

Our proposal for level-2 zeta generators is driven by the $\tau \rightarrow i\infty$ degeneration of the $\dd z$ component (\ref{dzexpanded}) of the CG connection $\mathbb K_{\rm CG}^{(z)}$. We shall study its $\tau\rightarrow i\infty$ limit in the coordinate
\beq
w(z)=e^{2\pi i z}\,,
\eeq
adapted to the nodal sphere which maps the torsion points at real values $z=0$ and $z=\frac{1}{2}$ of the coordinate $z$ on the torus (which is periodically identified with $z{+}1$ and $z{+}\tau$) to
\beq\label{sigmaplane}
w(0)=1\,,\quad w(\tfrac{1}{2})={-}1 \, .
\eeq
The limit $\dd z\, \mathbb K^{(z)}_\text{CG}(z,\tau \rightarrow i \infty)$ is obtained from $\dd z = \frac{\dd w}{2\pi i w}$ and the following degeneration of the Kronecker-Eisenstein kernels $ g^{(k)}$, $ \tilde g^{(k)}$ (see section \ref{defgtg}) at constant $z$, where for $k=1$, 
\begin{align}
\lim_{\tau\rightarrow i\infty} g^{(1)}(z,\tau)&=2\pi i w\biggl(\frac{B_1}{w}+\frac{1}{w{-}1}\biggr)\,,\\
\lim_{\tau\rightarrow i\infty} g^{(1)}(z{+}\tfrac{1}{2},\tau)&=2\pi i w\biggl(\frac{B_1}{w}+\frac{1}{w{+}1}\biggr)\,,
\notag\\
    \lim_{\tau\rightarrow i\infty} \tilde{g}^{(1)}(z{+}\tfrac{1}{2},\tfrac{\tau}{2},\tau) &=
 \tilde{g}^{(1)}(z,\tfrac{\tau}{2},\tau)=0\,,
 \notag
 \end{align}
with Bernoulli number $B_1=-\frac{1}{2}$, and for $k\geq 2$ we have,
\begin{align}
\lim_{\tau\rightarrow i\infty} g^{(k)}(z,\tau)&=\lim_{\tau\rightarrow i\infty} g^{(k)}(z{+}\tfrac{1}{2},\tau)=\frac{(2\pi i)^{k}B_k}{k!}\,, \label{degg1s}\\
\lim_{\tau\rightarrow i\infty} \tilde{g}^{(k)}(z{+}\tfrac{1}{2},\tfrac{\tau}{2},\tau) &=\lim_{\tau\rightarrow i\infty} \tilde{g}^{(k)}(z,\tfrac{\tau}{2},\tau)=\frac{(2\pi i)^{k}B_k(\tfrac{1}{2})}{k!}\,.
\notag 
\end{align}
These degeneration formul\ae\ give rise to a KZ connection
\beq\label{dsiggen}
\lim_{\tau\rightarrow i\infty}\dd z \,\mathbb K^{(z)}_\text{CG}(z,\tau)={-}\dd w\biggl(\frac{\egen_0}{w}+\frac{\egen_1}{w{-}1}+\frac{\egen_{-1}}{w{+}1}\biggr)\,,
\eeq
whose residues at the singular points $0$, $1$, $-1$ are given by the following combinations of the generators in (\ref{dzexpanded}):
\begin{align}
&\egen_0={-}\agen+\frac{1}{2}(t_{2|1}+t_{2|2})-\sum_{k=2}^\infty \BF_k \biggl(t_{k+1|1}+t_{k+1|2}+r_k (t_{k+1|3}+t_{k+1|4})\biggr)\, ,\notag\\
&\egen_1={-}t_{2|1}\,,\quad \egen_{-1}={-}t_{2|2}\, .
\label{mplapp.03}
\end{align}
The path-ordered exponential (\ref{mplapp.04}) of the KZ connection (\ref{dsiggen}) was studied as a generating series of level-2 MPLs in the previous appendix.
Up to the normalisation conventions for the generators, (\ref{dsiggen}) and (\ref{mplapp.03}) are equivalent to the degenerations of the level-2 connections in section 3.3 of \cite{Broedel:2017jdo} and Theorem 10.3 of \cite{Hopper:2021ell}.

\subsection{Bracket relations of level-1 zeta generators}
\label{sec:furth.1}

The first-principles definition of genus-1 zeta generators $\sigma_w^{\SLZ}$ for level 1 starts from the degeneration of the $\dd z$ component of the CEE connection (\ref{CEEexpanded}),
\begin{align}
&\lim_{\tau\rightarrow i\infty}\dd z \,\mathbb K^{(z)}_\text{CEE}(z,\tau)={-}\dd w\biggl(\frac{E_{0}}{w}+\frac{E_1}{w{-}1}\biggr)\,,
\label{lv1r.01}
\\
&\ \ E_1 = [a,b] \, , \ \ \ \ E_0 = - \sum_{k=0}^{\infty} \BF_k {\rm ad}_b^k(a)\, .
\notag
\end{align}
More precisely, level-1 zeta generators $\sigma_w^{\SLZ}$ with $w\geq 3$ odd are defined as Ihara derivations on the free algebra ${\rm Lie}[a,b]$ in two steps:
\begin{itemize}
\item[(i)] Determine their action on the free subalgebra ${\rm Lie}[E_0,E_1]$ of ${\rm Lie}[a,b]$ from \cite{EnriquezEllAss, Brown:depth3, Schneps:2015mzv, hain_matsumoto_2020, Dorigoni:2024iyt},
\beq
[\sigma_w^{\SLZ},E_1] = 0 \, , \ \ \ \ \ \ [\sigma_w^{\SLZ},E_0] = \big[ E_0,P_w(E_1,E_0)\big] \,,
\label{lv1r.02}
\eeq
which involve the same degree-$w$ Lie polynomials $P_w(E_1,E_0)$ of (\ref{revgg.06}) from the Drinfeld associator as seen in the brackets (\ref{revgg.04}) of the genus-0 zeta generators $M_w$. 
\item[(ii)] Construct the unique Ihara derivation $\sigma_w^{\SLZ}$ on ${\rm Lie}[a,b]$ from the action (\ref{lv1r.02}) on ${\rm Lie}[E_0,E_1]$ by means of the extension lemma of \cite{Schneps:2015mzv},
as detailed in section 5 of \cite{Dorigoni:2024iyt}.
\end{itemize}
Given that $E_0$, $E_1$ of (\ref{lv1r.01}) both have $a$-degree 1 and therefore yield $P_w(E_0,E_1)$ of $a$-degree $w$, the brackets (\ref{lv1r.02}) expose the $a$-degree $w$ of $\sigma_w^{\SLZ}$.
The expansion of $\sigma_w^{\SLZ}$ as a Lie-series in $\ep_{k}^{(j)}$ of different $b$-degrees and a single arithmetic term $z_w$ of $(a,b)$-degrees $(w,w)$ reviewed in section \ref{sec:lvl1} can then be checked by acting on $a$ and $b$. In particular, the extension of (\ref{lv1r.02}) to ${\rm Lie}[a,b]$ determines $z_w$ as an explicit degree-$(w,w)$ derivation on $a$ and $b$.

Note that the roles of the singular points of the underlying KZ connections are reversed in comparison to section \ref{sec:tmpl}:
While the generator $x$ annihilated by $M_w$ is associated with the base-point $\lambda'=0$ of the MPLs in (\ref{revgg.01}), the generator $E_1$ annihilated by $\sigma_w^{\SLZ}$ is associated with the image $w=1$ of the base-point $z=0$ of elliptic MPLs. More specifically, iterated integrals on the torus are conventionally defined with the marked point $z=0$ on the torus as a base-point \cite{BrownLev, Broedel:2014vla, Broedel:2017kkb} which upon translation to the nodal sphere with coordinate $w= e^{2\pi i z}$ results in the base-point $w=1$ for MPLs, which we have also adopted in (\ref{mplapp.04}).

\subsection{Bracket relations of level-2 zeta generators}
\label{sec:furth.12}

In the genus-0 setting of a free Lie algebra with generators $\egen_0$, $\egen_1$, $\egen_{-1}$ for singular points $0$, $1$, $-1$ on the Riemann sphere, the action of the level-2 zeta generators $\sigma_w$ on these generators is determined by (\ref{lv2.28}). In passing to genus 1, a complete characterisation of $\sigma_w$ requires a specification of their action on the algebra generators $a$, $b$, $t_{2|A}$ in $\mathbb K_{\rm CG}^{(z)}$ with $A=1,2,3,4$.

The dictionary (\ref{mplapp.03}) relating $\egen_1 = -t_{2|1}$, $\egen_{-1} = -t_{2|2}$ and $\egen_0$ to a Lie series in $a$, $b$, $t_{2|A}$ is now proposed to produce bracket relations of the genus-1 incarnation of $\sigma_w$: The expansion of $\sigma_w$ for level 2 as a Lie series in $\ep_{k|\mu}^{(j)}$ with unbounded $b$-degree and a single arithmetic derivation $z_w$ of $(a,b)$-degrees $(w,w)$ is claimed to respect the following genus-1 transcriptions of (\ref{lv2.28}),
\begin{align}
[\sigma_w,t_{2|1}] &= 0 \, ,\label{trsg2} \\
[\sigma_w,t_{2|2}] &= \big[ t_{2|2}, R_w(\egen_0,-t_{2|1},-t_{2|2}) \big]\, ,
\notag \\
[\sigma_w,\egen_{0}] &= \big[ \egen_{0}, Q_w(\egen_0,-t_{2|1},-t_{2|2}) \big]\, .
\notag
\end{align}
See (\ref{lv2.27}) for the definition of degree-$w$ Lie polynomials $Q_w$ and $R_w$ from level-2 associators.

By the $a$-degree 1 of $t_{2|A}$ and $\egen_0$ in  (\ref{mplapp.03}), both of $R_w$ and $Q_w$ have $a$-degree $w$ which ties in with the $a$-degree $w$ of $\sigma_w$ conjectured in (iii) below (\ref{eq:Scoc2}).
The expression (\ref{mplapp.03}) for $\egen_{0}$ explicitly determines the right-hand sides of (\ref{trsg2}) as Lie series in the generators $a$, $b$, $t_{2|A}$ of fixed $a$-degree $w{+}1$ and unbounded $b$-degree. On the left-hand sides of (\ref{trsg2}), the contribution from the geometric parts $\sigma^{\rm g}_w = \sigma_w - z_w$ can be determined from their expansion in terms of $\ep_{k|\mu}^{(j)} \in \mathfrak{u}$ (see sections \ref{sec: recursion} and \ref{sec: HW vectors}) and the action of $\mathfrak{u}$ on $a$, $b$, $t_{2|A}$ in section \ref{sec:actderiv}. The contributions from $\sigma_w \rightarrow z_w$ to the left-hand side are unknown at this stage, so one can in fact reverse the logic and solve (\ref{trsg2}) for $[z_w,t_{2|1}]$, $[z_w,t_{2|2}]$ and $[z_w,\egen_0]$ as a Lie series in $a,b,t_{2|A}$. 
It is highly non-trivial that the resulting expressions for these brackets of $z_w$ are compatible with their expected $(a,b)$-degrees $(w,w)$ and their $\mathfrak{sl}_2 \times {\cal S}_3$ invariance. We have tested up to and including degree 12 that this is the case for $z_1$ and $z_3$, given their geometric parts $\sigma^{\rm g}_1$ and $\sigma^{\rm g}_3$ in (\ref{exactz1}), (\ref{zet.61}) and (\ref{sysxi3.03}), which provides non-trivial support of (\ref{trsg2}).

Assuming the validity of (\ref{trsg2}) and again considering the expansion of $\sigma^{\rm g}_w$ as given, one can determine the explicit form of $[z_w,a]$ by isolating the degree-$(2w{+}1)$ terms of $[z_w,\egen_0]$ and then infer $[z_w,b]= [\ep^\vee_0,[z_w,a]]$. Similarly, $[z_w,t_{2|4}]$ and $[z_w,t_{2|3}]$ follow from the images of $[z_w,t_{2|2}]$ under $\theta_{S}^{-1}\bullet \theta_S$ and $\theta_T^{-1}\theta_{S}^{-1}\bullet \theta_S \theta_T$, respectively. This is how (\ref{trsg2}) together with the availability of $\sigma^{\rm g}_w$ fixes the action of $z_w$ on the complete algebra of $\mathbb K_{\rm CG}^{(z)}$ after using their $\mathfrak{sl}_2 \times {\cal S}_3$ invariance.

Without earlier knowledge of $\sigma_w^{\rm g}$, however, the three equations (\ref{trsg2}) by themselves cannot suffice to fix the action of $\sigma_w$ on all of the algebra generators $a$, $b$, $t_{2|A}$ with $A=1,2,3,4$. In view of the relation (\ref{brackab2}), one needs at least two additional conditions besides (\ref{trsg2}) -- for instance the action $[\sigma_w,t_{2|3}]$ and $[\sigma_w,t_{2|4}]$ in terms of known Lie series in $a$, $b$, $t_{2|A}$ -- which we shall investigate in future work. 

In the case of level 1, the link between the genus-0 and genus-1 incarnation of zeta generators can be understood from their role as generators of the motivic Galois group. At level $\geq 2$, the work of \cite{Hopper:2022ell, Hirose:2023} on the genus-1 incarnation of the motivic Galois group is expected to shed light on the mathematical foundations of the expansions and actions of $\sigma_w$ obtained in a rather empirical way in this work.

The vanishing of $[\sigma_w , t_{2|1}]$ in (\ref{trsg2}) and the genus-0 setup of appendix \ref{ass.lv2} make reference to the MPL integration base-point $w=1$ on the nodal sphere inherited from the conventional base point $z=0$ for iterated integrals on the torus. The relation $w=e^{2\pi i z}$ between the respective coordinates affects the tangent vectors in the regularisations at genus 0 and 1. The tangent vectors $-\vec{1}_1$, $\vec{1}_0$, $\vec{1}_{-1}$ on the sphere in appendix \ref{ass.lv2} descend from the regularisation prescriptions for iterated integrals on the torus below (2.26) of \cite{Broedel:2014vla} and in (3.5) of \cite{Broedel:2018iwv}, also see the discussion below (\ref{aemzv.02}).

\section{Iterated Eisenstein integrals for level 2 and twisted elliptic MZVs}
\label{app:eMZV}

The level-2 Pollack relations of section \ref{sec: pollack} project the iterated Eisenstein integrals (\ref{eq: eez}) in the degree $\geq 8$ terms of the expansion (\ref{eq: I2}) of $\mathbb I(\ep;\tau)$ to specific linear combinations. We shall here demonstrate that these combinations are expressible as twisted ($\mathfrak{A}$-cycle) elliptic MZVs  \cite{Broedel:2017jdo}, i.e., expansion coefficients of ellipsitomic associators for level 2 \cite{Calaque:2020ell}.

\subsection{Generating series of twisted elliptic MZVs}
\label{app:eMZV.1}

The ellipsitomic associator of our main interest is the regularised $\mathfrak{A}$-cycle monodromy \cite{Calaque:2020ell}
\beq
\Phi_{\mathfrak{A}}(\tau) = {\rm Pexp}\bigg( \int^{ \tvec_1}_{\tvec_0} \mathbb K_\text{CG}^{(z)}(z,\tau) \, \dd z\bigg)\,,
\label{aemzv.01}
\eeq
of the $\dd z$ component (\ref{dzexpanded}) of the CG connection, with tangent vectors $ \tvec_1$ and $ \tvec_0$ at the endpoints $z=1$ and $z=0$ of the $\mathfrak{A}$-cycle, respectively. We shall consider $\Phi_{\mathfrak{A}}(\tau)$ as a generating series of twisted elliptic MZVs for level 2, which are more generally defined by regularised iterated integrals over the $\mathfrak{A}$-cycle \cite{Broedel:2017jdo},\footnote{We have replaced the doubly-periodic integration kernels $f^{(n)}(z{-}v{+}u\tau,\tau)$ of \cite{Broedel:2017jdo} by meromorphic $\tilde{g}^{(n)}(z{-}v,u,\tau)$ which agree with $f^{(n)}(z{-}v{+}u\tau,\tau)$ for $z \in \mathbb R$, but additionally render the integrals (\ref{aemzv.02}) homotopy invariant. We have furthermore changed the parental letter of the twisted elliptic MZVs $\omega(\ldots)$ into $\varpi_{\mathfrak{A}}(\ldots)$ to avoid a clash of notation with the torsion points $\omega_A$ entering in (\ref{aemzv.03}) and below.}
\begin{align}
\varpi_{\mathfrak{A}}\Big( \smallmatrix n_1 &n_2 &\ldots &n_\ell \\ u_1 &u_2 &\ldots &u_\ell \\ v_1 &v_2 &\ldots &v_\ell \endsmallmatrix ;\tau \Big) &= \int^{ \tvec_1}_{\tvec_0} \dd z_\ell \, \tilde{g}^{(n_\ell)}(z_\ell{-}v_\ell,u_\ell ,\tau)  \int^{z_\ell}_{\tvec_0} \dd z_{\ell-1} \, \tilde{g}^{(n_{\ell-1})}(z_{\ell-1}{-}v_{\ell-1},u_{\ell-1},\tau)  \notag\\
&\quad \ldots \int^{z_3}_{\tvec_0} \dd z_2 \, \tilde{g}^{(n_2)}(z_2{-}v_2,u_2 ,\tau) \int^{z_2}_{\tvec_0} \dd z_1 \, \tilde{g}^{(n_1)}(z_1{-}v_1,u_1 ,\tau) \,,
\label{aemzv.02}
\end{align}
for $n_i \in \mathbb N_0 = \{ 0,1,2,\ldots\}$. The entries $u_i, v_i \in \mathbb Q$ are the co-moving coordinates of torsion points $u_i \tau {+} v_i$, and the meromorphic polynomials $\tilde{g}^{(n_i)}$ in Kronecker-Eisenstein kernels with $u_i$-dependent coefficients are defined by
(\ref{g kernel tilde}) and (\ref{moregtild}).

The tangent vectors $ \tvec_1$ and $ \tvec_0$ in (\ref{aemzv.01}) and (\ref{aemzv.02}) shuffle-regularise twisted elliptic MZVs in a way which is compatible with the invariance of the Kronecker-Eisenstein kernels under $z_i \rightarrow z_i{+}1$. A tempting choice is to set $p=1$, i.e.,  $\tvec_1 = \vec{1}_1$ and $ \tvec_0 = \vec{1}_0$, which regularises the simple pole of $g^{(1)}(z_1,\tau)= \frac{1}{z_1}+ {\cal O}(z_1)$ according to $\int_{\vec{1}_0}^{z} \frac{\dd z_1}{z_1} = \log(z)$. However, it is advantageous to set $p={-}\frac{1}{2\pi i}$ as the coordinate change $w=e^{2\pi iz}$ induced in the $\tau\rightarrow i\infty$ degeneration (see, e.g., section 2.2.1 of \cite{Broedel:2014vla} and (3.5) of \cite{Broedel:2018iwv}) of (\ref{aemzv.02}) sends the tangent vectors $\vec{p}_0$ and $\vec{p}_1$ to ${-}\vec{1}_1$ and $\vec{1}_1$, respectively.
In particular, this prevents the appearance of $\log(-2\pi i)$ in the aforementioned limit.
With $p=-\frac{1}{2\pi i}$, we recover the tangent vector $-\vec{1}_1$ of appendix \ref{ass.lv2} 
in passing to the coordinate $w=e^{2\pi i z}$ of the nodal sphere, see, e.g., section 2.2.1 of \cite{Broedel:2014vla} and (3.5) of \cite{Broedel:2018iwv}. We shall nevertheless leave $p$ generic throughout this appendix. 
In case of entries with $u_i=0$ and $n_i=1$, the singularities of (\ref{aemzv.02}) on the real line are bypassed by infinitesimally displacing the integration path into the positive imaginary direction, following section 2.1 of~\cite{Broedel:2017jdo}. 

Given the notation $\omega_1 = 0$ and the half-periods $\omega_\mu$ with $\mu=2,3,4$ in (\ref{eq: torsion}), we shall also denote the level-2 instances of twisted elliptic MZVs (\ref{aemzv.02})  more compactly~by
\beq
\varpi_{\mathfrak{A}}\Big( \smallmatrix n_1 &n_2 &\ldots &n_\ell \\   \big. \omega_{A_1} &\omega_{A_2} &\ldots &\omega_{A_\ell}  \big.\endsmallmatrix ;\tau \Big) = 
\varpi_{\mathfrak{A}}\Big( \smallmatrix n_1 &n_2 &\ldots &n_\ell \\ u_1 &u_2 &\ldots &u_\ell \\ v_1 &v_2 &\ldots &v_\ell \endsmallmatrix ;\tau \Big)\, , \ \ \ \ {\rm for} \ A_i\in \{ 1,2,3,4\} \ {\rm and} \ \omega_{A_i} = u_i \tau{+} v_i \, .
\label{aemzv.03}
\eeq
In this notation, the expansion of the path-ordered exponential (\ref{aemzv.01}) produces the series
\begin{align}
\Phi_{\mathfrak{A}}(\tau) &= 1 + \sum_{\ell=1}^\infty
\sum_{k_1,\ldots,k_\ell=0}^{\infty} \! \! \! (2\pi i)^{\ell-k_1-\ldots-k_\ell} \! \! \!  \sum_{A_1,\ldots,A_\ell=1}^4 
\! \! t_{k_\ell+1|A_\ell} \ldots t_{k_1+1|A_1}
\varpi_{\mathfrak{A}}\Big( \smallmatrix k_1   &\ldots &k_\ell \\   \big. \omega_{A_1}  &\ldots &\omega_{A_\ell}  \big.\endsmallmatrix ;\tau \Big)
\label{aemzv.04} \\
&= 1 + \sum_{k_1=0}^{\infty} (2\pi i)^{1-k_1}\sum_{A_1=1}^4 
t_{k_1+1|A_1}
\varpi_{\mathfrak{A}}\Big( \smallmatrix k_1 \\   \big. \omega_{A_1}   \big.\endsmallmatrix ;\tau \Big) \notag \\
&\quad
+ \sum_{k_1,k_2=0}^{\infty}  (2\pi i)^{2-k_1-k_2} \sum_{A_1,A_2=1}^4 
\! \! t_{k_2+1|A_2}   t_{k_1+1|A_1}
\varpi_{\mathfrak{A}}\Big( \smallmatrix k_1    &k_2 \\   \big. \omega_{A_1}   &\omega_{A_2}  \big.\endsmallmatrix ;\tau \Big) + \ldots
\, ,
\notag
\end{align}
in twisted elliptic MZVs and the generators $t_{k_i +1 | A_i}$ of $(a,b)$-degrees $(1,k_i)$ in (\ref{deftkmu}), with terms involving $\geq 3$ generators $t_{k_i +1 | A_i}$ in the ellipsis. We formally define $\sum_{A=1}^4 t_{1|A}= a$ for the summands with $k_i=0$ as suggested by the algebra relation (\ref{brackab2}), which for instance rewrites the terms with a single $t_{k_i +1 | A_i}$ in the second line of (\ref{aemzv.04}) as $ 2\pi i a \varpi_{\mathfrak{A}}\big( \smallmatrix 0 \\  0  \endsmallmatrix ;\tau \big) +\sum_{k_1=1}^{\infty} (2\pi i)^{1-k_1}\sum_{A_1=1}^4 
t_{k_1+1|A_1}
\varpi_{\mathfrak{A}}\big( \smallmatrix k_1 \\    \omega_{A_1}   \endsmallmatrix ;\tau \big)$. The generators $a$ and $t_{k_i+1|A_i}$ in the resulting expansion of (\ref{aemzv.04}) form a free Lie algebra such that all possible entries of twisted elliptic MZVs in (\ref{aemzv.02}) and (\ref{aemzv.03}) for level 2 are independently realised within the generating series~$\Phi_{\mathfrak{A}}(\tau)$. 

\subsection{\texorpdfstring{$\tau$-derivatives of twisted elliptic MZVs}{tau-derivatives of twisted elliptic MZVs}}
\label{app:eMZV.2}

We shall now derive a separate representation of the ellipsitomic associator $\Phi_{\mathfrak{A}}(\tau)$ where all the $\tau$-dependence is carried by iterated Eisenstein integrals (\ref{eq: eez}) and polynomials in $\tau$ and $\int^\tau_{i\infty} \frac{\dd \tau_1}{2\pi i} {\rm G}_2(\tau_1)$. The key idea is to use the flatness conditions of the CG connection, namely the separate vanishing of both sides of (\ref{flatnessCG}). Moreover, we shall use the differential equation
\beq
\partial_\tau \Phi_{\mathfrak{A}}(\tau) = \int^{ \tvec_1}_{\tvec_0} \dd s \, \Psi( \tvec_1,s;\tau) \, \partial_\tau \mathbb K^{(z)}_\text{CG}(s,\tau)\, \Psi(s,\tvec_0;\tau) \,,
\label{aemzv.05}
\eeq
of the path-ordered exponential in (\ref{aemzv.01}), where the quantities $\Psi(z_1,z_2;\tau)$ on the right-hand side generalise $\Phi_{\mathfrak{A}}(\tau)$ to more general endpoints of the integration path,
\beq
\Psi(z_1,z_2;\tau) = {\rm Pexp}\bigg( \int^{z_1}_{z_2} \mathbb K_\text{CG}^{(z)}(z,\tau) \, \dd z\bigg)\, .
\label{aemzv.06}
\eeq
In this way, $\Phi_{\mathfrak{A}}(\tau)$ is recovered from the regularised limit $(z_1,z_2) \rightarrow (\tvec_1,\tvec_0)$ of $\Psi(z_1,z_2;\tau)$.

We shall simplify the $\tau$-derivative (\ref{aemzv.05}) using the flatness condition $\partial_\tau \mathbb K^{(z)}_\text{CG}(s,\tau) = \partial_s \mathbb K^{(\tau)}_\text{CG}(s,\tau)$ followed by integration by parts in $s$,
\begin{align}
\partial_\tau \Phi_{\mathfrak{A}}(\tau) &= \int^{ \tvec_1}_{\tvec_0} \dd s \, \Psi( \tvec_1,s;\tau) \, \big[ \partial_s \mathbb K^{(\tau)}_\text{CG}(s,\tau) \big] \, \Psi(s,\tvec_0;\tau) 
\label{aemzv.07} \\
&= \int^{ \tvec_1}_{\tvec_0} \dd s \, \bigg\{ \partial_s \, \Big[ \Psi( \tvec_1,s;\tau) \,  \mathbb K^{(\tau)}_\text{CG}(s,\tau) \, \Psi(s,\tvec_0;\tau) \Big] \notag \\
&\quad - \big[ \partial_s \Psi(\tvec_1,s;\tau) \big] \,  \mathbb K^{(\tau)}_\text{CG}(s,\tau) \, \Psi(s,\tvec_0;\tau)
- \Psi(\tvec_1,s;\tau) \,  \mathbb K^{(\tau)}_\text{CG}(s,\tau) \, \partial_s \Psi(s,\tvec_0;\tau) \Bigg\} \notag \\
&= \Psi( \tvec_1,s;\tau) \,  \mathbb K^{(\tau)}_\text{CG}(s,\tau) \, \Psi(s,\tvec_0;\tau) \, \Big|^{s= \tvec_1}_{s=\tvec_0} \notag \\
&=\big[ \mathbb K^{(\tau)}_\text{CG}(0,\tau) , \Phi_{\mathfrak{A}}(\tau) \big] \, .\notag
\end{align}
We have used the differential equation
\beq
\dd \Psi(z_1,z_2;\tau) = \dd z_1\,  \mathbb K^{(z)}_\text{CG}(z_1,\tau) \Psi(z_1,z_2;\tau) -   \dd z_2  \, \Psi(z_1,z_2;\tau) \mathbb K^{(z)}_\text{CG}(z_2,\tau)\,,
\label{aemzv.00}
\eeq
of (\ref{aemzv.06}) to simplify the integrand in the third line of (\ref{aemzv.07}) to
\beq
\Psi(\tvec_1,s;\tau) \,  \big[ \mathbb K^{(z)}_\text{CG}(s,\tau) , \mathbb K^{(\tau)}_\text{CG}(s,\tau) \big]\,  \Psi(s,\tvec_0;\tau)
= 0\, ,
\label{aemzv.08}
\eeq
where the vanishing follows from the flatness condition $[ \mathbb K^{(z)}_\text{CG}(s,\tau) , \mathbb K^{(\tau)}_\text{CG}(s,\tau) ]=0$ of the CG connection. The last step of (\ref{aemzv.07}) uses the periodicity $ \mathbb K^{(\tau)}_\text{CG}(1,\tau)= \mathbb K^{(\tau)}_\text{CG}(0,\tau)$. With the representation (\ref{k0lv2}) of $\mathbb K^{(\tau)}_\text{CG}(0,\tau)$ in terms of $\tilde \ep_{2|\nu}$ and $\ep_{k|\mu}$ with $k\geq 4$, the conclusion of (\ref{aemzv.07}) matches the differential equation of ellipsitomic $\mathfrak{A}$-cycle associators in section 7 of \cite{Calaque:2020ell}. 

\subsection{Iterated Eisenstein integrals as twisted elliptic MZVs}
\label{app:eMZV.3}

We  solve the differential equation (\ref{aemzv.07}) in terms of an initial value at $\tau \rightarrow i \infty$ via
\begin{align}
\Phi_{\mathfrak{A}}(\tau) &= {\rm Pexp} \bigg( \int^\tau_{i\infty} \dd \tau' \, \mathbb K^{(\tau)}_\text{CG}(0,\tau') \bigg) \,  \Phi_{\mathfrak{A}}(i \infty) \, {\rm Pexp} \bigg( \int^{i\infty}_\tau \dd \tau' \, \mathbb K^{(\tau)}_\text{CG}(0,\tau') \bigg) \label{aemzv.09} \\
&= {\cal V}(\tau)^{-1}
\mathbb{I}(\ep;\tau)^{-1}
\Phi_{\mathfrak{A}}(i \infty)
\mathbb{I}(\ep;\tau)
{\cal V}(\tau) \, ,
\notag
\end{align}
where the gauge transformation ${\cal V}(\tau)$ is defined by (\ref{gaulv1.3}). In passing to the second line, we have used the gauge equivalence  (\ref{gaulv1.9}) of $\mathbb K^{(\tau)}_\text{CG}(0,\tau)$ to the level-2 connection $\mathbb A(\ep;\tau)$ of (\ref{eq: N2conn}). For the choice $\tvec_a = \overrightarrow{2\pi i}_a$ of the tangent vectors in (\ref{aemzv.01}) with $a=0,1$, the initial value is known from section 3.3 of \cite{Broedel:2017jdo},
\beq
\Phi_{\mathfrak{A}}(i \infty) = e^{i\pi t_{2|1}} Z_0(\egen_0,\egen_1,\egen_{-1}) e^{2\pi i \egen_0} Z_0(\egen_0,\egen_1,\egen_{-1})^{-1}\, ,
\label{aemzv.10}
\eeq
and expressed in terms of the genus-0 associator $Z_0$ for level 2 given by (\ref{mplapp.05}). The letters $\egen_a$ of $Z_0$ are $\egen_1 = -t_{2|1}$, $\egen_{-1} = -t_{2|2}$ and a Lie series $\egen_0$ in $a$, $b$, $t_{2|A}$ given by (\ref{mplapp.03}). The lowest-degree term $\egen_0 = -a+\ldots$ translates into $\Phi_{\mathfrak{A}}(i \infty) = -2\pi i a +\ldots$ with terms of degree $\geq 2$ of both ellipses and is insensitive to the choice of tangent vectors $\tvec_0$, $\tvec_1$ in (\ref{aemzv.01}). As a consequence, the second line of (\ref{aemzv.09}) features the iterated Eisenstein integral $\eez{j_1 &\ldots &j_r}{k_1 &\ldots &k_r}{\omega_{\mu_1} &\ldots &\omega_{\mu_r}}{\tau}$ from the expansion (\ref{eq: I2}) of $\mathbb{I}(\ep;\tau)$ along with a rational multiple of $2\pi i [\ep_{k_r | \mu_r}^{(j_r)},[ \ldots ,[\ep_{k_1 | \mu_1}^{(j_1)}, a] \ldots ]]$ plus terms of higher degree. For each word in $\ep_{k_1 | \mu_1}^{(j_1)} \ldots \ep_{k_r | \mu_r}^{(j_r)}$ that is linearly independent under level-2 Pollack relations, the action on $a$ in the leading-degree term along with $\eez{j_1 &\ldots &j_r}{k_1 &\ldots &k_r}{\omega_{\mu_1} &\ldots &\omega_{\mu_r}}{\tau}$ is linearly independent as well.\footnote{This can be seen by the following analogy with level 1: Since the Tsunogai derivations of section \ref{sec:221ts} annihilate the bracket $[b,a]$, they are entirely determined by their action on $a$ \cite{Schneps:2012}. The level-2 derivations $\ep_{k|\mu}^{(j)}$ entering $\mathbb{A}(\ep;\tau)$ in turn were found to all annihilate the generator $t_{2|1}$ in section \ref{sec:actderiv}. If we quotient out $t_{2| \mu} \rightarrow 0$ for $\mu=2,3,4$, then (\ref{brackab2}) identifies $t_{2| 1} \rightarrow [b,a]$ and reduces the level-2 Lie algebra to the free ${\rm Lie}[a,b]$. Hence, the quotient $t_{2| \mu} \rightarrow 0$ for $\mu=2,3,4$ makes all of $\ep_{k|\mu}^{(j)}$ annihilate the bracket $[b,a]$ such that they are determined by their action on $a$. Linear independence of words in derivations thus translates into linear independence in their action $[\ep_{k_r | \mu_r}^{(j_r)},[ \ldots ,[\ep_{k_1 | \mu_1}^{(j_1)}, a] \ldots ]]$ on $a$ within the quotient $t_{2| \mu} \rightarrow 0$ for $\mu=2,3,4$. Hence, when leaving the quotient and  passing back to non-trivial $t_{2| \mu}$ with $\mu=2,3,4$, linearly independent words in derivations must keep their linearly independent action on $a$.} As a consequence, already the lowest-degree term $\Phi_{\mathfrak{A}}(i \infty) = -2\pi i a +\ldots$ of the initial condition ensures that the associator (\ref{aemzv.09}) retains all linear combinations of iterated Eisenstein integrals present in the expansion of $\mathbb{I}(\ep;\tau)$. In other words, the linear combinations of iterated Eisenstein integrals which enter $\mathbb{I}(\ep;\tau)$ on the support of Pollack relations all appear as expansion coefficients of $\Phi_{\mathfrak{A}}(i \infty)$ and by (\ref{aemzv.04}) are realised as twisted ($\mathfrak{A}$-cycle) elliptic MZVs.

\section{Further examples}
\label{app: examples}

We shall here supplement the discussion of section \ref{sec: HW vectors} through additional examples of higher-order terms in the expansions of zeta generators, bracket relations of their arithmetic terms and the $S$-cocycle. Appendix \ref{appd.3} adds examples of equivariant iterated Eisenstein integrals to those of section \ref{ex:sec44}.

\subsection{\texorpdfstring{Degree-8 contributions with $\mathbb Q[\pi^{2}]$ coefficients to the $S$-cocycle}{Degree-8 contributions with Q[pi2k] coefficients to the S-cocycle}}
\label{appd.pisq}

The factorised form (\ref{ag:Scoc2}) of the $S$-cocycle features a series $\mathbb X(\ep)$ with $\mathbb Q[\pi^{2}]$ coefficients whose terms of modular depth 2 and degree $\leq 6$ can be found in (\ref{eq: rhoX}). We shall here supplement the degree-8 terms $\mathbb{X}^{[2]}_8(\ep)$ at modular depth 2, namely
\begin{align}
    \mathbb X(\ep)  &= 1 + \mathbb X^{[1]}(\ep) + \tfrac{1}{2}\big(\mathbb X^{[1]}(\ep)\big)^2 -\tfrac{1}{16} \zeta_2\,[\ep_{2|3},\ep_{2|4}]    \label{eq:morerho}  \\
    &\quad + \zeta_2\Big({-}\tfrac{1}{192}[\ep_{2|3},\ep_{4|2}] + \tfrac{13}{1280}[\ep_{2|3},\ep_{4|3}] + \tfrac{1}{1920}[\ep_{2|3},\ep_{4|4}] \nn\\
    &\qquad \quad + \tfrac{1}{960}[\ep_{2|4},\ep_{4|2}] + \tfrac{7}{7680}[\ep_{2|4},\ep_{4|3}] + \tfrac{31}{2560}[\ep_{2|4},\ep_{4|4}]\Big) \nn\\
    &\quad + \zeta_4\Big({-}\tfrac{89}{256}[\ep_{2|3},\ep_{4|2}^{(2)}] + \tfrac{71}{256}[\ep_{2|3},\ep_{4|3}^{(2)}] - \tfrac{3}{16}[\ep_{2|3},\ep_{4|4}^{(2)}] \nn\\
    &\qquad \quad - \tfrac{93}{256}[\ep_{2|4},\ep_{4|2}^{(2)}] - \tfrac{7}{256}[\ep_{2|4},\ep_{4|3}^{(2)}] - \tfrac{1}{32}[\ep_{2|4},\ep_{4|4}^{(2)}]\Big) + \mathbb{X}^{[2]}_8(\ep) + \ldots
    \nn\,,
    \end{align}
with
\begin{align}
\mathbb{X}^{[2]}_8(\ep) &= \zeta_2\Big(\tfrac{4247}{20321280}[\ep_{4|2},\ep_{4|3}] -\tfrac{169}{4064256}[\ep_{4|2},\ep_{4|4}] +\tfrac{3391}{40642560}[\ep_{4|3},\ep_{4|4}] \label{xx28}\\
    &\qquad \quad +\tfrac{7241}{20321280}[\ep_{2|3},\ep_{6|2}] -\tfrac{12167}{40642560}[\ep_{2|3},\ep_{6|3}] -\tfrac{389}{4064256}[\ep_{2|3},\ep_{6|4}] \nn\\
    &\qquad \quad +\tfrac{277}{6773760}[\ep_{2|4},\ep_{6|2}] -\tfrac{6649}{13547520}[\ep_{2|4},\ep_{6|4}]\Big) \nn\\
    &\quad + \zeta_4\Big(\tfrac{7}{7200}\big([\ep_{4|2},\ep_{4|2}^{(2)}]+[\ep_{4|4},\ep_{4|4}^{(2)}]\big) -\tfrac{4217}{460800}[\ep_{4|3},\ep_{4|3}^{(2)}] \nn\\
    &\qquad \quad +\tfrac{62057}{6451200}\big([\ep_{4|2},\ep_{4|3}^{(2)}]+[\ep_{4|3},\ep_{4|4}^{(2)}]\big) +\tfrac{6977}{6451200}[\ep_{4|2},\ep_{4|4}^{(2)}] \nn\\
    &\qquad \quad +\tfrac{7723}{6451200}\big([\ep_{4|2}^{(2)},\ep_{4|3}] + [\ep_{4|3}^{(2)},\ep_{4|4}]\big) +\tfrac{16483}{6451200}[\ep_{4|2}^{(2)},\ep_{4|4}] \nn\\
    &\qquad \quad -\tfrac{127}{143360}\big([\ep_{4|2}^{(1)},\ep_{4|3}^{(1)}]+[\ep_{4|3}^{(1)},\ep_{4|4}^{(1)}]\big) -\tfrac{559}{143360}[\ep_{4|2}^{(1)},\ep_{4|4}^{(1)}] \nn\\
    &\qquad \quad +\tfrac{3}{2048}[\ep_{2|3},\ep_{6|2}^{(2)}] -\tfrac{3}{7168}[\ep_{2|3},\ep_{6|4}^{(2)}] +\tfrac{15}{14336}\big([\ep_{2|4},\ep_{6|2}^{(2)}] +[\ep_{2|4},\ep_{6|4}^{(2)}]\big)\Big) \nn\\
    &\quad + \zeta_6\Big(\tfrac{3391}{64512}[\ep_{4|2}^{(2)},\ep_{4|3}^{(2)}] -\tfrac{845}{32256}[\ep_{4|2}^{(2)},\ep_{4|4}^{(2)}] +\tfrac{4247}{32256}[\ep_{4|3}^{(2)},\ep_{4|4}^{(2)}] \nn\\
    &\qquad \quad -\tfrac{16057}{387072}[\ep_{2|3},\ep_{6|2}^{(4)}] +\tfrac{12167}{387072}[\ep_{2|3},\ep_{6|3}^{(4)}] -\tfrac{3205}{96768}[\ep_{2|3},\ep_{6|4}^{(4)}] \nn\\
    &\qquad \quad -\tfrac{6649}{129024}[\ep_{2|4},\ep_{6|2}^{(4)}] +\tfrac{277}{64512}[\ep_{2|4},\ep_{6|4}^{(4)}]\Big)  \nn\,,
\end{align}
and terms of degree $\geq 10$ and/or modular depth $\geq 3$ in the ellipsis of (\ref{eq:morerho}).
The square $\frac{1}{2}(\mathbb X^{[1]}(\ep))^2$ of the modular depth 1 terms in (\ref{os:Xdepth1}) exhausts the contributions to (\ref{eq:morerho}) of modular depth 2 that are not Lie-algebra-valued, as required by $\mathbb X(\ep)$ being group-like. At degree 8 it is $\frac{5}{9216}\zeta_4\big(2\ep^{(1)}_{4|2}-\ep^{(1)}_{4|3}+2\ep^{(1)}_{4|4}\big)^2$ and includes the products $\ep^{(1)}_{4|\mu}\ep^{(1)}_{4|\nu}$ with $\mu\neq\nu$. The brackets of degree 8 are given modulo the $\mathfrak{sl}_2$ multiplet of the Pollack relation (\ref{eq: pollack8}), which we have used to eliminate the five brackets $[\ep_{2|4},\ep^{(p)}_{6|3}]$ with $0\leq p \leq 4$. 

The logarithm of $\mathbb X(\ep)$ can be found in the ancillary file at modular depth 1 through degree 16, following from (\ref{os:Xdepth1}), and at all modular depths through degree 10, with the terms of modular depth 2 and degree $\leq 8$ as given in (\ref{eq:morerho}). From degree 8 on, these terms are given modulo the Pollack relations.

\subsection{\texorpdfstring{Highest-weight vectors of $\sigma_9^{[2]}$}{Highest-weight vectors of sigma9[2]}}
\label{appd.1}

The highest-weight vectors in $\sigma_5^{[2]}$ and $\sigma_7^{[2]}$ given by (\ref{eq: HWd2}) generalise as follows to $\sigma_9^{[2]}$ 
\begin{align}
\sigma_9^{[2]}\Big\rvert_{\rm HW}&=\tfrac{1753}{13056} s_{4|2,\, 8|2}^{3} + \tfrac{5}{3072} (s_{4|2,\, 8|3}^{3} {+} s_{4|2,\, 8|4}^{3}) + \tfrac{35}{1024} (s_{4|3,\, 8|2}^{3} {+} s_{4|4,\, 8|2}^{3}) \label{eq:HWsig9}\\
&\quad - \tfrac{7}{68} (s_{4|3,\, 8|3}^{3} {+} s_{4|4,\, 8|4}^{3}) + \tfrac{105}{1024} (s_{4|3,\, 8|4}^{3} {+} s_{4|4,\, 8|3}^{3}) +\tfrac{1295}{17408} s_{6|2,\, 6|2}^{3} \nn\\
&\quad + \tfrac{35}{6144} (s_{6|2,\, 6|3}^{3} {+} s_{6|2,\, 6|4}^{3}) - \tfrac{35}{816} (s_{6|3,\, 6|3}^{3} {+} s_{6|4,\, 6|4}^{3}) + \tfrac{175}{2048} s_{6|4,\, 6|3}^{3} \nn\\
&\quad -\tfrac{77}{36864} s_{6|2,\, 8|2}^{5} - \tfrac{119}{589824} (s_{6|2,\, 8|3}^{5} {+} s_{6|2,\, 8|4}^{5}) - \tfrac{595}{589824} (s_{6|3,\, 8|2}^{5} {+} s_{6|4,\, 8|2}^{5}) \nn\\
&\quad + \tfrac{7}{2304} (s_{6|3,\, 8|3}^{5} {+} s_{6|4,\, 8|4}^{5}) - \tfrac{595}{196608} (s_{6|3,\, 8|4}^{5} {+} s_{6|4,\, 8|3}^{5}) +\tfrac{43}{2580480} s_{8|2,\, 8|2}^{7} \nn\\
&\quad + \tfrac{527}{16515072} (s_{8|3,\, 8|2}^{7} {+} s_{8|4,\, 8|2}^{7}) - \tfrac{31}{645120} (s_{8|3,\, 8|3}^{7} {+} s_{8|4,\, 8|4}^{7}) + \tfrac{527}{5505024} s_{8|4,\, 8|3}^{7} \, , \nn
\end{align}
where pairs of terms related by $\theta_T$-invariance are grouped into parentheses as in (\ref{eq: HWd2}). The terms with $s^3$, $s^5$ and $s^7$ are of degree 12, 14 and 16, respectively. We have used $s^d_{p|\mu,q|\nu}=s^d_{q|\nu,p|\mu}$ for odd $d$ to collect both orderings of $p|\mu$ and $q|\nu$, so that $s^3_{6|4,\,6|3}$ and $s^7_{8|4,\,8|3}$ are $\theta_T$-invariant by themselves. These coefficients agree with the closed formula (\ref{hwlevel2}), and at degree 12 the series (\ref{eq: eqvprop}) built from them has been checked numerically to be $S$-equivariant modulo Pollack relations.

\subsection{\texorpdfstring{Degree 10 terms of $\sigma^{[3]}_5$ and $\sigma^{[3]}_7$}{Degree 10 terms of sigma5[3] and sigma7[3]}}
\label{appd.2}

We use the combinations
\beq
\ep_{k|\pm}=\ep_{k|3}\pm\ep_{k|4}\,,
\label{eq:d3projD2}
\eeq
which are even and odd under $\theta_T$, respectively, together with the $\theta_T$-invariant $\ep_{k|2}$, where $\ep_{2|2}=-\ep_{2|+}$ by (\ref{fixeps}). A term $s^{d_2}\big(\ep_{k_3|c},t^{d_1}(\ep_{k_1|a},\ep_{k_2|b})\big)$ of the projectors (\ref{eq:eqv:d3sdef}) with $a,b,c\in\{2,+,-\}$ is then $\theta_T$-invariant, if it carries an even number of labels $-$, and $\theta_T$-odd otherwise.

The expansion of $\sigma^{[3]}_5$ starts with the degree $8$ terms (\ref{eq:eqv:lam58}), and its subleading terms of degree 10 are given by
\begin{align}
\sigma^{[3]}_5 \, \big|_{{\rm deg} \ 10} &= \tfrac{5}{8192}\big[\ep_{2|+},[\ep_{2|+},\ep^{(2)}_{6|2}]\big]
+ \tfrac{3}{1024}\big[\ep_{2|2},[\ep^{(2)}_{4|2},\ep_{4|2}]\big] \label{eq:sig53d10}\\
&\quad - \tfrac{1}{256}\Big(\big[\ep_{2|2},[\ep^{(2)}_{4|3},\ep_{4|3}]\big] + \big[\ep_{2|2},[\ep^{(2)}_{4|4},\ep_{4|4}]\big]\Big) \notag\\
&\quad + \tfrac{5}{4096}\Big(\big[\ep_{2|2},[\ep^{(2)}_{4|3},\ep_{4|2}]-[\ep_{4|3},\ep^{(2)}_{4|2}]\big] + \big[\ep_{2|2},[\ep^{(2)}_{4|4},\ep_{4|2}]-[\ep_{4|4},\ep^{(2)}_{4|2}]\big]\Big) \notag\\
&\quad + \tfrac{15}{4096}\big[\ep_{2|2},[\ep^{(2)}_{4|4},\ep_{4|3}]-[\ep_{4|4},\ep^{(2)}_{4|3}]\big] \notag\\
&\quad + \tfrac{811}{28800}\, s^4\big(\ep_{4|2},t^2(\ep_{2|+},\ep_{4|2})\big) - \tfrac{26201}{1843200}\, s^4\big(\ep_{4|2},t^2(\ep_{2|+},\ep_{4|+})\big) \notag\\
&\quad - \tfrac{26201}{1843200}\, s^4\big(\ep_{4|+},t^2(\ep_{2|+},\ep_{4|2})\big) + \tfrac{811}{460800}\, s^4\big(\ep_{4|+},t^2(\ep_{2|+},\ep_{4|+})\big) \notag\\
&\quad - \tfrac{1}{4096}\, s^4\big(\ep_{4|-},t^2(\ep_{2|+},\ep_{4|-})\big) - \tfrac{1157}{92160}\, s^4\big(\ep_{4|2},t^2(\ep_{2|-},\ep_{4|-})\big) \notag\\
&\quad + \tfrac{5909}{737280}\, s^4\big(\ep_{4|+},t^2(\ep_{2|-},\ep_{4|-})\big) + \tfrac{107}{61440}\, s^4\big(\ep_{4|-},t^2(\ep_{2|-},\ep_{4|2})\big) \notag\\
&\quad + \tfrac{71}{81920}\, s^4\big(\ep_{4|-},t^2(\ep_{2|-},\ep_{4|+})\big) \, .\notag
\end{align}
The terms with nested brackets in (\ref{eq:sig53d10}) are the unique solution of (\ref{eq: sigmarec}) in the image of ${\rm ad}_{\ep_0^\vee}$, given that $\sigma^{[2]}_5$ is the sum of (\ref{sysxi2.09}) and (\ref{eq: HWd2}). The terms with $s^4$ are highest-weight vectors at the key degree of $\sigma_5$ and therefore $\mathfrak{sl}_2$ singlets. Out of the nine $\theta_T$-invariant singlets at this degree, the three in the trivial representation of ${\cal S}_3$ are absorbed into $z_5$ by (\ref{def:split}), and the displayed highest-weight combination has no component in the trivial representation. The six coefficients in the standard representation $\mathbf{2}$ of ${\cal S}_3$ are determined as explained around (\ref{getvw}).

By the arguments of early section \ref{s43.3}, the expansion of 
$\sigma^{[3]}_7$ starts with highest-weight vectors at degree 10. Their explicit form is again obtained from the $S$-cocycle,
\begin{align}
\sigma^{[3]}_7 \, \big|_{{\rm deg} \ 10} &= \tfrac{151089}{50176}\, s^2\big(\ep_{2|+},t^2(\ep_{2|+},\ep_{6|2})\big) - \tfrac{19851}{401408}\, s^2\big(\ep_{2|+},t^2(\ep_{2|+},\ep_{6|+})\big) \label{eq:sig73d10} \\
&\quad - \tfrac{95965}{57344}\, s^2\big(\ep_{2|+},t^2(\ep_{2|-},\ep_{6|-})\big) + \tfrac{356089}{602112}\, s^2\big(\ep_{2|-},t^2(\ep_{2|-},\ep_{6|2})\big) \notag\\
&\quad - \tfrac{42151}{301056}\, s^2\big(\ep_{2|-},t^2(\ep_{2|-},\ep_{6|+})\big) - \tfrac{20523}{10240}\, s^2\big(\ep_{4|2},t^2(\ep_{2|+},\ep_{4|2})\big) \notag\\
&\quad + \tfrac{26013}{143360}\, s^2\big(\ep_{4|2},t^2(\ep_{2|+},\ep_{4|+})\big) - \tfrac{141}{8192}\, s^2\big(\ep_{4|+},t^2(\ep_{2|+},\ep_{4|2})\big) \notag\\
&\quad - \tfrac{39}{16384}\, s^2\big(\ep_{4|+},t^2(\ep_{2|+},\ep_{4|+})\big) - \tfrac{1905}{16384}\, s^2\big(\ep_{4|-},t^2(\ep_{2|+},\ep_{4|-})\big) \notag\\
&\quad + \tfrac{77773}{43008}\, s^2\big(\ep_{4|2},t^2(\ep_{2|-},\ep_{4|-})\big) - \tfrac{172213}{344064}\, s^2\big(\ep_{4|+},t^2(\ep_{2|-},\ep_{4|-})\big) \notag\\
&\quad - \tfrac{182881}{172032}\, s^2\big(\ep_{4|-},t^2(\ep_{2|-},\ep_{4|2})\big) + \tfrac{150877}{344064}\, s^2\big(\ep_{4|-},t^2(\ep_{2|-},\ep_{4|+})\big) \, . \notag
\end{align}
The projectors in (\ref{eq:sig73d10}) reduce to double brackets,
\begin{align}
s^2\big(\ep_{2|c},t^2(\ep_{2|a},\ep_{6|b})\big) &= \tfrac{1}{24}\big[\ep_{2|c},[\ep_{2|a},\ep^{(4)}_{6|b}]\big]\, , \notag \\
s^2\big(\ep_{4|c},t^2(\ep_{2|a},\ep_{4|b})\big) &= \tfrac{1}{4}\big[\ep^{(2)}_{4|c},[\ep_{2|a},\ep^{(2)}_{4|b}]\big]\, . \notag
\end{align}
The fifteen $\theta_T$-invariant terms of these two types are linearly independent in the free Lie algebra. In $\mathfrak u$, however, they obey one linear relation since the highest-weight vector $s^2(\ep_{2|-},P_8^2)=\frac{1}{4!}[\ep_{2|-},{\rm ad}_{\ep_0}^4 P_8^2]$ built from the Pollack relation (\ref{eq: pollack8}) is a combination of them with coefficient $\frac{3}{2}$ for $s^2\big(\ep_{2|-},t^2(\ep_{2|+},\ep_{6|-})\big)$. We have used this relation to eliminate $s^2\big(\ep_{2|-},t^2(\ep_{2|+},\ep_{6|-})\big)$ from (\ref{eq:sig73d10}).

\subsection{\texorpdfstring{Equivariant integrals over $f^{(2)}(\omega_\nu,\tau)$ at modular depth four and five}{Equivariant integrals over f2(omeganu,tau) at modular depth four and five}}
\label{appd.3}

We supplement the examples of section \ref{ex:sec44} by two equivariant integrals over $f^{(2)}(\omega_\nu,\tau)$ alone, of modular depth four and five and degree eight and ten:
\begin{align}
\eezeqv{0&0&0&0}{2&2&2&2}{\omega_3&\omega_3&\omega_3&\omega_4}{\tau} &= \eez{0&0&0&0}{2&2&2&2}{\omega_3&\omega_3&\omega_3&\omega_4}{\tau} + \overline{\eez{0&0&0&0}{2&2&2&2}{\omega_4&\omega_3&\omega_3&\omega_3}{\tau}} \label{eq:Eeqvex4} \\
&\quad + \eez{0&0&0}{2&2&2}{\omega_3&\omega_3&\omega_4}{\tau}\, \overline{\eez{0}{2}{\omega_3}{\tau}} + \eez{0&0}{2&2}{\omega_3&\omega_4}{\tau}\, \overline{\eez{0&0}{2&2}{\omega_3&\omega_3}{\tau}} \notag \\
&\quad + \eez{0}{2}{\omega_4}{\tau}\, \overline{\eez{0&0&0}{2&2&2}{\omega_3&\omega_3&\omega_3}{\tau}} \notag \\
&\quad - \tfrac{2}{3}\log(2) \Big( \eez{0&0&0}{2&2&2}{\omega_3&\omega_3&\omega_4}{\tau} + \overline{\eez{0&0&0}{2&2&2}{\omega_3&\omega_3&\omega_3}{\tau}} \notag \\
&\quad \phantom{{}-\tfrac{2}{3}\log(2)\Big(} + \eez{0&0}{2&2}{\omega_3&\omega_4}{\tau}\, \overline{\eez{0}{2}{\omega_3}{\tau}} + \eez{0}{2}{\omega_4}{\tau}\, \overline{\eez{0&0}{2&2}{\omega_3&\omega_3}{\tau}} \Big) \notag \\
&\quad + \tfrac{2}{9}\log^2(2) \Big( \eez{0&0}{2&2}{\omega_3&\omega_4}{\tau} + \overline{\eez{0&0}{2&2}{\omega_3&\omega_3}{\tau}} + \eez{0}{2}{\omega_4}{\tau}\, \overline{\eez{0}{2}{\omega_3}{\tau}} \Big) \notag \\
&\quad - \tfrac{4}{81}\log^3(2) \Big( \eez{0}{2}{\omega_4}{\tau} + \overline{\eez{0}{2}{\omega_3}{\tau}} \Big) + \tfrac{2}{243}\log^4(2) \notag \\
&\quad + \tfrac{1}{432}\,\zeta_3 \Big( \overline{\eez{0}{2}{\omega_3}{\tau}} - \overline{\eez{0}{2}{\omega_4}{\tau}} \Big) \, , \notag
\end{align}
and
\begin{align}
\eezeqv{0&0&0&0&0}{2&2&2&2&2}{\omega_3&\omega_3&\omega_3&\omega_3&\omega_4}{\tau} &= \eez{0&0&0&0&0}{2&2&2&2&2}{\omega_3&\omega_3&\omega_3&\omega_3&\omega_4}{\tau} + \overline{\eez{0&0&0&0&0}{2&2&2&2&2}{\omega_4&\omega_3&\omega_3&\omega_3&\omega_3}{\tau}} \label{eq:Eeqvex5} \\
&\quad + \eez{0&0&0&0}{2&2&2&2}{\omega_3&\omega_3&\omega_3&\omega_4}{\tau}\, \overline{\eez{0}{2}{\omega_3}{\tau}} + \eez{0&0&0}{2&2&2}{\omega_3&\omega_3&\omega_4}{\tau}\, \overline{\eez{0&0}{2&2}{\omega_3&\omega_3}{\tau}} \notag \\
&\quad + \eez{0&0}{2&2}{\omega_3&\omega_4}{\tau}\, \overline{\eez{0&0&0}{2&2&2}{\omega_3&\omega_3&\omega_3}{\tau}} + \eez{0}{2}{\omega_4}{\tau}\, \overline{\eez{0&0&0&0}{2&2&2&2}{\omega_3&\omega_3&\omega_3&\omega_3}{\tau}} \notag \\
&\quad - \tfrac{2}{3}\log(2) \Big( \eez{0&0&0&0}{2&2&2&2}{\omega_3&\omega_3&\omega_3&\omega_4}{\tau} + \overline{\eez{0&0&0&0}{2&2&2&2}{\omega_3&\omega_3&\omega_3&\omega_3}{\tau}} \notag \\
&\quad \phantom{{}-\tfrac{2}{3}\log(2)\Big(} + \eez{0&0&0}{2&2&2}{\omega_3&\omega_3&\omega_4}{\tau}\, \overline{\eez{0}{2}{\omega_3}{\tau}} + \eez{0&0}{2&2}{\omega_3&\omega_4}{\tau}\, \overline{\eez{0&0}{2&2}{\omega_3&\omega_3}{\tau}} \notag \\
&\quad \phantom{{}-\tfrac{2}{3}\log(2)\Big(} + \eez{0}{2}{\omega_4}{\tau}\, \overline{\eez{0&0&0}{2&2&2}{\omega_3&\omega_3&\omega_3}{\tau}} \Big) \notag \\
&\quad + \tfrac{2}{9}\log^2(2) \Big( \eez{0&0&0}{2&2&2}{\omega_3&\omega_3&\omega_4}{\tau} + \overline{\eez{0&0&0}{2&2&2}{\omega_3&\omega_3&\omega_3}{\tau}} \notag \\
&\quad \phantom{{}+\tfrac{2}{9}\log^2(2)\Big(} + \eez{0&0}{2&2}{\omega_3&\omega_4}{\tau}\, \overline{\eez{0}{2}{\omega_3}{\tau}} + \eez{0}{2}{\omega_4}{\tau}\, \overline{\eez{0&0}{2&2}{\omega_3&\omega_3}{\tau}} \Big) \notag \\
&\quad - \tfrac{4}{81}\log^3(2) \Big( \eez{0&0}{2&2}{\omega_3&\omega_4}{\tau} + \overline{\eez{0&0}{2&2}{\omega_3&\omega_3}{\tau}} + \eez{0}{2}{\omega_4}{\tau}\, \overline{\eez{0}{2}{\omega_3}{\tau}} \Big) \notag \\
&\quad + \tfrac{2}{243}\log^4(2) \Big( \eez{0}{2}{\omega_4}{\tau} + \overline{\eez{0}{2}{\omega_3}{\tau}} \Big) - \tfrac{4}{3645}\log^5(2) \notag \\
&\quad + \tfrac{1}{432}\,\zeta_3 \Big( \overline{\eez{0&0}{2&2}{\omega_3&\omega_3}{\tau}} - \overline{\eez{0&0}{2&2}{\omega_4&\omega_3}{\tau}} \Big) + \tfrac{1}{13824}\,\zeta_5 \, . \notag
\end{align}
In all integrals over $f^{(2)}(\omega_\nu,\tau)$ through degree ten, $\rho^{-1}(\sv(f_1f_1f_3))$ cancels between the cubic terms of $\mathbb M^\sv(\sigma)$ and the brackets (\ref{eq:z3tail4}).

Both examples are expressible via svMPLs in the Hauptmodul $\lambda=\lambda(\tau)$ given by (\ref{eq:hauptmodul}). By the translation (\ref{eq:xy.ep}) between $\tilde\ep_{2|\nu}$ and the generators $x,y$ of the MPL series in section \ref{sec:tmpl}, the coefficients of $\tilde\ep_{2|3}^3\tilde\ep_{2|4}$ and $\tilde\ep_{2|3}^4\tilde\ep_{2|4}$ in the series $\pi_2\left(\mathbb{I}^{\mathrm {eqv}}\right)=\mathbb G^\eqv=\Psi^\sv\mathbb G^\sv$ of (\ref{defgeqv}) lead us to identify 
\begin{align}
\eezeqv{0&0&0&0}{2&2&2&2}{\omega_3&\omega_3&\omega_3&\omega_4}{\tau} &= \tfrac{1}{12^4}  \! \sum_{a_i\in\{0,1\}}  \! \Big( G^{\sv}(0,a_1,a_2,a_3;\lambda) - 2\, G^{\sv}(1,a_1,a_2,a_3;\lambda) \Big) \, ,
\label{eq:Eeqvex5sv} \\
\eezeqv{0&0&0&0&0}{2&2&2&2&2}{\omega_3&\omega_3&\omega_3&\omega_3&\omega_4}{\tau} &= - \tfrac{1}{12^5} \! \sum_{a_i\in\{0,1\}} \! \Big( G^{\sv}(0,a_1,\ldots,a_4;\lambda) - 2\, G^{\sv}(1,a_1,\ldots,a_4;\lambda) \Big)  + \tfrac{\zeta_5}{13824} \, , \notag
\end{align}
where the svMPLs $G^{\rm sv}$ are defined in (\ref{GGsvdef}). The zeta value in the second line can be absorbed by passing to the equivariant version $G^{\rm eqv}$ of MPLs discussed in section \ref{sec:eqvmpl},
\beq
\eezeqv{0&0&0&0&0}{2&2&2&2&2}{\omega_3&\omega_3&\omega_3&\omega_3&\omega_4}{\tau} = - \tfrac{1}{12^5} \! \sum_{a_i\in\{0,1\}} \! \Big( G^{\rm eqv}(0,a_1,\ldots,a_4;\lambda) - 2\, G^{\rm eqv}(1,a_1,\ldots,a_4;\lambda) \Big) \, .
\label{eqvrocks}
\eeq

\section{Numerical evaluation of MMVs}
\label{sec:num.mmv}

The determination of highest-weight vectors in the expansion of $\sigma^{\rm g}_w$ in section \ref{sec: HW vectors} relies on numerical values of the MMVs (\ref{eq: mmvdef}) on the left-hand side of (\ref{ag:Scoc2}). We shall here comment on how we obtain these values for the iterated Eisenstein integrals for $\Gamma(2)$. Following the techniques at level 1 \cite{Broedel:2015hia, Broedel:2018izr, brown2017multiple, Dorigoni:2021ngn, Kleinschmidt:2025dtk}, we combine the $q$-expansions of the iterated integrals (\ref{eq: eez}) around the cusp $i\infty$ with the $S$-transformation (\ref{eq: IS}) of their generating series at the fixed point $\tau=i$ of $S$.

The $q$-expansions of the iterated integrals (\ref{eq: eez}) follow from those of their integration kernels. At the torsion points $\omega_\mu$, where the co-moving coordinates take the values $u,v\in\{0,\tfrac12\}$, the Fourier expansion (\ref{eq: FKroneis}) splits $f^{(k)}(\omega_\mu,\tau)$ into the constant $\frac{(2\pi i)^k}{k!}B_k(u)$ and a power series in $q_2=e^{\pi i\tau}$ without constant term whose coefficients are rational multiples of $(2\pi i)^k$. Odd powers of $q_2$ only arise from $\omega_3$ and $\omega_4$, where $u=\tfrac12$. Upon insertion into (\ref{eq: eez}), each Fourier mode $e^{2\pi i n\tau_r}$ with $n>0$ yields a convergent integral
\begin{align}
\int^{i\infty}_\tau \dd\tau_r\, \tau_r^{p}\, e^{2\pi i n\tau_r} = - e^{2\pi i n\tau} \sum_{a=0}^{p} \frac{(-1)^a\, p!}{(p{-}a)!}\, \frac{\tau^{p-a}}{(2\pi i n)^{a+1}} \, , \ \ \ \ n\in \tfrac12\mathbb N\,,
\label{eq:numint.q}
\end{align}
whereas the constant terms of the kernels lead to integrals over powers of $\tau_r$ which diverge at $\tau_r\rightarrow i\infty$. As stated below (\ref{eq: eez}) and further discussed in section \ref{sec:reg}, these divergences are regularised by tangential base-points \cite{brown2017multiple}, which assign the value
\begin{align}
\int^{i\infty}_\tau \dd\tau_r\, \tau_r^{p} = - \frac{\tau^{p+1}}{p{+}1}
\label{eq:numint.tbp}
\end{align}
to the integrals over powers of $\tau_r$. Iterating (\ref{eq:numint.q}) and (\ref{eq:numint.tbp}), every iterated integral (\ref{eq: eez}) becomes a power series in $q_2$ whose coefficients are polynomials in $\tau$. The part without any power of $q_2$ is a polynomial in $\tau$ without constant term, in line with the regularised boundary value $\lim_{\tau\rightarrow i\infty}\INN{\tau}=1$ below (\ref{prevpexp}).

Organising the nested sums over Fourier modes as in section 3.5 of \cite{Hidding:2022vjf} gives the coefficients of these $q$-series in closed form to all orders in $q_2$. The expansions of \cite{Hidding:2022vjf} were derived for $0<u<1$ and therefore apply directly at $\omega_3$ and $\omega_4$, where $u=\frac12$. In section 3.2.1 of \cite{Hidding:2022vjf}, the case $u=0$ was attributed to the holomorphic Eisenstein series for $\SLZ$, which is appropriate for $z=0$, but not for the torsion point $\omega_2=\frac12$ with $u=0$ and $v=\frac12$. At $\omega_2$, the term $u^{k-1}/(1{-}e^{-2\pi i(u\tau+v)})$ of (\ref{eq: FKroneis}) vanishes for $k\geq 2$. In \cite{Hidding:2022vjf}, this term is expanded as a geometric series, which gives rise to the terms with $n_i=0$ and $\sigma_i=1$ in their nested sums over $\sigma_i=\pm1$, $m_i\geq 1$ and $n_i\geq \frac12(1{-}\sigma_i)$, where $i=1,2,\ldots,r$ labels the integration kernels in (\ref{eq: eez}). After the integrations, these terms carry the factor $(n_i{+}\sigma_i u_i)^{k_i-1}$ in their numerators, which vanishes at $u_i=0$, and for $i=1$ also a vanishing denominator. We therefore omit them at $\omega_2$ by starting the sums at $n_i=1$ for both $\sigma_i=\pm1$, which reproduces the $q$-expansion of $f^{(k)}(\frac12,\tau)$ obtained from (\ref{eq: FKroneis}).

The MMVs (\ref{eq: mmvdef}) are regularised limits $\tau\rightarrow 0$ of these iterated integrals, where their $q$-series do not converge. We therefore evaluate the $S$-transformation (\ref{eq: IS}) at the fixed point $\tau=i$ of $S$, where $\tau$ and $-1/\tau$ coincide, and solve for the $S$-cocycle,
\begin{align}
\mathbb S(\ep) = \INN{i}\, U_S^{-1}\theta_S^{-1}\big(\INN{i}\big)^{-1}\theta_S U_S\,.
\label{eq:Sfixedpoint}
\end{align}
By the expansion (\ref{eq: Scocgen}) of its left-hand side, the coefficient of each word of modular depth $r$ in (\ref{eq:Sfixedpoint}) expresses an MMV of modular depth $r$ as a polynomial in iterated integrals (\ref{eq: eez}) of modular depth $\leq r$ at $\tau=i$.\footnote{One can even pass to a version of $\mathbb S(\ep)$ and $\INN{i}$ in (\ref{eq:Sfixedpoint}) where $\tilde \ep_{2| \nu} \rightarrow \tilde{ \mathrm{e}}_{2| \nu}$ and $\ep_{k| \mu}^{(j)} \rightarrow \mathrm{e}_{k| \mu}^{(j)}$ are promoted to free-Lie-algebra generators, i.e., without any Pollack relations, which results in numerical values for a larger class of MMVs (\ref{eq: mmvdef}) than those in $\INN{-\frac{1}{\tau}}$.} This is the level-2 counterpart of splitting the integration path of (\ref{eq: mmvdef}) at $\tau=i$ and mapping its part between $0$ and $i$ to the part between $i$ and $i\infty$ by $S$ \cite{Kleinschmidt:2025dtk}, where at level 2 the map $S$ in addition exchanges $\omega_2\leftrightarrow\omega_4$ through $\theta_S$. Hence, only the $q$-series around the cusp $i\infty$ are needed, while the regularisation of (\ref{eq: mmvdef}) at $\tau\rightarrow0$ enters through the $S$-transformation (\ref{eq: IS}). At $\tau=i$, we have $q_1=e^{-2\pi}\approx 1.9\cdot 10^{-3}$. In our tests, keeping the $q$-series up to $q_1^{40}$ (i.e.\ $q_2^{80}$) yields about 100 correct digits, compared with the closed form at modular depth 1 and with the series up to $q_1^{60}$ at modular depth 2, which is the precision of the MMVs of modular depth 2 used in this section.

The numerical MMVs enter the determination of the zeta generators and of $\mathbb X(\ep)$ through the factorised form (\ref{ag:Scoc2}) of the $S$-cocycle. As explained in section \ref{s43.2}, once the lower-degree terms of the zeta generators and of $\mathbb X(\ep)$ are known, the coefficient of each word in $\tilde\ep_{2|\nu}$ and $\ep_{k|\mu}^{(j)}$ that is independent under the Pollack relations of section \ref{sec: pollack} contains at most one unknown rational number, as at level 1 (see section 4.4 of \cite{Kleinschmidt:2025dtk}). For words of even $a$-degree $2w$, this is the coefficient of $\pi^{2w}$ in $\mathbb X(\ep)$, and for words of odd $a$-degree $2\ell{-}1$, it is the coefficient of $\zeta_{2\ell{-}1}$ from the highest-weight vectors of the zeta generator $\sigma_{2\ell{-}1}$, see the cases (i) to (iii) above. Since all other contributions are known exactly, this rational number is fixed by applying PSLQ to only two numbers, so that one rational number is determined per word that is independent under the Pollack relations, rather than per MMV. The periods of cusp forms beyond cyclotomic MZVs, which occur in individual MMVs, drop out of the coefficients of these words, in line with the factorised form (\ref{ag:Scoc2}).

The ancillary file lists the MMVs obtained in this way to about 100 digits, through degree 16 at modular depth 1 and 2, through degree 12 at modular depth 3 and through degree 10 at modular depth 4 and 5.

\vspace{5mm}
\hrule
\vspace{5mm}

\bibliographystyle{JHEP}
\bibliography{refs.bib}

\end{document}